\documentclass[english,11pt,a4paper]{article}
\usepackage[T1]{fontenc}
\usepackage[utf8]{inputenc}
\usepackage{geometry}
\usepackage{color}
\usepackage{babel}
\usepackage{amsmath}
\usepackage{amsthm}
\usepackage{amssymb}
\usepackage{wasysym}
\usepackage{mleftright}
\usepackage[unicode=true,pdfusetitle,
 bookmarks=true,bookmarksnumbered=false,bookmarksopen=false,
 breaklinks=false,pdfborder={0 0 0},pdfborderstyle={},backref=false,colorlinks=true]
 {hyperref}
\hypersetup{
 linkcolor=vblue, citecolor=vblue}

\makeatletter
\numberwithin{equation}{section}
\theoremstyle{plain}
\newtheorem{thm}{\protect\theoremname}[section]
\theoremstyle{plain}
\newtheorem{assumption}[thm]{\protect\assumptionname}
\theoremstyle{plain}
\newtheorem*{thm*}{\protect\theoremname}
\theoremstyle{plain}
\newtheorem{lem}[thm]{\protect\lemmaname}
\theoremstyle{plain}
\newtheorem{prop}[thm]{\protect\propositionname}
\theoremstyle{plain}
\newtheorem*{lem*}{\protect\lemmaname}
\newtheorem{cor}[thm]{\protect\corollaryname}
\theoremstyle{plain}
\newtheorem*{prop*}{\protect\propositionname}

\definecolor{vblue}{RGB}{0,0,255}
\allowdisplaybreaks

\makeatother

\providecommand{\assumptionname}{Assumption}
\providecommand{\lemmaname}{Lemma}
\providecommand{\propositionname}{Proposition}
\providecommand{\theoremname}{Theorem}
\providecommand{\corollaryname}{Corollary}

\newcommand{\eps}{\epsilon}

\newcommand{\nn}{\nonumber}

\newcommand{\cN}{\mathcal{N}}

\begin{document}
\title{The Correlation Energy of the Electron Gas in the Mean-Field Regime}
\author{Martin Ravn Christiansen, Christian Hainzl, Phan Th\`anh Nam\\
\\
{\footnotesize{}Department of Mathematics, Ludwig Maximilian University
of Munich, Germany}\\
{\footnotesize{}Emails: martin.christiansen@math.lmu.de, hainzl@math.lmu.de,
nam@math.lmu.de}}
\maketitle
\begin{abstract}
We prove a rigorous lower bound on the correlation energy of interacting
fermions in the mean-field regime for a wide class of singular interactions, including the Coulomb potential. Combined with the upper bound obtained in \cite{ChrHaiNam-23b}, our result establishes an analogue of the Gell-Mann--Brueckner formula $c_{1}\rho\log\left(\rho\right)+c_{2}\rho$ for the correlation energy of the electron gas 
in the high-density limit. Moreover, our analysis allows us to go beyond mean-field scaling while still covering the same class of  potentials. 
\end{abstract}
\tableofcontents{}

\section{Introduction}

A long-standing challenge in mathematical physics is the rigorous understanding of quantum correlations in interacting systems, starting from microscopic principles. For jellium--an idealized model of electrons moving in a uniform neutralizing background--this problem goes back to Wigner \cite{Wigner-34} and Heisenberg \cite{Heisenberg-47}, who already recognized the limitations of perturbative methods. A cornerstone in the subsequent development was the \textit{random phase approximation} (RPA) introduced by Bohm and Pines \cite{BohPin-51,BohPin-52,BohPin-53,Pines-53}. In this theory, electron correlations are described through the decoupling of collective plasmon excitations from quasi-electrons, which then interact in the plasmon background via a screened Coulomb potential. This leads to a prediction for the \emph{correlation energy}, i.e. the difference between the exact ground-state energy and the Hartree--Fock energy. 

\medskip
The justification of the RPA has been the subject of major theoretical works. In a seminal paper, Gell-Mann and Brueckner \cite{GelBru-57} reproduced the RPA by resumming classes of Feynman diagrams in the high-density electron gas, and predicted that the correlation energy can be approximated by
\begin{align}\label{eq:GB}
E_{\rm corr} = c_1 \rho \log(\rho) + c_2 \rho + o(\rho), \quad \rho \to \infty,
\end{align}
with density $\rho$, where $c_1,c_2$ are explicit constants. Here $E_{\rm corr}=E_N-E_{\rm FS}$ denotes the difference between the exact ground-state energy $E_N$ and the Fermi sea energy $E_{\rm FS}$ (the energy of the non-interacting ground-state).\footnote{Alternatively one could use the Hartree--Fock energy $E_{\rm HF}$, but optimizing over all Slater determinants only improves $E_{\rm FS}$ by an exponentially small term \cite{GonHaiLew-19}.} The leading contribution $c_1\rho \log \rho$ had already been obtained by Pines \cite{Pines-53} and by Macke \cite{Macke-50} via partial resummations with screened Coulomb interactions. The new insight of Gell-Mann and Brueckner was to identify the next-order term $c_2\rho$, which encodes the \emph{exchange contribution}. In diagrammatic terms this corresponds to two electrons being excited above the Fermi surface, interacting, and then recombining with exchanged momenta. This mechanism, absent from the original RPA of Bohm and Pines, is essential for a complete description of correlations and will also play a central role in the present paper. 

\medskip
Around the same time, Sawada \cite{Sawada-57} and Sawada--Brueckner--Fukuda--Brout \cite{SawBruFukBro-57} developed an alternative derivation of the RPA. In their approach, certain pairs of fermions are treated as effective bosons, yielding a quadratic bosonic Hamiltonian that can be diagonalized by a Bogolubov transformation. This bosonization method correctly reproduces the $c_1\rho \log\rho$ term, but the purely bosonic picture misses the exchange contribution of order $\rho$.

\medskip
While physicists developed these theories in the 1950s using heuristic or approximate arguments, rigorous results emerged later. The existence of the thermodynamic limit for jellium was first proved by Lieb and Narnhofer \cite{LiebNarn-75}, building on earlier foundational work on the thermodynamics of matter with Coulomb forces by Lebowitz and Lieb \cite{LebLie-69}, and on the stability of matter which was first established by Dyson and Lenard \cite{DysLen-67} and subsequently simplified with semiclassical bounds by Lieb and Thirring \cite{LieThi-75}.  This placed the jellium model on solid ground. The size of the correlation energy as a correction to mean-field theory was then addressed by Graf and Solovej \cite{GraSol-94}, who proved the upper bound $E_{\rm corr} \leq O(\rho^{4/3-\varepsilon})$ for some $\varepsilon>0$. In particular, they confirmed rigorously that the Hartree--Fock energy $c_{\rm TF}\rho^{5/3}+c_{\rm D}\rho^{4/3}+ o(\rho^{4/3})$ provides the leading and first subleading terms of the ground-state energy  (see \cite[Theorem~2]{GraSol-94}), where the constant $c_{\rm TF}$ comes from the Thomas--Fermi theory for the kinetic energy, and $c_{\rm D}$ is predicted by the Dirac exchange correction for the Coulomb interaction. Their method was based on correlation inequalities inspired by Bach \cite{Bach-92}, which were originally developed to treat large atoms (see also Fefferman--Seco \cite{FerSec-90} for related results in the atomic setting). However, their result still falls far short of the Gell-Mann--Brueckner prediction \eqref{eq:GB}, which remains unproven due to the combined challenges of the singular and long-range nature of the Coulomb potential. 

\medskip
In the last decade, major progress has been made in a \emph{mean-field regime}, where $N$ fermions are confined to a torus of fixed volume and interact through a potential scaled by $k_F^{-1}\sim N^{-1/3}$ as $N\to\infty$. This scaling ensures that kinetic and interaction energies are comparable, both of order $N^{5/3}$, and thus provides a mathematically tractable setting that still captures key features of the high-density limit. In this regime, Benedikter, Nam, Porta, Schlein, and Seiringer \cite{BNPSS-20,BNPSS-21} carried out the first rigorous computation of the correlation energy for smooth weak interactions. Their approach, inspired by the physical idea of mapping particle--hole excitations near the Fermi surface to bosonic modes, involved a careful patch decomposition of the Fermi surface and led to a quadratic bosonic Hamiltonian that could be diagonalized explicitly. They established both upper \cite{BNPSS-20} and matching lower bounds \cite{BNPSS-21}, confirming the accuracy of the RPA at leading order for weak interactions. Perturbative results were also obtained earlier by Hainzl, Porta, and Rexze  \cite{HaiPorRex-20}, where the potential is small,  leading to a further simplification of the leading-order term of the correlation energy.

\medskip
These results were later extended in two directions in parallel works: Benedikter, Porta, Schlein, and Seiringer \cite{BPSS-21} refined the original method in \cite{BNPSS-20,BNPSS-21} to treat a larger class of potentials with non-negative Fourier modes and $\sum_k |k| \hat V_k < \infty$; independently, in \cite{ChrHaiNam-23a}, we developed a new bosonization approach in which each particle--hole pair is treated individually as a quasi-bosonic operator. The refinement in \cite{ChrHaiNam-23a}, closer in spirit to Sawada’s original idea, lets us treat the same class of potentials and, at the same time, offers new insights into the bosonic excitation spectrum of the mean-field fermionic system. Moreover, the method in \cite{ChrHaiNam-23a} is crucial for subsequent work on singular potentials, such as the Coulomb case, since it allows one to keep track of fermionic corrections in addition to the dominant bosonic behavior.

\medskip
In a subsequent work \cite{ChrHaiNam-23b}, we applied this approach to initiate a program toward the full asymptotic expansion of the correlation energy for the mean-field Coulomb gas. As a first step we established a rigorous upper bound, showing that
\begin{align}\label{eq:GB-Vcoulomb}
E_{\rm corr}\leq \frac{1}{\pi}\sum_{k\in\mathbb{Z} _{\ast}^{3}}\int_{0}^{\infty}F\!\left(\frac{\hat{V}_{k}k_{F}^{-1}}{(2\pi)^{3}}\sum_{p\in L_{k}}\frac{\lambda_{k,p}}{\lambda_{k,p}^{2}+t^{2}}\right)dt + \frac{k_{F}^{-2}}{4(2\pi)^{6}}\sum_{k\in\mathbb{Z}_{\ast}^{3}}\sum_{p,q\in L_{k}}\frac{\hat{V}_{k}\hat{V}_{p+q-k}}{\lambda_{k,p}+\lambda_{k,q}} + o(k_F),
\end{align}
where $F(x)=\log(1+x)-x$, $L_k=\{ p\in\mathbb{Z}^{3}\mid\left|p-k\right|\leq k_{F}<\left|p\right|\}$, and $\lambda_{k,p}=\tfrac12(|p|^2-|p-k|^2)$. For the Coulomb potential $\hat V_k \sim |k|^{-2}$ this yields
\begin{align}\label{eq:GB-Vcoulomb-simplified}
E_{\rm corr} \leq \widetilde c_1 k_F \log k_F + \widetilde c_2 k_F + o(k_F),
\end{align}
with explicit constants $\widetilde c_1, \widetilde c_2$. The right-hand side of equation \eqref{eq:GB-Vcoulomb} can be seen as the mean-field analogue of the Gell-Mann--Brueckner formula \eqref{eq:GB}, and indeed formally reduces to it in the thermodynamic limit after rescaling. Importantly, it contains not only the bosonic contribution of Sawada et al.~\cite{SawBruFukBro-57}, but also the exchange contribution identified by Gell-Mann and Brueckner \cite{GelBru-57}, which is absent in a purely bosonic picture. Roughly speaking, the rigorous proof in \cite{ChrHaiNam-23b} follows the bosonization framework of \cite{ChrHaiNam-23a}, itself inspired by \cite{Sawada-57,SawBruFukBro-57}, but refined to retain the relevant fermionic corrections. 

\medskip
For less singular potentials, namely those with $\hat V_k \ge 0$ and $\sum_k \hat V_k^2 |k|<\infty$, earlier results based on \cite{BPSS-21,ChrHaiNam-23a} already provided rigorous upper bounds. In that case the bosonic contribution dominates, while the exchange term is negligible. This can be seen by expanding $F(x)$ for small $x$,
\begin{align}\label{eq:F-Taylor}
F(x)=\log(1+x)-x = -\tfrac12 x^2 + O(x^3), \quad x \to 0,
\end{align}
so that the bosonic correlation term in \eqref{eq:GB-Vcoulomb} can be approximated by
\begin{align} \label{eq:E-bos-expansion}
\frac{1}{\pi}\sum_{k\in\mathbb{Z} _{\ast}^{3}}\int_{0}^{\infty}F\!\left(\frac{\hat{V}_{k}k_{F}^{-1}}{(2\pi)^{3}}\sum_{p\in L_{k}}\frac{\lambda_{k,p}}{\lambda_{k,p}^{2}+t^{2}}\right)dt  
 \approx -\frac{1}{4(2\pi)^6}\sum_{k\in \mathbb{Z}^3_*}  ( \hat V_k k_F^{-1})^2 \sum_{p,q\in L_k} \frac{1}{\lambda_{k,p} + \lambda_{k,q}} .
\end{align}
Using that $|L_k|\sim k_F^2 \min\{|k|,k_F\}$ and $\lambda_{k,p}\sim |k|\max\{|k|,k_F\}$ on average, one finds that if $\sum_k \hat V_k^2 |k|<\infty$ then \eqref{eq:E-bos-expansion} is bounded by $O(k_F)$. The Coulomb potential is thus \emph{critical}: The series diverges logarithmically, reflecting the logarithmic term in both the original and mean-field Gell-Mann--Brueckner formulas. 

\medskip
\noindent\textbf{Present work.} 
The aim of this paper is to prove the corresponding lower bound, thereby fully establishing the right-hand side of \eqref{eq:GB-Vcoulomb} as the correlation energy of the electron gas in the mean-field regime. Compared with the upper bound analysis in \cite{ChrHaiNam-23b}, the main difficulty is that strong a priori estimates can not be established for the true ground state, as they can for the upper bound trial state. The weaker a priori estimates of the lower bound of \cite{ChrHaiNam-23a} also do not apply in the case of a singular interaction potential. This prevents the transformation-based approach of those papers from being effective.

Our strategy is instead to extract the correlation energy directly from the Hamiltonian by completing appropriate squares that mix bosonizable and non-bosonizable terms. This representation captures all leading contributions of Coulomb-type potentials, while leaving error terms that we control via new correlation inequalities. These tools may also be useful for treating  singular interactions in other contexts. 

\medskip
Furthermore, we extend our result to the weaker scaling $k_F^{-\beta} V$ with $\beta<1$. In this regime, kinetic and potential energies are no longer balanced, making the analysis considerably more delicate. Nonetheless, our method still yields the expansion \eqref{eq:GB-Vcoulomb}, thus providing, to our knowledge, the first rigorous result on correlation energies beyond the mean-field scaling. On the conceptual level, it shows that our method goes significantly beyond the standard perturbation theory, as we can deal with the case where the interaction is much stronger than the kinetic operator. We believe that this makes the method interesting from both mathematical and physical perspectives, and that the flexibility of our approach will be helpful for future applications.

\medskip
The precise statements of our main theorems and an outline of the proofs are given below.

\subsection{Main Result}

We consider for a given Fermi momentum $k_{F}>0$ and fixed $\beta>\frac{11}{12}$
the $\beta$-scaled Hamiltonian
\begin{equation}
H_{N}=H_{\mathrm{kin}}+k_{F}^{-\beta}H_{\mathrm{int}}=-\sum_{i=1}^{N}\Delta_{i}+k_{F}^{-\beta}\sum_{1\leq i<j\le N}V\mleft(x_{i}-x_{j}\mright)
\end{equation}
on $D\mleft(H_{N}\mright)=D\mleft(H_{\mathrm{kin}}\mright)=\bigwedge^{N}H^{2}\mleft(\mathbb{T}^{3}\mright)$
where $\mathbb{T}^{3}=\left[0,2\pi\right]^{3}$ with periodic boundary
conditions and
\begin{equation}
N=\left|B_{F}\right|,\quad B_{F}=\overline{B}\mleft(0,k_{F}\mright)\cap\mathbb{Z}^{3}.
\end{equation}
We take $V$ to admit the Fourier decomposition 
\begin{equation}
V\mleft(x\mright)=\mleft(2\pi\mright)^{-3}\sum_{k\in \mathbb{Z}_{\ast}^{3}}\hat{V}_{k}e^{ik\cdot x},\quad \mathbb{Z}_{\ast}^{3}=\mathbb{Z}^{3}\backslash\left\{ 0\right\},
\end{equation}
and make the following assumptions on the Fourier coefficients $\hat{V}_{k}$. 
\begin{assumption}
\label{Assumption:Potential}The Fourier coefficients $\hat{V}_{k}$
satisfy $\hat{V}_{k}=\hat{V}_{-k}\geq0$ for all $k\in\mathbb{Z}_{\ast}^{3}$,
are radially decreasing with respect to $k\in\mathbb{Z}_{\ast}^{3}$,
and there exists a constant $C_{V}>0$ such that
\[
\hat{V}_{k}\leq C_{V}\left|k\right|^{-2},\quad k\in\mathbb{Z}_{\ast}^{3}.
\]
\end{assumption}

\medskip

The leading order of the ground state energy of $H_{N}$ is given
by the Fermi state
\begin{equation}
\psi_{\mathrm{FS}}=\bigwedge_{p\in B_{F}}u_{p},\quad u_{p}\left(x\right)=\left(2\pi\right)^{-\frac{3}{2}}e^{ip\cdot x}, \label{eq:FermiStateDefinition}
\end{equation}
with the corresponding energy (see e.g. \cite[Eqs. (1.10) and (1.20)]{ChrHaiNam-23a})
\begin{equation} \label{eq:EPS}
E_{\mathrm{FS}}=\left\langle \psi_{\mathrm{FS}},H_{N}\psi_{\mathrm{FS}}\right\rangle = \sum_{p\in B_F} |p|^2 - \frac{k_F^{-\beta}}{2(2\pi)^3} \sum_{k\in \mathbb{Z}^3_*} \hat V(k) \left( N-|L_k|\right)
\end{equation}
where for every $k\in\mathbb{Z}_{\ast}^{3}$, we denote the \textit{lune} $L_{k}\subset\mathbb{Z}^{3}$ by
\begin{equation}
L_{k}=\left(B_{F}+k\right)\backslash B_{F}=\left\{ p\in\mathbb{Z}^{3}\mid\left|p-k\right|\leq k_{F}<\left|p\right|\right\}. 
\end{equation}

To describe the correlation energy, we set $\lambda_{k,p}=\frac{1}{2}(\left|p\right|^{2}-\left|p-k\right|^{2})$ and
 define the \textit{bosonic} and \textit{exchange
contributions} (to the correlation energy) by
\begin{equation}
E_{\mathrm{corr},\mathrm{bos}}=\frac{1}{\pi}\sum_{k\in\mathbb{Z}_{\ast}^{3}}\int_{0}^{\infty}F\mleft(\frac{\hat{V}_{k}k_{F}^{-\beta}}{\mleft(2\pi\mright)^{3}}\sum_{p\in L_{k}}\frac{\lambda_{k,p}}{\lambda_{k,p}^{2}+t^{2}}\mright)dt,\quad F\mleft(x\mright)=\log\mleft(1+x\mright)-x,
\end{equation}
and
\begin{equation}
E_{\mathrm{corr},\mathrm{ex}}=\frac{k_{F}^{-2\beta}}{4\mleft(2\pi\mright)^{6}}\sum_{k\in\mathbb{Z}_{\ast}^{3}}\sum_{p,q\in L_{k}}\frac{\hat{V}_{k}\hat{V}_{p+q-k}}{\lambda_{k,p}+\lambda_{k,q}}.
\end{equation}
Our main result can then be stated as follows:

\begin{thm}[Operator and a priori estimates]
\label{them:MainTheorem}Let $\frac{11}{12}<\beta\leq1$ and let $V$
obey Assumption \ref{Assumption:Potential}. Then it holds as $k_{F}\rightarrow\infty$
that
\[
H_{N}\geq E_{\mathrm{FS}}+E_{\mathrm{corr},\mathrm{bos}}+E_{\mathrm{corr},\mathrm{ex}}+\mathcal{E}
\]
where $\mathcal{E}$ is an operator obeying for any $\epsilon>0$ the lower bound 
\[
\mathcal{E}\geq-C_{V,\epsilon}k_{F}^{-\frac{1}{6}+2\mleft(1-\beta\mright)+\epsilon}\mleft(H_{\mathrm{kin}}^{\prime}+k_{F}\mright)\quad \text{ with }\quad H_{\mathrm{kin}}^{\prime}=H_{\mathrm{kin}}-\left\langle \psi_{\mathrm{FS}},H_{\mathrm{kin}}\psi_{\mathrm{FS}}\right\rangle .
\]
Furthermore, every state $\Psi\in D\mleft(H_{N}\mright)$ obeying $\left\langle \Psi,H_{N}\Psi\right\rangle \leq\left\langle \psi_{\mathrm{FS}},H_{N}\psi_{\mathrm{FS}}\right\rangle $
also satisfies
\[
\left\langle \Psi,H_{\mathrm{kin}}^{\prime}\Psi\right\rangle \leq C_{V,\epsilon} k_{F}^{3-2\beta+\epsilon}.
\]
Here $C_{V,\epsilon}$ denotes a general constant depending only on $C_{V}$ and $\epsilon$.
\end{thm}

In particular, applying this to a ground state of $H_N$ immediately implies the ground state energy lower bound
\begin{equation} \label{eq:corr-energy-lower-bound}
\inf \sigma(H_N) \geq E_{\mathrm{FS}}+E_{\mathrm{corr},\mathrm{bos}}+E_{\mathrm{corr},\mathrm{ex}}-O(k_{F}^{5/6+4\mleft(1-\beta\mright)+\epsilon}).
\end{equation}

Here are some remarks on our result. 

\medskip
{\bf 1.} In the mean-field case $\beta=1$, the lower bound \eqref{eq:corr-energy-lower-bound} matches the upper bound of
\cite{ChrHaiNam-23b}, leading to a complete justification of \eqref{eq:GB-Vcoulomb}:

\begin{cor}[Correlation energy in the mean-field regime] \label{cor:MainTheorem} Let $\beta=1$ and let $V$
obey Assumption \ref{Assumption:Potential}. Then it holds as $k_{F}\rightarrow\infty$ that, for every $\epsilon>0$, 
$$ \inf \sigma(H_N) = E_{\mathrm{FS}}+E_{\mathrm{corr},\mathrm{bos}}+E_{\mathrm{corr},\mathrm{ex}} + O(k_{F}^{5/6+\epsilon}).$$
\end{cor}

Additionally, when $\beta=1$, it always holds that $E_{\mathrm{corr},\mathrm{bos}}\leq-Ck_{F}$, so there is a $k_{F}^{-1/6+\epsilon}$ separation between
the error term and $E_{\mathrm{corr},\mathrm{bos}}$ (the order of
$E_{\mathrm{corr},\mathrm{ex}}$ depends on the particular potential). For regular potentials satisfying $\sum_{k\in\mathbb{Z}^3} \hat V_k |k|<\infty$,  similar results to that of Corollary  \ref{cor:MainTheorem} were previously proved in \cite{BNPSS-20,BNPSS-21,ChrHaiNam-23a,BPSS-21}.

\medskip

{\bf 2.} If there is equality in Assumption \ref{Assumption:Potential}, i.e.
if $\hat{V}_{k}\propto\left|k\right|^{-2}$ is the Coulomb potential,
then
\begin{equation}
E_{\mathrm{corr},\mathrm{bos}}=O(k_{F}^{3-2\beta}\log\mleft(k_{F}\mright)),\quad E_{\mathrm{corr},\mathrm{ex}}=O(k_{F}^{3-2\beta}). 
\end{equation}
Therefore, the lower bound \eqref{eq:corr-energy-lower-bound} is a non-trivial statement
for all $\frac{11}{12}<\beta\leq1$. The matching upper bound is open for $\beta<1$ (the upper bound analysis in \cite{ChrHaiNam-23b} requires mean-field scaling). 

\subsection{Outline of the Proof}

Our method is inspired by the idea of bosonization which goes back to Sawada \cite{Sawada-57} and Sawada--Brueckner--Fukuda--Brout \cite{SawBruFukBro-57}. The key observation is that after extracting the energy of the Fermi state, the main contribution of the Hamiltonian comes from certain "bosonizable" terms, which can be written as quasi-bosonic quadratic terms in which particular pairs of fermions behave as effective bosons.   As already explained in \cite{ChrHaiNam-23b}, for singular potentials this bosonization method has to be implemented carefully in order to capture a subtle correction which is missed in the purely bosonic picture of \cite{Sawada-57,SawBruFukBro-57}. For regular potentials studied in \cite{BNPSS-20,BNPSS-21,ChrHaiNam-23a,BPSS-21},  
the situation is conceptually simpler, since the purely bosonic computation proposed by Sawada is sufficient at the heuristic level, and  the main difficulty lies in obtaining precise error estimates.

\medskip

On the mathematical side, while we will start with the rigorous formulation of the bosonization method from  \cite{ChrHaiNam-23a,ChrHaiNam-23b}, the proof in the present paper proceeds in a very different way. Most notably, we will not use quasi-bosonic Bogolubov transformations as in \cite{ChrHaiNam-23a,ChrHaiNam-23b} since controlling the errors caused by these transformations would become extremely complicated due to the lack of strong a priori estimates. As a comparison, for regular potentials satisfying $\sum_{k \in \mathbb{Z}^3} \hat V_k |k|<\infty$ studied in \cite{ChrHaiNam-23a,BPSS-21}, the pointwise inequality 
\begin{align}\label{eq:Onsager-pointwise}
&\quad\,\sum_{1\le i<j \le N} V(x_i-x_j) -  \langle \psi_{\mathrm{FS}}, H_{\rm int} \psi_{\mathrm{FS}}\rangle +\frac{1}{2(2\pi)^3} \sum_{k \ne 0}\hat V_k |L_k| \\
&=\sum_{1\le i<j \le N} V(x_i-x_j) - \frac{1}{2\mleft(2\pi\mright)^{3}}( N^{2}\hat{V}_{0}-NV(0) ) = \frac{1}{2\mleft(2\pi\mright)^{3}} \sum_{k\ne 0} \hat V_k \left| \sum_{j=1}^N e^{i k \cdot x_j}\right|^2 \ge 0\nonumber
\end{align}
implies that 
\begin{align}\label{eq:Onsager-pointwise-2}
&k_F^{-1} H_{\rm int}   -  \langle \psi_{\mathrm{FS}}, k_F^{-1} H_{\rm int} \psi_{\mathrm{FS}}\rangle \ge -\frac{k_F^{-1}}{2(2\pi)^3}  \sum_{k \ne 0}\hat V_k |L_k|  \ge - C k_F \sum_{k \ne 0}\hat V_k |k|. 
\end{align}
Consequently, in the mean-field regime with $\sum_{k} \hat V_k |k|<\infty$, the correlation energy  is of order $O(k_F)$, leading to the a priori estimate  $
\left\langle \Psi,H_{\mathrm{kin}}^{\prime}\Psi\right\rangle \leq O(k_F)$ for every state satisfying 
$\left\langle \Psi,H_{N}\Psi\right\rangle \leq\left\langle \psi_{\mathrm{FS}},H_{N}\psi_{\mathrm{FS}}\right\rangle $. Unfortunately, this simple Onsager-type argument does not apply to singular potentials: From \eqref{eq:Onsager-pointwise-2} we see that even the case $\sum_{k\ne 0}\hat V_k |k|=\infty$ already causes difficulties, and not just the case where $V$ is unbounded. For Coulomb systems, an adaptation of the deeper 
techniques from \cite{Bach-92,GraSol-94} to our mean-field situation yields an a priori bound of order $O(k_F^{3-\eps})$ for the correlation energy, while the stronger bound $O(k_F^{1+\eps})$ is typically required to apply the bosonization method from \cite{ChrHaiNam-23a,ChrHaiNam-23b}. 

\medskip

To overcome this difficulty, we will derive a new representation of the Hamiltonian, wherein we extract the correlation energy directly by completing appropriate squares containing both bosonizable and non-bosonizable terms. In the bosonic picture, the realization that the ground state energy of a quadratic Hamiltonian can be extracted by completing suitable squares was first made by Bogolubov in 1947 \cite{Bogolubov-47}. Variations of this technique have been employed in various contexts, such as the proof of Foldy's formula for ``bosonic jellium'' \cite{LieSol-01}, the derivation of the Lee-Huang-Yang formula for dilute Bose gases  \cite{FouSol-20}, and recent work \cite{Brooks-23} on the diagonalization of Bose gases beyond the Gross-Pitaevskii regime.
It might therefore not seem surprising that attempting to replace quasi-bosonic Bogolubov transformations with the completion of squares should work, but the fact that the kinetic operator $H_{\mathrm{kin}}^\prime$ is not expressible in terms of pairs of fermions in the same sense as the interaction term prevents a \emph{naive} application of such an argument from working. This would also not explain why the non-bosonizable terms should be negligible.

\medskip

The significance of our new formula lies not only in being the first realization of such a factorization argument for a high-density fermion system (as opposed to the bosonic systems considered in the above works), but also in incorporating the most difficult non-bosonizable terms directly, removing the need to estimate these separately. Equipped with this representation we will then derive new correlation inequalities, which may be of independent interest, which allow us to estimate the remaining error terms as being small relative to the kinetic operator $H_{\mathrm{kin}}^{\prime}$. Further details of our proof are outlined as follows.

\subsection*{Second Quantization, Bosonizable and Non-bosonizable Terms}

The starting point of the analysis is the second quantized representation
of the Hamiltonian $H_{N}$, which can be decomposed as
\begin{equation}
H_{N}=E_{\mathrm{FS}}+H_{\mathrm{B}}+\mathcal{C}+\mathcal{Q}
\end{equation}
where $E_{\mathrm{FS}}=\left\langle \psi_{\mathrm{FS}},H_{N}\psi_{\mathrm{FS}}\right\rangle $ is
the energy of the Fermi state and the \textit{bosonizable,} \textit{cubic}
and \textit{quartic} terms\footnote{The names “cubic” and “quartic” terms come from the analogous decomposition for Bose gases, referring to the number of creation and annihilation operators in the interaction terms after extracting the condensate (see, e.g.,  \cite{BBCS-19}).} are given by
\begin{align}
H_{\mathrm{B}} & =H_{\mathrm{kin}}^{\prime}+\frac{k_{F}^{-\beta}}{2\mleft(2\pi\mright)^{3}}\sum_{k\in\mathbb{Z}_{\ast}^{3}}\hat{V}_{k}\mleft(2B_{k}^{\ast}B_{k}+B_{k}B_{-k}+B_{-k}^{\ast}B_{k}^{\ast}\mright),\nonumber \\
\mathcal{C} & =\frac{k_{F}^{-\beta}}{\mleft(2\pi\mright)^{3}}\,\mathrm{Re}\sum_{k\in\mathbb{Z}_{\ast}^{3}}\hat{V}_{k}\mleft(B_{k}+B_{-k}^{\ast}\mright)^{\ast}D_{k},\label{eq:DecompositionofHN}\\
\mathcal{Q} & =\frac{k_{F}^{-\beta}}{2\mleft(2\pi\mright)^{3}}\sum_{k\in\mathbb{Z}_{\ast}^{3}}\hat{V}_{k}\biggl(D_{k}^{\ast}D_{k}-\sum_{p\in L_{k}}\mleft(c_{p}^{\ast}c_{p}+c_{p-k}c_{p-k}^{\ast}\mright)\biggr),\nonumber 
\end{align}
respectively (see e.g. \cite[eqs. (1.16) - (1.26)]{ChrHaiNam-23b} for
the computation). Here $c_{p}^{\ast}$ and $c_{p}$ denote the creation
and annihilation operators associated with the plane wave states with
momenta $p\in\mathbb{Z}^{3}$, which satisfy the canonical anticommutation
relations (CAR)
\begin{equation}
\left\{ c_{p},c_{q}^{\ast}\right\} =\delta_{p,q},\quad\left\{ c_{p},c_{q}\right\} =0=\left\{ c_{p}^{\ast},c_{q}^{\ast}\right\} .
\end{equation}
In \eqref{eq:DecompositionofHN} above, $H_{\mathrm{kin}}^{\prime}$ denotes the localized kinetic operator,
which is
\begin{equation}
H_{\mathrm{kin}}^{\prime}=H_{\mathrm{kin}}-\left\langle \psi_{\mathrm{FS}},H_{\mathrm{kin}}\psi_{\mathrm{FS}}\right\rangle =\sum_{p\in B_{F}^{c}}\left|p\right|^{2}c_{p}^{\ast}c_{p}-\sum_{p\in B_{F}}\left|p\right|^{2}c_{p}c_{p}^{\ast},
\end{equation}
and $B_{k}$, $D_{k}$ are given by
\begin{equation}
B_{k}=\sum_{p\in L_{k}}c_{p-k}^{\ast}c_{p},\quad D_{k}=\sum_{p\in B_{F}^{c}\cap\mleft(B_{F}^{c}+k\mright)}c_{p-k}^{\ast}c_{p}+\sum_{p\in B_{F}\cap\mleft(B_{F}+k\mright)}c_{p-k}^{\ast}c_{p}.
\end{equation}

Here we refer to the operator $H'_{\rm kin}$ as being ``localized'' since we have already extracted the kinetic energy of the Fermi state $\psi_{\rm FS}$, and by changing the point of reference from the vacuum state to the Fermi state we may think of $H'_{\rm kin}$ as a version of $H_{\rm kin}$  localized around $\psi_{\rm FS}$.

\subsection*{Extraction of the Correlation Energy by Factorization}

The correlation energy (including $E_{\rm corr,ex}$) arises from the bosonizable terms $H_{\mathrm{B}}$, so we start by considering these in detail. Note that the Hamiltonian $H_{\mathrm{B}}$ is not equivalent to a bosonic Hamiltonian in a precise technical sense, and consequently the full correlation energy extracted from $H_{\mathrm{B}}$ is not exactly the same as the ground state energy of the bosonic analogue that we discuss below (indeed, this exactly bosonic analogue misses $E_{\mathrm{corr},\mathrm{ex}}$). Nevertheless, the formal analogy to a bosonic Hamiltonian is still very helpful, as we will explain.

The reason for its name
is the following: If we define the \textit{excitation operators} $b_{k,p}$
and $b_{k,p}^{\ast}$ by
\begin{equation}
b_{k,p}=c_{p-k}^{\ast}c_{p},\quad b_{k,p}^{\ast}=c_{p}^{\ast}c_{p-k},\quad k\in\mathbb{Z}_{\ast}^{3},\,p\in L_{k},
\end{equation}
then it follows immediately from the CAR that these obey commutation
relations of the form
\begin{equation}
[b_{k,p},b_{l,q}^{\ast}]=\delta_{k,l}\delta_{p,q}+\varepsilon_{k,l}(e_p;e_q),\quad[b_{k,p},b_{l,q}]=0=[b_{k,p}^{\ast},b_{l,q}^{\ast}],
\end{equation}
which are seen to be analogous to canonical commutation relations
up to a correction term $\varepsilon_{k,l}\mleft(e_{p};e_{q}\mright)$ (the precise form of which is not important for this outline). Here $\{e_p\}_{p\in L_k}$ is the standard orthonormal basis for the (real) Hilbert space  $\ell^2 (L_k)$. Furthermore,
there holds the exact commutator
\begin{equation}
\left[H_{\mathrm{kin}}^{\prime},b_{k,p}^{\ast}\right]=2\lambda_{k,p}b_{k,p}^{\ast}
\end{equation}
which given the \textit{quasi-bosonic} behaviour of the $b_{k,p}^{\ast}$
operators suggests an informal relation of the form
\begin{equation}
H_{\mathrm{kin}}^{\prime}\sim\sum_{k\in\mathbb{Z}_{\ast}^{3}}\sum_{p\in L_{k}}2\lambda_{k,p}b_{k,p}^{\ast}b_{k,p},\label{eq:InformalKineticRelation}
\end{equation}
and defining operators $h_{k},P_{k}:\ell^{2}\mleft(L_{k}\mright)\rightarrow\ell^{2}\mleft(L_{k}\mright)$
and a vector $v_{k}\in\ell^{2}\mleft(L_{k}\mright)$ by
\begin{equation}
\left\langle e_{p},h_{k}e_{q}\right\rangle =\lambda_{k,p}\delta_{p,q},\quad P_{k}=\left|v_{k}\right\rangle \left\langle v_{k}\right|,\quad\left\langle e_{p},v_{k}\right\rangle =\sqrt{\frac{\hat{V}_{k}k_{F}^{-\beta}}{2\mleft(2\pi\mright)^{3}}},\quad p,q\in L_{k},
\end{equation}
this suggests that
\begin{equation}\label{eq:HB-quasi-bosonic}
H_{\mathrm{B}}\sim\sum_{k\in\mathbb{Z}_{\ast}^{3}}\mleft(2\sum_{p,q\in L_{k}}\left\langle e_{p},\mleft(h_{k}+P_{k}\mright)e_{q}\right\rangle b_{k,p}^{\ast}b_{k,q}+2\,\mathrm{Re}\sum_{p,q\in L_{k}}\left\langle e_{p},P_{k}e_{q}\right\rangle b_{k,p}b_{-k,-q}\mright)
\end{equation}
which has the form of a quadratic Hamiltonian with respect to $b_{k,p}^{\ast}$.

\subsubsection*{Exactly Bosonic Bogolubov Factorization}

Now, \textit{if} \eqref{eq:HB-quasi-bosonic} were a genuine identity, and \textit{if} the
operators $b_{k,p}^{\ast}$ were genuinely bosonic (i.e. if $\varepsilon_{k,l}\mleft(e_p;e_q\mright)=0$),
then this would imply that $H_{\mathrm{B}}$ would be \textit{diagonalizable}
by a Bogolubov transformation $e^{\mathcal{K}}$, i.e. there would
exist an (explicit) Bogolubov kernel $\mathcal{K}$, which is an anti-symmetric operator used to construct the Bogoliubov unitary transformation,  such that
\begin{equation}
e^{\mathcal{K}}H_{\mathrm{B}}e^{-\mathcal{K}}=\sum_{k\in\mathbb{Z}_{\ast}^{3}}\mathrm{tr}\mleft(E_{k}-h_{k}-P_{k}\mright)+2\sum_{k\in\mathbb{Z}_{\ast}^{3}}\sum_{p,q\in L_{k}}\left\langle e_{p},E_{k}e_{q}\right\rangle b_{k,p}^{\ast}b_{k,q}\label{eq:ExactBosonicDiagonalization}
\end{equation}
with $E_{k}:\ell^{2}\mleft(L_{k}\mright)\rightarrow\ell^{2}\mleft(L_{k}\mright)$ given by
\begin{equation}\label{eq:Ek-def}
E_{k}=(h_{k}^{\frac{1}{2}}(h_{k}+2P_{k})h_{k}^{\frac{1}{2}})^{\frac{1}{2}}. 
\end{equation}

%
%

In fact (see e.g. \cite[Propositions 7.1, 7.6]{ChrHaiNam-23a})
\begin{equation}
E_{\mathrm{corr},\mathrm{bos}}=\sum_{k\in\mathbb{Z}_{\ast}^{3}}\mathrm{tr}\mleft(E_{k}-h_{k}-P_{k}\mright)
\end{equation}
which explains why we refer to this as the bosonic contribution to
the correlation energy.

In the exact bosonic case the transformation $e^{\mathcal{K}}$ would
(for a lower bound) technically be superfluous, since ``undoing''
the transformation shows that
\begin{equation}
H_{\mathrm{B}}=E_{\mathrm{corr},\mathrm{bos}}+2\sum_{k\in\mathbb{Z}_{\ast}^{3}}\sum_{p,q\in L_{k}}\left\langle e_{p},E_{k}e_{q}\right\rangle e^{-\mathcal{K}}b_{k,p}^{\ast}e^{\mathcal{K}}e^{-\mathcal{K}}b_{k,q}e^{\mathcal{K}}
\end{equation}
and the transformation $e^{\mathcal{K}}$ would additionally satisfy
\begin{align}
e^{-\mathcal{K}}b_{k,p}e^{\mathcal{K}} & =\sum_{q\in L_{k}}\left\langle C_{k}e_{p},e_{q}\right\rangle b_{k,q}+\sum_{q\in L_{k}}\left\langle e_{-q},S_{-k}e_{-p}\right\rangle b_{-k,-q}^{\ast}\\
 & =:b_{k}\mleft(C_{k}e_{p}\mright)+b_{-k}^{\ast}\mleft(S_{-k}e_{-p}\mright)\nonumber 
\end{align}
for operators $C_{k},S_{k}:\ell^{2}\mleft(L_{k}\mright)\rightarrow\ell^{2}\mleft(L_{k}\mright)$
given by
\begin{equation}
C_{k}=\frac{1}{2}(h_{k}^{-\frac{1}{2}}E_{k}^{\frac{1}{2}}+h_{k}^{\frac{1}{2}}E_{k}^{-\frac{1}{2}}),\quad S_{k}=\frac{1}{2}(h_{k}^{-\frac{1}{2}}E_{k}^{\frac{1}{2}}-h_{k}^{\frac{1}{2}}E_{k}^{-\frac{1}{2}}),
\end{equation}
i.e.
\begin{equation}
H_{\mathrm{B}}=E_{\mathrm{corr},\mathrm{bos}}+2\sum_{k\in\mathbb{Z}_{\ast}^{3}}\sum_{p,q\in L_{k}}\left\langle e_{p},E_{k}e_{q}\right\rangle \mleft(b_{k}\mleft(C_{k}e_{p}\mright)+b_{-k}^{\ast}\mleft(S_{-k}e_{-p}\mright)\mright)^{\ast}\mleft(b_{k}\mleft(C_{k}e_{q}\mright)+b_{-k}^{\ast}\mleft(S_{-k}e_{-q}\mright)\mright)\label{eq:ExactBosonicAlgebraicIdentity}
\end{equation}
which is simply an algebraic rewriting of $H_{\mathrm{B}}$, as can
be verified by expanding the expression and applying the definitions
of $E_{k}$, $C_{k}$ and $S_{k}$. Since $E_k \geq 0$, \eqref{eq:ExactBosonicAlgebraicIdentity} would immediately imply that $H_{\rm B} \geq E_{\rm{corr,bos}}$.

\subsubsection*{Quasi-Bosonic Bogolubov Factorization}

Returning to the non-exact case, a result to the effect of equation
(\ref{eq:ExactBosonicDiagonalization}) was established in \cite[Theorem 1.1]{ChrHaiNam-23a}
(for $\beta=1$ and potentials obeying $\sum_{k\in\mathbb{Z}_{\ast}^{3}}\left|k\right|\hat{V}_{k}<\infty$),
in which a unitary operator $\mathcal{U}$ (a product of two quasi-bosonic
Bogolubov transformations) was constructed such that
\begin{equation}
\mathcal{U}H_{\mathrm{B}}\mathcal{U}^{\ast}\sim E_{\mathrm{corr},\mathrm{bos}}+H_{\mathrm{kin}}^{\prime}+2\sum_{k\in\mathbb{Z}_{\ast}^{3}}\sum_{p,q\in L_{k}}\left\langle e_{p},\mleft(E_{k}-h_{k}\mright)e_{q}\right\rangle b_{k,p}^{\ast}b_{k,q}.\label{eq:OperatorStatement}
\end{equation}
Note the difference from equation (\ref{eq:ExactBosonicDiagonalization}):
We have the additional terms 
$$H_{\mathrm{kin}}^{\prime}-\sum_{k\in\mathbb{Z}_{\ast}^{3}}\sum_{p\in L_{k}}2\lambda_{k,p}b_{k,p}^{\ast}b_{k,q}$$
which reflects the fact that the relation of equation (\ref{eq:InformalKineticRelation})
only holds in an indirect sense. One could hope to make this more
direct, but in fact this is impossible, as it was also noted in \cite[Proposition 10.1]{ChrHaiNam-23a}
that
\begin{equation}
\sum_{k\in\mathbb{Z}_{\ast}^{3}}\sum_{p\in L_{k}}2\lambda_{k,p}b_{k,p}^{\ast}b_{k,q}=\mathcal{N}_{E}H_{\mathrm{kin}}^{\prime}
\end{equation}
where $\mathcal{N}_E$ denotes the \textit{excitation number operator}
$$
\mathcal{N}_{E} = \sum_{p\in B_F^c} a_p^* a_p =  \sum_{q\in B_F} a_q a_q^*. 
$$
Here the last identity, called the particle-hole relation, is valid on the $N$-particle space due to the fact that $N=\left|B_{F}\right|$. So 
$$H_{\mathrm{kin}}^{\prime}-\sum_{k\in\mathbb{Z}_{\ast}^{3}}\sum_{p\in L_{k}}2\lambda_{k,p}b_{k,p}^{\ast}b_{k,q}=-\mleft(\mathcal{N}_{E}-1\mright)H_{\mathrm{kin}}^{\prime}$$
can not be considered small on its own. It is nonetheless the case that 
$$2\sum_{k\in\mathbb{Z}_{\ast}^{3}}\sum_{p,q\in L_{k}}\left\langle e_{p},\mleft(E_{k}-h_{k}\mright)e_{q}\right\rangle b_{k,p}^{\ast}b_{k,q}\geq0,$$
so this does suffice to show that $H_{\mathrm{B}}\gtrsim E_{\mathrm{corr},\mathrm{bos}}$,
but it appears to preclude a transformation-free approach that could
yield something similar to equation (\ref{eq:ExactBosonicAlgebraicIdentity}).

By modifying the approach this is however possible: If we similarly
``undo'' the transformation of equation (\ref{eq:OperatorStatement})
we see that
\begin{equation}
H_{\mathrm{B}}\sim E_{\mathrm{corr},\mathrm{bos}}+\sum_{p\in\mathbb{Z}^{3}}\vert\left|p\right|^{2}-k_{F}^{2}\vert\left|\mathcal{U}^{\ast}\tilde{c}_{p}\mathcal{U}\right|^{2}+2\sum_{k\in\mathbb{Z}_{\ast}^{3}}\sum_{p,q\in L_{k}}\left\langle e_{p},\mleft(E_{k}-h_{k}\mright)e_{q}\right\rangle \mathcal{U}^{\ast}b_{k,p}^{\ast}\mathcal{U}\mathcal{U}^{\ast}b_{k,q}\mathcal{U}\label{eq:QuasiBosonicSuggestedIdentity}
\end{equation}
where we introduced the notation 
\begin{equation}
\tilde{c}_{p}=\begin{cases}
c_{p} & p\in B_{F}^{c}\\
c_{p}^{\ast} & p\in B_{F}
\end{cases}.
\end{equation}

Now, if $\mathcal{U}=e^{\mathcal{K}}$ for a quasi-bosonic kernel
$\mathcal{K}$ 
(defined as a slightly modified version of the kernels defining the two transformations used in \cite{ChrHaiNam-23a}) 
one finds similarly
to the exact case that
\begin{equation}
\mathcal{U}^{\ast}b_{k,p}\mathcal{U}\sim b_{k}\mleft(C_{k}e_{p}\mright)+b_{-k}^{\ast}\mleft(S_{-k}e_{-p}\mright)
\end{equation}
while the operators $\tilde{c}_{p}$ obey
\begin{equation}
\mathcal{U}^{\ast}\tilde{c}_{p}\mathcal{U}\sim\tilde{c}_{p}+d_{p}^{1}+d_{p}^{2}
\end{equation}
where $d_{p}^{1}$ and $d_{p}^{2}$ are given by
\begin{equation}
d_{p}^{1}=\begin{cases}
+\sum_{k\in\mathbb{Z}_{\ast}^{3}}1_{L_{k}}\mleft(p\mright)\tilde{c}_{p-k}^{\ast}b_{k}\mleft(\mleft(C_{k}-1\mright)e_{p}\mright) & p\in B_{F}^{c}\\
-\sum_{k\in\mathbb{Z}_{\ast}^{3}}1_{L_{k}-k}\mleft(p\mright)\tilde{c}_{p+k}^{\ast}b_{k}\mleft(\mleft(C_{k}-1\mright)e_{p+k}\mright) & p\in B_{F}
\end{cases}
\end{equation}
and
\begin{equation}
d_{p}^{2}=\begin{cases}
+\sum_{k\in\mathbb{Z}_{\ast}^{3}}1_{L_{k}}\mleft(p\mright)\tilde{c}_{p-k}^{\ast}b_{-k}^{\ast}\mleft(S_{-k}e_{-p}\mright) & p\in B_{F}^{c}\\
-\sum_{k\in\mathbb{Z}_{\ast}^{3}}1_{L_{k}-k}\mleft(p\mright)\tilde{c}_{p+k}^{\ast}b_{-k}^{\ast}\mleft(S_{-k}e_{-p-k}\mright) & p\in B_{F}
\end{cases},
\end{equation}
respectively (note that these are sums of triples of fermionic creation
and annihilation operators).

Equation (\ref{eq:QuasiBosonicSuggestedIdentity}) consequently suggests
an identity of the form\footnote{A factorization of a similar form was recently used in \cite{Brooks-23},
which inspired this approach.}
\begin{align}
H_{\mathrm{\mathrm{B}}} & \sim E_{\mathrm{corr},\mathrm{bos}}+\sum_{p\in\mathbb{Z}^{3}}\vert\left|p\right|^{2}-k_{F}^{2}\vert\left|\tilde{c}_{p}+d_{p}^{1}+d_{p}^{2}\right|^{2}\label{eq:HBMotivatingRelation}\\
 & +2\sum_{k\in\mathbb{Z}_{\ast}^{3}}\sum_{p,q\in L_{k}}\left\langle e_{p},\mleft(E_{k}-h_{k}\mright)e_{q}\right\rangle \mleft(b_{k}\mleft(C_{k}e_{p}\mright)+b_{-k}^{\ast}\mleft(S_{-k}e_{-p}\mright)\mright)^{\ast}\mleft(b_{k}\mleft(C_{k}e_{q}\mright)+b_{-k}^{\ast}\mleft(S_{-k}e_{-q}\mright)\mright)\nonumber 
\end{align}
which \textit{is} a purely algebraic statement. This is of course
not exact, but the crucial point is that we can simply take the right-hand
side as an \textit{ansatz} and expand it to obtain a genuine identity
for $H_{\mathrm{B}}$. This is precisely what we will do in Sections
\ref{sec:ExtractionoftheCorrelationEnergybyFactorization} and \ref{sec:EstimationofEB} (see Theorems \ref{them:QuasibosonicBogolubovFactorization} and \ref{them:EBEstimate})
to obtain the following:

\begin{thm} 
It holds that
\begin{align*}
H_{\mathrm{B}} & =\sum_{p\in\mathbb{Z}^{3}}\vert\left|p\right|^{2}-k_{F}^{2}\vert\mleft(\left|\tilde{c}_{p}+d_{p}^{1}+d_{p}^{2}\right|^{2}+\left|(d_{p}^{1}+d_{p}^{2})^{\ast}\right|^{2}\mright)-2\sum_{k\in\mathbb{Z}_{\ast}^{3}}\sum_{p,q\in L_{k}}\left\langle e_{p},S_{k}E_{k}S_{k}^{\ast}e_{q}\right\rangle \varepsilon_{k,k}\mleft(e_{p};e_{q}\mright)\\
 & +2\sum_{k\in\mathbb{Z}_{\ast}^{3}}\sum_{p,q\in L_{k}}\left\langle e_{p},\mleft(E_{k}-h_{k}\mright)e_{q}\right\rangle \mleft(b_{k}\mleft(C_{k}e_{p}\mright)+b_{-k}^{\ast}\mleft(S_{-k}e_{-p}\mright)\mright)^{\ast}\mleft(b_{k}\mleft(C_{k}e_{q}\mright)+b_{-k}^{\ast}\mleft(S_{-k}e_{-q}\mright)\mright)\\
 & +E_{\mathrm{corr},\mathrm{bos}}+E_{\mathrm{corr},\mathrm{ex}}+\mathcal{E}_{\mathrm{B}}
\end{align*}
for an operator $\mathcal{\mathcal{E}_{\mathrm{B}}}$ which under
Assumption \ref{Assumption:Potential} obeys
\[
\pm\mathcal{E}_{\mathrm{B}}\leq C_{V,\epsilon}k_{F}^{-\frac{1}{6}+2\mleft(1-\beta\mright)+\epsilon}   \mleft(H_{\mathrm{kin}}^{\prime}+k_{F}\mright),\quad k_{F}\rightarrow\infty.
\]
\end{thm}

There are two things to remark about this identity: The first is that
all terms on the first two lines of the right-hand side are manifestly
non-negative (since $E_{k}-h_{k},S_{k}E_{k}S_{k}^{\ast}\geq0$ and
$\varepsilon_{k,k}\mleft(e_{p};e_{q}\mright)=\delta_{p,q}\varepsilon_{k,k}\mleft(e_{p};e_{p}\mright)\leq0$),
and so despite their apparent complexity these terms can be ignored
for a lower bound. This includes in particular all terms with $6$
creation and annihilation operators ($c$ and $c^*$).

The second is that although not anticipated by the motivating relation
of equation (\ref{eq:HBMotivatingRelation}), the exchange contribution
$E_{\mathrm{corr},\mathrm{ex}}$ automatically appears during the
expansion procedure. This identity thus accounts for the full correlation
energy.

\subsection*{Handling the Cubic and Quartic Terms}

The identity for the bosonizable terms essentially suffices to prove
a version of Theorem \ref{them:MainTheorem} for $H_{\mathrm{B}}$,
but the full Hamiltonian $H_{N}$ also contains the cubic and quartic
terms $\mathcal{C}$ and $\mathcal{Q}$; see \eqref{eq:DecompositionofHN}. The quartic terms are in
a sense ``mostly positive'', but the non-definite cubic terms are
difficult to estimate directly.

\subsubsection*{Incorporation of the Small $k$ Cubic Terms}

We will deal with this issue by partially including them in the factorization
identity above. To motivate this, let us note that $\mathcal{C}$
can be written as
\begin{equation}
\mathcal{C}=4\,\mathrm{Re}\sum_{k\in\mathbb{Z}_{\ast}^{3}}\frac{\hat{V}_{k}k_{F}^{-\beta}}{2\mleft(2\pi\mright)^{3}}B_{k}^{\ast}D_{k}=4\,\mathrm{Re}\sum_{k\in\mathbb{Z}_{\ast}^{3}}\biggl(\sum_{p\in L_{k}}\frac{\hat{V}_{k}k_{F}^{-\beta}}{2\mleft(2\pi\mright)^{3}}b_{k,p}^{\ast}\biggr)D_{k}
\end{equation}
where the first equality follows from the observations that $D_{k}^{\ast}=D_{-k}$
and $\left[B_{k},D_{k}^{\ast}\right]=0$.

If we define $w_{k}\in\ell^{2}\mleft(L_{k}\mright)$ by $\left\langle e_{p},w_{k}\right\rangle =2^{-1}\mleft(2\pi\mright)^{-3}\hat{V}_{k}k_{F}^{-\beta}$
we can express this as
\begin{equation}
\mathcal{C}=4\,\mathrm{Re}\sum_{k\in\mathbb{Z}_{\ast}^{3}}\biggl(\sum_{p\in L_{k}}\left\langle e_{p},w_{k}\right\rangle b_{k,p}^{\ast}\biggr)D_{k}
\end{equation}
which suggests how we should modify the \textit{ansatz} we used for
$H_{\mathrm{B}}$: To generate expressions of the form $\sum_{p\in L_{k}}\left\langle e_{p},\mleft(\cdot\mright)\right\rangle b_{k,p}^{\ast}D_{k}$
we can modify the quadratic part accordingly by adding $\langle e_p,\eta_k\rangle D_k$ to $b_{k}\mleft(C_{k}e_{p}\mright)+b_{-k}^{\ast}\mleft(S_{-k}e_{-p}\mright)$, for some  $\eta_{k}\in\ell^{2}\mleft(L_{k}\mright)$ (to be fixed at
the end), and correspondingly include an additional term $d_{p}^{3}$
in the kinetic factorization, where
\begin{equation}
d_{p}^{3}=\begin{cases}
+\sum_{k\in\mathbb{Z}_{\ast}^{3}}1_{L_{k}}\mleft(p\mright)\left\langle e_{p},\eta_{k}\right\rangle \tilde{c}_{p-k}^{\ast}D_{k} & p\in B_{F}^{c}\\
-\sum_{k\in\mathbb{Z}_{\ast}^{3}}1_{L_{k}-k}\mleft(p\mright)\left\langle e_{p+k},\eta_{k}\right\rangle \tilde{c}_{p+k}^{\ast}D_{k} & p\in B_{F}
\end{cases}.
\end{equation}
In Sections \ref{sec:InclusionfoftheSmallkCubicTerms} and \ref{sec:EstimationofEC} (see Theorems \ref{them:BigFactorization} and \ref{them:ECBound}) 
we show that the specific choice 
\begin{equation}
\eta_{k}=\begin{cases}
E_{k}^{-\frac{3}{2}}h_{k}^{\frac{1}{2}}w_{k} & \left|k\right|<k_{F}^{1/3}\\
0 & \text{otherwise}
\end{cases}
\end{equation}
(the choice of cut-off $|k|<k_F^{1/3}$ comes from optimizing the final estimates; see \eqref{eq:optimizing-cut-off-end}) yields the following:

\begin{thm} 
It holds that
\begin{align*}
 & \quad\;\;\,H_{\mathrm{B}}+4\,\mathrm{Re}\sum_{k\in B(0,k_{F}^{1/3})\cap\mathbb{Z}_{\ast}^{3}}\frac{\hat{V}_{k}k_{F}^{-\beta}}{2\mleft(2\pi\mright)^{3}}B_{k}^{\ast}D_{k}+\frac{k_{F}^{-\beta}}{2\mleft(2\pi\mright)^{3}}\sum_{k\in B(0,k_{F}^{1/3})\cap\mathbb{Z}_{\ast}^{3}}\hat{V}_{k}\frac{2\left\langle v_{k},h_{k}^{-1}v_{k}\right\rangle }{1+2\left\langle v_{k},h_{k}^{-1}v_{k}\right\rangle }D_{k}^{\ast}D_{k}\\
 & =\sum_{p\in\mathbb{Z}^{3}}\vert\left|p\right|^{2}-k_{F}^{2}\vert\mleft(\left|\tilde{c}_{p}+d_{p}^{1}+d_{p}^{2}+d_{p}^{3}\right|^{2}+\left|(d_{p}^{1}+d_{p}^{2}+d_{p}^{3})^{\ast}\right|^{2}\mright)-2\sum_{k\in\mathbb{Z}_{\ast}^{3}}\sum_{p\in L_{k}}\varepsilon_{k,k}\mleft(e_{p};S_{k}E_{k}S_{k}^{\ast}e_{p}\mright)\\
 & \quad +\sum_{k\in\mathbb{Z}_{\ast}^{3}}\sum_{p,q\in L_{k}}2\left\langle e_{p},\mleft(E_{k}-h_{k}\mright)e_{q}\right\rangle \mleft(b_{k}\mleft(C_{k}e_{p}\mright)+b_{-k}^{\ast}\mleft(S_{-k}e_{-p}\mright)+\left\langle e_{p},\eta_{k}\right\rangle D_{k}\mright)^{\ast} \\
 &\qquad\qquad\qquad\qquad\qquad\qquad\qquad\qquad \times \mleft(b_{k}\mleft(C_{k}e_{q}\mright)+b_{-k}^{\ast}\mleft(S_{-k}e_{-q}\mright)+\left\langle e_{q},\eta_{k}\right\rangle D_{k}\mright)\\
 &\quad +E_{\mathrm{corr},\mathrm{bos}}+E_{\mathrm{corr},\mathrm{ex}}+\mathcal{E}_{\mathrm{B}}+\mathcal{E}_{\mathcal{C}}
\end{align*}
for an operator $\mathcal{\mathcal{E}_{\mathrm{\mathcal{C}}}}$ which
under Assumption \ref{Assumption:Potential} obeys
\[
\pm\mathcal{E}_{\mathcal{C}}\leq C_{V,\epsilon}k_{F}^{-\frac{1}{6}+2\mleft(1-\beta\mright)+\epsilon} \mleft(H_{\mathrm{kin}}^{\prime}+k_{F}\mright),\quad k_{F}\rightarrow\infty.
\]
\end{thm}

This identity only includes the ``small $k$'' part of $\mathcal{C}$,
i.e. the sum over $k\in B(0,k_{F}^{1/3})\cap\mathbb{Z}_{\ast}^{3}$.
This of course leaves the ``large $k$'' terms unaccounted for,
but these \textit{can} be estimated directly.

Note also the additional sum involving $D_{k}^{\ast}D_{k}$ terms,
reminiscent of the quartic terms. Such expressions are unavoidable
when attempting to include the cubic terms by factorization, but the
crucial point here is the obvious inequality
\begin{equation}
\frac{2\left\langle v_{k},h_{k}^{-1}v_{k}\right\rangle }{1+2\left\langle v_{k},h_{k}^{-1}v_{k}\right\rangle }\leq1.
\end{equation}
That this factor is always less than $1$ means that we can use the
``almost positivity'' of the quartic terms $\mathcal{Q}$ in \eqref{eq:DecompositionofHN} to partially cancel these
terms.

\subsubsection*{Estimation of the Remaining Terms}

The parts of $H_{N}$ which remain unaccounted for are the ``large
$k$'' cubic and quartic terms, which we bound in Section \ref{sec:EstimationoftheRemainingTerms}.
To illustrate how to estimate these, consider for definiteness the
cubic terms
\begin{equation}
\mathcal{E}_{\mathcal{C},\mathrm{large}}=\frac{2k_{F}^{-\beta}}{\mleft(2\pi\mright)^{3}}\mathrm{Re}\sum_{k\in\mathbb{Z}_{\ast}^{3}\backslash B(0,k_{F}^{1/3})}\hat{V}_{k}B_{k}^{\ast}D_{k}.
\end{equation}
The key observation is that if one expands $B_{k}^{\ast}$, one can
write the sum as
\begin{align}
\sum_{k\in\mathbb{Z}_{\ast}^{3}\backslash B(0,k_{F}^{1/3})}\hat{V}_{k}B_{k}^{\ast}D_{k} & =\sum_{k\in\mathbb{Z}_{\ast}^{3}\backslash B(0,k_{F}^{1/3})}\sum_{p\in L_{k}}\hat{V}_{k}\tilde{c}_{p}^{\ast}\tilde{c}_{p-k}^{\ast}D_{k}\\
 & =\sum_{p\in B_{F}^{c}}\tilde{c}_{p}^{\ast}\biggl(\sum_{k\in\mathbb{Z}_{\ast}^{3}\backslash B(0,k_{F}^{1/3})}1_{L_{k}}\mleft(p\mright)\hat{V}_{k}\tilde{c}_{p-k}^{\ast}D_{k}\biggr)\nonumber 
\end{align}
and so, by the identity\footnote{This is a consequence of particle-hole symmetry.}
\begin{equation}\label{eq:H-kin-localized-def}
H_{\mathrm{kin}}^{\prime}=\sum_{p\in B_{F}^{c}}(\left|p\right|^{2}-\zeta)\,c_{p}^{\ast}c_{p}+\sum_{p\in B_{F}}(\zeta-\left|p\right|^{2})\,c_{p}c_{p}^{\ast}=\sum_{p\in\mathbb{Z}^{3}}\vert\left|p\right|^{2}-\zeta\vert\,\tilde{c}_{p}^{\ast}\tilde{c}_{p}
\end{equation}
which is valid for any $\zeta\in[\sup_{p\in B_{F}}\left|p\right|^{2},\inf_{p\in B_{F}^{c}}\left|p\right|^{2}]$,
one can estimate
\begin{align}
&\vert\langle\Psi,\mathcal{E}_{\mathcal{C},\mathrm{large}}\Psi\rangle\vert \nn \\
& \leq Ck_{F}^{-\beta}\sqrt{\sum_{p\in B_{F}^{c}}\vert\left|p\right|^{2}-\zeta\vert\Vert\tilde{c}_{p}\Psi\Vert^{2}}\sqrt{\sum_{p\in B_{F}^{c}}\vert\left|p\right|^{2}-\zeta\vert^{-1}\biggl\Vert\sum_{k\in\mathbb{Z}_{\ast}^{3}\backslash B(0,k_{F}^{1/3})}1_{L_{k}}\mleft(p\mright)\hat{V}_{k}\tilde{c}_{p-k}^{\ast}D_{k}\Psi\biggr\Vert^{2}}\nonumber \\
 & \leq Ck_{F}^{-\beta}\sqrt{\left\langle \Psi,H_{\mathrm{kin}}^{\prime}\Psi\right\rangle \sum_{p\in B_{F}^{c}}\vert\left|p\right|^{2}-\zeta\vert^{-1}\left\langle \Psi,A_{p}^{\ast}A_{p}\Psi\right\rangle }
\end{align}
for
\begin{equation}
A_{p}=\sum_{k\in\mathbb{Z}_{\ast}^{3}\backslash B(0,k_{F}^{1/3})}1_{L_{k}}\mleft(p\mright)\hat{V}_{k}\tilde{c}_{p-k}^{\ast}D_{k}.
\end{equation}
Clearly $A_{p}^{\ast}A_{p}\leq A_{p}^{\ast}A_{p}+A_{p}A_{p}^{\ast}=\{A_{p}^{\ast},A_{p}\}$,
and the point is that $A_{p}$ is a sum of triples of fermionic creation
and annihilation operators. As a consequence, the anticommutator $\{A_{p}^{\ast},A_{p}\}$ consists
only of sums of 4 or less creation and annihilation operators, which can be controlled by the kinetic operator $H_{\mathrm{kin}}^{\prime}$. In fact, combining this observation and the fact that $\zeta$ can be chosen such that
\begin{equation}
\sum_{p\in L_{k}}\vert\left|p\right|^{2}-\zeta\vert^{-1}\leq C_{\epsilon}k_{F}^{1+\epsilon}
\end{equation}
(which also enters in the estimation of $\mathcal{E}_{\mathrm{B}}$
and $\mathcal{E}_{\mathcal{C}}$ from the previous steps) we eventually
arrive at the bound
\begin{equation}\label{eq:cE-C-large-k}
\pm\mathcal{E}_{\mathcal{C},\mathrm{large}}\leq C_{\epsilon}k_{F}^{1-\beta+\epsilon}\sqrt{\sum_{k\in\mathbb{Z}_{\ast}^{3}\backslash B(0,k_{F}^{1/3})}\hat{V}_{k}^{2}}H_{\mathrm{kin}}^{\prime}.
\end{equation}
The large $k$ quartic terms can be estimated in a similar fashion,
with one exception: There remains the term
\begin{equation}\label{eq:Q5-large-k}
\frac{k_{F}^{-\beta}}{2\mleft(2\pi\mright)^{3}}\sum_{k\in\mathbb{Z}_{\ast}^{3}\backslash B(0,k_{F}^{1/3})}\hat{V}_{k}\sum_{p,q\in A\cap(A+k)}c_{p}^{\ast}c_{q-k}^{\ast}c_{q}c_{p-k}
\end{equation}
where $A=\mathbb{Z}^{3}\backslash\overline{B}(0,2k_{F})$, which is
to say the part of the interaction which involves momenta exclusively
``far away'' from the Fermi ball. This condition can however be
exploited to also control this term in the same form as the other
terms. In particular, on the set $A$ we have the improved spectral gap $||k|^2-k_F^2| \ge k_F^2$ and hence $\sum_{p\in A} c_p^* c_p \le k_F^{-2} H_{\mathrm{kin}}^\prime$ which is significantly stronger than the simple bound $\cN_E\le  H_{\mathrm{kin}}^\prime$.


\subsubsection*{Concluding Theorem \ref{them:MainTheorem}}

%
%

With all the estimates in place we thus obtain the first part of Theorem
\ref{them:MainTheorem}, i.e. the inequality
\begin{equation}\label{eq:lower-bound-V-2V-a}
H_N -  E_{\mathrm{FS}} = H_{\rm B}+ \mathcal{C} + \mathcal{Q} \ge E_{\mathrm{corr},\mathrm{bos}}+E_{\mathrm{corr},\mathrm{ex}}+\mathcal{E}
\end{equation}
where $\mathcal{E}$ obeys $\mathcal{E}\geq-C_{V,\epsilon}k_{F}^{-\frac{1}{6}+2\mleft(1-\beta\mright)+\epsilon}\mleft(H_{\mathrm{kin}}^{\prime}+k_{F}\mright)$,
but not the second part, i.e. the estimate
\begin{equation}
\left\langle \Psi,H_{\mathrm{kin}}^{\prime}\Psi\right\rangle \leq C_{V,\epsilon}k_{F}^{3-2\beta+\epsilon}
\end{equation}
for low-lying states $\Psi$. This however follows as a simple consequence
of the first inequality, since we can write
\begin{equation}
2\mleft(H_{N}-E_{\mathrm{FS}}\mright)=H_{\mathrm{kin}}^{\prime}+(\tilde{H}_{\mathrm{B}}+\tilde{\mathcal{C}}+\tilde{\mathcal{Q}})
\end{equation}
where the tilde quantities are the same as those of equation (\ref{eq:DecompositionofHN})
up to the replacement $\hat{V}_{k}\rightarrow2\,\hat{V}_{k}$. By using the first part, namely by applying \eqref{eq:lower-bound-V-2V-a} with $(H_B + \mathcal{C} + \mathcal{Q})$ replaced by $(\tilde{H}_{\mathrm{B}}+\tilde{\mathcal{C}}+\tilde{\mathcal{Q}})$, we have 
\begin{align}
\tilde{H}_{\mathrm{B}}+\tilde{\mathcal{C}}+\tilde{\mathcal{Q}} & \geq\tilde{E}_{\mathrm{corr},\mathrm{bos}}+\tilde{E}_{\mathrm{corr},\mathrm{ex}}-\tilde{C}_{V,\epsilon}k_{F}^{-\frac{1}{6}+2\mleft(1-\beta\mright)+\epsilon}\mleft(H_{\mathrm{kin}}^{\prime}+k_{F}\mright)\\
 & \geq\tilde{E}_{\mathrm{corr},\mathrm{bos}}-o\mleft(1\mright)\mleft(H_{\mathrm{kin}}^{\prime}+k_{F}\mright),\quad k_{F}\rightarrow\infty.\nonumber 
\end{align}
Here we also used $\tilde{E}_{\mathrm{corr},\mathrm{ex}}\geq0$. Thus 
\begin{equation}
\mleft(1-o\mleft(1\mright)\mright)H_{\mathrm{kin}}^{\prime}\leq2\mleft(H_{N}-E_{\mathrm{FS}}\mright)-\tilde{E}_{\mathrm{corr},\mathrm{bos}}+C_{V,\epsilon}k_{F},\quad k_{F}\rightarrow\infty,
\end{equation}
from which the second part follows by proving that $-\tilde{E}_{\mathrm{corr},\mathrm{bos}}\leq C_{V,\epsilon}k_{F}^{3-2\beta+\epsilon}$.

\bigskip
{\bf Organization of the paper.} In Section \ref{sec:ExtractionoftheCorrelationEnergybyFactorization} we will extract the correlation energy from $H_{\rm B}$ by an explicit factorization. The error $\mathcal{E}_{\rm B}$ of this step is estimated in Section \ref{sec:EstimationofEB}.  In Section \ref{sec:InclusionfoftheSmallkCubicTerms} we extend the exact factorization to include also the low-momentum part of the cubic terms $\mathcal{C}$. The error $\mathcal{E}_{\mathcal{C}}$ of this step is estimated in Section \ref{sec:EstimationofEC}. All of the remaining
terms are estimated in Section \ref{sec:EstimationoftheRemainingTerms}, leading to the conclusion of Theorem \ref{them:MainTheorem}.

\bigskip
{\bf Acknowledgements.} This work was partially funded by the Deutsche Forschungsgemeinschaft (DFG,
German Research Foundation) via the TRR 352 -- Project-ID 470903074.  PTN
was partially supported by the European Research Council via the ERC CoG RAMBAS -- Project-Nr. 101044249.

\section{\label{sec:ExtractionoftheCorrelationEnergybyFactorization}Extraction
of the Correlation Energy by Factorization}

In this section we perform the computations leading to the factorized
expression for $H_{\mathrm{B}}$.

For convenience we recall that the operators $E_{k},C_{k},S_{k}:\ell^{2}\mleft(L_{k}\mright)\rightarrow\ell^{2}\mleft(L_{k}\mright)$
are defined by
\begin{equation}
E_{k}=(h_{k}^{\frac{1}{2}}(h_{k}+2P_{k})h_{k}^{\frac{1}{2}})^{\frac{1}{2}}\label{eq:EkDefinition}
\end{equation}
and
\begin{equation}
C_{k}=\frac{1}{2}(h_{k}^{-\frac{1}{2}}E_{k}^{\frac{1}{2}}+h_{k}^{\frac{1}{2}}E_{k}^{-\frac{1}{2}}),\quad S_{k}=\frac{1}{2}(h_{k}^{-\frac{1}{2}}E_{k}^{\frac{1}{2}}-h_{k}^{\frac{1}{2}}E_{k}^{-\frac{1}{2}}),\label{eq:CkSkDefinition}
\end{equation}
while the operators $d_{p}^{1}$ and $d_{p}^{2}$ are defined by
\begin{equation}
d_{p}^{1}=\begin{cases}
+\sum_{k\in\mathbb{Z}_{\ast}^{3}}1_{L_{k}}\mleft(p\mright)\tilde{c}_{p-k}^{\ast}b_{k}\mleft(\mleft(C_{k}-1\mright)e_{p}\mright) & \text{ for }p\in B_{F}^{c}\\
-\sum_{k\in\mathbb{Z}_{\ast}^{3}}1_{L_{k}-k}\mleft(p\mright)\tilde{c}_{p+k}^{\ast}b_{k}\mleft(\mleft(C_{k}-1\mright)e_{p+k}\mright) & \text{ for } p\in B_{F}
\end{cases}\label{eq:dp1Definition}
\end{equation}
and
\begin{equation}
d_{p}^{2}=\begin{cases}
+\sum_{k\in\mathbb{Z}_{\ast}^{3}}1_{L_{k}}\mleft(p\mright)\tilde{c}_{p-k}^{\ast}b_{-k}^{\ast}\mleft(S_{-k}e_{-p}\mright) & \text{ for } p\in B_{F}^{c}\\
-\sum_{k\in\mathbb{Z}_{\ast}^{3}}1_{L_{k}-k}\mleft(p\mright)\tilde{c}_{p+k}^{\ast}b_{-k}^{\ast}\mleft(S_{-k}e_{-p-k}\mright) & \text{ for } p\in B_{F}
\end{cases},\label{eq:dp2Definition}
\end{equation}
respectively. Our goal is the following:
\begin{thm}
\label{them:QuasibosonicBogolubovFactorization}It holds that
\begin{align*}
H_{\mathrm{B}} & =\sum_{p\in\mathbb{Z}^{3}}\vert\left|p\right|^{2}-k_{F}^{2}\vert\mleft(\left|\tilde{c}_{p}+d_{p}^{1}+d_{p}^{2}\right|^{2}+\left|(d_{p}^{1}+d_{p}^{2})^{\ast}\right|^{2}\mright)-2\sum_{k\in\mathbb{Z}_{\ast}^{3}}\sum_{p\in L_{k}}\varepsilon_{k,k}\mleft(e_{p};S_{k}E_{k}S_{k}^{\ast}e_{p}\mright)\\
 & +2\sum_{k\in\mathbb{Z}_{\ast}^{3}}\sum_{p,q\in L_{k}}\left\langle e_{p},\mleft(E_{k}-h_{k}\mright)e_{q}\right\rangle \mleft(b_{k}\mleft(C_{k}e_{p}\mright)+b_{-k}^{\ast}\mleft(S_{-k}e_{-p}\mright)\mright)^{\ast}\mleft(b_{k}\mleft(C_{k}e_{q}\mright)+b_{-k}^{\ast}\mleft(S_{-k}e_{-q}\mright)\mright)\\
 & +E_{\mathrm{corr},\mathrm{bos}}+E_{\mathrm{corr},\mathrm{ex}}+\mathcal{E}_{\mathrm{B}}
\end{align*}
for an operator $\mathcal{\mathcal{E}_{\mathrm{B}}}$ defined below.
\end{thm}

\subsubsection*{Quasi-Bosonic Operators}

Before we start in earnest we will recall some properties of the quasi-bosonic
operators we must consider.

First, we define for general symmetric operators $A_{k},B_{k}:\ell^{2}\mleft(L_{k}\mright)\rightarrow\ell^{2}\mleft(L_{k}\mright)$
the expressions
\begin{equation}
Q_{1}^{k}\mleft(A_{k}\mright)=\sum_{p,q\in L_{k}}\left\langle e_{p},A_{k}e_{q}\right\rangle b_{k,p}^{\ast}b_{k,q},\quad Q_{2}^{k}\mleft(B_{k}\mright)=2\,\mathrm{Re}\sum_{p,q\in L_{k}}\left\langle e_{p},B_{k}e_{q}\right\rangle b_{k,p}b_{-k,-q},
\end{equation}
in terms of which the interaction part of $H_{N}$ can be written
\begin{equation}
\frac{k_{F}^{-\beta}}{2\mleft(2\pi\mright)^{3}}\sum_{k\in\mathbb{Z}_{\ast}^{3}}\hat{V}_{k}\mleft(2B_{k}^{\ast}B_{k}+B_{k}B_{-k}+B_{-k}^{\ast}B_{k}^{\ast}\mright)=\sum_{k\in\mathbb{Z}_{\ast}^{3}}\mleft(2\,Q_{1}^{k}\mleft(P_{k}\mright)+Q_{2}^{k}\mleft(P_{k}\mright)\mright)
\end{equation}
for $P_{k}=\left|v_{k}\right\rangle \left\langle v_{k}\right|$ with
$v_{k}\in\ell^{2}\mleft(L_{k}\mright)$ defined by $\left\langle e_{p},v_{k}\right\rangle =\sqrt{2^{-1}\mleft(2\pi\mright)^{-3}\hat{V}_{k}k_{F}^{-\beta}}$.

For any $\varphi\in\ell^{2}\mleft(L_{k}\mright)$ we also define the
generalized excitation operators
\begin{equation}
b_{k}\mleft(\varphi\mright)=\sum_{p\in L_{k}}\left\langle \varphi,e_{p}\right\rangle b_{k,p},\quad b_{k}^{\ast}\mleft(\varphi\mright)=\sum_{p\in L_{k}}\left\langle e_{p},\varphi\right\rangle b_{k,p}^{\ast},
\end{equation}
which lets us write $Q_{1}^{k}\mleft(A_{k}\mright)$ and $Q_{2}^{k}\mleft(B_{k}\mright)$
as
\begin{equation}
Q_{1}^{k}\mleft(A_{k}\mright)=\sum_{p\in L_{k}}b_{k}^{\ast}\mleft(A_{k}e_{p}\mright)b_{k,p},\quad Q_{2}^{k}\mleft(B_{k}\mright)=2\,\mathrm{Re}\sum_{p\in L_{k}}b_{k}\mleft(B_{k}e_{p}\mright)b_{-k,-p}.
\end{equation}
The generalized excitation operators obey the quasi-bosonic commutation
relations
\begin{align}
\left[b_{k}\mleft(\varphi\mright),b_{l}\mleft(\psi\mright)\right] & =\left[b_{k}^{\ast}\mleft(\varphi\mright),b_{l}^{\ast}\mleft(\psi\mright)\right]=0\\
\left[b_{k}\mleft(\varphi\mright),b_{l}^{\ast}\mleft(\psi\mright)\right] & =\delta_{k,l}\left\langle \varphi,\psi\right\rangle +\varepsilon_{k,l}\mleft(\varphi;\psi\mright)\nonumber 
\end{align}
where the \textit{exchange correction} $\varepsilon_{k,l}\mleft(\varphi;\psi\mright)$
is given by
\begin{equation}
\varepsilon_{k,l}\mleft(\varphi;\psi\mright)=-\sum_{q\in L_{k}\cap L_{l}}\left\langle \varphi,e_{q}\right\rangle \left\langle e_{q},\psi\right\rangle \tilde{c}_{q-l}^{\ast}\tilde{c}_{q-k}-\sum_{q\in\mleft(L_{k}-k\mright)\cap\mleft(L_{l}-l\mright)}\left\langle \varphi,e_{q+k}\right\rangle \left\langle e_{q+l},\psi\right\rangle \tilde{c}_{q+l}^{\ast}\tilde{c}_{q+k}.
\end{equation}

Below we will often encounter expressions of the ``trace form''
$\sum_{i=1}^{n}q\mleft(Se_{i},Te_{i}\mright)$ for some bilinear mapping
$q$, for example
\begin{equation}
Q_{1}^{k}\mleft(A_{k}\mright)=\sum_{p\in L_{k}}b_{k}^{\ast}\mleft(A_{k}e_{p}\mright)b_{k}\mleft(e_{p}\mright)=\sum_{p\in L_{k}}q\mleft(A_{k}e_{p},e_{p}\mright),\quad q\mleft(\varphi,\psi\mright)=b_{k}^{\ast}\mleft(\varphi\mright)b_{k}\mleft(\psi\mright).
\end{equation}
(Here we drop the $k$-dependence in $q$ as we will use the above formula for each $k$ separately.) For that reason we recall the following lemma which simplifies the
calculations with these significantly:
\begin{lem}
\label{lemma:TraceFormLemma}Let $V$ be an $n$-dimensional Hilbert
space and let $q:V\times V\rightarrow W$ be a sesquilinear mapping
into a vector space $W$. Then for any orthonormal basis $\mleft(e_{i}\mright)_{i=1}^{n}$
of $V$ and operators $S,T:V\rightarrow V$ it holds that
\[
\sum_{i=1}^{n}q\mleft(Se_{i},Te_{i}\mright)=\sum_{i=1}^{n}q\mleft(ST^{\ast}e_{i},e_{i}\mright).
\]
\end{lem}

The lemma is immediate by orthonormal expansion.

We remark that we will only consider $\ell^{2}\mleft(L_{k}\mright)$
as a \textit{real} vector space (so sesquilinearity is simply bilinearity).

Finally we point out that the operators $E_{k},C_{k}$ and $S_{k}$
all obey a symmetry condition of the form
\begin{equation}
\left\langle e_{p},E_{k}e_{q}\right\rangle =\left\langle e_{-p},E_{-k}e_{-q}\right\rangle ,\quad p,q\in L_{k},
\end{equation}
since these are directly determined by $h_{k}$ and $P_{k}$ which
also satisfy this.

\subsection{Factorization of the Interaction Terms}

We begin with the terms $\sum_{k\in\mathbb{Z}_{\ast}^{3}}\mleft(2\,Q_{1}^{k}\mleft(P_{k}\mright)+Q_{2}^{k}\mleft(P_{k}\mright)\mright)$
which come from the interaction. Since we will also need this for
the kinetic terms below, we state a general identity:
\begin{prop}
\label{prop:QuadraticExpansionIdentity}For symmetric operators $A_{k}:\ell^{2}\mleft(L_{k}\mright)\rightarrow\ell^{2}\mleft(L_{k}\mright)$,
$k\in\mathbb{Z}_{\ast}^{3}$, obeying
\[
\left\langle e_{p},A_{k}e_{q}\right\rangle =\left\langle e_{-p},A_{-k}e_{-q}\right\rangle ,\quad p,q\in L_{k},
\]
it holds that
\begin{align*}
 & \quad\,\sum_{k\in\mathbb{Z}_{\ast}^{3}}\sum_{p,q\in L_{k}}2\left\langle e_{p},A_{k}e_{q}\right\rangle \mleft(b_{k}\mleft(C_{k}e_{p}\mright)+b_{-k}^{\ast}\mleft(S_{-k}e_{-p}\mright)\mright)^{\ast}\mleft(b_{k}\mleft(C_{k}e_{q}\mright)+b_{-k}^{\ast}\mleft(S_{-k}e_{-q}\mright)\mright)\\
 & =\sum_{k\in\mathbb{Z}_{\ast}^{3}}\mleft(2\,Q_{1}^{k}\mleft(C_{k}A_{k}C_{k}^{\ast}+S_{k}A_{k}S_{k}^{\ast}\mright)+Q_{2}^{k}\mleft(C_{k}A_{k}S_{k}^{\ast}+S_{k}A_{k}C_{k}^{\ast}\mright)\mright)\\
 & +\sum_{k\in\mathbb{Z}_{\ast}^{3}}2\,\mathrm{tr}\mleft(S_{k}A_{k}S_{k}^{\ast}\mright)+2\sum_{k\in\mathbb{Z}_{\ast}^{3}}\sum_{p\in L_{k}}\varepsilon_{k,k}\mleft(e_{p};S_{k}A_{k}S_{k}^{\ast}e_{p}\mright).
\end{align*}
\end{prop}

\textbf{Proof:} By expanding the terms and applying Lemma \ref{lemma:TraceFormLemma}
we see that
\begin{align}
 & \quad\;\;\;\sum_{k\in\mathbb{Z}_{\ast}^{3}}\sum_{p,q\in L_{k}}2\left\langle e_{p},A_{k}e_{q}\right\rangle \mleft(b_{k}\mleft(C_{k}e_{p}\mright)+b_{-k}^{\ast}\mleft(S_{-k}e_{-p}\mright)\mright)^{\ast}\mleft(b_{k}\mleft(C_{k}e_{q}\mright)+b_{-k}^{\ast}\mleft(S_{-k}e_{-q}\mright)\mright)\nonumber \\
 & =2\sum_{k\in\mathbb{Z}_{\ast}^{3}}\sum_{p,q\in L_{k}}\left\langle e_{p},A_{k}e_{q}\right\rangle b_{k}^{\ast}\mleft(C_{k}e_{p}\mright)b_{k}\mleft(C_{k}e_{q}\mright)+2\sum_{k\in\mathbb{Z}_{\ast}^{3}}\sum_{p,q\in L_{k}}\left\langle e_{p},A_{k}e_{q}\right\rangle b_{-k}\mleft(S_{-k}e_{-p}\mright)b_{-k}^{\ast}\mleft(S_{-k}e_{-q}\mright)\nonumber \\
 & +4\,\mathrm{Re}\sum_{k\in\mathbb{Z}_{\ast}^{3}}\sum_{p,q\in L_{k}}\left\langle e_{p},A_{k}e_{q}\right\rangle b_{-k}\mleft(S_{-k}e_{-p}\mright)b_{k}\mleft(C_{k}e_{q}\mright)\\
 & =2\sum_{k\in\mathbb{Z}_{\ast}^{3}}\sum_{p,q\in L_{k}}\left\langle e_{p},C_{k}A_{k}C_{k}^{\ast}e_{q}\right\rangle b_{k,p}^{\ast}b_{k,q}+2\sum_{k\in\mathbb{Z}_{\ast}^{3}}\sum_{p,q\in L_{k}}\left\langle e_{p},S_{k}A_{k}S_{k}^{\ast}e_{q}\right\rangle b_{k,p}b_{k,q}^{\ast}\nonumber \\
 & +4\,\mathrm{Re}\sum_{k\in\mathbb{Z}_{\ast}^{3}}\sum_{p,q\in L_{k}}\left\langle e_{p},S_{k}A_{k}C_{k}^{\ast}e_{q}\right\rangle b_{k,p}b_{-k,-q}\nonumber 
\end{align}
where we also took advantage of the symmetry of $A_{k},C_{k}$ and
$S_{k}$ under $\mleft(k,p,q\mright)\rightarrow\mleft(-k,-p,-q\mright)$.
Now
\begin{equation}
\sum_{p,q\in L_{k}}\left\langle e_{p},C_{k}A_{k}C_{k}^{\ast}e_{q}\right\rangle b_{k,p}^{\ast}b_{k,q}=Q_{1}^{k}\mleft(C_{k}A_{k}C_{k}^{\ast}\mright)
\end{equation}
by definition, while
\begin{align}
\sum_{k\in\mathbb{Z}_{\ast}^{3}}\sum_{p,q\in L_{k}}\left\langle e_{p},S_{k}A_{k}S_{k}^{\ast}e_{q}\right\rangle b_{k,p}b_{k,q}^{\ast} & =\sum_{k\in\mathbb{Z}_{\ast}^{3}}\sum_{p,q\in L_{k}}\left\langle e_{p},S_{k}A_{k}S_{k}^{\ast}e_{q}\right\rangle b_{k,q}^{\ast}b_{k,p}+\sum_{k\in\mathbb{Z}_{\ast}^{3}}\sum_{p\in L_{k}}\left\langle e_{p},S_{k}A_{k}S_{k}^{\ast}e_{p}\right\rangle \nonumber \\
 & +\sum_{k\in\mathbb{Z}_{\ast}^{3}}\sum_{p,q\in L_{k}}\left\langle e_{p},S_{k}A_{k}S_{k}^{\ast}e_{q}\right\rangle \varepsilon_{k,k}\mleft(e_{p};e_{q}\mright)\label{eq:ReOrderingQ1Term}\\
 & =\sum_{k\in\mathbb{Z}_{\ast}^{3}}Q_{1}^{k}\mleft(S_{k}A_{k}S_{k}^{\ast}\mright)+\sum_{k\in\mathbb{Z}_{\ast}^{3}}\mathrm{tr}\mleft(S_{k}A_{k}S_{k}^{\ast}\mright)+\sum_{k\in\mathbb{Z}_{\ast}^{3}}\sum_{p\in L_{k}}\varepsilon_{k,k}\mleft(e_{p};S_{k}A_{k}S_{k}^{\ast}e_{p}\mright)\nonumber 
\end{align}
by symmetry of $A$ and the fact that the matrix elements are real-valued.
Similarly, renaming variables and using the symmetries involved once
more
\begin{align}
\sum_{k\in\mathbb{Z}_{\ast}^{3}}\sum_{p,q\in L_{k}}\left\langle e_{p},S_{k}A_{k}C_{k}^{\ast}e_{q}\right\rangle b_{k,p}b_{-k,-q} & =\sum_{k\in\mathbb{Z}_{\ast}^{3}}\sum_{p,q\in L_{k}}\left\langle C_{k}A_{k}S_{k}^{\ast}e_{p},e_{q}\right\rangle b_{-k,-q}b_{k,p}\\
 & =\sum_{k\in\mathbb{Z}_{\ast}^{3}}\sum_{p,q\in L_{k}}\left\langle e_{p},C_{k}A_{k}S_{k}^{\ast}e_{q}\right\rangle b_{k,p}b_{-k,-q}\nonumber 
\end{align}
which implies
\begin{align}
4\,\mathrm{Re}\sum_{k\in\mathbb{Z}_{\ast}^{3}}\sum_{p,q\in L_{k}}\left\langle e_{p},S_{k}A_{k}C_{k}^{\ast}e_{q}\right\rangle b_{k,p}b_{-k,-q} & =2\,\mathrm{Re}\sum_{k\in\mathbb{Z}_{\ast}^{3}}\sum_{p,q\in L_{k}}\left\langle e_{p},\mleft(C_{k}A_{k}S_{k}^{\ast}+S_{k}A_{k}C_{k}^{\ast}\mright)e_{q}\right\rangle b_{k,p}b_{-k,-q}\nonumber \\
 & =\sum_{k\in\mathbb{Z}_{\ast}^{3}}Q_{2}^{k}\mleft(C_{k}A_{k}S_{k}^{\ast}+S_{k}A_{k}C_{k}^{\ast}\mright).
\end{align}
$\hfill\square$

This yields the following identity for the interaction terms:
\begin{prop}
\label{prop:InteractionIdentity}It holds that
\begin{align*}
 & \sum_{k\in\mathbb{Z}_{\ast}^{3}}\mleft(2\,Q_{1}^{k}\mleft(P_{k}\mright)+Q_{2}^{k}\mleft(P_{k}\mright)\mright)=-\sum_{k\in\mathbb{Z}_{\ast}^{3}}2\,Q_{1}^{k}\mleft(h_{k}\mright)+E_{\mathrm{corr},\mathrm{bos}}-2\sum_{k\in\mathbb{Z}_{\ast}^{3}}\varepsilon_{k,k}\mleft(e_{p};S_{k}E_{k}S_{k}^{\ast}e_{p}\mright)\\
 & \;\,+\sum_{k\in\mathbb{Z}_{\ast}^{3}}\sum_{p,q\in L_{k}}2\left\langle e_{p},E_{k}e_{q}\right\rangle \mleft(b_{k}\mleft(C_{k}e_{p}\mright)+b_{-k}^{\ast}\mleft(S_{-k}e_{-p}\mright)\mright)^{\ast}\mleft(b_{k}\mleft(C_{k}e_{q}\mright)+b_{-k}^{\ast}\mleft(S_{-k}e_{-q}\mright)\mright).
\end{align*}
\end{prop}

\textbf{Proof:} From the definitions of the equations (\ref{eq:EkDefinition})
and (\ref{eq:CkSkDefinition}) it readily follows that
\begin{align}
C_{k}E_{k}C_{k}^{\ast}+S_{k}E_{k}S_{k}^{\ast} & =\frac{1}{2}\mleft(h_{k}+2P_{k}+h_{k}\mright)=h_{k}+P_{k}\\
C_{k}E_{k}S_{k}^{\ast}+S_{k}E_{k}C_{k}^{\ast} & =\frac{1}{2}\mleft(h_{k}+2P_{k}-h_{k}\mright)=P_{k}\nonumber 
\end{align}
and so the previous proposition tells us that
\begin{align}
 & \quad\,\sum_{k\in\mathbb{Z}_{\ast}^{3}}\sum_{p,q\in L_{k}}2\left\langle e_{p},E_{k}e_{q}\right\rangle \mleft(b_{k}\mleft(C_{k}e_{p}\mright)+b_{-k}^{\ast}\mleft(S_{-k}e_{-p}\mright)\mright)^{\ast}\mleft(b_{k}\mleft(C_{k}e_{q}\mright)+b_{-k}^{\ast}\mleft(S_{-k}e_{-q}\mright)\mright)\\
 & =\sum_{k\in\mathbb{Z}_{\ast}^{3}}\mleft(2\,Q_{1}^{k}\mleft(h_{k}+P_{k}\mright)+Q_{2}^{k}\mleft(P_{k}\mright)\mright)+\sum_{k\in\mathbb{Z}_{\ast}^{3}}2\,\mathrm{tr}\mleft(S_{k}E_{k}S_{k}^{\ast}\mright)+2\sum_{k\in\mathbb{Z}_{\ast}^{3}}\sum_{p\in L_{k}}\varepsilon_{k,k}\mleft(e_{p};S_{k}E_{k}S_{k}^{\ast}e_{p}\mright)\nonumber 
\end{align}
which can be rearranged for the claim since
\begin{equation}
2\,S_{k}E_{k}S_{k}^{\ast}=h_{k}+P_{k}-\frac{1}{2}\mleft(h_{k}^{-\frac{1}{2}}E_{k}h_{k}^{\frac{1}{2}}+h_{k}^{\frac{1}{2}}E_{k}h_{k}^{-\frac{1}{2}}\mright)
\end{equation}
whence
\begin{equation}
-\sum_{k\in\mathbb{Z}_{\ast}^{3}}2\,\mathrm{tr}\mleft(S_{k}E_{k}S_{k}^{\ast}\mright)=\sum_{k\in\mathbb{Z}_{\ast}^{3}}\mathrm{tr}\mleft(E_{k}-h_{k}-P_{k}\mright)=E_{\mathrm{corr},\mathrm{bos}}
\end{equation}
as calculated in \cite[Propositions 7.1, 7.6]{ChrHaiNam-23a}.

$\hfill\square$

\subsection{Factorization of the Kinetic Terms}

Recall the kinetic operator $H_{\mathrm{kin}}^{\prime}$ in \eqref{eq:H-kin-localized-def}. Clearly
\begin{align}
\sum_{p\in\mathbb{Z}^{3}}\vert\left|p\right|^{2}-k_{F}^{2}\vert\left|\tilde{c}_{p}+d_{p}^{1}+d_{p}^{2}\right|^{2} & =H_{\mathrm{kin}}^{\prime}+\sum_{p\in\mathbb{Z}^{3}}2\,\vert\left|p\right|^{2}-k_{F}^{2}\vert\,\mathrm{Re}\mleft(\tilde{c}_{p}^{\ast}d_{p}^{1}+\tilde{c}_{p}^{\ast}d_{p}^{2}\mright)\label{eq:KineticExpansion}\\
 & +\sum_{p\in\mathbb{Z}^{3}}\vert\left|p\right|^{2}-k_{F}^{2}\vert\mleft((d_{p}^{1})^{\ast}d_{p}^{1}+2\,\mathrm{Re}\mleft((d_{p}^{1})^{\ast}d_{p}^{2}\mright)+(d_{p}^{2})^{\ast}d_{p}^{2}\mright)\nonumber 
\end{align}
so we consider the sums on the right-hand side in order. First the
simplest:
\begin{prop}
It holds that
\begin{align*}
\sum_{p\in\mathbb{Z}^{3}}2\,\vert\left|p\right|^{2}-k_{F}^{2}\vert\,\mathrm{Re}\mleft(\tilde{c}_{p}^{\ast}d_{p}^{1}\mright) & =\sum_{k\in\mathbb{Z}_{\ast}^{3}}2\,Q_{1}^{k}\mleft(\mleft(C_{k}-1\mright)h_{k}+h_{k}\mleft(C_{k}^{\ast}-1\mright)\mright)\\
\sum_{p\in\mathbb{Z}^{3}}2\,\vert\left|p\right|^{2}-k_{F}^{2}\vert\,\mathrm{Re}\mleft(\tilde{c}_{p}^{\ast}d_{p}^{2}\mright) & =\sum_{k\in\mathbb{Z}_{\ast}^{3}}Q_{2}^{k}\mleft(S_{k}h_{k}+h_{k}S_{k}^{\ast}\mright).
\end{align*}
\end{prop}

\textbf{Proof:} For $p\in B_{F}^{c}$ we have from the definitions
of the equations (\ref{eq:dp1Definition}) and (\ref{eq:dp2Definition})
that
\begin{align}
\sum_{p\in B_{F}^{c}}\vert\left|p\right|^{2}-k_{F}^{2}\vert\,\tilde{c}_{p}^{\ast}d_{p}^{1} & =\sum_{k\in\mathbb{Z}_{\ast}^{3}}\sum_{p\in L_{k}}\vert\left|p\right|^{2}-k_{F}^{2}\vert\,\tilde{c}_{p}^{\ast}\tilde{c}_{p-k}^{\ast}b_{k}\mleft(\mleft(C_{k}-1\mright)e_{p}\mright)\\
\sum_{p\in B_{F}^{c}}\vert\left|p\right|^{2}-k_{F}^{2}\vert\,\tilde{c}_{p}^{\ast}d_{p}^{2} & =\sum_{k\in\mathbb{Z}_{\ast}^{3}}\sum_{p\in L_{k}}\vert\left|p\right|^{2}-k_{F}^{2}\vert\,\tilde{c}_{p}^{\ast}\tilde{c}_{p-k}^{\ast}b_{-k}^{\ast}\mleft(S_{-k}e_{-p}\mright)\nonumber 
\end{align}
while for $p\in B_{F}$ (after substituting $p\rightarrow p-k$ in
the inner sums)
\begin{align}
\sum_{p\in B_{F}}\vert\left|p\right|^{2}-k_{F}^{2}\vert\,\tilde{c}_{p}^{\ast}d_{p}^{1} & =-\sum_{k\in\mathbb{Z}_{\ast}^{3}}\sum_{p\in L_{k}}\vert\left|p-k\right|^{2}-k_{F}^{2}\vert\,\tilde{c}_{p-k}^{\ast}\tilde{c}_{p}^{\ast}b_{k}\mleft(\mleft(C_{k}-1\mright)e_{p}\mright)\\
\sum_{p\in B_{F}}\vert\left|p\right|^{2}-k_{F}^{2}\vert\,\tilde{c}_{p}^{\ast}d_{p}^{2} & =-\sum_{k\in\mathbb{Z}_{\ast}^{3}}\sum_{p\in L_{k}}\vert\left|p-k\right|^{2}-k_{F}^{2}\vert\,\tilde{c}_{p-k}^{\ast}\tilde{c}_{p}^{\ast}b_{-k}^{\ast}\mleft(S_{-k}e_{-p}\mright).\nonumber 
\end{align}
In the $d_{p}^{1}$ case this implies that together
\begin{align}
\sum_{p\in\mathbb{Z}^{3}}\vert\left|p\right|^{2}-k_{F}^{2}\vert\,\tilde{c}_{p}^{\ast}d_{p}^{1} & =\sum_{k\in\mathbb{Z}_{\ast}^{3}}\sum_{p\in L_{k}}\mleft(\vert\left|p\right|^{2}-k_{F}^{2}\vert+\vert\left|p-k\right|^{2}-k_{F}^{2}\vert\mright)\tilde{c}_{p}^{\ast}\tilde{c}_{p-k}^{\ast}b_{k}\mleft(\mleft(C_{k}-1\mright)e_{p}\mright)\nonumber \\
 & =\sum_{k\in\mathbb{Z}_{\ast}^{3}}\sum_{p\in L_{k}}2\lambda_{k,p}b_{k,p}^{\ast}b_{k}\mleft(\mleft(C_{k}-1\mright)e_{p}\mright)=\sum_{k\in\mathbb{Z}_{\ast}^{3}}\sum_{p\in L_{k}}2\,b_{k,p}^{\ast}b_{k}\mleft(\mleft(C_{k}-1\mright)h_{k}e_{p}\mright)\\
 & =\sum_{k\in\mathbb{Z}_{\ast}^{3}}\sum_{p\in L_{k}}2\,b_{k}^{\ast}\mleft(h_{k}\mleft(C_{k}^{\ast}-1\mright)e_{p}\mright)b_{k,p}\nonumber 
\end{align}
whence
\begin{align}
2\,\mathrm{Re}\sum_{p\in\mathbb{Z}^{3}}\vert\left|p\right|^{2}-k_{F}^{2}\vert\,\tilde{c}_{p}^{\ast}d_{p}^{1} & =\sum_{k\in\mathbb{Z}_{\ast}^{3}}\sum_{p\in L_{k}}2\mleft(b_{k}^{\ast}\mleft(h_{k}\mleft(C_{k}^{\ast}-1\mright)e_{p}\mright)b_{k,p}+b_{k,p}^{\ast}b_{k}\mleft(h_{k}\mleft(C_{k}^{\ast}-1\mright)e_{p}\mright)\mright)\nonumber \\
 & =\sum_{k\in\mathbb{Z}_{\ast}^{3}}\sum_{p\in L_{k}}2\,b_{k}^{\ast}\mleft(\mleft(\mleft(C_{k}-1\mright)h_{k}+h_{k}\mleft(C_{k}^{\ast}-1\mright)\mright)e_{p}\mright)b_{k,p}\\
 & =\sum_{k\in\mathbb{Z}_{\ast}^{3}}\sum_{p\in L_{k}}2\,Q_{1}^{k}\mleft(\mleft(C_{k}-1\mright)h_{k}+h_{k}\mleft(C_{k}^{\ast}-1\mright)\mright)\nonumber 
\end{align}
and similarly, using the $\mleft(k,p,q\mright)\rightarrow\mleft(-k,-p,-q\mright)$
symmetry,
\begin{align}
\sum_{p\in\mathbb{Z}^{3}}\vert\left|p\right|^{2}-k_{F}^{2}\vert\,\tilde{c}_{p}^{\ast}d_{p}^{2} & =\sum_{k\in\mathbb{Z}_{\ast}^{3}}\sum_{p\in L_{k}}2\lambda_{k,p}b_{k,p}^{\ast}b_{-k}^{\ast}\mleft(S_{-k}e_{-p}\mright)=\sum_{k\in\mathbb{Z}_{\ast}^{3}}\sum_{p\in L_{k}}2\,b_{k,p}^{\ast}b_{-k}^{\ast}\mleft(S_{-k}h_{-k}e_{-p}\mright)\nonumber \\
 & =\sum_{k\in\mathbb{Z}_{\ast}^{3}}\sum_{p\in L_{k}}\mleft(b_{k}^{\ast}\mleft(h_{k}S_{k}^{\ast}e_{p}\mright)b_{-k,-p}^{\ast}+b_{-k,-p}^{\ast}b_{k}^{\ast}\mleft(S_{k}h_{k}e_{p}\mright)\mright)\\
 & =\sum_{k\in\mathbb{Z}_{\ast}^{3}}\sum_{p\in L_{k}}b_{k}^{\ast}\mleft(\mleft(S_{k}h_{k}+h_{k}S_{k}^{\ast}\mright)e_{p}\mright)b_{-k,-p}^{\ast}\nonumber 
\end{align}
yielding
\begin{equation}
2\,\mathrm{Re}\sum_{p\in\mathbb{Z}^{3}}\vert\left|p\right|^{2}-k_{F}^{2}\vert\,\tilde{c}_{p}^{\ast}d_{p}^{2}=\sum_{k\in\mathbb{Z}_{\ast}^{3}}Q_{2}^{k}\mleft(S_{k}h_{k}+h_{k}S_{k}^{\ast}\mright).
\end{equation}
$\hfill\square$

To state the identities for terms of the form $\sum_{p\in\mathbb{Z}^{3}}\vert\left|p\right|^{2}-k_{F}^{2}\vert\,(d_{p})^{\ast}d_{p}$
we must define some error terms. The first is
\begin{align}
\mathcal{E}_{\mathrm{B},1} & =\sum_{k,l\in\mathbb{Z}_{\ast}^{3}}\sum_{p\in L_{k}\cap L_{l}}\vert\left|p\right|^{2}-k_{F}^{2}\vert\,\tilde{c}_{p-l}^{\ast}\left[b_{k}^{\ast}\mleft(\mleft(C_{k}-1\mright)e_{p}\mright),b_{l}\mleft(\mleft(C_{l}-1\mright)e_{p}\mright)\right]\tilde{c}_{p-k}\\
 & +\sum_{k,l\in\mathbb{Z}_{\ast}^{3}}\sum_{p\in\mleft(L_{k}-k\mright)\cap\mleft(L_{l}-l\mright)}\vert\left|p\right|^{2}-k_{F}^{2}\vert\,\tilde{c}_{p+l}^{\ast}\left[b_{k}^{\ast}\mleft(\mleft(C_{k}-1\mright)e_{p+k}\mright),b_{l}\mleft(\mleft(C_{l}-1\mright)e_{p+l}\mright)\right]\tilde{c}_{p+k}\nonumber 
\end{align}
and we note that the two sums are of a similar form, in that the second
can be obtained from the first by the substitions $(L_{k},\tilde{c}_{p-k},e_{p})\rightarrow(L_{k}-k,\tilde{c}_{p+k},e_{p+k})$,
which reflects the fact that the definitions of $d_{p}^{1}$ and $d_{p}^{2}$ differ
in this way depending on whether $p\in B_{F}^{c}$ or $p\in B_{F}$.
We can thus write $\mathcal{E}_{\mathrm{B},1}=\mathcal{E}_{\mathrm{B},1}^{\mleft(1\mright)}+\mathcal{E}_{\mathrm{B},1}^{\mleft(2\mright)}$
for
\begin{equation}
\mathcal{E}_{\mathrm{B},1}^{\mleft(1\mright)}=\sum_{k,l\in\mathbb{Z}_{\ast}^{3}}\sum_{p\in L_{k}\cap L_{l}}\vert\left|p\right|^{2}-k_{F}^{2}\vert\,\tilde{c}_{p-l}^{\ast}\left[b_{k}^{\ast}\mleft(\mleft(C_{k}-1\mright)e_{p}\mright),b_{l}\mleft(\mleft(C_{l}-1\mright)e_{p}\mright)\right]\tilde{c}_{p-k}\label{eq:EB1Definition}
\end{equation}
with $\mathcal{E}_{\mathrm{B},1}^{\mleft(2\mright)}$ obtained from
this by the above substitution. In this notation we similarly define
$\mathcal{E}_{\mathrm{B},m}=\mathcal{E}_{\mathrm{B},m}^{\mleft(1\mright)}+\mathcal{E}_{\mathrm{B},m}^{\mleft(2\mright)}$
for $m=2,\ldots,5$ by
\begin{align}
\mathcal{E}_{\mathrm{B},2}^{\mleft(1\mright)} & =\sum_{k,l\in\mathbb{Z}_{\ast}^{3}}\sum_{p\in L_{k}\cap L_{l}}\vert\left|p\right|^{2}-k_{F}^{2}\vert\,\tilde{c}_{p-l}^{\ast}b_{k}^{\ast}\mleft(\mleft(C_{k}-1\mright)e_{p}\mright)\left[\tilde{c}_{p-k},b_{-l}^{\ast}\mleft(S_{-l}e_{-p}\mright)\right]\nonumber \\
\mathcal{E}_{\mathrm{B},3}^{\mleft(1\mright)} & =\sum_{k,l\in\mathbb{Z}_{\ast}^{3}}\sum_{p\in L_{k}\cap L_{l}}\vert\left|p\right|^{2}-k_{F}^{2}\vert\tilde{c}_{p-l}^{\ast}\left[b_{-k}\mleft(S_{-k}e_{-p}\mright),b_{-l}^{\ast}\mleft(S_{-l}e_{-p}\mright)\right]\tilde{c}_{p-k}\label{eq:EB2345Definition}\\
\mathcal{E}_{\mathrm{B},4}^{\mleft(1\mright)} & =\sum_{k,l\in\mathbb{Z}_{\ast}^{3}}\sum_{p\in L_{k}\cap L_{l}}\vert\left|p\right|^{2}-k_{F}^{2}\vert\tilde{c}_{p-l}^{\ast}b_{-k}\mleft(S_{-k}e_{-p}\mright)\left[\tilde{c}_{p-k},b_{-l}^{\ast}\mleft(S_{-l}e_{-p}\mright)\right]\nonumber \\
\mathcal{E}_{\mathrm{B},5}^{\mleft(1\mright)} & =\sum_{k,l\in\mathbb{Z}_{\ast}^{3}}\sum_{p\in L_{k}\cap L_{l}}\vert\left|p\right|^{2}-k_{F}^{2}\vert\left[\tilde{c}_{p-l},b_{-k}^{\ast}\mleft(S_{-k}e_{-p}\mright)\right]^{\ast}\left[\tilde{c}_{p-k},b_{-l}^{\ast}\mleft(S_{-l}e_{-p}\mright)\right]\nonumber 
\end{align}
The identities then take the following forms:
\begin{prop}
It holds that
\begin{align*}
\sum_{p\in\mathbb{Z}^{3}}\vert\left|p\right|^{2}-k_{F}^{2}\vert\,(d_{p}^{1})^{\ast}d_{p}^{1} & =\sum_{k\in\mathbb{Z}_{\ast}^{3}}2\,Q_{1}^{k}\mleft(\mleft(C_{k}-1\mright)h_{k}\mleft(C_{k}^{\ast}-1\mright)\mright)-\sum_{p\in\mathbb{Z}^{3}}\vert\left|p\right|^{2}-k_{F}^{2}\vert\,d_{p}^{1}(d_{p}^{1})^{\ast}-\mathcal{E}_{\mathrm{B},1}\\
2\,\mathrm{Re}\sum_{p\in\mathbb{Z}^{3}}\vert\left|p\right|^{2}-k_{F}^{2}\vert\,(d_{p}^{1})^{\ast}d_{p}^{2} & =\sum_{k\in\mathbb{Z}_{\ast}^{3}}Q_{2}^{k}\mleft(\mleft(C_{k}-1\mright)h_{k}S_{k}^{\ast}+S_{k}h_{k}\mleft(C_{k}^{\ast}-1\mright)\mright)\\
 & -2\,\mathrm{Re}\sum_{p\in\mathbb{Z}^{3}}\vert\left|p\right|^{2}-k_{F}^{2}\vert\,d_{p}^{2}(d_{p}^{1})^{\ast}-2\,\mathrm{Re}\mleft(\mathcal{E}_{\mathrm{B},2}\mright)\\
\sum_{p\in\mathbb{Z}^{3}}\vert\left|p\right|^{2}-k_{F}^{2}\vert\,(d_{p}^{2})^{\ast}d_{p}^{2} & =\sum_{k\in\mathbb{Z}_{\ast}^{3}}2\,Q_{1}^{k}\mleft(S_{k}h_{k}S_{k}^{\ast}\mright)+2\sum_{k\in\mathbb{Z}_{\ast}^{3}}\mathrm{tr}\mleft(S_{k}h_{k}S_{k}^{\ast}\mright)+2\sum_{k\in\mathbb{Z}_{\ast}^{3}}\varepsilon_{k,k}\mleft(e_{p};S_{k}h_{k}S_{k}^{\ast}e_{p}\mright)\\
 & -\sum_{p\in\mathbb{Z}^{3}}\vert\left|p\right|^{2}-k_{F}^{2}\vert\,d_{p}^{2}(d_{p}^{2})^{\ast}-\mathcal{E}_{\mathrm{B},3}-2\,\mathrm{Re}\mleft(\mathcal{E}_{\mathrm{B},4}\mright)-\mathcal{E}_{\mathrm{B},5}.
\end{align*}
\end{prop}

\textbf{Proof:} The first part of the derivation is similar for all
three terms, so we focus on $\sum_{p\in\mathbb{Z}^{3}}\vert\left|p\right|^{2}-k_{F}^{2}\vert\,(d_{p}^{1})^{\ast}d_{p}^{1}$.

By equation (\ref{eq:dp1Definition}) we have for $p\in B_{F}^{c}$
that
\begin{align}
\sum_{p\in B_{F}^{c}}\vert\left|p\right|^{2}-k_{F}^{2}\vert\,(d_{p}^{1})^{\ast}d_{p}^{1} & =\sum_{p\in B_{F}^{c}}\sum_{k,l\in\mathbb{Z}_{\ast}^{3}}1_{L_{k}\cap L_{l}}\mleft(p\mright)\vert\left|p\right|^{2}-k_{F}^{2}\vert\,b_{k}^{\ast}\mleft(\mleft(C_{k}-1\mright)e_{p}\mright)\tilde{c}_{p-k}\tilde{c}_{p-l}^{\ast}b_{l}\mleft(\mleft(C_{l}-1\mright)e_{p}\mright)\nonumber \\
 & =\sum_{k\in\mathbb{Z}_{\ast}^{3}}\sum_{p\in L_{k}}\vert\left|p\right|^{2}-k_{F}^{2}\vert\,b_{k}^{\ast}\mleft(\mleft(C_{k}-1\mright)e_{p}\mright)b_{k}\mleft(\mleft(C_{k}-1\mright)e_{p}\mright)\\
 & -\sum_{k,l\in\mathbb{Z}_{\ast}^{3}}\sum_{p\in L_{k}\cap L_{l}}\vert\left|p\right|^{2}-k_{F}^{2}\vert\,b_{k}^{\ast}\mleft(\mleft(C_{k}-1\mright)e_{p}\mright)\tilde{c}_{p-l}^{\ast}\tilde{c}_{p-k}b_{l}\mleft(\mleft(C_{l}-1\mright)e_{p}\mright)\nonumber 
\end{align}
and for $p\in B_{F}$ that
\begin{align}
\sum_{p\in B_{F}}\vert\left|p\right|^{2}-k_{F}^{2}\vert\,(d_{p}^{1})^{\ast}d_{p}^{1} & =\sum_{p\in B_{F}}\sum_{k,l\in\mathbb{Z}_{\ast}^{3}}1_{\mleft(L_{k}-k\mright)\cap\mleft(L_{l}-l\mright)}\mleft(p\mright)\vert\left|p\right|^{2}-k_{F}^{2}\vert\,b_{k}^{\ast}\mleft(\mleft(C_{k}-1\mright)e_{p+k}\mright)\tilde{c}_{p+k}\tilde{c}_{p+l}^{\ast}b_{l}\mleft(\mleft(C_{l}-1\mright)e_{p+l}\mright)\nonumber \\
 & =\sum_{k\in\mathbb{Z}_{\ast}^{3}}\sum_{p\in L_{k}}\vert\left|p-k\right|^{2}-k_{F}^{2}\vert\,b_{k}^{\ast}\mleft(\mleft(C_{k}-1\mright)e_{p}\mright)b_{k}\mleft(\mleft(C_{k}-1\mright)e_{p}\mright)\\
 & -\sum_{k,l\in\mathbb{Z}_{\ast}^{3}}\sum_{p\in\mleft(L_{k}-k\mright)\cap\mleft(L_{l}-l\mright)}\vert\left|p\right|^{2}-k_{F}^{2}\vert\,b_{k}^{\ast}\mleft(\mleft(C_{k}-1\mright)e_{p+k}\mright)\tilde{c}_{p+l}^{\ast}\tilde{c}_{p+k}b_{l}\mleft(\mleft(C_{l}-1\mright)e_{p+l}\mright).\nonumber 
\end{align}
When summing over all $p\in\mathbb{Z}^{3}$ the first terms combine
to form
\begin{align}
 & \quad\,\sum_{k\in\mathbb{Z}_{\ast}^{3}}\sum_{p\in L_{k}}\mleft(\vert\left|p\right|^{2}-k_{F}^{2}\vert+\vert\left|p-k\right|^{2}-k_{F}^{2}\vert\mright)\,b_{k}^{\ast}\mleft(\mleft(C_{k}-1\mright)e_{p}\mright)b_{k}\mleft(\mleft(C_{k}-1\mright)e_{p}\mright)\\
 & =\sum_{k\in\mathbb{Z}_{\ast}^{3}}\sum_{p\in L_{k}}2\lambda_{k,p}b_{k}^{\ast}\mleft(\mleft(C_{k}-1\mright)e_{p}\mright)b_{k}\mleft(\mleft(C_{k}-1\mright)e_{p}\mright)=\sum_{k\in\mathbb{Z}_{\ast}^{3}}\sum_{p\in L_{k}}2\,Q_{1}^{k}\mleft(\mleft(C_{k}-1\mright)h_{k}\mleft(C_{k}^{\ast}-1\mright)\mright)\nonumber 
\end{align}
while for the second we have e.g. (using that $\left[b^{\ast}\mleft(\cdot\mright),\tilde{c}^{\ast}\right]=0$)
\begin{align}
 & \quad\,\sum_{k,l\in\mathbb{Z}_{\ast}^{3}}\sum_{p\in L_{k}\cap L_{l}}\vert\left|p\right|^{2}-k_{F}^{2}\vert\,b_{k}^{\ast}\mleft(\mleft(C_{k}-1\mright)e_{p}\mright)\tilde{c}_{p-l}^{\ast}\tilde{c}_{p-k}b_{l}\mleft(\mleft(C_{l}-1\mright)e_{p}\mright)\nonumber \\
 & =\sum_{k,l\in\mathbb{Z}_{\ast}^{3}}\sum_{p\in L_{k}\cap L_{l}}\vert\left|p\right|^{2}-k_{F}^{2}\vert\,\tilde{c}_{p-l}^{\ast}b_{k}^{\ast}\mleft(\mleft(C_{k}-1\mright)e_{p}\mright)b_{l}\mleft(\mleft(C_{l}-1\mright)e_{p}\mright)\tilde{c}_{p-k}\nonumber \\
 & =\sum_{k,l\in\mathbb{Z}_{\ast}^{3}}\sum_{p\in L_{k}\cap L_{l}}\vert\left|p\right|^{2}-k_{F}^{2}\vert\,\tilde{c}_{p-l}^{\ast}b_{l}\mleft(\mleft(C_{l}-1\mright)e_{p}\mright)b_{k}^{\ast}\mleft(\mleft(C_{k}-1\mright)e_{p}\mright)\tilde{c}_{p-k}\\
 & +\sum_{k,l\in\mathbb{Z}_{\ast}^{3}}\sum_{p\in L_{k}\cap L_{l}}\vert\left|p\right|^{2}-k_{F}^{2}\vert\,\tilde{c}_{p-l}^{\ast}\left[b_{k}^{\ast}\mleft(\mleft(C_{k}-1\mright)e_{p}\mright),b_{l}\mleft(\mleft(C_{l}-1\mright)e_{p}\mright)\right]\tilde{c}_{p-k}\nonumber \\
 & =\sum_{p\in B_{F}^{c}}\vert\left|p\right|^{2}-k_{F}^{2}\vert\,d_{p}^{1}(d_{p}^{1})^{\ast}+\mathcal{E}_{\mathrm{B},1}^{\mleft(1\mright)}.\nonumber 
\end{align}
For $\sum_{p\in\mathbb{Z}^{3}}\vert\left|p\right|^{2}-k_{F}^{2}\vert\,(d_{p}^{1})^{\ast}d_{p}^{2}$
one likewise finds terms combining to form
\begin{equation}
\sum_{k\in\mathbb{Z}_{\ast}^{3}}\sum_{p\in L_{k}}b_{-k,-p}^{\ast}b_{k}^{\ast}\mleft(\mleft(\mleft(C_{k}-1\mright)h_{k}S_{k}^{\ast}+S_{k}h_{k}\mleft(C_{k}^{\ast}-1\mright)\mright)e_{p}\mright),
\end{equation}
yielding the corresponding $Q_{2}^{k}$ terms when taking $2\,\mathrm{Re}$,
and additional terms of the form
\begin{align}
 & \quad\,\sum_{k,l\in\mathbb{Z}_{\ast}^{3}}\sum_{p\in L_{k}\cap L_{l}}\vert\left|p\right|^{2}-k_{F}^{2}\vert\,b_{k}^{\ast}\mleft(\mleft(C_{k}-1\mright)e_{p}\mright)\tilde{c}_{p-l}^{\ast}\tilde{c}_{p-k}b_{-l}^{\ast}\mleft(S_{-l}e_{-p}\mright)\nonumber \\
 & =\sum_{k,l\in\mathbb{Z}_{\ast}^{3}}\sum_{p\in L_{k}\cap L_{l}}\vert\left|p\right|^{2}-k_{F}^{2}\vert\,\tilde{c}_{p-l}^{\ast}b_{k}^{\ast}\mleft(\mleft(C_{k}-1\mright)e_{p}\mright)b_{-l}^{\ast}\mleft(S_{-l}e_{-p}\mright)\tilde{c}_{p-k}\\
 & +\sum_{k,l\in\mathbb{Z}_{\ast}^{3}}\sum_{p\in L_{k}\cap L_{l}}\vert\left|p\right|^{2}-k_{F}^{2}\vert\,\tilde{c}_{p-l}^{\ast}b_{k}^{\ast}\mleft(\mleft(C_{k}-1\mright)e_{p}\mright)\left[\tilde{c}_{p-k},b_{-l}^{\ast}\mleft(S_{-l}e_{-p}\mright)\right]\nonumber \\
 & =\sum_{p\in B_{F}^{c}}\vert\left|p\right|^{2}-k_{F}^{2}\vert\,d_{p}^{2}(d_{p}^{1})^{\ast}+\mathcal{E}_{\mathrm{B},2}^{\mleft(1\mright)}\nonumber 
\end{align}
where we also used that $\left[b_{k}^{\ast}\mleft(\cdot\mright),b_{l}^{\ast}\mleft(\cdot\mright)\right]=0$.

Lastly one has for $\sum_{p\in\mathbb{Z}^{3}}\vert\left|p\right|^{2}-k_{F}^{2}\vert\,(d_{p}^{2})^{\ast}d_{p}^{2}$
terms combining to yield
\begin{align}
 & \quad\,\sum_{k\in\mathbb{Z}_{\ast}^{3}}\sum_{p\in L_{k}}2\lambda_{k,p}b_{-k}\mleft(S_{-k}e_{-p}\mright)b_{-k}^{\ast}\mleft(S_{-k}e_{-p}\mright)\\
 & =\sum_{k\in\mathbb{Z}_{\ast}^{3}}2\,Q_{1}^{k}\mleft(S_{k}h_{k}S_{k}^{\ast}\mright)+2\sum_{k\in\mathbb{Z}_{\ast}^{3}}\mathrm{tr}\mleft(S_{k}h_{k}S_{k}^{\ast}\mright)+2\sum_{k\in\mathbb{Z}_{\ast}^{3}}\varepsilon_{k,k}\mleft(e_{p};S_{k}h_{k}S_{k}^{\ast}e_{p}\mright),\nonumber 
\end{align}
the right-hand side following as in equation (\ref{eq:ReOrderingQ1Term}),
and terms of the form
\begin{align}
 & \quad\,\sum_{k,l\in\mathbb{Z}_{\ast}^{3}}\sum_{p\in L_{k}\cap L_{l}}\vert\left|p\right|^{2}-k_{F}^{2}\vert\,b_{-k}\mleft(S_{-k}e_{-p}\mright)\tilde{c}_{p-l}^{\ast}\tilde{c}_{p-k}b_{-l}^{\ast}\mleft(S_{-l}e_{-p}\mright)\nonumber \\
 & =\sum_{k,l\in\mathbb{Z}_{\ast}^{3}}\sum_{p\in L_{k}\cap L_{l}}\vert\left|p\right|^{2}-k_{F}^{2}\vert\tilde{c}_{p-l}^{\ast}b_{-k}\mleft(S_{-k}e_{-p}\mright)b_{-l}^{\ast}\mleft(S_{-l}e_{-p}\mright)\tilde{c}_{p-k}\nonumber \\
 & +2\,\mathrm{Re}\sum_{k,l\in\mathbb{Z}_{\ast}^{3}}\sum_{p\in L_{k}\cap L_{l}}\vert\left|p\right|^{2}-k_{F}^{2}\vert\tilde{c}_{p-l}^{\ast}b_{-k}\mleft(S_{-k}e_{-p}\mright)\left[\tilde{c}_{p-k},b_{-l}^{\ast}\mleft(S_{-l}e_{-p}\mright)\right]\\
 & +\sum_{k,l\in\mathbb{Z}_{\ast}^{3}}\sum_{p\in L_{k}\cap L_{l}}\vert\left|p\right|^{2}-k_{F}^{2}\vert\left[\tilde{c}_{p-l},b_{-k}^{\ast}\mleft(S_{-k}e_{-p}\mright)\right]^{\ast}\left[\tilde{c}_{p-k},b_{-l}^{\ast}\mleft(S_{-l}e_{-p}\mright)\right]\nonumber \\
 & =\sum_{p\in B_{F}^{c}}\vert\left|p\right|^{2}-k_{F}^{2}\vert\,d_{p}^{2}(d_{p}^{2})^{\ast}+\mathcal{E}_{\mathrm{B},3}^{\mleft(1\mright)}+2\,\mathrm{Re}\mleft(\mathcal{E}_{\mathrm{B},4}^{\mleft(1\mright)}\mright)+\mathcal{E}_{\mathrm{B},5}^{\mleft(1\mright)}.\nonumber 
\end{align}
$\hfill\square$

We can now conclude the following identity for $H_{\mathrm{kin}}^{\prime}$:
\begin{prop}
\label{prop:KineticIdentity}It holds that
\begin{align*}
H_{\mathrm{kin}}^{\prime} & =\sum_{k\in\mathbb{Z}_{\ast}^{3}}2\,Q_{1}^{k}\mleft(h_{k}\mright)+\sum_{p\in\mathbb{Z}^{3}}\vert\left|p\right|^{2}-k_{F}^{2}\vert\mleft(\left|\tilde{c}_{p}+d_{p}^{1}+d_{p}^{2}\right|^{2}+\left|(d_{p}^{1}+d_{p}^{2})^{\ast}\right|^{2}\mright)\\
 & -\sum_{k\in\mathbb{Z}_{\ast}^{3}}\sum_{p,q\in L_{k}}2\left\langle e_{p},h_{k}e_{q}\right\rangle \mleft(b_{k}\mleft(C_{k}e_{p}\mright)+b_{-k}^{\ast}\mleft(S_{-k}e_{-p}\mright)\mright)^{\ast}\mleft(b_{k}\mleft(C_{k}e_{q}\mright)+b_{-k}^{\ast}\mleft(S_{-k}e_{-q}\mright)\mright)\\
 & +\mathcal{E}_{\mathrm{B},1}+2\,\mathrm{Re}\mleft(\mathcal{E}_{\mathrm{B},2}\mright)+\mathcal{E}_{\mathrm{B},3}+2\,\mathrm{Re}\mleft(\mathcal{E}_{\mathrm{B},4}\mright)+\mathcal{E}_{\mathrm{B},5}
\end{align*}
with $\mathcal{E}_{\mathrm{B},m}$ defined in \eqref{eq:EB2345Definition}. 
\end{prop}

\textbf{Proof:} By rearranging the terms of equation (\ref{eq:KineticExpansion})
and inserting the identities we have derived we find
\begin{align}
H_{\mathrm{kin}}^{\prime} & =\sum_{k\in\mathbb{Z}_{\ast}^{3}}2\,Q_{1}^{k}\mleft(h_{k}\mright)+\sum_{p\in\mathbb{Z}^{3}}\vert\left|p\right|^{2}-k_{F}^{2}\vert\mleft(\left|\tilde{c}_{p}+d_{p}^{1}+d_{p}^{2}\right|^{2}+\left|(d_{p}^{1}+d_{p}^{2})^{\ast}\right|^{2}\mright)\nonumber \\
 & +\mathcal{E}_{\mathrm{B},1}+2\,\mathrm{Re}\mleft(\mathcal{E}_{\mathrm{B},2}\mright)+\mathcal{E}_{\mathrm{B},3}+2\,\mathrm{Re}\mleft(\mathcal{E}_{\mathrm{B},4}\mright)+\mathcal{E}_{\mathrm{B},5}\\
 & -\sum_{k\in\mathbb{Z}_{\ast}^{3}}\mleft(2\,Q_{1}^{k}\mleft(C_{k}h_{k}C_{k}^{\ast}+S_{k}h_{k}S_{k}^{\ast}\mright)+Q_{2}^{k}\mleft(C_{k}h_{k}S_{k}^{\ast}+S_{k}h_{k}C_{k}^{\ast}\mright)\mright)\nonumber \\
 & -2\sum_{k\in\mathbb{Z}_{\ast}^{3}}\mathrm{tr}\mleft(S_{k}h_{k}S_{k}^{\ast}\mright)-2\sum_{k\in\mathbb{Z}_{\ast}^{3}}\varepsilon_{k,k}\mleft(e_{p};S_{k}h_{k}S_{k}^{\ast}e_{p}\mright)\nonumber 
\end{align}
and by Proposition \ref{prop:QuadraticExpansionIdentity} the terms
on the two final lines combine to form
\begin{equation}
-\sum_{k\in\mathbb{Z}_{\ast}^{3}}\sum_{p,q\in L_{k}}2\left\langle e_{p},h_{k}e_{q}\right\rangle \mleft(b_{k}\mleft(C_{k}e_{p}\mright)+b_{-k}^{\ast}\mleft(S_{-k}e_{-p}\mright)\mright)^{\ast}\mleft(b_{k}\mleft(C_{k}e_{q}\mright)+b_{-k}^{\ast}\mleft(S_{-k}e_{-q}\mright)\mright).
\end{equation}
$\hfill\square$

\subsection{Extraction of $E_{\mathrm{corr},\mathrm{ex}}$}

To conclude Theorem \ref{them:QuasibosonicBogolubovFactorization}
it essentially only remains to identify $E_{\mathrm{corr},\mathrm{ex}}$. In the following, we will again decompose $\mathcal{E}_{{\rm B},i} = \mathcal{E}_{{\rm B},i}^{(1)} + \mathcal{E}_{{\rm B},i}^{(2)}$ with  $\mathcal{E}_{{\rm B},i}^{(2)}$ obtained from $\mathcal{E}_{{\rm B},i}^{(1)} $ by the substitution $(L_k, \tilde{c}_{p-k}, e_p) \to (L_k-k, \tilde{c}_{p+k}, e_{p+k})$.  Then by anticommuting
the commutators we can write $\mathcal{E}_{\mathrm{B},5}=-\mathcal{E}_{\mathrm{B},5}^{\prime}+\mathcal{E}_{\mathrm{B},6}$
where e.g.
\begin{align}
\mathcal{E}_{\mathrm{B},5}^{\prime\mleft(1\mright)} & =\sum_{k,l\in\mathbb{Z}_{\ast}^{3}}\sum_{p\in L_{k}\cap L_{l}}\vert\left|p\right|^{2}-k_{F}^{2}\vert\left[\tilde{c}_{p-k},b_{-l}^{\ast}\mleft(S_{-l}e_{-p}\mright)\right]\left[\tilde{c}_{p-l},b_{-k}^{\ast}\mleft(S_{-k}e_{-p}\mright)\right]^{\ast}\label{eq:EBPrime56Definition}\\
\mathcal{E}_{\mathrm{B},6}^{\mleft(1\mright)} & =\sum_{k,l\in\mathbb{Z}_{\ast}^{3}}\sum_{p\in L_{k}\cap L_{l}}\vert\left|p\right|^{2}-k_{F}^{2}\vert\left\{ \left[\tilde{c}_{p-l},b_{-k}^{\ast}\mleft(S_{-k}e_{-p}\mright)\right]^{\ast},\left[\tilde{c}_{p-k},b_{-l}^{\ast}\mleft(S_{-l}e_{-p}\mright)\right]\right\} ,\nonumber 
\end{align}
and noting that
\begin{align}
\left[\tilde{c}_{p-k},b_{-l}^{\ast}\mleft(S_{-l}e_{-p}\mright)\right] & =\sum_{q\in L_{l}}\left\langle e_{-q},S_{-l}e_{-p}\right\rangle \left[\tilde{c}_{p-k},\tilde{c}_{-q}^{\ast}\tilde{c}_{-q+l}^{\ast}\right]\\
 & =-\sum_{q\in L_{l}}\delta_{p-k,-q+l}\left\langle e_{q},S_{l}e_{p}\right\rangle \tilde{c}_{-q}^{\ast}\nonumber 
\end{align}
for $p\in L_{k}$, we have
\begin{align}
\mathcal{E}_{\mathrm{B},6}^{\mleft(1\mright)} & =\sum_{k,l\in\mathbb{Z}_{\ast}^{3}}\sum_{p\in L_{k}\cap L_{l}}\vert\left|p\right|^{2}-k_{F}^{2}\vert\left\{ \sum_{q\in L_{k}}\delta_{p-l,-q+k}\left\langle S_{k}e_{p},e_{q}\right\rangle \tilde{c}_{-q},\sum_{q'\in L_{l}}\delta_{p-k,-q'+l}\left\langle e_{q'},S_{l}e_{p}\right\rangle \tilde{c}_{-q'}^{\ast}\right\} \\
 & =\sum_{k,l\in\mathbb{Z}_{\ast}^{3}}\sum_{p,q\in L_{k}\cap L_{l}}\delta_{p-k,-q+l}\vert\left|p\right|^{2}-k_{F}^{2}\vert\left\langle S_{k}e_{p},e_{q}\right\rangle \left\langle e_{q},S_{l}e_{p}\right\rangle \nonumber 
\end{align}
which is simply a constant. A similar calculation shows that
\begin{equation}
\mathcal{E}_{\mathrm{B},6}^{\mleft(2\mright)}=\sum_{k,l\in\mathbb{Z}_{\ast}^{3}}\sum_{p,q\in\mleft(L_{k}-k\mright)\cap\mleft(L_{l}-l\mright)}\delta_{p+k,-q-l}\vert\left|p\right|^{2}-k_{F}^{2}\vert\left\langle S_{k}e_{p+k},e_{q+k}\right\rangle \left\langle e_{q+l},S_{l}e_{p+l}\right\rangle .
\end{equation}
The point is that $\mathcal{E}_{\mathrm{B},6}$ is, to leading order
in $k_{F}$, $E_{\mathrm{corr},\mathrm{ex}}$. To see this we first
rewrite the expressions:
\begin{prop}
It holds that
\[
\mathcal{E}_{\mathrm{B},6}=\sum_{k\in\mathbb{Z}_{\ast}^{3}}\sum_{p,q\in L_{k}}2\lambda_{k,p}\left\langle S_{k}e_{p},e_{q}\right\rangle \left\langle e_{q},S_{p+q-k}e_{p}\right\rangle .
\]
\end{prop}

\textbf{Proof:} We begin by noting that the Kronecker delta $\delta_{p-k,-q+l}$
implies that
\begin{align}
\mathcal{E}_{\mathrm{B},6}^{\mleft(1\mright)} & =\sum_{k,l\in\mathbb{Z}_{\ast}^{3}}\sum_{p,q\in L_{k}}\delta_{p-k,-q+l}\vert\left|p\right|^{2}-k_{F}^{2}\vert\left\langle S_{k}e_{p},e_{q}\right\rangle \left\langle e_{q},S_{l}e_{p}\right\rangle \\
 & =\sum_{k\in\mathbb{Z}_{\ast}^{3}}\sum_{p,q\in L_{k}}\vert\left|p\right|^{2}-k_{F}^{2}\vert\left\langle S_{k}e_{p},e_{q}\right\rangle \left\langle e_{q},S_{p+q-k}e_{p}\right\rangle \nonumber 
\end{align}
since, as observed in \cite[eq. 4.69]{ChrHaiNam-23b}, $p,q\in L_{p+q-k}\Leftrightarrow p,q\in L_{k}$.
Likewise $p,q\in\mleft(L_{-p-q-k}+p+q+k\mright)\Leftrightarrow p,q\in L_{k}-k$,
so (using also the $\mleft(k,p,q\mright)\rightarrow\mleft(-k,-p,-q\mright)$
symmetry of the matrix elements)
\begin{align}
\mathcal{E}_{\mathrm{B},6}^{\mleft(2\mright)} & =\sum_{k,l\in\mathbb{Z}_{\ast}^{3}}\sum_{p,q\in\mleft(L_{k}-k\mright)}\delta_{p+k,-q-l}\vert\left|p\right|^{2}-k_{F}^{2}\vert\left\langle S_{k}e_{p+k},e_{q+k}\right\rangle \left\langle e_{q+l},S_{l}e_{p+l}\right\rangle \nonumber \\
 & =\sum_{k\in\mathbb{Z}_{\ast}^{3}}\sum_{p,q\in\mleft(L_{k}-k\mright)}\vert\left|p\right|^{2}-k_{F}^{2}\vert\left\langle S_{k}e_{p+k},e_{q+k}\right\rangle \left\langle e_{q-\mleft(p+q+k\mright)},S_{-\mleft(p+q+k\mright)}e_{p-\mleft(p+q+k\mright)}\right\rangle \\
 & =\sum_{k\in\mathbb{Z}_{\ast}^{3}}\sum_{p,q\in\mleft(L_{k}-k\mright)}\vert\left|p\right|^{2}-k_{F}^{2}\vert\left\langle S_{k}e_{p+k},e_{q+k}\right\rangle \left\langle e_{p+k},S_{p+q+k}e_{q+k}\right\rangle \nonumber \\
 & =\sum_{k\in\mathbb{Z}_{\ast}^{3}}\sum_{p,q\in L_{k}}\vert\left|p-k\right|^{2}-k_{F}^{2}\vert\left\langle S_{k}e_{p},e_{q}\right\rangle \left\langle e_{p},S_{p+q-k}e_{q}\right\rangle \nonumber 
\end{align}
whence
\begin{align}
\mathcal{E}_{\mathrm{B},6} & =\sum_{k\in\mathbb{Z}_{\ast}^{3}}\sum_{p,q\in L_{k}}\mleft(\vert\left|p\right|^{2}-k_{F}^{2}\vert+\vert\left|p-k\right|^{2}-k_{F}^{2}\vert\mright)\left\langle S_{k}e_{p},e_{q}\right\rangle \left\langle e_{q},S_{p+q-k}e_{p}\right\rangle \\
 & =\sum_{k\in\mathbb{Z}_{\ast}^{3}}\sum_{p,q\in L_{k}}2\lambda_{k,p}\left\langle S_{k}e_{p},e_{q}\right\rangle \left\langle e_{q},S_{p+q-k}e_{p}\right\rangle .\nonumber 
\end{align}
$\hfill\square$

We show in appendix Section \ref{subsec:One-BodyOperatorEstimates}
that $\left\langle e_{p},S_{k}e_{q}\right\rangle \approx\frac{\hat{V}_{k}k_{F}^{-\beta}}{2\mleft(2\pi\mright)^{3}}\frac{1}{\lambda_{k,p}+\lambda_{k,q}}$,
suggesting that
\begin{align}
\mathcal{E}_{\mathrm{B},6} & \approx\frac{k_{F}^{-2\beta}}{4\mleft(2\pi\mright)^{6}}\sum_{k\in\mathbb{Z}_{\ast}^{3}}\sum_{p,q\in L_{k}}2\lambda_{k,p}\frac{\hat{V}_{k}}{\lambda_{k,p}+\lambda_{k,q}}\frac{\hat{V}_{p+q-k}}{\lambda_{p+q-k,p}+\lambda_{p+q-k,q}}\label{eq:EB6Approximation}\\
 & =\frac{k_{F}^{-2\beta}}{4\mleft(2\pi\mright)^{6}}\sum_{k\in\mathbb{Z}_{\ast}^{3}}\sum_{p,q\in L_{k}}\frac{\hat{V}_{k}\hat{V}_{p+q-k}}{\lambda_{k,p}+\lambda_{k,q}}=E_{\mathrm{corr},\mathrm{ex}}\nonumber 
\end{align}
where we used that $\lambda_{p+q-k,p}+\lambda_{p+q-k,q}=\lambda_{k,p}+\lambda_{k,q}$
and the fact that the summand on the right-hand side is symmetric
in $p$ and $q$. We leave the estimates to the next section, but
this justifies defining $\mathcal{E}_{\mathrm{B},6}^{\prime}=\mathcal{E}_{\mathrm{B},6}-E_{\mathrm{corr},\mathrm{ex}}$
to write
\begin{equation}
\mathcal{E}_{\mathrm{B},5}=E_{\mathrm{corr},\mathrm{ex}}-\mathcal{E}_{\mathrm{B},5}^{\prime}+\mathcal{E}_{\mathrm{B},6}^{\prime}
\end{equation}
and Theorem \ref{them:QuasibosonicBogolubovFactorization} now follows
from the Propositions \ref{prop:InteractionIdentity} and \ref{prop:KineticIdentity}
with
\begin{equation}
\mathcal{E}_{\mathrm{B}}=\mathcal{E}_{\mathrm{B},1}+2\,\mathrm{Re}\mleft(\mathcal{E}_{\mathrm{B},2}\mright)+\mathcal{E}_{\mathrm{B},3}+2\,\mathrm{Re}\mleft(\mathcal{E}_{\mathrm{B},4}\mright)-\mathcal{E}_{\mathrm{B},5}^{\prime}+\mathcal{E}_{\mathrm{B},6}^{\prime}.
\end{equation}

\section{\label{sec:EstimationofEB}Estimation of $\mathcal{E}_{\mathrm{B}}$}

In this section we bound the error term $\mathcal{E}_{\mathrm{B}}$
appearing in Theorem \ref{them:QuasibosonicBogolubovFactorization},
obtaining the following estimate:
\begin{thm}
\label{them:EBEstimate}For any symmetric\footnote{in the sense that $S = -S$.} set $S\subset\mathbb{Z}_{\ast}^{3}$  
and $\epsilon>0$ it holds as $k_{F}\rightarrow\infty$ that
\begin{align*}
\pm\mathcal{E}_{\mathrm{B}} & \leq  C_{\epsilon}k_{F}^{2\mleft(1-\beta\mright)+\epsilon}\mleft(\sqrt{\sum_{k\in\mathbb{Z}_{\ast}^{3}\backslash S}\hat{V}_{k}^{2}}+k_{F}^{-\frac{1}{2}}\sum_{k\in S}\hat{V}_{k}\mright)\sqrt{\sum_{k\in\mathbb{Z}_{\ast}^{3}}\hat{V}_{k}^{2}\min\left\{ \left|k\right|,k_{F}\right\} }\mleft(H_{\mathrm{kin}}^{\prime}
+k_{F}\mright)\\&+C k_F^{3(1-\beta)} \sqrt{\sum_{k\in\mathbb{Z}_{\ast}^{3}}\hat{V}_{k}^{2}}\sum_{k\in\mathbb{Z}_{\ast}^{3}}\hat{V}_{k}^{2}\left|k\right|^{\frac{1}{2}}
\end{align*}
for constants $C,C_{\epsilon}>0$ with $C$ independent of all quantities
and $C_{\epsilon}$ depending only on $\epsilon$.
\end{thm}

\subsubsection*{Reduction to Schematic Forms}

Let us describe the main features of the error terms $\mathcal{E}_{\mathrm{B}}$. Recall that $\mathcal{E}_{\mathrm{B}}$ was defined to be
\begin{equation}
\mathcal{E}_{\mathrm{B}}=\mathcal{E}_{\mathrm{B},1}+2\,\mathrm{Re}\mleft(\mathcal{E}_{\mathrm{B},2}\mright)+\mathcal{E}_{\mathrm{B},3}+2\,\mathrm{Re}\mleft(\mathcal{E}_{\mathrm{B},4}\mright)-\mathcal{E}_{\mathrm{B},5}^{\prime}+\mathcal{E}_{\mathrm{B},6}^{\prime},
\end{equation}
the sub-terms $\mathcal{E}_{\mathrm{B},1},\ldots,\mathcal{E}_{\mathrm{B},5}^{\prime}$
being defined in the equations (\ref{eq:EB1Definition}), (\ref{eq:EB2345Definition}),
(\ref{eq:EBPrime56Definition}) and $\mathcal{E}_{\mathrm{B},6}^{\prime}$
being
\begin{equation}
\mathcal{E}_{\mathrm{B},6}^{\prime}=\sum_{k\in\mathbb{Z}_{\ast}^{3}}\sum_{p,q\in L_{k}}2\lambda_{k,p}\left\langle S_{k}e_{p},e_{q}\right\rangle \left\langle e_{q},S_{p+q-k}e_{p}\right\rangle -E_{\mathrm{corr},\mathrm{ex}}.
\end{equation}
Consider $\mathcal{E}_{\mathrm{B},1}$, which is the sum of the two
terms
\begin{align}
\mathcal{E}_{\mathrm{B},1}^{\mleft(1\mright)} & =\sum_{k,l\in\mathbb{Z}_{\ast}^{3}}\sum_{p\in L_{k}\cap L_{l}}\vert\left|p\right|^{2}-k_{F}^{2}\vert\,\tilde{c}_{p-l}^{\ast}\left[b_{k}^{\ast}\mleft(\mleft(C_{k}-1\mright)e_{p}\mright),b_{l}\mleft(\mleft(C_{l}-1\mright)e_{p}\mright)\right]\tilde{c}_{p-k}\\
\mathcal{E}_{\mathrm{B},1}^{\mleft(2\mright)} & =\sum_{k,l\in\mathbb{Z}_{\ast}^{3}}\sum_{p\in\mleft(L_{k}-k\mright)\cap\mleft(L_{l}-l\mright)}\vert\left|p\right|^{2}-k_{F}^{2}\vert\,\tilde{c}_{p+l}^{\ast}\left[b_{k}^{\ast}\mleft(\mleft(C_{k}-1\mright)e_{p+k}\mright),b_{l}\mleft(\mleft(C_{l}-1\mright)e_{p+l}\mright)\right]\tilde{c}_{p+k}.\nonumber 
\end{align}
As already noted, these terms are clearly similar. Indeed, they are
both of the schematic form
\begin{equation}
\tilde{\mathcal{E}}_{\mathrm{B},1}=\sum_{k,l\in\mathbb{Z}_{\ast}^{3}}\sum_{p\in M_{k}\cap M_{l}}\tilde{c}_{p\mp l}^{\ast}\left[b_{k}^{\ast}\mleft(\varphi_{k,p}\mright),b_{l}\mleft(\varphi_{l,p}\mright)\right]\tilde{c}_{p\mp k}\label{eq:EB1SchematicForm}
\end{equation}
where the sets $M_{k}$, the signs $p\mp k$ and $\varphi_{k,p}\in\ell^{2}\mleft(L_{k}\mright)$
are given by
\begin{equation}
\mleft(M_{k},p\mp k,\varphi_{k,p}\mright)=\begin{cases}
\Bigl(L_{k},p-k,\sqrt{\vert\left|p\right|^{2}-k_{F}^{2}\vert}\mleft(C_{k}-1\mright)e_{p}\Bigr) & \text{for }\mathcal{E}_{\mathrm{B},1}^{\mleft(1\mright)}\\
\Bigl(L_{k}-k,p+k,\sqrt{\vert\left|p\right|^{2}-k_{F}^{2}\vert}\mleft(C_{k}-1\mright)e_{p+k}\Bigr) & \text{for }\mathcal{E}_{\mathrm{B},1}^{\mleft(2\mright)}
\end{cases}.\label{eq:MkPkPhikDefinition}
\end{equation}
It thus suffices to obtain estimates for the schematic form of equation
(\ref{eq:EB1SchematicForm}) rather than the specific terms $\mathcal{E}_{\mathrm{B},1}^{\mleft(1\mright)}$
and $\mathcal{E}_{\mathrm{B},1}^{\mleft(2\mright)}$. The same is true
of the other error terms: $\mathcal{E}_{\mathrm{B},2}$, for instance,
consists of the terms
\begin{align}
\mathcal{E}_{\mathrm{B},2}^{\mleft(1\mright)} & =\sum_{k,l\in\mathbb{Z}_{\ast}^{3}}\sum_{p\in L_{k}\cap L_{l}}\vert\left|p\right|^{2}-k_{F}^{2}\vert\,\tilde{c}_{p-l}^{\ast}b_{k}^{\ast}\mleft(\mleft(C_{k}-1\mright)e_{p}\mright)\left[\tilde{c}_{p-k},b_{-l}^{\ast}\mleft(S_{-l}e_{-p}\mright)\right]\\
\mathcal{E}_{\mathrm{B},2}^{\mleft(2\mright)} & =\sum_{k,l\in\mathbb{Z}_{\ast}^{3}}\sum_{p\in\mleft(L_{k}-k\mright)\cap\mleft(L_{l}-l\mright)}\vert\left|p\right|^{2}-k_{F}^{2}\vert\,\tilde{c}_{p+l}^{\ast}b_{k}^{\ast}\mleft(\mleft(C_{k}-1\mright)e_{p+k}\mright)\left[\tilde{c}_{p+k},b_{-l}^{\ast}\mleft(S_{-l}e_{-p-l}\mright)\right]\nonumber 
\end{align}
which we can likewise summarize in the schematic form
\begin{equation}
\tilde{\mathcal{E}}_{\mathrm{B},2}=\sum_{k,l\in\mathbb{Z}_{\ast}^{3}}\sum_{p\in M_{k}\cap M_{l}}\tilde{c}_{p\mp l}^{\ast}b_{k}^{\ast}\mleft(\varphi_{k,p}\mright)\left[\tilde{c}_{p\mp k},b_{-l}^{\ast}\mleft(\psi_{-l,-p}\mright)\right]
\end{equation}
provided we also define $\psi_{l,p}\in\ell^{2}\mleft(L_{l}\mright)$
by
\begin{equation}
\psi_{l,p}=\begin{cases}
\sqrt{\vert\left|p\right|^{2}-k_{F}^{2}\vert}S_{l}e_{p} & \text{for }\mathcal{E}_{\mathrm{B},2}^{\mleft(1\mright)}\\
\sqrt{\vert\left|p\right|^{2}-k_{F}^{2}\vert}S_{l}e_{p+l} & \text{for }\mathcal{E}_{\mathrm{B},2}^{\mleft(2\mright)}
\end{cases}.\label{eq:PsikDefinition}
\end{equation}
The quantities of the equations (\ref{eq:MkPkPhikDefinition}) and
(\ref{eq:PsikDefinition}) suffice to write all error terms schematically
as
\begin{align}
\tilde{\mathcal{E}}_{\mathrm{B},1} & =\sum_{k,l\in\mathbb{Z}_{\ast}^{3}}\sum_{p\in M_{k}\cap M_{l}}\tilde{c}_{p\mp l}^{\ast}\left[b_{k}^{\ast}\mleft(\varphi_{k,p}\mright),b_{l}\mleft(\varphi_{l,p}\mright)\right]\tilde{c}_{p\mp k}\nonumber \\
\tilde{\mathcal{E}}_{\mathrm{B},2} & =\sum_{k,l\in\mathbb{Z}_{\ast}^{3}}\sum_{p\in M_{k}\cap M_{l}}\tilde{c}_{p\mp l}^{\ast}b_{k}^{\ast}\mleft(\varphi_{k,p}\mright)\left[\tilde{c}_{p\mp k},b_{-l}^{\ast}\mleft(\psi_{-l,-p}\mright)\right]\nonumber \\
\tilde{\mathcal{E}}_{\mathrm{B},3} & =\sum_{k,l\in\mathbb{Z}_{\ast}^{3}}\sum_{p\in M_{k}\cap M_{l}}\tilde{c}_{p\mp l}^{\ast}\left[b_{-k}\mleft(\psi_{-k,-p}\mright),b_{-l}^{\ast}\mleft(\psi_{-l,-p}\mright)\right]\tilde{c}_{p\mp k}\label{eq:EBSchematicForms}\\
\tilde{\mathcal{E}}_{\mathrm{B},4} & =\sum_{k,l\in\mathbb{Z}_{\ast}^{3}}\sum_{p\in M_{k}\cap M_{l}}\tilde{c}_{p\mp l}^{\ast}b_{-k}\mleft(\psi_{-k,-p}\mright)\left[\tilde{c}_{p\mp k},b_{-l}^{\ast}\mleft(\psi_{-l,-p}\mright)\right]\nonumber \\
\tilde{\mathcal{E}}_{\mathrm{B},5}^{\prime} & =\sum_{k,l\in\mathbb{Z}_{\ast}^{3}}\sum_{p\in M_{k}\cap M_{l}}\left[\tilde{c}_{p\mp k},b_{-l}^{\ast}\mleft(\psi_{-l,-p}\mright)\right]\left[\tilde{c}_{p\mp l},b_{-k}^{\ast}\mleft(\psi_{-k,-p}\mright)\right]^{\ast}\nonumber 
\end{align}
and it is these general forms which we will estimate. We will then
insert the particular expressions for $\varphi_{k,p}$ and $\psi_{l,p}$
at the end to obtain Theorem \ref{them:EBEstimate}.

\subsection{Estimation of $\tilde{\mathcal{E}}_{\mathrm{B},1}$ and $\tilde{\mathcal{E}}_{\mathrm{B},2}$}

The schematic forms of $\tilde{\mathcal{E}}_{\mathrm{B},1}$ and $\tilde{\mathcal{E}}_{\mathrm{B},2}$
display the typical structure we will need to consider, so we first
consider these in detail.

We begin with $\tilde{\mathcal{E}}_{\mathrm{B},1}$, which since $[b_{k}^{\ast}\mleft(\varphi_{k,p}\mright),b_{l}\mleft(\varphi_{l,p}\mright)]=-\delta_{k,l}\left\Vert \varphi_{k,p}\right\Vert ^{2}-\varepsilon_{l,k}\mleft(\varphi_{l,p};\varphi_{k,p}\mright)$
can be further decomposed as
\begin{align}
\tilde{\mathcal{E}}_{\mathrm{B},1} & =\sum_{k,l\in\mathbb{Z}_{\ast}^{3}}\sum_{p\in M_{k}\cap M_{l}}\tilde{c}_{p\mp l}^{\ast}\left[b_{k}^{\ast}\mleft(\varphi_{k,p}\mright),b_{l}\mleft(\varphi_{l,p}\mright)\right]\tilde{c}_{p\mp k}\nonumber \\
 & =-\sum_{k\in\mathbb{Z}_{\ast}^{3}}\sum_{p\in M_{k}}\left\Vert \varphi_{k,p}\right\Vert ^{2}\tilde{c}_{p\mp k}^{\ast}\tilde{c}_{p\mp k}+\sum_{k,l\in\mathbb{Z}_{\ast}^{3}}\sum_{p\in M_{k}\cap M_{l}}\sum_{q\in L_{k}\cap L_{l}}\left\langle \varphi_{l,p},e_{q}\right\rangle \left\langle e_{q},\varphi_{k,p}\right\rangle \tilde{c}_{p\mp l}^{\ast}\tilde{c}_{q-k}^{\ast}\tilde{c}_{q-l}\tilde{c}_{p\mp k}\nonumber \\
 & +\sum_{k,l\in\mathbb{Z}_{\ast}^{3}}\sum_{p\in M_{k}\cap M_{l}}\sum_{q\in\mleft(L_{k}-k\mright)\cap\mleft(L_{l}-l\mright)}\left\langle \varphi_{l,p},e_{q+l}\right\rangle \left\langle e_{q+k},\varphi_{k,p}\right\rangle \tilde{c}_{p\mp l}^{\ast}\tilde{c}_{q+k}^{\ast}\tilde{c}_{q+l}\tilde{c}_{p\mp k}\\
 & =:-\tilde{\mathcal{E}}_{\mathrm{B},1,1}+\tilde{\mathcal{E}}_{\mathrm{B},1,2}+\tilde{\mathcal{E}}_{\mathrm{B},1,3}.\nonumber 
\end{align}
To control this we will use the following:
\begin{lem}
\label{lemma:GeneralKineticSum}For any $A\subset\mathbb{Z}^{3}$
with $\left|A\right|\leq\left|\overline{B}\mleft(0,2k_{F}\mright)\cap\mathbb{Z}^{3}\right|$
and any $\epsilon>0$ it holds that
\[
\sum_{p\in A}\frac{1}{\vert\left|p\right|^{2}-\zeta\vert}\leq C_{\epsilon}k_{F}^{1+\epsilon}
\]
for a constant $C_{\epsilon}>0$ depending only on $\epsilon$.
\end{lem}

For the proof see appendix section \ref{subsec:KineticSumEstimates}.

We can now prove the following:
\begin{prop}
\label{prop:EB1Bound}For any symmetric set $S\subset\mathbb{Z}_{\ast}^{3}$
and $\epsilon>0$ it holds as $k_{F}\rightarrow\infty$ that
\begin{align*}
\pm\tilde{\mathcal{E}}_{\mathrm{B},1,1} & \leq\sum_{k\in\mathbb{Z}_{\ast}^{3}}\max_{p\in M_{k}}\left\Vert \varphi_{k,p}\right\Vert ^{2}H_{\mathrm{kin}}^{\prime}\\
\pm\tilde{\mathcal{E}}_{\mathrm{B},1,2},\pm\tilde{\mathcal{E}}_{\mathrm{B},1,3} & \leq C_{\epsilon}\mleft(k_{F}^{1+\epsilon}\sum_{k\in\mathbb{Z}_{\ast}^{3}\backslash S}\sum_{p\in M_{k}}\max_{q\in L_{k}}\left|\left\langle e_{q},\varphi_{k,p}\right\rangle \right|^{2}+\mleft(\sum_{k\in S}\sqrt{\sum_{p\in M_{k}}\max_{q\in L_{k}}\left|\left\langle e_{q},\varphi_{k,p}\right\rangle \right|^{2}}\mright)^{2}\mright)H_{\mathrm{kin}}^{\prime}
\end{align*}
for a constant $C_{\epsilon}>0$ depending only on $\epsilon$.
\end{prop}

Note that $\max_{p\in M_{k}}\left\Vert \varphi_{k,p}\right\Vert ^{2}\le Ck_{F}^{1-2\beta}\hat{V}_{k}^{2}$ and $\sum_{p\in M_{k}}\max_{q\in L_{k}}\left|\left\langle e_{q},\varphi_{k,p}\right\rangle \right|^{2}\le Ck_{F}^{1-2\beta}\hat{V}_{k}^{2}$, see \eqref{eq:max-varphi} below. 


\textbf{Proof:} The estimate for $\tilde{\mathcal{E}}_{\mathrm{B},1,1}$
is immediate, since
\begin{align}
\tilde{\mathcal{E}}_{\mathrm{B},1,1} & =\sum_{k\in\mathbb{Z}_{\ast}^{3}}\sum_{p\in M_{k}}\left\Vert \varphi_{k,p}\right\Vert ^{2}\tilde{c}_{p\mp k}^{\ast}\tilde{c}_{p\mp k}\leq\sum_{k\in\mathbb{Z}_{\ast}^{3}}\max_{p\in M_{k}}\left\Vert \varphi_{k,p}\right\Vert ^{2}\sum_{p\in M_{k}}\tilde{c}_{p\mp k}^{\ast}\tilde{c}_{p\mp k}\\
 & \leq\mleft(\sum_{k\in\mathbb{Z}_{\ast}^{3}}\max_{p\in M_{k}}\left\Vert \varphi_{k,p}\right\Vert ^{2}\mright)\mathcal{N}_{E}\leq\mleft(\sum_{k\in\mathbb{Z}_{\ast}^{3}}\max_{p\in M_{k}}\left\Vert \varphi_{k,p}\right\Vert ^{2}\mright)H_{\mathrm{kin}}^{\prime},\nonumber 
\end{align}
where we used that $\mathcal{N}_{E}\leq H_{\mathrm{kin}}^{\prime}$
at the end (a consequence of the representation $H_{\mathrm{kin}}^{\prime}=\sum_{p\in\mathbb{Z}^{3}}\vert\left|p\right|^{2}-\zeta\vert\,\tilde{c}_{p}^{\ast}\tilde{c}_{p}$).

The terms $\tilde{\mathcal{E}}_{\mathrm{B},1,2}$ and $\tilde{\mathcal{E}}_{\mathrm{B},1,3}$
are similar, so we focus on $\tilde{\mathcal{E}}_{\mathrm{B},1,2}$.
For this we note that for any $\Psi\in D\mleft(H_{\mathrm{kin}}^{\prime}\mright)$
\begin{align}
\left|\left\langle \Psi,\tilde{\mathcal{E}}_{\mathrm{B},1,2}\Psi\right\rangle \right| & \leq\sum_{k,l\in\mathbb{Z}_{\ast}^{3}}\sum_{p\in M_{k}\cap M_{l}}\sum_{q\in L_{k}\cap L_{l}}\left|\left\langle \varphi_{l,p},e_{q}\right\rangle \right|\left|\left\langle e_{q},\varphi_{k,p}\right\rangle \right|\left\Vert \tilde{c}_{p\mp l}\tilde{c}_{q-k}\Psi\right\Vert \left\Vert \tilde{c}_{p\mp k}\tilde{c}_{q-l}\Psi\right\Vert \nonumber \\
 & \leq\sum_{p,q\in\mathbb{Z}^{3}}\mleft(\sum_{k\in\mathbb{Z}_{\ast}^{3}}1_{M_{k}}\mleft(p\mright)1_{L_{k}}\mleft(q\mright)\left|\left\langle e_{q},\varphi_{k,p}\right\rangle \right|\left\Vert \tilde{c}_{q-k}\Psi\right\Vert \mright)^{2}\label{eq:EB12MainEstimate}\\
 & \leq2\sum_{p,q\in\mathbb{Z}^{3}}\mleft(\sum_{k\in\mathbb{Z}_{\ast}^{3}\backslash S}1_{M_{k}}\mleft(p\mright)1_{L_{k}}\mleft(q\mright)\left|\left\langle e_{q},\varphi_{k,p}\right\rangle \right|\left\Vert \tilde{c}_{q-k}\Psi\right\Vert \mright)^{2}\nonumber \\
 & +2\sum_{p,q\in\mathbb{Z}^{3}}\mleft(\sum_{k\in S}1_{M_{k}}\mleft(p\mright)1_{L_{k}}\mleft(q\mright)\left|\left\langle e_{q},\varphi_{k,p}\right\rangle \right|\left\Vert \tilde{c}_{q-k}\Psi\right\Vert \mright)^{2}\nonumber 
\end{align}
where we used the triangle and Cauchy-Schwarz inequalities and that $\Vert\tilde{c}_{p\pm l}\Vert_{\mathrm{op}}=1$. The first term
on the right-hand side can be further estimated as
\begin{align}
 & \quad\,\sum_{p,q\in\mathbb{Z}^{3}}\mleft(\sum_{k\in\mathbb{Z}_{\ast}^{3}\backslash S}1_{M_{k}}\mleft(p\mright)1_{L_{k}}\mleft(q\mright)\left|\left\langle e_{q},\varphi_{k,p}\right\rangle \right|\left\Vert \tilde{c}_{q-k}\Psi\right\Vert \mright)^{2}\nonumber \\
 & \leq\sum_{p,q\in\mathbb{Z}^{3}}\mleft(\sum_{k\in\mathbb{Z}_{\ast}^{3}\backslash S}1_{M_{k}}\mleft(p\mright)\frac{1_{L_{k}}\mleft(q\mright)}{\vert\left|q-k\right|^{2}-\zeta\vert}\left|\left\langle e_{q},\varphi_{k,p}\right\rangle \right|^{2}\mright)\mleft(\sum_{k\in\mathbb{Z}_{\ast}^{3}\backslash S}1_{L_{k}}\mleft(q\mright)\vert\left|q-k\right|^{2}-\zeta\vert\left\Vert \tilde{c}_{q-k}\Psi\right\Vert ^{2}\mright)\\
 & \leq\sum_{k\in\mathbb{Z}_{\ast}^{3}\backslash S}\sum_{p\in M_{k}}\max_{q\in L_{k}}\left|\left\langle e_{q},\varphi_{k,p}\right\rangle \right|^{2}\sum_{q\in L_{k}}\frac{1_{L_{k}}\mleft(q\mright)}{\vert\left|q-k\right|^{2}-\zeta\vert}\left\langle \Psi,H_{\mathrm{kin}}^{\prime}\Psi\right\rangle \nonumber \\
 & \leq C_{\epsilon}k_{F}^{1+\epsilon}\sum_{k\in\mathbb{Z}_{\ast}^{3}\backslash S}\sum_{p\in M_{k}}\max_{q\in L_{k}}\left|\left\langle e_{q},\varphi_{k,p}\right\rangle \right|^{2}\left\langle \Psi,H_{\mathrm{kin}}^{\prime}\Psi\right\rangle \nonumber 
\end{align}
where we could apply Lemma \ref{lemma:GeneralKineticSum} since $\left|L_{k}\right|\leq\left|B_{F}\right|\leq\left|\overline{B}\mleft(0,2k_{F}\mright)\cap\mathbb{Z}^{3}\right|$.

For the second we instead expand and bound
\begin{align}
 & \quad\sum_{p,q\in\mathbb{Z}^{3}}\mleft(\sum_{k\in S}1_{M_{k}}\mleft(p\mright)1_{L_{k}}\mleft(q\mright)\left|\left\langle e_{q},\varphi_{k,p}\right\rangle \right|\left\Vert \tilde{c}_{q-k}\Psi\right\Vert \mright)^{2}\nonumber \\
 & =\sum_{k,l\in S}\sum_{p\in M_{k}\cap M_{l}}\sum_{q\in L_{k}\cap L_{l}}\left|\left\langle e_{q},\varphi_{k,p}\right\rangle \right|\left|\left\langle e_{q},\varphi_{l,p}\right\rangle \right|\left\Vert \tilde{c}_{q-k}\Psi\right\Vert \left\Vert \tilde{c}_{q-l}\Psi\right\Vert \nonumber \\
 & \leq\sum_{k,l\in S}\sum_{p\in M_{k}\cap M_{l}}\mleft(\max_{q\in L_{k}}\left|\left\langle e_{q},\varphi_{k,p}\right\rangle \right|\mright)\mleft(\max_{q\in L_{l}}\left|\left\langle e_{q},\varphi_{l,p}\right\rangle \right|\mright)\sqrt{\sum_{q\in L_{k}\cap L_{l}}\left\Vert \tilde{c}_{q-k}\Psi\right\Vert ^{2}}\sqrt{\sum_{q\in L_{k}\cap L_{l}}\left\Vert \tilde{c}_{q-l}\Psi\right\Vert ^{2}}\\
 & \leq\sum_{k,l\in S}\sqrt{\sum_{p\in M_{k}\cap M_{l}}\max_{q\in L_{k}}\left|\left\langle e_{q},\varphi_{k,p}\right\rangle \right|^{2}}\sqrt{\sum_{p\in M_{k}\cap M_{l}}\max_{q\in L_{l}}\left|\left\langle e_{q},\varphi_{l,p}\right\rangle \right|^{2}}\left\langle \Psi,\mathcal{N}_{E}\Psi\right\rangle \nonumber \\
 & \leq\mleft(\sum_{k\in S}\sqrt{\sum_{p\in M_{k}}\max_{q\in L_{k}}\left|\left\langle e_{q},\varphi_{k,p}\right\rangle \right|^{2}}\mright)^{2}\left\langle \Psi,H_{\mathrm{kin}}^{\prime}\Psi\right\rangle .\nonumber 
\end{align}
$\hfill\square$

For $\tilde{\mathcal{E}}_{\mathrm{B},2}$ we compute that when $p\in M_{k}\cap M_{l}$
with $M_{k}=L_{k}$
\begin{align}
\left[\tilde{c}_{p\mp k},b_{-l}^{\ast}\mleft(\psi_{-l,-p}\mright)\right] & =\sum_{q\in L_{-l}}\left\langle e_{q},\psi_{-l,-p}\right\rangle \left[c_{p-k}^{\ast},c_{q}^{\ast}c_{q+l}\right]=-\sum_{q\in L_{-l}}\delta_{p-k,q+l}\left\langle e_{q},\psi_{-l,-p}\right\rangle c_{q}^{\ast}\\
 & =-1_{L_{-l}}\mleft(p-k-l\mright)\left\langle e_{p-k-l},\psi_{-l,-p}\right\rangle \tilde{c}_{p-k-l}^{\ast}\nonumber 
\end{align}
and likewise when $p\in M_{k}\cap M_{l}$ with $M_{k}=\mleft(L_{k}-k\mright)$
\begin{align}
\left[\tilde{c}_{p\mp k},b_{-l}^{\ast}\mleft(\psi_{-l,-p}\mright)\right] & =\sum_{q\in L_{-l}}\left\langle e_{q},\psi_{-l,-p}\right\rangle \left[c_{p+k},c_{q}^{\ast}c_{q+l}\right]=\sum_{q\in L_{-l}}\delta_{p+k,q}\left\langle e_{q},\psi_{-l,-p}\right\rangle c_{q+l}\\
 & =1_{L_{-l}}\mleft(p+k\mright)\left\langle e_{p+k},\psi_{-l,-p}\right\rangle \tilde{c}_{p+k+l}^{\ast}=1_{L_{-l}+l}\mleft(p+k+l\mright)\left\langle e_{p+k},\psi_{-l,-p}\right\rangle \tilde{c}_{p+k+l}^{\ast}.\nonumber 
\end{align}
We can summarize this as
\begin{equation}
\left[\tilde{c}_{p\mp k},b_{-l}^{\ast}\mleft(\psi_{-l,-p}\mright)\right]=\mp1_{M_{-l}}\mleft(p\mp k\mp l\mright)\left\langle e_{p\mp k\mp l'},\psi_{-l,-p}\right\rangle \tilde{c}_{p\mp k\mp l}^{\ast}\label{eq:c-pkblpsiCommutator}
\end{equation}
where $l'=l$ when $M_{k}=L_{k}$ and $l'=0$ when $M_{k}=L_{k}-k$
(the presence or absence of this will not make a difference to the
estimation below, so this definition is convenient).

Using also that $\left[\tilde{c}^{\ast},b^{\ast}\mleft(\cdot\mright)\right]=0$
we can then write $\tilde{\mathcal{E}}_{\mathrm{B},2}$ as
\begin{align}
\tilde{\mathcal{E}}_{\mathrm{B},2} & =\mp\sum_{k,l\in\mathbb{Z}_{\ast}^{3}}\sum_{p\in M_{k}\cap M_{l}}1_{M_{-l}}\mleft(p\mp k\mp l\mright)\left\langle e_{p\mp k\mp l'},\psi_{-l,-p}\right\rangle \tilde{c}_{p\mp l}^{\ast}b_{k}^{\ast}\mleft(\varphi_{k,p}\mright)\tilde{c}_{p\mp k\mp l}^{\ast}\\
 & =\pm\sum_{k\in\mathbb{Z}_{\ast}^{3}}\sum_{p\in M_{k}}\mleft(\sum_{l\in\mathbb{Z}_{\ast}^{3}}1_{M_{l}}\mleft(p\mright)1_{M_{-l}}\mleft(p\mp k\mp l\mright)\left\langle \psi_{-l,-p},e_{p\mp k\mp l'}\right\rangle \tilde{c}_{p\mp l}\tilde{c}_{p\mp k\mp l}\mright)^{\ast}b_{k}^{\ast}\mleft(\varphi_{k,p}\mright).\nonumber 
\end{align}
To control this we note the following bounds from \cite[Propositions 4.4, A.1, A.2]{ChrHaiNam-23a} (the idea of which  originates from \cite{HaiPorRex-20}):

\begin{prop}
\label{prop:QuasiBosonicKineticEstimates}For any $k\in\mathbb{Z}_{\ast}^{3}$
and $\varphi\in\ell^{2}\mleft(L_{k}\mright)$ it holds that
\[
b_{k}^{\ast}\mleft(\varphi\mright)b_{k}\mleft(\varphi\mright)\leq\left\langle \varphi,h_{k}^{-1}\varphi\right\rangle H_{\mathrm{kin}}^{\prime},\quad b_{k}\mleft(\varphi\mright)b_{k}^{\ast}\mleft(\varphi\mright)\leq\left\langle \varphi,h_{k}^{-1}\varphi\right\rangle H_{\mathrm{kin}}^{\prime}+\left\Vert \varphi\right\Vert ^{2}.
\]
\end{prop}

\begin{prop}
\label{prop:SimpleLuneRiemannSums}For any $k\in\mathbb{Z}_{\ast}^{3}$
it holds as $k_{F}\rightarrow\infty$ that
\[
\sum_{p\in L_{k}}\lambda_{k,p}^{-1}\leq Ck_{F},\quad\left|L_{k}\right|\leq Ck_{F}^{2}\min\left\{ \left|k\right|,k_{F}\right\} ,
\]
for a constant $C>0$ independent of all quantities.
\end{prop}

With this we can prove the following:
\begin{prop}
\label{prop:EB2Bound}For any symmetric set $S\subset\mathbb{Z}_{\ast}^{3}$
and $\epsilon>0$ it holds as $k_{F}\rightarrow\infty$ that
\begin{align*}
\pm\tilde{\mathcal{E}}_{\mathrm{B},2} & \leq C_{\epsilon}\mleft(\sqrt{k_{F}^{1+\epsilon}\sum_{k\in\mathbb{Z}_{\ast}^{3}\backslash S}\sum_{p\in M_{k}}\max_{q\in L_{k}}\left|\left\langle e_{q},\psi_{k,p}\right\rangle \right|^{2}}+\sum_{k\in S}\sqrt{\sum_{p\in M_{k}}\max_{q\in L_{k}}\left|\left\langle e_{q},\psi_{k,p}\right\rangle \right|^{2}}\mright)\\
 & \qquad\qquad\qquad\qquad\qquad\quad\,\cdot\sqrt{k_{F}\sum_{k\in\mathbb{Z}_{\ast}^{3}}\min\left\{ \left|k\right|,k_{F}\right\} \sum_{p\in M_{k}}\max_{q\in L_{k}}\left|\left\langle e_{q},\varphi_{k,p}\right\rangle \right|^{2}}\mleft(H_{\mathrm{kin}}^{\prime}+k_{F}\mright)
\end{align*}
for a constant $C_{\epsilon}>0$ depending only on $\epsilon$.
\end{prop}

Note that $\max_{p\in M_{k}}\left\Vert \psi_{k,p}\right\Vert ^{2}\le Ck_{F}^{1-2\beta}\hat{V}_{k}^{2}$ and $\sum_{p\in M_{k}}\max_{q\in L_{k}}\left|\left\langle e_{q},\psi_{k,p}\right\rangle \right|^{2}\le Ck_{F}^{1-2\beta}\hat{V}_{k}^{2}$, namely the $\psi_{k,p}$'s satisfy the same bounds as the $\varphi_{k,p}$'s, see \eqref{eq:max-varphi} below.

\textbf{Proof:} By the computation above we can for any $\Psi\in D\mleft(H_{\mathrm{kin}}^{\prime}\mright)$
estimate
\begin{align}
\left\langle \Psi,\tilde{\mathcal{E}}_{\mathrm{B},2}\Psi\right\rangle  & \leq\sum_{k\in\mathbb{Z}_{\ast}^{3}}\sum_{p\in M_{k}}\left\Vert \sum_{l\in\mathbb{Z}_{\ast}^{3}\backslash S}1_{M_{l}}\mleft(p\mright)1_{M_{-l}}\mleft(p\mp k\mp l\mright)\left\langle \psi_{-l,-p},e_{p\mp k\mp l'}\right\rangle \tilde{c}_{p\mp l}\tilde{c}_{p\mp k\mp l}\Psi\right\Vert \left\Vert b_{k}^{\ast}\mleft(\varphi_{k,p}\mright)\Psi\right\Vert \\
 & +\sum_{l\in S}\sum_{k\in\mathbb{Z}_{\ast}^{3}}\sum_{p\in M_{k}}1_{M_{l}}\mleft(p\mright)1_{M_{-l}}\mleft(p\mp k\mp l\mright)\left|\left\langle \psi_{-l,-p},e_{p\mp k\mp l'}\right\rangle \right|\left\Vert \tilde{c}_{p\mp l}\tilde{c}_{p\mp k\mp l}\Psi\right\Vert \left\Vert b_{k}^{\ast}\mleft(\varphi_{k,p}\mright)\Psi\right\Vert \nonumber 
\end{align}
and by Cauchy-Schwarz the terms on the right-hand side can be bounded
by the terms
\begin{align}
 & \sqrt{\sum_{k\in\mathbb{Z}_{\ast}^{3}}\sum_{p\in M_{k}}\left\Vert \sum_{l\in\mathbb{Z}_{\ast}^{3}\backslash S}1_{M_{l}}\mleft(p\mright)1_{M_{-l}}\mleft(p\mp k\mp l\mright)\left\langle \psi_{-l,-p},e_{p\mp k\mp l'}\right\rangle \tilde{c}_{p\mp l}\tilde{c}_{p\mp k\mp l}\Psi\right\Vert ^{2}}\sqrt{\sum_{k\in\mathbb{Z}_{\ast}^{3}}\sum_{p\in M_{k}}\left\Vert b_{k}^{\ast}\mleft(\varphi_{k,p}\mright)\Psi\right\Vert ^{2}}, \nonumber \\
 & \sum_{l\in S}\sqrt{\sum_{k\in\mathbb{Z}_{\ast}^{3}}\sum_{p\in M_{k}}1_{M_{l}}\mleft(p\mright)1_{M_{-l}}\mleft(p\mp k\mp l\mright)\left|\left\langle \psi_{-l,-p},e_{p\mp k\mp l'}\right\rangle \right|^{2}\left\Vert \tilde{c}_{p\mp l}\tilde{c}_{p\mp k\mp l}\Psi\right\Vert ^{2}}\sqrt{\sum_{k\in\mathbb{Z}_{\ast}^{3}}\sum_{p\in M_{k}}\left\Vert b_{k}^{\ast}\mleft(\varphi_{k,p}\mright)\Psi\right\Vert ^{2}}.\label{eq:EB2EstimateSplit}
\end{align}
Beginning with the common factor $\sum_{k\in\mathbb{Z}_{\ast}^{3}}\sum_{p\in M_{k}}\left\Vert b_{k}^{\ast}\mleft(\varphi_{k,p}\mright)\Psi\right\Vert ^{2}$
we can apply the Propositions \ref{prop:QuasiBosonicKineticEstimates}
and \ref{prop:SimpleLuneRiemannSums} to see that
\begin{align}
\left\Vert b_{k}^{\ast}\mleft(\varphi_{k,p}\mright)\Psi\right\Vert ^{2} & \leq\left\langle \varphi_{k,p},h_{k}^{-1}\varphi_{k,p}\right\rangle \left\langle \Psi,H_{\mathrm{kin}}^{\prime}\Psi\right\rangle +\left\Vert \varphi_{k,p}\right\Vert ^{2}\left\Vert \Psi\right\Vert ^{2}\nonumber \\
 & =\sum_{q\in L_{k}}\frac{1}{\lambda_{k,q}}\left|\left\langle e_{q},\varphi_{k,p}\right\rangle \right|^{2}\left\langle \Psi,H_{\mathrm{kin}}^{\prime}\Psi\right\rangle +\sum_{q\in L_{k}}\left|\left\langle e_{q},\varphi_{k,p}\right\rangle \right|^{2}\left\Vert \Psi\right\Vert ^{2}\label{eq:bastphikpBound}\\
 & \leq Ck_{F}\max_{q\in L_{k}}\left|\left\langle e_{q},\varphi_{k,p}\right\rangle \right|^{2}\left\langle \Psi,H_{\mathrm{kin}}^{\prime}\Psi\right\rangle +Ck_{F}^{2}\min\left\{ \left|k\right|,k_{F}\right\} \max_{q\in L_{k}}\left|\left\langle e_{q},\varphi_{k,p}\right\rangle \right|^{2}\left\Vert \Psi\right\Vert ^{2}\nonumber \\
 & \leq Ck_{F}\min\left\{ \left|k\right|,k_{F}\right\} \max_{q\in L_{k}}\left|\left\langle e_{q},\varphi_{k,p}\right\rangle \right|^{2}\left\langle \Psi,\mleft(H_{\mathrm{kin}}^{\prime}+k_{F}\mright)\Psi\right\rangle \nonumber 
\end{align}
for any $k\in\mathbb{Z}_{\ast}^{3}$ and $p\in M_{k}$, whence
\begin{equation}
\sum_{k\in\mathbb{Z}_{\ast}^{3}}\sum_{p\in M_{k}}\left\Vert b_{k}^{\ast}\mleft(\varphi_{k,p}\mright)\Psi\right\Vert ^{2}\leq Ck_{F}\sum_{k\in\mathbb{Z}_{\ast}^{3}}\min\left\{ \left|k\right|,k_{F}\right\} \sum_{p\in M_{k}}\max_{q\in L_{k}}\left|\left\langle e_{q},\varphi_{k,p}\right\rangle \right|^{2}\left\langle \Psi,\mleft(H_{\mathrm{kin}}^{\prime}+k_{F}\mright)\Psi\right\rangle .
\end{equation}
For the remaining factors of equation (\ref{eq:EB2EstimateSplit})
we begin with
\begin{align}
 & \quad\,\sum_{k\in\mathbb{Z}_{\ast}^{3}}\sum_{p\in M_{k}}\left\Vert \sum_{l\in\mathbb{Z}_{\ast}^{3}\backslash S}1_{M_{l}}\mleft(p\mright)1_{M_{-l}}\mleft(p\mp k\mp l\mright)\left\langle \psi_{-l,-p},e_{p\mp k\mp l'}\right\rangle \tilde{c}_{p\mp l}\tilde{c}_{p\mp k\mp l}\Psi\right\Vert ^{2}\nonumber \\
 & \leq\sum_{k\in\mathbb{Z}_{\ast}^{3}}\sum_{p\in M_{k}}\mleft(\sum_{l\in\mathbb{Z}_{\ast}^{3}\backslash S}1_{M_{l}}\mleft(p\mright)\frac{1_{M_{-l}}\mleft(p\mp k\mp l\mright)}{\vert\left|p\mp k\mp l\right|^{2}-\zeta\vert}\left|\left\langle e_{p\mp k\mp l'},\psi_{-l,-p}\right\rangle \right|^{2}\mright)\nonumber \\
 & \qquad\qquad\;\;\;\times\mleft(\sum_{l\in\mathbb{Z}_{\ast}^{3}\backslash S}1_{M_{-l}}\mleft(p\mp k\mp l\mright)\vert\left|p\mp k\mp l\right|^{2}-\zeta\vert\left\Vert \tilde{c}_{p\mp l}\tilde{c}_{p\mp k\mp l}\Psi\right\Vert ^{2}\mright)\\
 & \leq\sum_{l\in\mathbb{Z}_{\ast}^{3}\backslash S}\sum_{p\in M_{l}}\max_{q\in L_{l}}\left|\left\langle e_{-q},\psi_{-l,-p}\right\rangle \right|^{2}\sum_{k\in\mathbb{Z}_{\ast}^{3}}\frac{1_{M_{-l}}\mleft(p\mp k\mp l\mright)}{\vert\left|p\mp k\mp l\right|^{2}-\zeta\vert}\left\langle \Psi,H_{\mathrm{kin}}^{\prime}\Psi\right\rangle \nonumber \\
 & \leq C_{\epsilon}k_{F}^{1+\epsilon}\sum_{l\in\mathbb{Z}_{\ast}^{3}\backslash S}\sum_{p\in M_{l}}\max_{q\in L_{l}}\left|\left\langle e_{q},\psi_{l,p}\right\rangle \right|^{2}\left\langle \Psi,H_{\mathrm{kin}}^{\prime}\Psi\right\rangle \nonumber 
\end{align}
where we used that the presence of the indicator function $1_{M_{-l}}\mleft(p\mp k\mp l\mright)$
restricts the $k$ summation to a set of cardinality at most $\left|M_{-l}\right|=\left|L_{l}\right|\leq\left|B_{F}\right|\leq\left|\overline{B}\mleft(0,2k_{F}\mright)\cap\mathbb{Z}^{3}\right|$,
so Lemma \ref{lemma:GeneralKineticSum} applies.

For the last factor of equation (\ref{eq:EB2EstimateSplit}) we simply
note that
\begin{align}
 & \quad\,\sum_{l\in S}\sqrt{\sum_{k\in\mathbb{Z}_{\ast}^{3}}\sum_{p\in M_{k}}1_{M_{l}}\mleft(p\mright)1_{M_{-l}}\mleft(p\mp k\mp l\mright)\left|\left\langle \psi_{-l,-p},e_{p\mp k\mp l'}\right\rangle \right|^{2}\left\Vert \tilde{c}_{p\mp l}\tilde{c}_{p\mp k\mp l}\Psi\right\Vert ^{2}}\nonumber \\
 & \leq\sum_{l\in S}\sqrt{\sum_{p\in M_{l}}\max_{q\in L_{l}}\left|\left\langle e_{-q},\psi_{-l,-p}\right\rangle \right|^{2}\sum_{k\in\mathbb{Z}_{\ast}^{3}}1_{M_{-l}}\mleft(p\mp k\mp l\mright)\left\Vert \tilde{c}_{p\mp k\mp l}\Psi\right\Vert ^{2}}\\
 & \leq\sum_{l\in S}\sqrt{\sum_{p\in M_{l}}\max_{q\in L_{l}}\left|\left\langle e_{-q},\psi_{-l,-p}\right\rangle \right|^{2}}\sqrt{\left\langle \Psi,\mathcal{N}_{E}\Psi\right\rangle }\leq\sum_{l\in S}\sqrt{\sum_{p\in M_{l}}\max_{q\in L_{l}}\left|\left\langle e_{q},\psi_{l,p}\right\rangle \right|^{2}}\sqrt{\left\langle \Psi,H_{\mathrm{kin}}^{\prime}\Psi\right\rangle }.\nonumber 
\end{align}
The proposition now follows by combining the estimates.

$\hfill\square$

\subsection{Estimation of $\tilde{\mathcal{E}}_{\mathrm{B},3}$, $\tilde{\mathcal{E}}_{\mathrm{B},4}$
and $\tilde{\mathcal{E}}_{\mathrm{B},5}^{\prime}$}

We now bound the remaining forms of equation (\ref{eq:EBSchematicForms}).
First is $\mathcal{\tilde{E}}_{\mathrm{B},3}$, which is an analog
of $\mathcal{\tilde{E}}_{\mathrm{B},1}$: We can write it as (substituting
$\mleft(k,l,p\mright)\rightarrow\mleft(-k,-l,-p\mright)$ first)
\begin{align}
\tilde{\mathcal{E}}_{\mathrm{B},3} & =\sum_{k,l\in\mathbb{Z}_{\ast}^{3}}\sum_{p\in M_{k}\cap M_{l}}\tilde{c}_{-p\pm l}^{\ast}\left[b_{k}\mleft(\psi_{k,p}\mright),b_{l}^{\ast}\mleft(\psi_{l,p}\mright)\right]\tilde{c}_{-p\pm k}\nonumber \\
 & =\sum_{k\in\mathbb{Z}_{\ast}^{3}}\sum_{p\in M_{k}}\left\Vert \psi_{k,p}\right\Vert ^{2}\tilde{c}_{-p\pm k}^{\ast}\tilde{c}_{-p\pm k}-\sum_{k,l\in\mathbb{Z}_{\ast}^{3}}\sum_{p\in M_{k}\cap M_{l}}\sum_{q\in L_{k}\cap L_{l}}\left\langle \psi_{k,p},e_{q}\right\rangle \left\langle e_{q},\psi_{l,p}\right\rangle \tilde{c}_{-p\pm l}^{\ast}\tilde{c}_{q-l}^{\ast}\tilde{c}_{q-k}\tilde{c}_{-p\pm k}\nonumber \\
 & -\sum_{k,l\in\mathbb{Z}_{\ast}^{3}}\sum_{p\in M_{k}\cap M_{l}}\sum_{q\in\mleft(L_{k}-k\mright)\cap\mleft(L_{l}-l\mright)}\left\langle \psi_{k,p},e_{q+k}\right\rangle \left\langle e_{q+l},\psi_{l,p}\right\rangle \tilde{c}_{-p\pm l}^{\ast}\tilde{c}_{q+l}^{\ast}\tilde{c}_{q+k}\tilde{c}_{-p\pm k}\\
 & :=\tilde{\mathcal{E}}_{\mathrm{B},3,1}-\tilde{\mathcal{E}}_{\mathrm{B},3,2}-\tilde{\mathcal{E}}_{\mathrm{B},3,3}.\nonumber 
\end{align}
The following can now be concluded exactly as we did in Proposition \ref{prop:EB1Bound}:
\begin{prop}
\label{prop:EB3Bound}For any symmetric set $S\subset\mathbb{Z}_{\ast}^{3}$
and $\epsilon>0$ it holds as $k_{F}\rightarrow\infty$ that
\begin{align*}
\pm\tilde{\mathcal{E}}_{\mathrm{B},3,1} & \leq\sum_{k\in\mathbb{Z}_{\ast}^{3}}\max_{p\in M_{k}}\left\Vert \psi_{k,p}\right\Vert ^{2}H_{\mathrm{kin}}^{\prime}\\
\pm\tilde{\mathcal{E}}_{\mathrm{B},3,2},\,\pm\tilde{\mathcal{E}}_{\mathrm{B},3,3} & \leq C_{\epsilon}\mleft(k_{F}^{1+\epsilon}\sum_{k\in\mathbb{Z}_{\ast}^{3}\backslash S}\sum_{p\in M_{k}}\max_{q\in L_{k}}\left|\left\langle e_{q},\psi_{k,p}\right\rangle \right|^{2}+\mleft(\sum_{k\in S}\sqrt{\sum_{p\in M_{k}}\max_{q\in L_{k}}\left|\left\langle e_{q},\psi_{k,p}\right\rangle \right|^{2}}\mright)^{2}\mright)H_{\mathrm{kin}}^{\prime}
\end{align*}
for a constant $C_{\epsilon}>0$ depending only on $\epsilon$.
\end{prop}

\textbf{Proof:} $\tilde{\mathcal{E}}_{\mathrm{B},3,1}$ is of the
exact same form as $\tilde{\mathcal{E}}_{\mathrm{B},1,1}$, the only
difference being the substitution $\mleft(\varphi_{k,p},p\mp k\mright)\rightarrow\mleft(\psi_{k,p},-p\pm k\mright)$,
so the first estimate follows exactly as in Proposition \ref{prop:EB1Bound}.

$\tilde{\mathcal{E}}_{\mathrm{B},3,2}$ and $\tilde{\mathcal{E}}_{\mathrm{B},3,3}$
are similar, so we consider $\tilde{\mathcal{E}}_{\mathrm{B},3,2}$.
This immediately factorizes as
\begin{equation}
\tilde{\mathcal{E}}_{\mathrm{B},3,2}=\sum_{p,q\in\mathbb{Z}^{3}}\left|\sum_{k\in\mathbb{Z}_{\ast}^{3}}1_{M_{k}}\mleft(p\mright)1_{L_{k}}\mleft(q\mright)\left\langle \psi_{k,p},e_{q}\right\rangle \tilde{c}_{-p\pm k}\tilde{c}_{q-k}\right|^{2}
\end{equation}
so for any $\Psi\in D\mleft(H_{\mathrm{kin}}^{\prime}\mright)$
\begin{equation}
\left|\left\langle \Psi,\tilde{\mathcal{E}}_{\mathrm{B},3,2}\Psi\right\rangle \right|\leq\sum_{p,q\in\mathbb{Z}^{3}}\mleft(\sum_{k\in\mathbb{Z}_{\ast}^{3}}1_{M_{k}}\mleft(p\mright)1_{L_{k}}\mleft(q\mright)\left|\left\langle \psi_{k,p},e_{q}\right\rangle \right|\left\Vert \tilde{c}_{q-k}\Psi\right\Vert \mright)^{2}
\end{equation}
which subject to the substitution $\varphi_{k,p}\rightarrow\psi_{k,p}$
is the same as that of equation (\ref{eq:EB12MainEstimate}), whence
the second estimate follows.

$\hfill\square$

Using equation (\ref{eq:c-pkblpsiCommutator}) we can write $\tilde{\mathcal{E}}_{\mathrm{B},4}$
as
\begin{align}
\tilde{\mathcal{E}}_{\mathrm{B},4} & =\mp\sum_{k,l\in\mathbb{Z}_{\ast}^{3}}\sum_{p\in M_{k}\cap M_{l}}1_{M_{-l}}\mleft(p\mp k\mp l\mright)\left\langle e_{p\mp k\mp l'},\psi_{-l,-p}\right\rangle \tilde{c}_{p\mp l}^{\ast}b_{-k}\mleft(\psi_{-k,-p}\mright)\tilde{c}_{p\mp k\mp l}^{\ast}\nonumber \\
 & =\pm\sum_{k\in\mathbb{Z}_{\ast}^{3}}\sum_{p\in M_{k}}\mleft(\sum_{l\in\mathbb{Z}_{\ast}^{3}}1_{M_{l}}\mleft(p\mright)1_{M_{-l}}\mleft(p\mp k\mp l\mright)\left\langle \psi_{-l,-p},e_{p\mp k\mp l'}\right\rangle \tilde{c}_{p\mp l}\tilde{c}_{p\mp k\mp l}\mright)^{\ast}b_{-k}\mleft(\psi_{-k,-p}\mright)\\
 & \mp\sum_{k,l\in\mathbb{Z}_{\ast}^{3}}\sum_{p\in M_{k}\cap M_{l}}1_{M_{-l}}\mleft(p\mp k\mp l\mright)\left\langle e_{p\mp k\mp l'},\psi_{-l,-p}\right\rangle \tilde{c}_{p\mp l}^{\ast}\left[b_{-k}\mleft(\psi_{-k,-p}\mright),\tilde{c}_{p\mp k\mp l}^{\ast}\right]\nonumber \\
 & =:\tilde{\mathcal{E}}_{\mathrm{B},4,1}+\tilde{\mathcal{E}}_{\mathrm{B},4,2}\nonumber 
\end{align}

and these terms can be bounded in the following manner:
\begin{prop}
\label{prop:EB4Bound}For any symmetric set $S\subset\mathbb{Z}_{\ast}^{3}$
and $\epsilon>0$ it holds as $k_{F}\rightarrow\infty$ that
\begin{align*}
\pm\tilde{\mathcal{E}}_{\mathrm{B},4,1} & \leq C_{\epsilon}\mleft(\sqrt{k_{F}^{1+\epsilon}\sum_{k\in\mathbb{Z}_{\ast}^{3}}\sum_{p\in M_{k}}\max_{q\in L_{k}}\left|\left\langle e_{q},\psi_{k,p}\right\rangle \right|^{2}}+\sum_{k\in S}\sqrt{\sum_{p\in M_{k}}\max_{q\in L_{k}}\left|\left\langle e_{q},\psi_{k,p}\right\rangle \right|^{2}}\mright)\\
 & \qquad\qquad\qquad\qquad\qquad\qquad\qquad\qquad\times\sqrt{k_{F}\sum_{k\in\mathbb{Z}_{\ast}^{3}}\sum_{p\in M_{k}}\max_{q\in L_{k}}\left|\left\langle e_{q},\psi_{k,p}\right\rangle \right|^{2}}H_{\mathrm{kin}}^{\prime}\\
\pm\tilde{\mathcal{E}}_{\mathrm{B},4,2} & \leq\sqrt{\sum_{k\in\mathbb{Z}_{\ast}^{3}}\max_{p\in M_{k}}\left\Vert \psi_{k,p}\right\Vert ^{2}}\sqrt{\sum_{k\in\mathbb{Z}_{\ast}^{3}}\sum_{p\in M_{k}}\max_{q\in L_{k}}\left|\left\langle e_{q},\psi_{k,p}\right\rangle \right|^{2}}H_{\mathrm{kin}}^{\prime}
\end{align*}
for a constant $C_{\epsilon}>0$ depending only on $\epsilon$.
\end{prop}

\textbf{Proof:} $\tilde{\mathcal{E}}_{\mathrm{B},4,1}$ is of the
same form as $\tilde{\mathcal{E}}_{\mathrm{B},2}$ up to the substitution
$b^{\ast}\mleft(\varphi_{k,p}\mright)\rightarrow b_{-k}\mleft(\psi_{-k,-p}\mright)$,
so the first estimate follows as in \ref{prop:EB2Bound}, using also Proposition \ref{prop:SimpleLuneRiemannSums} and noting that we now simply have
\begin{align}
\left\Vert b_{-k}\mleft(\psi_{-k,-p}\mright)\Psi\right\Vert  & \leq\left\langle \psi_{-k,-p},h_{-k}^{-1}\psi_{-k,-p}\right\rangle \left\langle \Psi,H_{\mathrm{kin}}^{\prime}\Psi\right\rangle \\
 & \leq Ck_{F}\max_{q\in L_{k}}\left|\left\langle e_{-q},\psi_{-k,-p}\right\rangle \right|^{2}\left\langle \Psi,H_{\mathrm{kin}}^{\prime}\Psi\right\rangle \nonumber 
\end{align}
by Proposition \ref{prop:QuasiBosonicKineticEstimates}, rather than
the more complicated bound of equation (\ref{eq:bastphikpBound})
which was needed for $\tilde{\mathcal{E}}_{\mathrm{B},2}$.

For $\tilde{\mathcal{E}}_{\mathrm{B},4,2}$ we compute that when $p\in M_{k}=L_{k}$
\begin{align}
\left[b_{-k}\mleft(\psi_{-k,-p}\mright),\tilde{c}_{p\mp k\mp l}^{\ast}\right] & =\sum_{q\in L_{-k}}\left\langle \psi_{-k,-p},e_{q}\right\rangle \left[c_{q+k}^{\ast}c_{q},c_{p-k-l}^{\ast}\right]=\sum_{q\in L_{-k}}\delta_{q,p-k-l}\left\langle \psi_{-k,-p},e_{q}\right\rangle c_{q+k}^{\ast}\\
 & =1_{L_{-k}}\mleft(p-k-l\mright)\left\langle \psi_{-k,-p},e_{p-k-l}\right\rangle \tilde{c}_{p-l}\nonumber 
\end{align}
while when $p\in M_{k}=\mleft(L_{k}-k\mright)$
\begin{align}
\left[b_{-k}\mleft(\psi_{-k,-p}\mright),\tilde{c}_{p\mp k\mp l}^{\ast}\right] & =\sum_{q\in L_{-k}}\left\langle \psi_{-k,-p},e_{q}\right\rangle \left[c_{q+k}^{\ast}c_{q},c_{p+k+l}\right]=-\sum_{q\in L_{-k}}\delta_{q,p+l}\left\langle \psi_{-k,-p},e_{q}\right\rangle c_{q}\\
 & =-1_{L_{-k}}\mleft(p+l\mright)\left\langle \psi_{-k,-p},e_{p+l}\right\rangle \tilde{c}_{p+l}=-1_{L_{-k}+k}\mleft(p+k+l\mright)\left\langle \psi_{-k,-p},e_{p+l}\right\rangle \tilde{c}_{p+l}\nonumber 
\end{align}
which we can summarize as
\begin{equation}
\left[b_{-k}\mleft(\psi_{-k,-p}\mright),\tilde{c}_{p\mp k\mp l}^{\ast}\right]=\pm1_{M_{-k}}\mleft(p\mp k\mp l\mright)\left\langle \psi_{-k,-p},e_{p\mp k'\mp l}\right\rangle \tilde{c}_{p\mp l}.
\end{equation}
$\tilde{\mathcal{E}}_{\mathrm{B},4,2}$ thus takes the form
\begin{equation}
\tilde{\mathcal{E}}_{\mathrm{B},4,2}=-\sum_{k,l\in\mathbb{Z}_{\ast}^{3}}\sum_{p\in M_{k}\cap M_{l}}1_{M_{-k}}\mleft(p\mp k\mp l\mright)1_{M_{-l}}\mleft(p\mp k\mp l\mright)\left\langle e_{p\mp k\mp l'},\psi_{-l,-p}\right\rangle \left\langle \psi_{-k,-p},e_{p\mp k'\mp l}\right\rangle \tilde{c}_{p\mp l}^{\ast}\tilde{c}_{p\mp l}
\end{equation}
so by Cauchy-Schwarz we may for any $\Psi\in D\mleft(H_{\mathrm{kin}}^{\prime}\mright)$
estimate that
\begin{align}
\left|\left\langle \Psi,\tilde{\mathcal{E}}_{\mathrm{B},4,2}\Psi\right\rangle \right| & \leq\sqrt{\sum_{k,l\in\mathbb{Z}_{\ast}^{3}}\sum_{p\in M_{k}\cap M_{l}}1_{M_{-k}}\mleft(p\mp k\mp l\mright)1_{M_{-l}}\mleft(p\mp k\mp l\mright)\left|\left\langle e_{p\mp k\mp l'},\psi_{-l,-p}\right\rangle \right|^{2}\left\Vert \tilde{c}_{p\mp l}\Psi\right\Vert ^{2}}\\
 & \times \sqrt{\sum_{k,l\in\mathbb{Z}_{\ast}^{3}}\sum_{p\in M_{k}\cap M_{l}}1_{M_{-k}}\mleft(p\mp k\mp l\mright)1_{M_{-l}}\mleft(p\mp k\mp l\mright)\left|\left\langle \psi_{-k,-p},e_{p\mp k'\mp l}\right\rangle \right|^{2}\left\Vert \tilde{c}_{p\mp l}\Psi\right\Vert ^{2}}.\nonumber 
\end{align}
The first quantity can be controlled by writing it as
\begin{align}
 & \quad\,\sum_{l\in\mathbb{Z}_{\ast}^{3}}\sum_{p\in M_{l}}\mleft(\sum_{k\in\mathbb{Z}_{\ast}^{3}}1_{M_{k}}\mleft(p\mright)1_{M_{-k}}\mleft(p\mp k\mp l\mright)1_{M_{-l}}\mleft(p\mp k\mp l\mright)\left|\left\langle e_{p\mp k\mp l'},\psi_{-l,-p}\right\rangle \right|^{2}\mright)\left\Vert \tilde{c}_{p\mp l}\Psi\right\Vert ^{2}\nonumber \\
 & \leq\sum_{l\in\mathbb{Z}_{\ast}^{3}}\sum_{p\in M_{l}}\left\Vert \psi_{-l,-p}\right\Vert ^{2}\left\Vert \tilde{c}_{p\mp l}\Psi\right\Vert ^{2}\leq\sum_{l\in\mathbb{Z}_{\ast}^{3}}\max_{p\in M_{l}}\left\Vert \psi_{-l,-p}\right\Vert ^{2}\sum_{p\in M_{l}}\left\Vert \tilde{c}_{p\mp l}\Psi\right\Vert ^{2}\\
 & \leq\sum_{l\in\mathbb{Z}_{\ast}^{3}}\max_{p\in M_{l}}\left\Vert \psi_{-l,-p}\right\Vert ^{2}\left\langle \Psi,\mathcal{N}_{E}\Psi\right\rangle \leq\sum_{l\in\mathbb{Z}_{\ast}^{3}}\max_{p\in M_{l}}\left\Vert \psi_{l,p}\right\Vert ^{2}\left\langle \Psi,H_{\mathrm{kin}}^{\prime}\Psi\right\rangle .\nonumber 
\end{align}
For the second we instead estimate
\begin{align}
 & \quad\,\sum_{k\in\mathbb{Z}_{\ast}^{3}}\sum_{p\in M_{k}}\sum_{l\in\mathbb{Z}_{\ast}^{3}} 1_{M_{l}}\mleft(p\mright) 1_{M_{-k}}\mleft(p\mp k\mp l\mright)1_{M_{-l}}\mleft(p\mp k\mp l\mright)\left|\left\langle \psi_{-k,-p},e_{p\mp k'\mp l}\right\rangle \right|^{2}\left\Vert \tilde{c}_{p\mp l}\Psi\right\Vert ^{2}\nonumber \\
 & \leq\sum_{k\in\mathbb{Z}_{\ast}^{3}}\sum_{p\in M_{k}}\max_{q\in L_{k}}\left|\left\langle e_{-q},\psi_{-k,-p}\right\rangle \right|^{2}\mleft(\sum_{l\in\mathbb{Z}_{\ast}^{3}} 1_{M_{l}}\mleft(p\mright) 1_{M_{-k}}\mleft(p\mp k\mp l\mright)1_{M_{-l}}\mleft(p\mp k\mp l\mright)\left\Vert \tilde{c}_{p\mp l}\Psi\right\Vert ^{2}\mright)\\
 & \leq\sum_{k\in\mathbb{Z}_{\ast}^{3}}\sum_{p\in M_{k}}\max_{q\in L_{k}}\left|\left\langle e_{-q},\psi_{-k,-p}\right\rangle \right|^{2}\left\langle \Psi,\mathcal{N}_{E}\Psi\right\rangle \leq\sum_{k\in\mathbb{Z}_{\ast}^{3}}\sum_{p\in M_{k}}\max_{q\in L_{k}}\left|\left\langle e_{q},\psi_{k,p}\right\rangle \right|^{2}\left\langle \Psi,H_{\mathrm{kin}}^{\prime}\Psi\right\rangle \nonumber 
\end{align}
and the claim follows.

$\hfill\square$

Finally we have $\tilde{\mathcal{E}}_{\mathrm{B},5}^{\prime}$:
\begin{prop}
\label{prop:EB5Bound}It holds as $k_{F}\rightarrow\infty$ that
\[
\pm\tilde{\mathcal{E}}_{\mathrm{B},5}^{\prime}\leq\sum_{k\in\mathbb{Z}_{\ast}^{3}}\sum_{p\in M_{k}}\max_{q\in L_{k}}\left|\left\langle e_{q},\psi_{k,p}\right\rangle \right|^{2}H_{\mathrm{kin}}^{\prime}.
\]
\end{prop}

\textbf{Proof:} From equation (\ref{eq:c-pkblpsiCommutator}) we have
\[
\tilde{\mathcal{E}}_{\mathrm{B},5}^{\prime}=\sum_{k,l\in\mathbb{Z}_{\ast}^{3}}\sum_{p\in M_{k}\cap M_{l}}1_{M_{-l}}\mleft(p\mp k\mp l\mright)1_{M_{-k}}\mleft(p\mp k\mp l\mright)\left\langle e_{p\mp k\mp l'},\psi_{-l,-p}\right\rangle \left\langle \psi_{-k,-p},e_{p\mp k'\mp l}\right\rangle \tilde{c}_{p\mp k\mp l}^{\ast}\tilde{c}_{p\mp k\mp l}
\]
and since the summand is symmetric in $k$ and $l$ we can for any
$\Psi\in D\mleft(H_{\mathrm{kin}}^{\prime}\mright)$ estimate using
Cauchy-Schwarz
\begin{align}
\left|\left\langle \Psi,\tilde{\mathcal{E}}_{\mathrm{B},5}^{\prime}\Psi\right\rangle \right| & \leq\sum_{k,l\in\mathbb{Z}_{\ast}^{3}}\sum_{p\in M_{k}\cap M_{l}}1_{M_{-l}}\mleft(p\mp k\mp l\mright)1_{M_{-k}}\mleft(p\mp k\mp l\mright)\left|\left\langle \psi_{-k,-p},e_{p\mp k'\mp l}\right\rangle \right|^{2}\left\Vert \tilde{c}_{p\mp k\mp l}\Psi\right\Vert ^{2}\\
 & \leq\sum_{k\in\mathbb{Z}_{\ast}^{3}}\sum_{p\in M_{k}}\max_{q\in L_{k}}\left|\left\langle e_{-q},\psi_{-k,-p}\right\rangle \right|^{2}\mleft(\sum_{l\in\mathbb{Z}_{\ast}^{3}}1_{M_{l}}\mleft(p\mright)1_{M_{-l}}\mleft(p\mp k\mp l\mright)1_{M_{-k}}\mleft(p\mp k\mp l\mright)\left\Vert \tilde{c}_{p\mp k\mp l}\Psi\right\Vert ^{2}\mright)\nonumber \\
 & \leq\sum_{k\in\mathbb{Z}_{\ast}^{3}}\sum_{p\in M_{k}}\max_{q\in L_{k}}\left|\left\langle e_{-q},\psi_{-k,-p}\right\rangle \right|^{2}\left\langle \Psi,\mathcal{N}_{E}\Psi\right\rangle \leq\sum_{k\in\mathbb{Z}_{\ast}^{3}}\sum_{p\in M_{k}}\max_{q\in L_{k}}\left|\left\langle e_{q},\psi_{k,p}\right\rangle \right|^{2}\left\langle \Psi,H_{\mathrm{kin}}^{\prime}\Psi\right\rangle .\nonumber 
\end{align}

$\hfill\square$

\subsection{Proof of Theorem \ref{them:EBEstimate}}

We are now ready to insert the particular $\varphi_{k,p}$'s and $\psi_{k,p}$'s
of our problem to conclude Theorem \ref{them:EBEstimate}. To estimate
the relevant quantities we will need the following matrix element
estimates on the one-body operators $C_{k}$ and $S_{k}$:
\begin{prop}
\label{prop:OneBodyOperatorEstimates}For any $k\in\mathbb{Z}_{\ast}^{3}$
and $p,q\in L_{k}$ it holds that
\begin{align*}
\left|\left\langle e_{p},\mleft(C_{k}-1\mright)e_{q}\right\rangle \right|,\left|\left\langle e_{p},S_{k}e_{q}\right\rangle \right| & \leq C\frac{\hat{V}_{k}k_{F}^{-\beta}}{\lambda_{k,p}+\lambda_{k,q}}\\
\left|\left\langle e_{p},S_{k}e_{q}\right\rangle -\frac{\hat{V}_{k}k_{F}^{-\beta}}{2\mleft(2\pi\mright)^{3}}\frac{1}{\lambda_{k,p}+\lambda_{k,q}}\right| & \leq C\frac{\hat{V}_{k}^{2}k_{F}^{1-2\beta}}{\lambda_{k,p}+\lambda_{k,q}}
\end{align*}
for a constant $C>0$ independent of all quantities.
\end{prop}

The proof of these estimates is similar to that of the one-body estimates
of \cite[Section 7]{ChrHaiNam-23a} so we leave this to appendix section
\ref{subsec:One-BodyOperatorEstimates}.

With these estimates we can also bound $\mathcal{E}_{\mathrm{B},6}^{\prime}$:
\begin{prop}\label{prop:EB6'}
It holds as $k_{F}\rightarrow\infty$ that

$$
\pm\mathcal{E}_{\mathrm{B},6}^{\prime} \le C k_F^{3(1-\beta)} \sqrt{\sum_{k\in\mathbb{Z}_{\ast}^{3}}\hat{V}_{k}^{2}}\sum_{k\in\mathbb{Z}_{\ast}^{3}}\hat{V}_{k}^{2}\left|k\right|^{\frac{1}{2}}
$$
for a constant $C>0$ independent of all quantities.
\end{prop}

\textbf{Proof:} As in equation (\ref{eq:EB6Approximation}), the fact
that $\lambda_{k,p}+\lambda_{k,q}=\lambda_{l,p}+\lambda_{l,q}$ when
there is a Kronecker delta $\delta_{p+q,k+l}$ means that we can write
\begin{align}
\mathcal{E}_{\mathrm{B},6}^{\prime} & =\sum_{k,l\in\mathbb{Z}_{\ast}^{3}}\sum_{p,q\in L_{k}\cap L_{l}}\delta_{p+q,k+l}\mleft(\lambda_{k,p}+\lambda_{k,q}\mright)\left\langle S_{k}e_{p},e_{q}\right\rangle \left\langle e_{q},S_{l}e_{p}\right\rangle -E_{\mathrm{corr},\mathrm{ex}}\\
E_{\mathrm{corr},\mathrm{ex}} & =\sum_{k,l\in\mathbb{Z}_{\ast}^{3}}\sum_{p,q\in L_{k}\cap L_{l}}\delta_{p+q,k+l}\mleft(\lambda_{k,p}+\lambda_{k,q}\mright)\mleft(\frac{\hat{V}_{k}k_{F}^{-\beta}}{2\mleft(2\pi\mright)^{3}}\frac{1}{\lambda_{k,p}+\lambda_{k,q}}\mright)\mleft(\frac{\hat{V}_{l}k_{F}^{-\beta}}{2\mleft(2\pi\mright)^{3}}\frac{1}{\lambda_{l,p}+\lambda_{l,q}}\mright)\nonumber 
\end{align}
so $\mathcal{E}_{\mathrm{B},6}^{\prime}$ can be written as the sum
of two terms
\begin{align}
\mathcal{E}_{\mathrm{B},6}^{\prime} & =\sum_{k,l\in\mathbb{Z}_{\ast}^{3}}\sum_{p,q\in L_{k}\cap L_{l}}\delta_{p+q,k+l}\mleft(\lambda_{l,p}+\lambda_{l,q}\mright)\mleft(\left\langle S_{k}e_{p},e_{q}\right\rangle -\frac{\hat{V}_{k}k_{F}^{-\beta}}{2\mleft(2\pi\mright)^{3}}\frac{1}{\lambda_{k,p}+\lambda_{k,q}}\mright)\left\langle e_{q},S_{l}e_{p}\right\rangle \\
 & +\sum_{k,l\in\mathbb{Z}_{\ast}^{3}}\sum_{p,q\in L_{k}\cap L_{l}}\delta_{p+q,k+l}\mleft(\lambda_{k,p}+\lambda_{k,q}\mright)\mleft(\frac{\hat{V}_{k}k_{F}^{-\beta}}{2\mleft(2\pi\mright)^{3}}\frac{1}{\lambda_{k,p}+\lambda_{k,q}}\mright)\mleft(\left\langle e_{q},S_{l}e_{p}\right\rangle -\frac{\hat{V}_{l}k_{F}^{-\beta}}{2\mleft(2\pi\mright)^{3}}\frac{1}{\lambda_{l,p}+\lambda_{l,q}}\mright).\nonumber 
\end{align}
By the estimates of Proposition \ref{prop:OneBodyOperatorEstimates}
these terms can be estimated in a similar form for
\begin{align}\label{eq:CB6-main-new}
\left|\mathcal{E}_{\mathrm{B},6}^{\prime}\right| & \leq C\sum_{k,l\in\mathbb{Z}_{\ast}^{3}}\sum_{p,q\in L_{k}\cap L_{l}}\delta_{p+q,k+l}\mleft(\lambda_{l,p}+\lambda_{l,q}\mright)\frac{\hat{V}_{k}^{2}k_{F}^{1-2\beta}}{\lambda_{k,p}+\lambda_{k,q}}\frac{\hat{V}_{l}k_{F}^{-\beta}}{\lambda_{l,p}+\lambda_{l,q}}\\
 & =Ck_{F}^{1-3\beta}\sum_{k,l\in\mathbb{Z}_{\ast}^{3}}\hat{V}_{k}^{2}\hat{V}_{l}\sum_{p,q\in L_{k}\cap L_{l}}\frac{\delta_{p+q,k+l}}{\lambda_{k,p}+\lambda_{k,q}}\nonumber.
\end{align}
The sum on the right-hand side of equation \eqref{eq:CB6-main-new} can now be estimated along the same lines as \cite[equation (4.80)]{ChrHaiNam-23b} by (using again the observation that $p,q\in L_{p+q-k}\Leftrightarrow p,q\in L_{k}$)
\begin{align}\label{eq:CB6-main-new_T1}
\sum_{k,l\in\mathbb{Z}_{\ast}^{3}}\hat{V}_{k}^{2}\hat{V}_{l}\sum_{p,q\in L_{k}\cap L_{l}}\frac{\delta_{p+q,k+l}}{\lambda_{k,p}+\lambda_{k,q}}\nonumber&=\sum_{k\in\mathbb{Z}_{\ast}^{3}}\hat{V}_{k}^{2}\sum_{p,q\in L_{k}}\frac{\hat{V}_{p+q-k}}{\lambda_{k,p}+\lambda_{k,q}}\\\nonumber&\leq\sum_{k\in\mathbb{Z}_{\ast}^{3}}\hat{V}_{k}^{2}\sqrt{\sum_{p\in L_{k}}\sum_{q\in L_{k}}\hat{V}_{p+q-k}^{2}}\sqrt{\sum_{p,q\in L_{k}}\frac{1}{\left(\lambda_{k,p}+\lambda_{k,q}\right)^{2}}}\\
&\leq\sqrt{\sum_{k'\in\mathbb{Z}_{\ast}^{3}}\hat{V}_{k'}^{2}}\sum_{k\in\mathbb{Z}_{\ast}^{3}}\hat{V}_{k}^{2}\sqrt{\left|L_{k}\right|}\sqrt{\sum_{p,q\in L_{k}}\frac{1}{\lambda_{k,p}}\frac{1}{\lambda_{k,q}}}\\\nonumber
&\leq Ck_{F}\sqrt{\sum_{k'\in\mathbb{Z}_{\ast}^{3}}\hat{V}_{k'}^{2}}\sum_{k\in\mathbb{Z}_{\ast}^{3}}\hat{V}_{k}^{2}\sqrt{k_{F}^{2}\min\left\{ \left|k\right|,k_{F}\right\} }\\\nonumber
&=Ck_{F}^{2}\sqrt{\sum_{k'\in\mathbb{Z}_{\ast}^{3}}\hat{V}_{k'}^{2}}\sum_{k\in\mathbb{Z}_{\ast}^{3}}\hat{V}_{k}^{2}\left|k\right|^{\frac{1}{2}}
\end{align}
where we also used the bound of Proposition \ref{prop:SimpleLuneRiemannSums}. 
$\hfill\square$

We now conclude the main result of this section:

\noindent
\textbf{Proof of Theorem \ref{them:EBEstimate}:} Recalling the definition of equation (\ref{eq:MkPkPhikDefinition}),
we can use Proposition \ref{prop:OneBodyOperatorEstimates} to estimate
that
\begin{align}\label{eq:max-varphi}
\max_{p\in M_{k}}\left\Vert \varphi_{k,p}\right\Vert ^{2} & =\max_{p\in L_{k}}\sum_{q\in L_{k}}\vert\left|p\right|^{2}-k_{F}^{2}\vert\left|\left\langle e_{q},\mleft(C_{k}-1\mright)e_{p}\right\rangle \right|^{2}\leq Ck_{F}^{-2\beta}\hat{V}_{k}^{2}\max_{p\in L_{k}}\sum_{q\in L_{k}}\frac{\vert\left|p\right|^{2}-k_{F}^{2}\vert}{\mleft(\lambda_{k,p}+\lambda_{k,q}\mright)^{2}}\nonumber \\
 & \leq Ck_{F}^{-2\beta}\hat{V}_{k}^{2}\sum_{q\in L_{k}}\lambda_{k,q}^{-1}\leq Ck_{F}^{1-2\beta}\hat{V}_{k}^{2}\\
\sum_{p\in M_{k}}\max_{q\in L_{k}}\left|\left\langle e_{q},\varphi_{k,p}\right\rangle \right|^{2} & =\sum_{p\in L_{k}}\max_{q\in L_{k}}\vert\left|p\right|^{2}-k_{F}^{2}\vert\left|\left\langle e_{q},\mleft(C_{k}-1\mright)e_{p}\right\rangle \right|^{2}\leq Ck_{F}^{-2\beta}\hat{V}_{k}^{2}\sum_{p\in L_{k}}\max_{q\in L_{k}}\frac{\vert\left|p\right|^{2}-k_{F}^{2}\vert}{\mleft(\lambda_{k,p}+\lambda_{k,q}\mright)^{2}}\nonumber \\
 & \leq Ck_{F}^{-2\beta}\hat{V}_{k}^{2}\sum_{p\in L_{k}}\lambda_{k,p}^{-1}\leq Ck_{F}^{1-2\beta}\hat{V}_{k}^{2}\nonumber 
\end{align}
when $p\in M_{k}=L_{k}$, where we used that $\vert\left|p\right|^{2}-k_{F}^{2}\vert\leq\vert\left|p\right|^{2}-k_{F}^{2}\vert+\vert\left|p-k\right|^{2}-k_{F}^{2}\vert=2\lambda_{k,p}$.
This is also true when $p\in M_{k}=\mleft(L_{k}-k\mright)$ (the only
difference being the substitition $\vert\left|p\right|^{2}-k_{F}^{2}\vert\rightarrow\vert\left|p-k\right|^{2}-k_{F}^{2}\vert$
in the formulas above) and, since the estimate for $\mleft(C_{k}-1\mright)$
is also valid for $S_{k}$, the same estimates hold when $\psi_{k,p}$
is substituted for $\varphi_{k,p}$.

Consequently all the estimates for $\tilde{\mathcal{E}}_{\mathrm{B},1},\ldots,\tilde{\mathcal{E}}_{\mathrm{B},5}^{\prime}$
of the Propositions \ref{prop:EB1Bound}, \ref{prop:EB2Bound}, \ref{prop:EB3Bound},
\ref{prop:EB4Bound} and \ref{prop:EB5Bound} can be dominated by
\begin{equation}
C_{\epsilon}k_{F}^{2\mleft(1-\beta\mright)+\epsilon}\mleft(\sqrt{\sum_{k\in\mathbb{Z}_{\ast}^{3}\backslash S}\hat{V}_{k}^{2}}+k_{F}^{-\frac{1}{2}}\sum_{k\in S}\hat{V}_{k}\mright)\sqrt{\sum_{k\in\mathbb{Z}_{\ast}^{3}}\hat{V}_{k}^{2}\min\left\{ \left|k\right|,k_{F}\right\} }\mleft(H_{\mathrm{kin}}^{\prime}+k_{F}\mright).
\end{equation}

Most of these follow directly by basic inequalities, but to reduce the bounds of the Propositions \ref{prop:EB1Bound} and \ref{prop:EB3Bound} to this form one should first exploit the fact that the estimate are valid for any $S \in \mathbb{Z}^3_\ast$ to minimize over such sets, and then estimate this minimum by a specific $S$ and $S = \emptyset$ separately.

Combining this with our estimate for $\mathcal{E}_{\mathrm{B},6}^{\prime}$ of Proposition \ref{prop:EB6'} we obtain the desired claim.
$\hfill\square$

\section{\label{sec:InclusionfoftheSmallkCubicTerms}Inclusion of the ``Small
$k$'' Cubic Terms}

In this section we perform the computations leading to the incorporation
of the ``small $k$'' cubic terms into the factorization of $H_{\mathrm{B}}$.

For convenience we recall that the (full) cubic terms can be written
\begin{equation}
\mathcal{C}=4\,\mathrm{Re}\sum_{k\in\mathbb{Z}_{\ast}^{3}}b_{k}^{\ast}\mleft(w_{k}\mright)D_{k}
\end{equation}
where
\begin{equation}
w_{k}=\frac{\hat{V}_{k}k_{F}^{-\beta}}{2\mleft(2\pi\mright)^{3}}\sum_{p\in L_{k}}e_{p}=\sqrt{\frac{\hat{V}_{k}k_{F}^{-\beta}}{2\mleft(2\pi\mright)^{3}}}v_{k}.
\end{equation}
We furthermore define
\begin{equation}
\eta_{k}=\begin{cases}
E_{k}^{-\frac{3}{2}}h_{k}^{\frac{1}{2}}w_{k} & k\in S\\
0 & \text{otherwise}
\end{cases}
\end{equation}
for a fixed symmetric subset $S\subset\mathbb{Z}_{\ast}^{3}$ (to
be optimized over at the end) and
\begin{equation}
d_{p}^{3}=\begin{cases}
+\sum_{k\in S}1_{L_{k}}\mleft(p\mright)\left\langle e_{p},\eta_{k}\right\rangle \tilde{c}_{p-k}^{\ast}D_{k} & \text{ for } p\in B_{F}^{c}\\
-\sum_{k\in S}1_{L_{k}-k}\mleft(p\mright)\left\langle e_{p+k},\eta_{k}\right\rangle \tilde{c}_{p+k}^{\ast}D_{k} & \text{ for } p\in B_{F}
\end{cases}.\label{eq:dp3Definition}
\end{equation}
We will prove the following:
\begin{thm}
\label{them:BigFactorization}It holds that
\begin{align*}
 & \quad\;\;H_{\mathrm{B}}+4\,\mathrm{Re}\sum_{k\in S}b_{k}^{\ast}\mleft(w_{k}\mright)D_{k}+\frac{k_{F}^{-\beta}}{2\mleft(2\pi\mright)^{3}}\sum_{k\in S}\hat{V}_{k}\frac{2\left\langle v_{k},h_{k}^{-1}v_{k}\right\rangle }{1+2\left\langle v_{k},h_{k}^{-1}v_{k}\right\rangle }D_{k}^{\ast}D_{k}\\
 & =\sum_{p\in\mathbb{Z}^{3}}\vert\left|p\right|^{2}-k_{F}^{2}\vert\mleft(\left|\tilde{c}_{p}+d_{p}^{1}+d_{p}^{2}+d_{p}^{3}\right|^{2}+\left|(d_{p}^{1}+d_{p}^{2}+d_{p}^{3})^{\ast}\right|^{2}\mright)-2\sum_{k\in\mathbb{Z}_{\ast}^{3}}\sum_{p\in L_{k}}\varepsilon_{k,k}\mleft(e_{p};S_{k}E_{k}S_{k}^{\ast}e_{p}\mright)\\
 & +\sum_{k\in\mathbb{Z}_{\ast}^{3}}\sum_{p,q\in L_{k}}2\left\langle e_{p},\mleft(E_{k}-h_{k}\mright)e_{q}\right\rangle \mleft(b_{k}\mleft(C_{k}e_{p}\mright)+b_{-k}^{\ast}\mleft(S_{-k}e_{-p}\mright)+\left\langle e_{p},\eta_{k}\right\rangle D_{k}\mright)^{\ast}\mleft(b_{k}\mleft(C_{k}e_{q}\mright)+b_{-k}^{\ast}\mleft(S_{-k}e_{-q}\mright)+\left\langle e_{q},\eta_{k}\right\rangle D_{k}\mright)\\
 & +E_{\mathrm{corr},\mathrm{bos}}+E_{\mathrm{corr},\mathrm{ex}}+\mathcal{E}_{\mathrm{B}}+\mathcal{E}_{\mathcal{C}}
\end{align*}
for an operator $\mathcal{E}_{\mathcal{C}}$ defined below.
\end{thm}
We stress that throughout the remainder of the paper $\eta_{k}$ and $\mathcal{E}_{\mathcal{C}}$ (and its related subexpressions) depend implicitly on the choice of the fixed set $S\subset\mathbb{Z}_{\ast}^{3}$.

\subsection{Expansion of the Potential Terms}

As in Section \ref{sec:ExtractionoftheCorrelationEnergybyFactorization}
we first consider the potential part of the factorization. For that
we first have the following:
\begin{prop}
\label{prop:ModifiedQuadraticExpansionIdentity}For any symmetric
operators $A_{k}:\ell^{2}\mleft(L_{k}\mright)\rightarrow\ell^{2}\mleft(L_{k}\mright)$,
$k\in\mathbb{Z}_{\ast}^{3}$, obeying
\[
\left\langle e_{p},A_{k}e_{q}\right\rangle =\left\langle e_{-p},A_{-k}e_{-q}\right\rangle ,\quad p,q\in L_{k},
\]
it holds that
\begin{align*}
 & \quad\,\sum_{k\in\mathbb{Z}_{\ast}^{3}}\sum_{p,q\in L_{k}}2\left\langle e_{p},A_{k}e_{q}\right\rangle \mleft(b_{k}\mleft(C_{k}e_{p}\mright)+b_{-k}^{\ast}\mleft(S_{-k}e_{-p}\mright)+\left\langle e_{p},\eta_{k}\right\rangle D_{k}\mright)^{\ast}\mleft(b_{k}\mleft(C_{k}e_{q}\mright)+b_{-k}^{\ast}\mleft(S_{-k}e_{-q}\mright)+\left\langle e_{q},\eta_{k}\right\rangle D_{k}\mright)\\
 & =\sum_{k\in\mathbb{Z}_{\ast}^{3}}\sum_{p,q\in L_{k}}2\left\langle e_{p},A_{k}e_{q}\right\rangle \mleft(b_{k}\mleft(C_{k}e_{p}\mright)+b_{-k}^{\ast}\mleft(S_{-k}e_{-p}\mright)\mright)^{\ast}\mleft(b_{k}\mleft(C_{k}e_{q}\mright)+b_{-k}^{\ast}\mleft(S_{-k}e_{-q}\mright)\mright)\\
 & +4\,\mathrm{Re}\sum_{k\in S}b_{k}^{\ast}\mleft(\mleft(C_{k}+S_{k}\mright)A_{k}\eta_{k}\mright)D_{k}+\sum_{k\in S}2\left\langle \eta_{k},A_{k}\eta_{k}\right\rangle D_{k}^{\ast}D_{k}. 
\end{align*}
\end{prop}

\textbf{Proof:} This is immediate by expansion upon noting that
\begin{align}
\sum_{k\in\mathbb{Z}_{\ast}^{3}}\sum_{p,q\in L_{k}}2\left\langle e_{p},A_{k}e_{q}\right\rangle \left\langle e_{q},\eta_{k}\right\rangle b_{k}^{\ast}\mleft(C_{k}e_{p}\mright)D_{k} & =2\sum_{k\in S} b_{k}^{\ast}\mleft(C_{k}A_{k}\eta_{k}\mright)D_{k},\\
\sum_{k\in\mathbb{Z}_{\ast}^{3}}\sum_{p,q\in L_{k}}2\left\langle e_{p},A_{k}e_{q}\right\rangle \left\langle \eta_{k},e_{p}\right\rangle \left\langle e_{q},\eta_{k}\right\rangle D_{k}^{\ast}D_{k} & =\sum_{k\in S}2\left\langle \eta_{k},A_{k}\eta_{k}\right\rangle D_{k}^{\ast}D_{k},\nonumber 
\end{align}
and (using also that the quantities $\left\langle e_{q},\eta_{k}\right\rangle $
are real and obey $\left\langle e_{-q},\eta_{-k}\right\rangle =\left\langle e_{q},\eta_{k}\right\rangle $)
\begin{align}
\sum_{k\in\mathbb{Z}_{\ast}^{3}}\sum_{p,q\in L_{k}}2\left\langle e_{p},A_{k}e_{q}\right\rangle \left\langle e_{q},\eta_{k}\right\rangle b_{-k}\mleft(S_{-k}e_{-p}\mright)D_{k} & =2\sum_{k\in S}b_{-k}\mleft(S_{-k}A_{-k}\eta_{-k}\mright)D_{-k}^{\ast}\label{eq:SkDkRewrite}\\
 & =2\sum_{k\in S}D_{k}^{\ast}b_{k}\mleft(S_{k}A_{k}\eta_{k}\mright)\nonumber 
\end{align}
as it holds in general that $\left[b_{k}\mleft(\cdot\mright),D_{k}^{\ast}\right]=0$.

$\hfill\square$

This allows us to conclude a generalization of Proposition \ref{prop:InteractionIdentity}:
\begin{prop}
\label{prop:ModifiedInteractionIdentity}It holds that
\begin{align*}
 & \qquad\,\sum_{k\in\mathbb{Z}_{\ast}^{3}}\mleft(2\,Q_{1}^{k}\mleft(P_{k}\mright)+Q_{2}^{k}\mleft(P_{k}\mright)\mright)+4\,\mathrm{Re}\sum_{k\in S}b_{k}^{\ast}\mleft(w_{k}\mright)D_{k}+\frac{k_{F}^{-\beta}}{2\mleft(2\pi\mright)^{3}}\sum_{k\in S}\hat{V}_{k}\frac{2\left\langle v_{k},h_{k}^{-1}v_{k}\right\rangle }{1+2\left\langle v_{k},h_{k}^{-1}v_{k}\right\rangle }D_{k}^{\ast}D_{k}\\
 & =-\sum_{k\in\mathbb{Z}_{\ast}^{3}}2\,Q_{1}^{k}\mleft(h_{k}\mright)+E_{\mathrm{corr},\mathrm{bos}}-2\sum_{k\in\mathbb{Z}_{\ast}^{3}}\varepsilon_{k,k}\mleft(e_{p};S_{k}E_{k}S_{k}^{\ast}e_{p}\mright)\\
 & +\sum_{k\in\mathbb{Z}_{\ast}^{3}}\sum_{p,q\in L_{k}}2\left\langle e_{p},E_{k}e_{q}\right\rangle \mleft(b_{k}\mleft(C_{k}e_{p}\mright)+b_{-k}^{\ast}\mleft(S_{-k}e_{-p}\mright)+\left\langle e_{p},\eta_{k}\right\rangle D_{k}\mright)^{\ast}\mleft(b_{k}\mleft(C_{k}e_{q}\mright)+b_{-k}^{\ast}\mleft(S_{-k}e_{-q}\mright)+\left\langle e_{q},\eta_{k}\right\rangle D_{k}\mright).
\end{align*}
\end{prop}

\textbf{Proof:} The only terms above which are not accounted for by
Proposition \ref{prop:InteractionIdentity} after applying the previous
proposition are the final two terms on the left-hand side. These arise
since $\eta_{k}$ obeys (for $k\in S$)
\begin{equation}
\mleft(C_{k}+S_{k}\mright)E_{k}\eta_{k}=h_{k}^{-\frac{1}{2}}E_{k}^{\frac{1}{2}}E_{k}E_{k}^{-\frac{3}{2}}h_{k}^{\frac{1}{2}}w_{k}=w_{k}
\end{equation}
whence
\begin{equation}
4\,\mathrm{Re}\sum_{k\in S}b_{k}^{\ast}\mleft(\mleft(C_{k}+S_{k}\mright)E_{k}\eta_{k}\mright)D_{k}=4\,\mathrm{Re}\sum_{k\in S}b_{k}^{\ast}\mleft(w_{k}\mright)D_{k},
\end{equation}
while by the definition of $w_{k}$ and $\eta_{k}$
\begin{equation}
\left\langle \eta_{k},E_{k}\eta_{k}\right\rangle =\left\langle w_{k},h_{k}^{\frac{1}{2}}E_{k}^{-2}h_{k}^{\frac{1}{2}}w_{k}\right\rangle =\frac{\hat{V}_{k}k_{F}^{-\beta}}{2\mleft(2\pi\mright)^{3}}\left\langle v_{k},\mleft(h_{k}+2P_{k}\mright)^{-1}v_{k}\right\rangle 
\end{equation}
and by the Sherman-Morrison formula
\begin{equation}
\mleft(h_{k}+2P_{k}\mright)^{-1}=h_{k}^{-1}-\frac{2}{1+2\left\langle v_{k},h_{k}^{-1}v_{k}\right\rangle }h_{k}^{-1}P_{k}h_{k}^{-1}
\end{equation}
so
\begin{equation}
\left\langle \eta_{k},E_{k}\eta_{k}\right\rangle =\frac{\hat{V}_{k}k_{F}^{-\beta}}{2\mleft(2\pi\mright)^{3}}\mleft(\left\langle v_{k},h_{k}^{-1}v_{k}\right\rangle -\frac{2\left\langle v_{k},h_{k}^{-1}v_{k}\right\rangle ^{2}}{1+2\left\langle v_{k},h_{k}^{-1}v_{k}\right\rangle }\mright)=\frac{\hat{V}_{k}k_{F}^{-\beta}}{2\mleft(2\pi\mright)^{3}}\frac{\left\langle v_{k},h_{k}^{-1}v_{k}\right\rangle }{1+2\left\langle v_{k},h_{k}^{-1}v_{k}\right\rangle }\label{eq:etaEketa}
\end{equation}
hence
\begin{equation}
\sum_{k\in S}2\left\langle \eta_{k},E_{k}\eta_{k}\right\rangle D_{k}^{\ast}D_{k}=\sum_{k\in S}\frac{\hat{V}_{k}k_{F}^{-\beta}}{2\mleft(2\pi\mright)^{3}}\frac{2\left\langle v_{k},h_{k}^{-1}v_{k}\right\rangle }{1+2\left\langle v_{k},h_{k}^{-1}v_{k}\right\rangle }D_{k}^{\ast}D_{k}.
\end{equation}
$\hfill\square$

\subsection{Expansion of the Kinetic Terms}

Obviously
\begin{align}
 & \quad\,\sum_{p\in\mathbb{Z}^{3}}\vert\left|p\right|^{2}-k_{F}^{2}\vert\left|\tilde{c}_{p}+d_{p}^{1}+d_{p}^{2}+d_{p}^{3}\right|^{2}\nonumber \\
 & =\sum_{p\in\mathbb{Z}^{3}}\vert\left|p\right|^{2}-k_{F}^{2}\vert\left|\tilde{c}_{p}+d_{p}^{1}+d_{p}^{2}\right|^{2}+2\,\mathrm{Re}\sum_{p\in\mathbb{Z}^{3}}\vert\left|p\right|^{2}-k_{F}^{2}\vert\,\tilde{c}_{p}^{\ast}d_{p}^{3}\label{eq:ModifiedKineticExpansion}\\
 & +\sum_{p\in\mathbb{Z}^{3}}\vert\left|p\right|^{2}-k_{F}^{2}\vert\,(d_{p}^{3})^{\ast}d_{p}^{3}+2\,\mathrm{Re}\sum_{p\in\mathbb{Z}^{3}}\vert\left|p\right|^{2}-k_{F}^{2}\vert\,(d_{p}^{1})^{\ast}d_{p}^{3}+2\,\mathrm{Re}\sum_{p\in\mathbb{Z}^{3}}\vert\left|p\right|^{2}-k_{F}^{2}\vert\,(d_{p}^{2})^{\ast}d_{p}^{3}\nonumber 
\end{align}
and the term $\sum_{p\in\mathbb{Z}^{3}}\vert\left|p\right|^{2}-k_{F}^{2}\vert\left|\tilde{c}_{p}+d_{p}^{1}+d_{p}^{2}\right|^{2}$
is what we considered in Section \ref{sec:ExtractionoftheCorrelationEnergybyFactorization},
so we examine the remaining expressions. First the simplest:
\begin{prop}
It holds that
\[
2\,\mathrm{Re}\sum_{p\in\mathbb{Z}^{3}}\vert\left|p\right|^{2}-k_{F}^{2}\vert\,\tilde{c}_{p}^{\ast}d_{p}^{3}=4\,\mathrm{Re}\sum_{k\in S}b_{k}^{\ast}\mleft(h_{k}\eta_{k}\mright)D_{k}.
\]
\end{prop}

\textbf{Proof:} It follows directly from equation (\ref{eq:dp3Definition})
that
\begin{align}
\sum_{p\in B_{F}^{c}}\vert\left|p\right|^{2}-k_{F}^{2}\vert\,\tilde{c}_{p}^{\ast}d_{p}^{3} & =+\sum_{k\in S}\sum_{p\in L_{k}}\vert\left|p\right|^{2}-k_{F}^{2}\vert\,\left\langle e_{p},\eta_{k}\right\rangle \tilde{c}_{p}^{\ast}\tilde{c}_{p-k}^{\ast}D_{k}\\
\sum_{p\in B_{F}}\vert\left|p\right|^{2}-k_{F}^{2}\vert\,\tilde{c}_{p}^{\ast}d_{p}^{3} & =-\sum_{k\in S}\sum_{p\in L_{k}}\vert\left|p-k\right|^{2}-k_{F}^{2}\vert\,\left\langle e_{p},\eta_{k}\right\rangle \tilde{c}_{p-k}^{\ast}\tilde{c}_{p}^{\ast}D_{k}\nonumber 
\end{align}
so
\begin{align}
\sum_{p\in\mathbb{Z}_{\ast}^{3}}\vert\left|p\right|^{2}-k_{F}^{2}\vert\,\tilde{c}_{p}^{\ast}d_{p}^{3} & =\sum_{k\in S}\sum_{p\in L_{k}}\mleft(\vert\left|p\right|^{2}-k_{F}^{2}\vert+\vert\left|p-k\right|^{2}-k_{F}^{2}\vert\mright)\left\langle e_{p},\eta_{k}\right\rangle \tilde{c}_{p}^{\ast}\tilde{c}_{p-k}^{\ast}D_{k}\\
 & =\sum_{k\in S}\sum_{p\in L_{k}}2\lambda_{k,p}\left\langle e_{p},\eta_{k}\right\rangle b_{k,p}^{\ast}D_{k}=2\sum_{k\in\mathbb{Z}_{\ast}^{3}}b_{k}^{\ast}\mleft(h_{k}\eta_{k}\mright)D_{k}\nonumber 
\end{align}
which implies the claim.

$\hfill\square$

For the remaining terms of equation (\ref{eq:ModifiedKineticExpansion})
we must again define a number of error terms. In the notation of Section
\ref{sec:ExtractionoftheCorrelationEnergybyFactorization} the first
of these are $\mathcal{E}_{\mathcal{C},m}=\mathcal{E}_{\mathcal{C},m}^{\mleft(1\mright)}+\mathcal{E}_{\mathcal{C},m}^{\mleft(2\mright)}$
where
\begin{align}
\mathcal{E}_{\mathcal{C},1}^{\mleft(1\mright)} & =\sum_{k,l\in S}\sum_{p\in L_{k}\cap L_{l}}\vert\left|p\right|^{2}-k_{F}^{2}\vert\,\left\langle \eta_{k},e_{p}\right\rangle \left\langle e_{p},\eta_{l}\right\rangle \tilde{c}_{p-l}^{\ast}\left[D_{k}^{\ast},D_{l}\right]\tilde{c}_{p-k}\nonumber \\
\mathcal{E}_{\mathcal{C},2}^{\mleft(1\mright)} & =\sum_{k,l\in S}\sum_{p\in L_{k}\cap L_{l}}\vert\left|p\right|^{2}-k_{F}^{2}\vert\,\left\langle \eta_{k},e_{p}\right\rangle \left\langle e_{p},\eta_{l}\right\rangle \tilde{c}_{p-l}^{\ast}D_{k}^{\ast}\left[\tilde{c}_{p-k},D_{l}\right]\label{eq:ECPureErrorTerms}\\
\mathcal{E}_{\mathcal{C},3}^{\mleft(1\mright)} & =\sum_{k,l\in S}\sum_{p\in L_{k}\cap L_{l}}\vert\left|p\right|^{2}-k_{F}^{2}\vert\,\left\langle \eta_{k},e_{p}\right\rangle \left\langle e_{p},\eta_{l}\right\rangle \left[\tilde{c}_{p-l},D_{k}\right]^{\ast}\left[\tilde{c}_{p-k},D_{l}\right]\nonumber 
\end{align}
and the substitutions in going from $\mathcal{E}_{\mathcal{C},m}^{\mleft(1\mright)}$
to $\mathcal{E}_{\mathcal{C},m}^{\mleft(2\mright)}$ now also includes
$\left\langle \eta_{k},e_{p}\right\rangle \rightarrow\left\langle \eta_{k},e_{p+k}\right\rangle $.

\bigskip
We can then state
\begin{prop}
It holds that
\[
\sum_{p\in\mathbb{Z}^{3}}\vert\left|p\right|^{2}-k_{F}^{2}\vert\,(d_{p}^{3})^{\ast}d_{p}^{3}=2\sum_{k\in S}\left\langle \eta_{k},h_{k}\eta_{k}\right\rangle D_{k}^{\ast}D_{k}-\sum_{p\in\mathbb{Z}^{3}}\vert\left|p\right|^{2}-k_{F}^{2}\vert\,d_{p}^{3}(d_{p}^{3})^{\ast}-\mathcal{E}_{\mathcal{C},1}-2\,\mathrm{Re}\mleft(\mathcal{E}_{\mathcal{C},2}\mright)-\mathcal{E}_{\mathcal{C},3}.
\]
\end{prop}

\textbf{Proof:} By the definition of equation (\ref{eq:dp3Definition})
we have that for $p\in B_{F}^{c}$
\newpage
\begin{align}
\sum_{p\in B_{F}^{c}}\vert\left|p\right|^{2}-k_{F}^{2}\vert\,(d_{p}^{3})^{\ast}d_{p}^{3} & =\sum_{k,l\in S}\sum_{p\in B_{F}^{c}}1_{L_{k}\cap L_{l}}\mleft(p\mright)\vert\left|p\right|^{2}-k_{F}^{2}\vert\,\left\langle \eta_{k},e_{p}\right\rangle \left\langle e_{p},\eta_{l}\right\rangle D_{k}^{\ast}\tilde{c}_{p-k}\tilde{c}_{p-l}^{\ast}D_{l}\nonumber \\
 & =\sum_{k\in S}\sum_{p\in L_{k}}\vert\left|p\right|^{2}-k_{F}^{2}\vert\,\left|\left\langle e_{p},\eta_{k}\right\rangle \right|^{2}D_{k}^{\ast}D_{k}\\
 & -\sum_{k,l\in S}\sum_{p\in L_{k}\cap L_{l}}\vert\left|p\right|^{2}-k_{F}^{2}\vert\,\left\langle \eta_{k},e_{p}\right\rangle \left\langle e_{p},\eta_{l}\right\rangle D_{k}^{\ast}\tilde{c}_{p-l}^{\ast}\tilde{c}_{p-k}D_{l}\nonumber 
\end{align}
and similarly, when $p\in B_{F}$,
\begin{align}
\sum_{p\in B_{F}}\vert\left|p\right|^{2}-k_{F}^{2}\vert\,(d_{p}^{3})^{\ast}d_{p}^{3} & =\sum_{k\in S}\sum_{p\in L_{k}}\vert\left|p-k\right|^{2}-k_{F}^{2}\vert\,\left|\left\langle e_{p},\eta_{k}\right\rangle \right|^{2}D_{k}^{\ast}D_{k}\\
 & -\sum_{k,l\in S}\sum_{p\in\mleft(L_{k}-k\mright)\cap\mleft(L_{l}-l\mright)}\vert\left|p\right|^{2}-k_{F}^{2}\vert\,\left\langle \eta_{k},e_{p+k}\right\rangle \left\langle e_{p+l},\eta_{l}\right\rangle D_{k}^{\ast}\tilde{c}_{p+l}^{\ast}\tilde{c}_{p+k}D_{l}.\nonumber 
\end{align}
The leading terms combine to form
\begin{align}
 & \quad\,\sum_{k\in S}\sum_{p\in L_{k}}\mleft(\vert\left|p\right|^{2}-k_{F}^{2}\vert+\vert\left|p-k\right|^{2}-k_{F}^{2}\vert\mright)\left|\left\langle e_{p},\eta_{k}\right\rangle \right|^{2}D_{k}^{\ast}D_{k}\\
 & =\sum_{k\in S}\mleft(\sum_{p\in L_{k}}2\lambda_{k,p}\left|\left\langle e_{p},\eta_{k}\right\rangle \right|^{2}\mright)D_{k}^{\ast}D_{k}=2\sum_{k\in S}\left\langle \eta_{k},h_{k}\eta_{k}\right\rangle D_{k}^{\ast}D_{k}\nonumber 
\end{align}
while the remaining terms obey e.g.
\begin{align}
 & \quad\,\sum_{k,l\in S}\sum_{p\in L_{k}\cap L_{l}}\vert\left|p\right|^{2}-k_{F}^{2}\vert\,\left\langle \eta_{k},e_{p}\right\rangle \left\langle e_{p},\eta_{l}\right\rangle D_{k}^{\ast}\tilde{c}_{p-l}^{\ast}\tilde{c}_{p-k}D_{l}\nonumber \\
 & =\sum_{k,l\in S}\sum_{p\in L_{k}\cap L_{l}}\vert\left|p\right|^{2}-k_{F}^{2}\vert\,\left\langle \eta_{k},e_{p}\right\rangle \left\langle e_{p},\eta_{l}\right\rangle \tilde{c}_{p-l}^{\ast}D_{k}^{\ast}D_{l}\tilde{c}_{p-k}\nonumber \\
 & +2\,\mathrm{Re}\sum_{k,l\in S}\sum_{p\in L_{k}\cap L_{l}}\vert\left|p\right|^{2}-k_{F}^{2}\vert\,\left\langle \eta_{k},e_{p}\right\rangle \left\langle e_{p},\eta_{l}\right\rangle \tilde{c}_{p-l}^{\ast}D_{k}^{\ast}\left[\tilde{c}_{p-k},D_{l}\right]\\
 & +\sum_{k,l\in S}\sum_{p\in L_{k}\cap L_{l}}\vert\left|p\right|^{2}-k_{F}^{2}\vert\,\left\langle \eta_{k},e_{p}\right\rangle \left\langle e_{p},\eta_{l}\right\rangle \left[\tilde{c}_{p-l},D_{k}\right]^{\ast}\left[\tilde{c}_{p-k},D_{l}\right]\nonumber \\
 & =\sum_{p\in B_{F}^{c}}\vert\left|p\right|^{2}-k_{F}^{2}\vert\,d_{p}^{3}(d_{p}^{3})^{\ast}+\mathcal{E}_{\mathcal{C},1}^{\mleft(1\mright)}+2\,\mathrm{Re}\mleft(\mathcal{E}_{\mathcal{C},2}^{\mleft(1\mright)}\mright)+\mathcal{E}_{\mathcal{C},3}^{\mleft(1\mright)}.\nonumber 
\end{align}
$\hfill\square$

For the last terms of equation (\ref{eq:ModifiedKineticExpansion})
we define the final error terms by
\begin{align}
\mathcal{E}_{\mathcal{C},4}^{\mleft(1\mright)} & =\sum_{k\in\mathbb{Z}_{\ast}^{3}}\sum_{l\in S}\sum_{p\in L_{k}\cap L_{l}}\vert\left|p\right|^{2}-k_{F}^{2}\vert\,\left\langle e_{p},\eta_{l}\right\rangle \tilde{c}_{p-l}^{\ast}\left[b_{k}^{\ast}\mleft(\mleft(C_{k}-1\mright)e_{p}\mright)\tilde{c}_{p-k},D_{l}\right]\nonumber \\
\mathcal{E}_{\mathcal{C},5}^{\mleft(1\mright)} & =\sum_{k\in\mathbb{Z}_{\ast}^{3}}\sum_{l\in S}\sum_{p\in L_{k}\cap L_{l}}\vert\left|p\right|^{2}-k_{F}^{2}\vert\,\left\langle e_{p},\eta_{l}\right\rangle \tilde{c}_{p-l}^{\ast}\left[b_{-k}\mleft(S_{-k}e_{-p}\mright)\tilde{c}_{p-k},D_{l}\right]\label{eq:ECMixedErrorTerms}\\
\mathcal{E}_{\mathcal{C},6}^{\mleft(1\mright)} & =\sum_{k\in\mathbb{Z}_{\ast}^{3}}\sum_{l\in S}\sum_{p\in L_{k}\cap L_{l}}\vert\left|p\right|^{2}-k_{F}^{2}\vert\,\left\langle e_{p},\eta_{l}\right\rangle \left[b_{-k}\mleft(S_{-k}e_{-p}\mright),\tilde{c}_{p-l}^{\ast}\right]\tilde{c}_{p-k}D_{l}\nonumber 
\end{align}
and compute the following:
\begin{prop}
It holds that
\begin{align*}
2\,\mathrm{Re}\sum_{p\in\mathbb{Z}^{3}}\vert\left|p\right|^{2}-k_{F}^{2}\vert\,(d_{p}^{1})^{\ast}d_{p}^{3} & =4\,\mathrm{Re}\sum_{k\in S}b_{k}^{\ast}\mleft(\mleft(C_{k}-1\mright)h_{k}\eta_{k}\mright)D_{k}- 2\,\mathrm{Re}\sum_{p\in\mathbb{Z}^{3}}\vert\left|p\right|^{2}-k_{F}^{2}\vert\,d_{p}^{3}(d_{p}^{1})^{\ast}-2\,\mathrm{Re}\mleft(\mathcal{E}_{\mathcal{C},4}\mright)\\
2\,\mathrm{Re}\sum_{p\in\mathbb{Z}^{3}}\vert\left|p\right|^{2}-k_{F}^{2}\vert\,(d_{p}^{2})^{\ast}d_{p}^{3} & =4\,\mathrm{Re}\sum_{k\in S}b_{k}^{\ast}\mleft(S_{k}h_{k}\eta_{k}\mright)D_{k}- 2\,\mathrm{Re}\sum_{p\in\mathbb{Z}^{3}}\vert\left|p\right|^{2}-k_{F}^{2}\vert\,d_{p}^{3}(d_{p}^{2})^{\ast}-2\,\mathrm{Re}\mleft(\mathcal{E}_{\mathcal{C},5}+\mathcal{E}_{\mathcal{C},6}\mright).
\end{align*}
\end{prop}

\textbf{Proof:} As in the previous proposition it is easily verified
that
\begin{align}
\sum_{p\in B_{F}^{c}}\vert\left|p\right|^{2}-k_{F}^{2}\vert\,(d_{p}^{1})^{\ast}d_{p}^{3} & =\sum_{k\in S}\sum_{p\in L_{k}}\vert\left|p\right|^{2}-k_{F}^{2}\vert\,\left\langle e_{p},\eta_{k}\right\rangle b_{k}^{\ast}\mleft(\mleft(C_{k}-1\mright)e_{p}\mright)D_{k}\nonumber \\
 & -\sum_{k\in\mathbb{Z}_{\ast}^{3}}\sum_{l\in S}\sum_{p\in L_{k}\cap L_{l}}\vert\left|p\right|^{2}-k_{F}^{2}\vert\,\left\langle e_{p},\eta_{l}\right\rangle b_{k}^{\ast}\mleft(\mleft(C_{k}-1\mright)e_{p}\mright)\tilde{c}_{p-l}^{\ast}\tilde{c}_{p-k}D_{l}\\
\sum_{p\in B_{F}}\vert\left|p\right|^{2}-k_{F}^{2}\vert\,(d_{p}^{1})^{\ast}d_{p}^{3} & =\sum_{k\in S}\sum_{p\in L_{k}}\vert\left|p-k\right|^{2}-k_{F}^{2}\vert\,\left\langle e_{p},\eta_{k}\right\rangle b_{k}^{\ast}\mleft(\mleft(C_{k}-1\mright)e_{p}\mright)D_{k}\nonumber \\
 & -\sum_{k\in\mathbb{Z}_{\ast}^{3}}\sum_{l\in S}\sum_{p\in\mleft(L_{k}-k\mright)\cap\mleft(L_{l}-l\mright)}\vert\left|p\right|^{2}-k_{F}^{2}\vert\,\left\langle e_{p+l},\eta_{l}\right\rangle b_{k}^{\ast}\mleft(\mleft(C_{k}-1\mright)e_{p+k}\mright)\tilde{c}_{p+l}^{\ast}\tilde{c}_{p+k}D_{l}\nonumber 
\end{align}
and the first terms form
\begin{align}
 & \quad\;\;\;\sum_{k\in S}\mleft(\sum_{p\in L_{k}}\mleft(\vert\left|p\right|^{2}-k_{F}^{2}\vert+\vert\left|p-k\right|^{2}-k_{F}^{2}\vert\mright)\left\langle e_{p},\eta_{k}\right\rangle b_{k}^{\ast}\mleft(\mleft(C_{k}-1\mright)e_{p}\mright)\mright)D_{k}\\
 & =2\sum_{k\in S}\mleft(\sum_{p\in L_{k}}\left\langle e_{p},h_{k}\eta_{k}\right\rangle b_{k}^{\ast}\mleft(\mleft(C_{k}-1\mright)e_{p}\mright)\mright)D_{k}=2\sum_{k\in S}b_{k}^{\ast}\mleft(\mleft(C_{k}-1\mright)h_{k}\eta_{k}\mright)D_{k}\nonumber 
\end{align}
whereas the second terms obey
\begin{align}
 & \quad\,\sum_{k\in\mathbb{Z}_{\ast}^{3}}\sum_{l\in S}\sum_{p\in L_{k}\cap L_{l}}\vert\left|p\right|^{2}-k_{F}^{2}\vert\,\left\langle e_{p},\eta_{l}\right\rangle b_{k}^{\ast}\mleft(\mleft(C_{k}-1\mright)e_{p}\mright)\tilde{c}_{p-l}^{\ast}\tilde{c}_{p-k}D_{l}\nonumber \\
 & =\sum_{k\in\mathbb{Z}_{\ast}^{3}}\sum_{l\in S}\sum_{p\in L_{k}\cap L_{l}}\vert\left|p\right|^{2}-k_{F}^{2}\vert\,\left\langle e_{p},\eta_{l}\right\rangle \tilde{c}_{p-l}^{\ast}D_{l}b_{k}^{\ast}\mleft(\mleft(C_{k}-1\mright)e_{p}\mright)\tilde{c}_{p-k}\\
 & +\sum_{k\in\mathbb{Z}_{\ast}^{3}}\sum_{l\in S}\sum_{p\in L_{k}\cap L_{l}}\vert\left|p\right|^{2}-k_{F}^{2}\vert\,\left\langle e_{p},\eta_{l}\right\rangle \tilde{c}_{p-l}^{\ast}\left[b_{k}^{\ast}\mleft(\mleft(C_{k}-1\mright)e_{p}\mright)\tilde{c}_{p-k},D_{l}\right]\nonumber \\
 & =\sum_{p\in B_{F}^{c}}\vert\left|p\right|^{2}-k_{F}^{2}\vert\,d_{p}^{3}(d_{p}^{1})^{\ast}+\mathcal{E}_{\mathcal{C},4}^{\mleft(1\mright)}\nonumber 
\end{align}
where we also used that $\left[b_{k}^{\ast}\mleft(\cdot\mright),\tilde{c}^{\ast}\right]=0$.

For the $(d_{p}^{2})^{\ast}d_{p}^{3}$ sum one similarly finds terms
combining to yield
\begin{align}
 & \quad\;\;\;\sum_{k\in S}\sum_{p\in L_{k}}\mleft(\vert\left|p\right|^{2}-k_{F}^{2}\vert+\vert\left|p-k\right|^{2}-k_{F}^{2}\vert\mright)\left\langle e_{p},\eta_{k}\right\rangle b_{-k}\mleft(S_{-k}e_{-p}\mright)D_{k}\\
 & =2\sum_{k\in S}b_{-k}\mleft(S_{-k}h_{-k}\eta_{-k}\mright)D_{-k}^{\ast}=2\sum_{k\in S}D_{k}^{\ast}b_{k}\mleft(S_{k}h_{k}\eta_{k}\mright)\nonumber 
\end{align}
as in equation (\ref{eq:SkDkRewrite}), and additional terms of the
form
\begin{align}
 & \quad\,\sum_{k\in\mathbb{Z}_{\ast}^{3}}\sum_{l\in S}\sum_{p\in L_{k}\cap L_{l}}\vert\left|p\right|^{2}-k_{F}^{2}\vert\,\left\langle e_{p},\eta_{l}\right\rangle b_{-k}\mleft(S_{-k}e_{-p}\mright)\tilde{c}_{p-l}^{\ast}\tilde{c}_{p-k}D_{l}\nonumber \\
 & =\sum_{k\in\mathbb{Z}_{\ast}^{3}}\sum_{l\in S}\sum_{p\in L_{k}\cap L_{l}}\vert\left|p\right|^{2}-k_{F}^{2}\vert\,\left\langle e_{p},\eta_{l}\right\rangle \tilde{c}_{p-l}^{\ast}D_{l}b_{-k}\mleft(S_{-k}e_{-p}\mright)\tilde{c}_{p-k}\nonumber \\
 & +\sum_{k\in\mathbb{Z}_{\ast}^{3}}\sum_{l\in S}\sum_{p\in L_{k}\cap L_{l}}\vert\left|p\right|^{2}-k_{F}^{2}\vert\,\left\langle e_{p},\eta_{l}\right\rangle \tilde{c}_{p-l}^{\ast}\left[b_{-k}\mleft(S_{-k}e_{-p}\mright)\tilde{c}_{p-k},D_{l}\right]\\
 & +\sum_{k\in\mathbb{Z}_{\ast}^{3}}\sum_{l\in S}\sum_{p\in L_{k}\cap L_{l}}\vert\left|p\right|^{2}-k_{F}^{2}\vert\,\left\langle e_{p},\eta_{l}\right\rangle \left[b_{-k}\mleft(S_{-k}e_{-p}\mright),\tilde{c}_{p-l}^{\ast}\right]\tilde{c}_{p-k}D_{l}\nonumber \\
 & =\sum_{p\in B_{F}^{c}}\vert\left|p\right|^{2}-k_{F}^{2}\vert\,d_{p}^{3}(d_{p}^{2})^{\ast}+\mathcal{E}_{\mathcal{C},5}^{\mleft(1\mright)}+\mathcal{E}_{\mathcal{C},6}^{\mleft(1\mright)}.\nonumber 
\end{align}
$\hfill\square$

We can now conclude the generalization of Proposition \ref{prop:KineticIdentity}:

\begin{prop}
\label{prop:ModifiedKineticIdentity}It holds that
\begin{align*}
H_{\mathrm{kin}}^{\prime} & =\sum_{k\in\mathbb{Z}_{\ast}^{3}}2\,Q_{1}^{k}\mleft(h_{k}\mright)+\sum_{p\in\mathbb{Z}^{3}}\vert\left|p\right|^{2}-k_{F}^{2}\vert\mleft(\left|\tilde{c}_{p}+d_{p}^{1}+d_{p}^{2}+d_{p}^{3}\right|^{2}+\left|(d_{p}^{1}+d_{p}^{2}+d_{p}^{3})^{\ast}\right|^{2}\mright)\\
 & -\sum_{k\in\mathbb{Z}_{\ast}^{3}}\sum_{p,q\in L_{k}}2\left\langle e_{p},h_{k}e_{q}\right\rangle \mleft(b_{k}\mleft(C_{k}e_{p}\mright)+b_{-k}^{\ast}\mleft(S_{-k}e_{-p}\mright)+\left\langle e_{p},\eta_{k}\right\rangle D_{k}\mright)^{\ast}\mleft(b_{k}\mleft(C_{k}e_{q}\mright)+b_{-k}^{\ast}\mleft(S_{-k}e_{-q}\mright)+\left\langle e_{q},\eta_{k}\right\rangle D_{k}\mright)\\
 & +E_{\mathrm{corr},\mathrm{ex}}+\mathcal{E}_{\mathrm{B}}+\mathcal{E}_{\mathcal{C}}
\end{align*}
for
\[
\mathcal{E}_{\mathcal{C}}=\mathcal{E}_{\mathcal{C},1}+2\,\mathrm{Re}\mleft(\mathcal{E}_{\mathcal{C},2}\mright)+\mathcal{E}_{\mathcal{C},3}+2\,\mathrm{Re}\mleft(\mathcal{E}_{\mathcal{C},4}+\mathcal{E}_{\mathcal{C},5}+\mathcal{E}_{\mathcal{C},6}\mright).
\]
\end{prop}

\textbf{Proof:} From equation (\ref{eq:ModifiedKineticExpansion}),
the propositions above and the computation of Section \ref{sec:ExtractionoftheCorrelationEnergybyFactorization}
we have
\begin{align}
 & \quad\,\sum_{p\in\mathbb{Z}^{3}}\vert\left|p\right|^{2}-k_{F}^{2}\vert\left|\tilde{c}_{p}+d_{p}^{1}+d_{p}^{2}+d_{p}^{3}\right|^{2}\nonumber \\
 & =H_{\mathrm{kin}}^{\prime}-\sum_{k\in\mathbb{Z}_{\ast}^{3}}2\,Q_{1}^{k}\mleft(h_{k}\mright)-E_{\mathrm{corr},\mathrm{ex}}-\mathcal{E}_{\mathrm{B}}- \sum_{p\in\mathbb{Z}^{3}}\vert\left|p\right|^{2}-k_{F}^{2}\vert\left|(d_{p}^{1}+d_{p}^{2})^{\ast}\right|^{2}\nonumber \\
 & -2\,\mathrm{Re}\sum_{p\in\mathbb{Z}^{3}}\vert\left|p\right|^{2}-k_{F}^{2}\vert\,d_{p}^{3}(d_{p}^{1}+d_{p}^{2})^{\ast} - \sum_{p\in\mathbb{Z}^{3}}\vert\left|p\right|^{2}-k_{F}^{2}\vert\,d_{p}^{3}(d_{p}^{3})^{\ast}\\
 & -\mathcal{E}_{\mathcal{C},1}-2\,\mathrm{Re}\mleft(\mathcal{E}_{\mathcal{C},2}\mright)-\mathcal{E}_{\mathcal{C},3}-2\,\mathrm{Re}\mleft(\mathcal{E}_{\mathcal{C},4}+\mathcal{E}_{\mathcal{C},5}+\mathcal{E}_{\mathcal{C},6}\mright)\nonumber \\
 & +\sum_{k\in\mathbb{Z}_{\ast}^{3}}\sum_{p,q\in L_{k}}2\left\langle e_{p},h_{k}e_{q}\right\rangle \mleft(b_{k}\mleft(C_{k}e_{p}\mright)+b_{-k}^{\ast}\mleft(C_{-k}e_{-p}\mright)\mright)^{\ast}\mleft(b_{k}\mleft(C_{k}e_{q}\mright)+b_{-k}^{\ast}\mleft(C_{-k}e_{-q}\mright)\mright)\nonumber \\
 & +4\,\mathrm{Re}\sum_{k\in S}b_{k}^{\ast}\mleft(\mleft(C_{k}+S_{k}\mright)h_{k}\eta_{k}\mright)D_{k}+2\sum_{k\in S}\left\langle \eta_{k},h_{k}\eta_{k}\right\rangle D_{k}^{\ast}D_{k}.\nonumber 
\end{align}
By Proposition \ref{prop:ModifiedQuadraticExpansionIdentity} the
terms on the two final lines combine to form
\[
\sum_{k\in\mathbb{Z}_{\ast}^{3}}\sum_{p,q\in L_{k}}2\left\langle e_{p},h_{k}e_{q}\right\rangle \mleft(b_{k}\mleft(C_{k}e_{p}\mright)+b_{-k}^{\ast}\mleft(S_{-k}e_{-p}\mright)+\left\langle e_{p},\eta_{k}\right\rangle D_{k}\mright)^{\ast}\mleft(b_{k}\mleft(C_{k}e_{q}\mright)+b_{-k}^{\ast}\mleft(S_{-k}e_{-q}\mright)+\left\langle e_{q},\eta_{k}\right\rangle D_{k}\mright)
\]
whereupon the claim follows by rearranging the equation.

$\hfill\square$

Theorem \ref{them:BigFactorization} now follows by combining Proposition
\ref{prop:ModifiedInteractionIdentity} and Proposition \ref{prop:ModifiedKineticIdentity}.

\section{\label{sec:EstimationofEC}Estimation of $\mathcal{E}_{\mathcal{C}}$}

In this section we bound the new error term $\mathcal{E}_{\mathcal{C}}$
of Theorem \ref{them:BigFactorization}, which consists of six sub-terms
\begin{equation}
\mathcal{E}_{\mathcal{C}}=\mathcal{E}_{\mathcal{C},1}+2\,\mathrm{Re}\mleft(\mathcal{E}_{\mathcal{C},2}\mright)+\mathcal{E}_{\mathcal{C},3}+2\,\mathrm{Re}\mleft(\mathcal{E}_{\mathcal{C},4}+\mathcal{E}_{\mathcal{C},5}+\mathcal{E}_{\mathcal{C},6}\mright)
\end{equation}
which are given by the equations (\ref{eq:ECPureErrorTerms}) and
(\ref{eq:ECMixedErrorTerms}). Recall that these depend implicitly upon the fixed $S\subset\mathbb{Z}_{\ast}^{3}$.

We will prove the following:
\begin{thm}
\label{them:ECBound}For any $\epsilon>0$ it holds as $k_{F}\rightarrow\infty$ that
\[
\pm\mathcal{E}_{\mathcal{C}}\leq C_{\epsilon}k_{F}^{2\mleft(1-\beta\mright)+\epsilon}\mleft(k_{F}^{-\frac{1}{2}}\sum_{k\in S}\hat{V}_{k}\mright)\mleft(\sqrt{\sum_{k\in\mathbb{Z}_{\ast}^{3}}\hat{V}_{k}^{2}\min\left\{ \left|k\right|,k_{F}\right\} }+k_{F}^{-\frac{1}{2}}\sum_{k\in S}\hat{V}_{k}\mright)\mleft(H_{\mathrm{kin}}^{\prime}+k_{F}\mright)
\]
for a constant $C_{\epsilon}>0$ depending only on $\epsilon$.
\end{thm}

Before we begin the estimation, we write the terms of $\mathcal{E}_{\mathcal{C}}$
more conveniently by introducing the quantity
\begin{equation}
\tilde{\eta}_{k,p}=\begin{cases}
\sqrt{\vert\left|p\right|^{2}-k_{F}^{2}\vert}\left\langle e_{p},\eta_{k}\right\rangle  & \text{ for } p\in B_{F}^{c}\\
\sqrt{\vert\left|p\right|^{2}-k_{F}^{2}\vert}\left\langle e_{p+k},\eta_{k}\right\rangle  & \text{ for } p\in B_{F}
\end{cases},
\end{equation}
which recalling also the definitions of Section \ref{sec:EstimationofEB}
lets us represent the different expressions defining $\mathcal{E}_{\mathcal{C}}$
by the schematic forms
\begin{align}
\tilde{\mathcal{E}}_{\mathcal{C},1} & =\sum_{k,l\in S}\sum_{p\in M_{k}\cap M_{l}}\tilde{\eta}_{k,p}\tilde{\eta}_{l,p}\tilde{c}_{p\mp l}^{\ast}\left[D_{k}^{\ast},D_{l}\right]\tilde{c}_{p\mp k},\nonumber \\
\tilde{\mathcal{E}}_{\mathcal{C},2} & =\sum_{k,l\in S}\sum_{p\in M_{k}\cap M_{l}}\tilde{\eta}_{k,p}\tilde{\eta}_{l,p}\tilde{c}_{p\mp l}^{\ast}D_{k}^{\ast}\left[\tilde{c}_{p\mp k},D_{l}\right],\nonumber \\
\tilde{\mathcal{E}}_{\mathcal{C},3} & =\sum_{k,l\in S}\sum_{p\in M_{k}\cap M_{l}}\tilde{\eta}_{k,p}\tilde{\eta}_{l,p}\left[\tilde{c}_{p\mp l},D_{k}\right]^{\ast}\left[\tilde{c}_{p\mp k},D_{l}\right],\\
\tilde{\mathcal{E}}_{\mathcal{C},4} & =\sum_{k\in\mathbb{Z}_{\ast}^{3}}\sum_{l\in S}\sum_{p\in M_{k}\cap M_{l}}\tilde{\eta}_{l,p}\tilde{c}_{p\mp l}^{\ast}\left[b_{k}^{\ast}\mleft(\varphi_{k,p}\mright)\tilde{c}_{p\mp k},D_{l}\right],\nonumber \\
\tilde{\mathcal{E}}_{\mathcal{C},5} & =\sum_{k\in\mathbb{Z}_{\ast}^{3}}\sum_{l\in S}\sum_{p\in M_{k}\cap M_{l}}\tilde{\eta}_{l,p}\tilde{c}_{p\mp l}^{\ast}\left[b_{-k}\mleft(\psi_{-k,-p}\mright)\tilde{c}_{p\mp k},D_{l}\right],\nonumber \\
\tilde{\mathcal{E}}_{\mathcal{C},6} & =\sum_{k\in\mathbb{Z}_{\ast}^{3}}\sum_{l\in S}\sum_{p\in M_{k}\cap M_{l}}\tilde{\eta}_{l,p}\left[b_{-k}\mleft(\psi_{-k,-p}\mright),\tilde{c}_{p\mp l}^{\ast}\right]\tilde{c}_{p\mp k}D_{l}.\nonumber 
\end{align}
We also recall that $D_{k}=D_{1,k}+D_{2,k}$ where
\begin{align}
D_{1,k} & =\sum_{p\in B_{F}^{c}\cap\mleft(B_{F}^{c}+k\mright)}c_{p-k}^{\ast}c_{p}=+\sum_{p\in B_{F}^{c}\cap\mleft(B_{F}^{c}-k\mright)}\tilde{c}_{p}^{\ast}\tilde{c}_{p+k},\\
D_{2,k} & =\sum_{p\in B_{F}\cap\mleft(B_{F}+k\mright)}c_{p-k}^{\ast}c_{p}=-\sum_{p\in B_{F}\cap\mleft(B_{F}+k\mright)}\tilde{c}_{p}^{\ast}\tilde{c}_{p-k}\nonumber 
\end{align}
which we can abbreviate as
\begin{equation}
D_{j,k}=\pm\sum_{p\in B_{F}^{\circ}\cap\mleft(B_{F}^{\circ}\mp k\mright)}\tilde{c}_{p}^{\ast}\tilde{c}_{p\pm k},\quad\mleft(B_{F}^{\circ},\pm\mright)=\begin{cases}
\mleft(B_{F}^{c},+\mright) & \text{ for } j=1\\
\mleft(B_{F},-\mright) & \text{ for } j=2
\end{cases}.
\end{equation}

\subsection{Estimation of $\mathcal{\tilde{E}}_{\mathcal{C},1}$, $\mathcal{\tilde{E}}_{\mathcal{C},2}$
and $\mathcal{\tilde{E}}_{\mathcal{C},3}$}

We begin with the error terms arising from the $(d_{p}^{3})^{\ast}d_{p}^{3}$
part of the factorization. For $\tilde{\mathcal{E}}_{\mathcal{C},1}$
we need to calculate the commutator $\left[D_{k}^{\ast},D_{l}\right]$.
Since $[D_{1,k},D_{2,l}^{\ast}]=0$ we need only consider the commutator
$[D_{j,k}^{\ast},D_{j,l}]$. This we compute to be
\begin{align}
\left[D_{j,k}^{\ast},D_{j,l}\right] & =\sum_{p\in B_{F}^{\circ}\cap\mleft(B_{F}^{\circ}\mp k\mright)}\sum_{q\in B_{F}^{\circ}\cap\mleft(B_{F}^{\circ}\mp l\mright)}\left[\tilde{c}_{p\pm k}^{\ast}\tilde{c}_{p},\tilde{c}_{q}^{\ast}\tilde{c}_{q\pm l}\right]\label{eq:DjkDjlCommutator}\\
 & =\sum_{p\in B_{F}^{\circ}\cap\mleft(B_{F}^{\circ}\mp k\mright)}\sum_{q\in B_{F}^{\circ}\cap\mleft(B_{F}^{\circ}\mp l\mright)}\tilde{c}_{p\pm k}^{\ast}\left\{ \tilde{c}_{p},\tilde{c}_{q}^{\ast}\right\} \tilde{c}_{q\pm l}-\sum_{p\in B_{F}^{\circ}\cap\mleft(B_{F}^{\circ}\mp k\mright)}\sum_{q\in B_{F}^{\circ}\cap\mleft(B_{F}^{\circ}\mp l\mright)}\tilde{c}_{q}^{\ast}\left\{ \tilde{c}_{p\pm k}^{\ast},\tilde{c}_{q\pm l}\right\} \tilde{c}_{p}\nonumber \\
 & =\sum_{q\in B_{F}^{\circ}\cap\mleft(B_{F}^{\circ}\mp k\mright)\cap\mleft(B_{F}^{\circ}\mp l\mright)}\tilde{c}_{q\pm k}^{\ast}\tilde{c}_{q\pm l}-\sum_{q\in B_{F}^{\circ}\cap\mleft(B_{F}^{\circ}\pm k\mright)\cap\mleft(B_{F}^{\circ}\pm l\mright)}\tilde{c}_{q\mp l}^{\ast}\tilde{c}_{q\mp k}.\nonumber 
\end{align}
We can now estimate $\tilde{\mathcal{E}}_{\mathcal{C},1}$ as follows:
\begin{prop}
\label{prop:EC1Bound}It holds as $k_{F}\rightarrow\infty$ that
\[
\pm\tilde{\mathcal{E}}_{\mathcal{C},1}\leq2\mleft(\sum_{k\in S}\sqrt{\sum_{p\in M_{k}}\tilde{\eta}_{k,p}^{2}}\mright)^{2}H_{\mathrm{kin}}^{\prime}.
\]
\end{prop}

\textbf{Proof:} For any $\Psi\in D\mleft(H_{\mathrm{kin}}^{\prime}\mright)$
we can estimate
\begin{align}
\left|\left\langle \Psi,\tilde{\mathcal{E}}_{\mathcal{C},1}\Psi\right\rangle \right| & \leq\sum_{k,l\in S}\sum_{p\in M_{k}\cap M_{l}}\sum_{q\in B_{F}^{\circ}\cap\mleft(B_{F}^{\circ}\mp k\mright)\cap\mleft(B_{F}^{\circ}\mp l\mright)}\left|\tilde{\eta}_{k,p}\tilde{\eta}_{l,p}\right|\left\Vert \tilde{c}_{q\pm k}\tilde{c}_{p\mp l}\Psi\right\Vert \left\Vert \tilde{c}_{q\pm l}\tilde{c}_{p\mp k}\Psi\right\Vert \\
 & +\sum_{k,l\in S}\sum_{p\in M_{k}\cap M_{l}}\sum_{q\in B_{F}^{\circ}\cap\mleft(B_{F}^{\circ}\pm k\mright)\cap\mleft(B_{F}^{\circ}\pm l\mright)}\left|\tilde{\eta}_{k,p}\tilde{\eta}_{l,p}\right|\left\Vert \tilde{c}_{q\mp l}\tilde{c}_{p\mp l}\Psi\right\Vert \left\Vert \tilde{c}_{q\mp k}\tilde{c}_{p\mp k}\Psi\right\Vert \nonumber 
\end{align}
and we focus on the first sum. Using that $\Vert c_{p\mp k}\Vert_{\mathrm{op}}\leq1$
we can bound this by
\begin{equation}
\mleft(\sum_{k\in S}\sqrt{\sum_{p\in M_{k}}\sum_{q\in\mleft(B_{F}^{\circ}\mp k\mright)}\tilde{\eta}_{k,p}^{2}\left\Vert \tilde{c}_{q\pm k}\Psi\right\Vert ^{2}}\mright)^{2}\leq\mleft(\sum_{k\in S}\sqrt{\sum_{p\in M_{k}}\tilde{\eta}_{k,p}^{2}}\mright)^{2}\left\langle \Psi,\mathcal{N}_{E}\Psi\right\rangle 
\end{equation}
whence the claim follows since $\mathcal{N}_{E}\leq H_{\mathrm{kin}}^{\prime}$.

$\hfill\square$

For $\tilde{\mathcal{E}}_{\mathcal{C},2}$ and $\tilde{\mathcal{E}}_{\mathcal{C},3}$
we need the commutator $\left[\tilde{c}_{p\mp k},D_{l}\right]$. When
$M_{k}=L_{k}$ (so $p\mp k=p-k\in B_{F}$) this is
\begin{align}
\left[\tilde{c}_{p\mp k},D_{l}\right] & =-\sum_{q\in B_{F}\cap\mleft(B_{F}+l\mright)}\left[\tilde{c}_{p-k},\tilde{c}_{q}^{\ast}\tilde{c}_{q-l}\right]=-\sum_{q\in B_{F}\cap\mleft(B_{F}+l\mright)}\delta_{p-k,q}\tilde{c}_{q-l}\\
 & =-1_{B_{F}}\mleft(p-k-l\mright)\tilde{c}_{p-k-l}\nonumber 
\end{align}
and likewise when $M_{k}=L_{k}-k$ (so $p\mp k=p+k\in B_{F}^{c}$)
\begin{align}
\left[\tilde{c}_{p\mp k},D_{l}\right] & =\sum_{q\in B_{F}^{c}\cap\mleft(B_{F}^{c}-l\mright)}\left[\tilde{c}_{p+k},\tilde{c}_{q}^{\ast}\tilde{c}_{q+l}\right]=\sum_{q\in B_{F}^{c}\cap\mleft(B_{F}^{c}-l\mright)}\delta_{p+k,q}\tilde{c}_{q+l}\\
 & =1_{B_{F}^{c}}\mleft(p+k+l\mright)\tilde{c}_{p+k+l}.\nonumber 
\end{align}
We can summarize these in the common expression
\begin{equation}
\left[\tilde{c}_{p\mp k},D_{l}\right]=\mp1_{B_{F}^{\circ}}\mleft(p\mp k\mp l\mright)\tilde{c}_{p\mp k\mp l},\quad B_{F}^{\circ}=\begin{cases}
B_{F} & M_{k}=L_{k}\\
B_{F}^{c} & M_{k}=L_{k}-k
\end{cases},\label{eq:cpkDlCommutator}
\end{equation}
and write
\begin{align}
\tilde{\mathcal{E}}_{\mathcal{C},2} & =\sum_{k,l\in S}\sum_{p\in M_{k}\cap M_{l}}\tilde{\eta}_{k,p}\tilde{\eta}_{l,p}\tilde{c}_{p\mp l}^{\ast}D_{k}^{\ast}\left[\tilde{c}_{p\mp k},D_{l}\right]\\
 & =\sum_{k,l\in S}\sum_{p\in M_{k}\cap M_{l}}\tilde{\eta}_{k,p}\tilde{\eta}_{l,p}D_{k}^{\ast}\tilde{c}_{p\mp l}^{\ast}\left[\tilde{c}_{p\mp k},D_{l}\right]+\tilde{\mathcal{E}}_{\mathcal{C},3}=\tilde{\mathcal{E}}_{\mathcal{C},2,2}+\tilde{\mathcal{E}}_{\mathcal{C},3}\nonumber 
\end{align}
where $\tilde{\mathcal{E}}_{\mathcal{C},2,2}$ is then
\begin{align}
\tilde{\mathcal{E}}_{\mathcal{C},2,2} & =\sum_{k,l\in S}\sum_{p\in M_{k}\cap M_{l}}\tilde{\eta}_{k,p}\tilde{\eta}_{l,p}D_{k}^{\ast}\tilde{c}_{p\mp l}^{\ast}\left[\tilde{c}_{p\mp k},D_{l}\right]\label{eq:EC22Definition}\\
 & =\mp\sum_{k\in S}\sum_{p\in M_{k}}\tilde{\eta}_{k,p}D_{k}^{\ast}\mleft(\sum_{l\in S}1_{M_{l}}\mleft(p\mright)1_{B_{F}^{\circ}}\mleft(p\mp k\mp l\mright)\tilde{\eta}_{l,p}\tilde{c}_{p\mp l}^{\ast}\tilde{c}_{p\mp k\mp l}\mright).\nonumber 
\end{align}
To handle the presence of the $D_{k}^{\ast}$ factor we need the following:
\begin{prop}
\label{prop:DkDkKineticEstimate}For any $k\in\mathbb{Z}_{\ast}^{3}$
and $\epsilon>0$ it holds that
\[
D_{1,k}^{\ast}D_{1,k},\,D_{2,k}^{\ast}D_{2,k}\leq C_{\epsilon}k_{F}^{1+\epsilon}H_{\mathrm{kin}}^{\prime}
\]
for a constant $C_{\epsilon}>0$ depending only on $\epsilon$.
\end{prop}

\textbf{Proof:} The bound for $D_{2,k}$ follows immediately from
Lemma \ref{lemma:GeneralKineticSum} as
\begin{align}
\left\Vert D_{2,k}\Psi\right\Vert  & \leq\sum_{p\in B_{F}\cap\mleft(B_{F}+k\mright)}\left\Vert \tilde{c}_{p}^{\ast}\tilde{c}_{p-k}\Psi\right\Vert \leq\sqrt{\sum_{p\in B_{F}\cap\mleft(B_{F}+k\mright)}\frac{1}{\vert\left|p-k\right|^{2}-\zeta\vert}}\sqrt{\sum_{p\in B_{F}\cap\mleft(B_{F}+k\mright)}\vert\left|p-k\right|^{2}-\zeta\vert\left\Vert \tilde{c}_{p-k}\Psi\right\Vert ^{2}}\nonumber \\
 & \leq\sqrt{C_{\epsilon}k_{F}^{1+\epsilon}\left\langle \Psi,H_{\mathrm{kin}}^{\prime}\Psi\right\rangle }.
\end{align}
For $D_{1,k}$ we define the sets
\begin{equation}
A_{1}=\left\{ p\in B_{F}^{c}\vert\left|p\right|\leq2k_{F}\right\} ,\quad A_{2}=\left\{ p\in B_{F}^{c}\vert\left|p\right|>2k_{F}\right\} ,
\end{equation}
and use the triangle inequality to see that
\begin{equation}
\left\Vert D_{1,k}\Psi\right\Vert \leq\mleft(\sum_{p\in A_{1}\cap\mleft(A_{1}-k\mright)}+\sum_{p\in A_{1}\cap\mleft(A_{2}-k\mright)}+\sum_{p\in A_{2}\cap\mleft(A_{1}-k\mright)}\mright)\left\Vert \tilde{c}_{p}^{\ast}\tilde{c}_{p+k}\Psi\right\Vert +\left\Vert D_{0,k}\Psi\right\Vert 
\end{equation}
where
\begin{equation}
D_{0,k}=\sum_{p\in A_{2}\cap\mleft(A_{2}-k\mright)}\tilde{c}_{p}^{\ast}\tilde{c}_{p+k}.
\end{equation}
The first three sums can be estimated in the same manner as we did
$D_{2,k}^{\ast}D_{2,k}$, so we need only consider $D_{0,k}$ further.
For this we note that
\begin{align}
D_{0,k}^{\ast}D_{0,k} & =\sum_{p,q\in A_{2}\cap\mleft(A_{2}-k\mright)}\tilde{c}_{p+k}^{\ast}\tilde{c}_{p}\tilde{c}_{q}^{\ast}\tilde{c}_{q+k}\\
 & =\sum_{p,q\in A_{2}\cap\mleft(A_{2}-k\mright)}\tilde{c}_{q}^{\ast}\tilde{c}_{p+k}^{\ast}\tilde{c}_{p}\tilde{c}_{q+k}+\sum_{p\in A_{2}\cap\mleft(A_{2}-k\mright)}\tilde{c}_{p+k}^{\ast}\tilde{c}_{p+k}\leq(\mathcal{N}_{E}^{\prime})^{2}\nonumber 
\end{align}
where $\mathcal{N}_{E}^{\prime}=\sum_{p\in A_{2}}\tilde{c}_{p}^{\ast}\tilde{c}_{p}$,
since
\begin{align}
\sum_{p,q\in A_{2}\cap\mleft(A_{2}-k\mright)}\left\langle \Psi,\tilde{c}_{q}^{\ast}\tilde{c}_{p+k}^{\ast}\tilde{c}_{p}\tilde{c}_{q+k}\Psi\right\rangle  & \leq\sqrt{\sum_{p,q\in A_{2}\cap\mleft(A_{2}-k\mright)}\left\Vert \tilde{c}_{p+k}\tilde{c}_{q}\Psi\right\Vert ^{2}}\sqrt{\sum_{p,q\in A_{2}\cap\mleft(A_{2}-k\mright)}\left\Vert \tilde{c}_{p}\tilde{c}_{q+k}\Psi\right\Vert ^{2}}\\
 & \leq\sum_{p,q\in A_{2}}\left\Vert \tilde{c}_{p}\tilde{c}_{q}\Psi\right\Vert ^{2}=\left\langle \Psi,\mathcal{N}_{E}^{\prime}\mleft(\mathcal{N}_{E}^{\prime}-1\mright)\Psi\right\rangle .\nonumber 
\end{align}
Now, $\mathcal{N}_{E}^{\prime}$ can be estimated in two different
ways. First we clearly have that
\begin{equation}
\mathcal{N}_{E}^{\prime}\leq\mathcal{N}_{E}\leq\left|B_{F}\right|\leq Ck_{F}^{3},
\end{equation}
but the condition $p\in A_{2}$ also lets us estimate
\begin{equation}
\mathcal{N}_{E}^{\prime}=\sum_{p\in A_{2}}\frac{\left|p\right|^{2}-k_{F}^{2}}{\left|p\right|^{2}-k_{F}^{2}}\tilde{c}_{p}^{\ast}\tilde{c}_{p}\leq\frac{1}{3k_{F}^{2}}\sum_{p\in A_{2}}\mleft(\left|p\right|^{2}-k_{F}^{2}\mright)\tilde{c}_{p}^{\ast}\tilde{c}_{p}\leq\frac{1}{3}k_{F}^{-2}H_{\mathrm{kin}}^{\prime}
\end{equation}
and combining the two we conclude that
\begin{equation}
D_{0,k}^{\ast}D_{0,k}\leq(\mathcal{N}_{E}^{\prime})^{2}\leq Ck_{F}H_{\mathrm{kin}}^{\prime}.
\end{equation}
$\hfill\square$

We can now estimate $\tilde{\mathcal{E}}_{\mathcal{C},2,2}$ and $\tilde{\mathcal{E}}_{\mathcal{C},3}$:
\begin{prop}
\label{prop:EC23Bound}For any $\epsilon>0$ it holds as $k_{F}\rightarrow\infty$ that
\begin{align*}
\pm\tilde{\mathcal{E}}_{\mathcal{C},2,2} & \leq C_{\epsilon}\mleft(\sum_{k\in S}\sqrt{\sum_{p\in M_{k}}\tilde{\eta}_{k,p}^{2}}\mright)\sqrt{k_{F}^{1+\epsilon}\sum_{k\in\mathbb{Z}_{\ast}^{3}}\sum_{p\in M_{k}}\tilde{\eta}_{k,p}^{2}}H_{\mathrm{kin}}^{\prime}\\
\pm\tilde{\mathcal{E}}_{\mathcal{C},3} & \leq\sum_{k\in S}\sum_{p\in M_{k}}\tilde{\eta}_{k,p}^{2}H_{\mathrm{kin}}^{\prime}
\end{align*}
for a constant $C_{\epsilon}>0$ depending only on $\epsilon$.
\end{prop}

\textbf{Proof:} From equation (\ref{eq:EC22Definition}) and Cauchy-Schwarz
we see that we can for any $\Psi\in D\mleft(H_{\mathrm{kin}}^{\prime}\mright)$
estimate
\begin{equation}
\left|\left\langle \Psi,\tilde{\mathcal{E}}_{\mathcal{C},2,2}\Psi\right\rangle \right|\leq\sqrt{\sum_{k\in\mathbb{Z}_{\ast}^{3}}\sum_{p\in M_{k}}\tilde{\eta}_{k,p}^{2}\left\Vert D_{k}\Psi\right\Vert ^{2}}\sum_{l\in S}\sqrt{\sum_{k\in\mathbb{Z}_{\ast}^{3}}\sum_{p\in M_{k}}1_{M_{l}}\mleft(p\mright)1_{B_{F}^{\circ}}\mleft(p\mp k\mp l\mright)\tilde{\eta}_{l,p}^{2}\Vert\tilde{c}_{p\mp l}^{\ast}\tilde{c}_{p\mp k\mp l}\Psi\Vert^{2}}.
\end{equation}
It is immediate from Proposition \ref{prop:DkDkKineticEstimate} that
the first factor can be bounded as
\begin{equation}
\sum_{k\in\mathbb{Z}_{\ast}^{3}}\sum_{p\in M_{k}}\tilde{\eta}_{k,p}^{2}\left\Vert D_{k}\Psi\right\Vert ^{2}\leq C_{\epsilon}k_{F}^{1+\epsilon}\sum_{k\in\mathbb{Z}_{\ast}^{3}}\sum_{p\in M_{k}}\tilde{\eta}_{k,p}^{2}\left\langle \Psi,H_{\mathrm{kin}}^{\prime}\Psi\right\rangle 
\end{equation}
so we turn to the latter. For this we simply bound
\begin{align}
 & \quad\,\sum_{l\in S}\sqrt{\sum_{k\in\mathbb{Z}_{\ast}^{3}}\sum_{p\in M_{k}}1_{M_{l}}\mleft(p\mright)1_{B_{F}^{\circ}}\mleft(p\mp k\mp l\mright)\tilde{\eta}_{l,p}^{2}\Vert\tilde{c}_{p\mp l}^{\ast}\tilde{c}_{p\mp k\mp l}\Psi\Vert^{2}}\\
 & \leq\sum_{l\in S}\sqrt{\sum_{p\in M_{l}}\tilde{\eta}_{l,p}^{2}\sum_{k\in\mathbb{Z}_{\ast}^{3}}1_{B_{F}^{\circ}}\mleft(p\mp k\mp l\mright)\left\Vert \tilde{c}_{p\mp k\mp l}\Psi\right\Vert ^{2}}\leq\sum_{l\in S}\sqrt{\sum_{p\in M_{l}}\tilde{\eta}_{l,p}^{2}}\sqrt{\left\langle \Psi,\mathcal{N}_{E}\Psi\right\rangle }\nonumber 
\end{align}
and use that $\mathcal{N}_{E}\leq H_{\mathrm{kin}}^{\prime}$.

For $\tilde{\mathcal{E}}_{\mathcal{C},3}$ we note that by equation
(\ref{eq:cpkDlCommutator}) this is
\begin{equation}
\tilde{\mathcal{E}}_{\mathcal{C},3}=\sum_{k,l\in S}\sum_{p\in M_{k}\cap M_{l}}1_{B_{F}^{\circ}}\mleft(p\mp k\mp l\mright)\tilde{\eta}_{k,p}\tilde{\eta}_{l,p}\tilde{c}_{p\mp k\mp l}^{\ast}\tilde{c}_{p\mp k\mp l}
\end{equation}
whence
\begin{align}
\left|\left\langle \Psi,\tilde{\mathcal{E}}_{\mathcal{C},3}\Psi\right\rangle \right| & \leq\sum_{k,l\in S}\sum_{p\in M_{k}\cap M_{l}}1_{B_{F}^{\circ}}\mleft(p\mp k\mp l\mright)\tilde{\eta}_{k,p}^{2}\left\Vert \tilde{c}_{p\mp k\mp l}\Psi\right\Vert ^{2}\nonumber \\
 & \leq\sum_{k\in S}\sum_{p\in M_{k}}\tilde{\eta}_{k,p}^{2}\sum_{l\in S}1_{B_{F}^{\circ}}\mleft(p\mp k\mp l\mright)\left\Vert \tilde{c}_{p\mp k\mp l}\Psi\right\Vert ^{2}\\
 & \leq\sum_{k\in S}\sum_{p\in M_{k}}\tilde{\eta}_{k,p}^{2}\left\langle \Psi,\mathcal{N}_{E}\Psi\right\rangle \leq\sum_{k\in S}\sum_{p\in M_{k}}\tilde{\eta}_{k,p}^{2}\left\langle \Psi,H_{\mathrm{kin}}^{\prime}\Psi\right\rangle .\nonumber 
\end{align}
$\hfill\square$

\subsection{Estimation of $\mathcal{\tilde{E}}_{\mathcal{C},4}$, $\mathcal{\tilde{E}}_{\mathcal{C},5}$
and $\mathcal{\tilde{E}}_{\mathcal{C},6}$}

Now we come to the ``mixed'' terms of $\mathcal{E}_{\mathcal{C}}$,
which include also $b_{k}\mleft(\cdot\mright)$ expressions. The first
of these, $\mathcal{\tilde{E}}_{\mathcal{C},4}$, is
\begin{align}
\tilde{\mathcal{E}}_{\mathcal{C},4} & =\sum_{k\in\mathbb{Z}_{\ast}^{3}}\sum_{l\in S}\sum_{p\in M_{k}\cap M_{l}}\tilde{\eta}_{l,p}\tilde{c}_{p\mp l}^{\ast}\left[b_{k}^{\ast}\mleft(\varphi_{k,p}\mright)\tilde{c}_{p\mp k},D_{l}\right]\nonumber \\
 & =\sum_{k\in\mathbb{Z}_{\ast}^{3}}\sum_{l\in S}\sum_{p\in M_{k}\cap M_{l}}\tilde{\eta}_{l,p}\tilde{c}_{p\mp l}^{\ast}\left[b_{k}^{\ast}\mleft(\varphi_{k,p}\mright),D_{l}\right]\tilde{c}_{p\mp k}\\
 & +\sum_{k\in\mathbb{Z}_{\ast}^{3}}\sum_{l\in S}\sum_{p\in M_{k}\cap M_{l}}\tilde{\eta}_{l,p}\tilde{c}_{p\mp l}^{\ast}b_{k}^{\ast}\mleft(\varphi_{k,p}\mright)\left[\tilde{c}_{p\mp k},D_{l}\right]=:\tilde{\mathcal{E}}_{\mathcal{C},4,1}+\tilde{\mathcal{E}}_{\mathcal{C},4,2}\nonumber 
\end{align}
and we can write the second, $\mathcal{\tilde{E}}_{\mathcal{C},5}$,
in the similar form
\begin{align}
\tilde{\mathcal{E}}_{\mathcal{C},5} & =\sum_{k\in\mathbb{Z}_{\ast}^{3}}\sum_{l\in S}\sum_{p\in M_{k}\cap M_{l}}\tilde{\eta}_{l,p}\tilde{c}_{p\mp l}^{\ast}\left[b_{-k}\mleft(\psi_{-k,-p}\mright),D_{l}\right]\tilde{c}_{p\mp k}\\
 & +\sum_{k\in\mathbb{Z}_{\ast}^{3}}\sum_{l\in S}\sum_{p\in M_{k}\cap M_{l}}\tilde{\eta}_{l,p}\tilde{c}_{p\mp l}^{\ast}b_{-k}\mleft(\psi_{-k,-p}\mright)\left[\tilde{c}_{p\mp k},D_{l}\right]=:\tilde{\mathcal{E}}_{\mathcal{C},5,1}+\tilde{\mathcal{E}}_{\mathcal{C},5,2}.\nonumber 
\end{align}
To bound the commutators of the form $\left[b_{k}^{\ast}\mleft(\cdot\mright),D_{l}\right]$
we prove the following:
\begin{prop}
For any $k,l\in\mathbb{Z}_{\ast}^{3}$, $p\in M_{k}$ and $\epsilon>0$
it holds as $k_{F}\rightarrow\infty$ that
\begin{align*}
\left|\left[b_{k}^{\ast}\mleft(\varphi_{k,p}\mright),D_{l}\right]^{\ast}\tilde{c}_{p\mp l}\right|^{2} & \leq C_{\epsilon}k_{F}^{1+\epsilon}\max_{q\in L_{k}}\left|\left\langle e_{q},\varphi_{k,p}\right\rangle \right|^{2}H_{\mathrm{kin}}^{\prime}\\
\left|\left[b_{-k}\mleft(\psi_{-k,-p}\mright),D_{l}\right]^{\ast}\tilde{c}_{p\mp l}\right|^{2} & \leq C_{\epsilon}k_{F}^{1+\epsilon}\max_{q\in L_{k}}\left|\left\langle e_{-q},\psi_{-k,-p}\right\rangle \right|^{2}H_{\mathrm{kin}}^{\prime}+2\left\Vert \psi_{-k,-p}\right\Vert ^{2}\tilde{c}_{p\mp l}^{\ast}\tilde{c}_{p\mp l}
\end{align*}
for a constant $C_{\epsilon}>0$ depending only on $\epsilon$.
\end{prop}

\textbf{Proof:} Computing that
\begin{align}
\left[b_{k,q}^{\ast},D_{l}\right] & =\sum_{p\in B_{F}^{c}\cap\mleft(B_{F}^{c}-l\mright)}\left[\tilde{c}_{q}^{\ast}\tilde{c}_{q-k}^{\ast},\tilde{c}_{p}^{\ast}\tilde{c}_{p+l}\right]-\sum_{p\in B_{F}\cap\mleft(B_{F}+l\mright)}\left[\tilde{c}_{q}^{\ast}\tilde{c}_{q-k}^{\ast},\tilde{c}_{p}^{\ast}\tilde{c}_{p-l}\right]\nonumber \\
 & =-\sum_{p\in B_{F}^{c}\cap\mleft(B_{F}^{c}-l\mright)}\tilde{c}_{p}^{\ast}\left\{ \tilde{c}_{q}^{\ast},\tilde{c}_{p+l}\right\} \tilde{c}_{q-k}^{\ast}+\sum_{p\in B_{F}\cap\mleft(B_{F}+l\mright)}\tilde{c}_{q}^{\ast}\tilde{c}_{p}^{\ast}\left\{ \tilde{c}_{q-k}^{\ast},\tilde{c}_{p-l}\right\} \\
 & =-1_{B_{F}^{c}}\mleft(q-l\mright)\tilde{c}_{q-l}^{\ast}\tilde{c}_{q-k}^{\ast}+1_{B_{F}}\mleft(q-k+l\mright)\tilde{c}_{q}^{\ast}\tilde{c}_{q-k+l}^{\ast}\nonumber 
\end{align}
we see that the commutator $\left[b_{k}^{\ast}\mleft(\varphi_{k,p}\mright),D_{l}\right]$
can be written as
\begin{align}
 & \;\left[b_{k}^{\ast}\mleft(\varphi_{k,p}\mright),D_{l}\right]=\sum_{q\in L_{k}}\left\langle e_{q},\varphi_{k,p}\right\rangle \left[b_{k,q}^{\ast},D_{l}\right]\nonumber \\
 & =-\sum_{q\in L_{k}}1_{B_{F}^{c}}\mleft(q-l\mright)\left\langle e_{q},\varphi_{k,p}\right\rangle \tilde{c}_{q-l}^{\ast}\tilde{c}_{q-k}^{\ast}+\sum_{q\in L_{k}}1_{B_{F}}\mleft(q-k+l\mright)\left\langle e_{q},\varphi_{k,p}\right\rangle \tilde{c}_{q}^{\ast}\tilde{c}_{q-k+l}^{\ast}\label{eq:bastDlcommutator}\\
 & =\sum_{q\in L_{k}}1_{B_{F}^{c}}\mleft(q-l\mright)\left\langle e_{q},\varphi_{k,p}\right\rangle \tilde{c}_{q-k}^{\ast}\tilde{c}_{q-l}^{\ast}+\sum_{q\in\mleft(L_{k}-k\mright)}1_{B_{F}}\mleft(q+l\mright)\left\langle e_{q+k},\varphi_{k,p}\right\rangle \tilde{c}_{q+k}^{\ast}\tilde{c}_{q+l}^{\ast}.\nonumber 
\end{align}
Consequently, for any $\Psi\in D\mleft(H_{\mathrm{kin}}^{\prime}\mright)$,
\begin{align}
\left\Vert \left[b_{k}^{\ast}\mleft(\varphi_{k,p}\mright),D_{l}\right]^{\ast}\tilde{c}_{p\mp l}\Psi\right\Vert  & \leq\sum_{q\in L_{k}}1_{B_{F}^{c}}\mleft(q-l\mright)\left|\left\langle e_{q},\varphi_{k,p}\right\rangle \right|\left\Vert \tilde{c}_{q-l}\tilde{c}_{q-k}\tilde{c}_{p\mp l}\Psi\right\Vert \nonumber \\
 & +\sum_{q\in\mleft(L_{k}-k\mright)}1_{B_{F}}\mleft(q+l\mright)\left|\left\langle e_{q+k},\varphi_{k,p}\right\rangle \right|\left\Vert \tilde{c}_{q+l}\tilde{c}_{q+k}\tilde{c}_{p\mp l}\Psi\right\Vert \nonumber \\
 & \leq\max_{q\in L_{k}}\left|\left\langle e_{q},\varphi_{k,p}\right\rangle \right|\sqrt{\sum_{q\in L_{k}}\frac{1}{\vert\left|q-k\right|^{2}-\zeta\vert}}\sqrt{\sum_{q\in L_{k}}\vert\left|q-k\right|^{2}-\zeta\vert\left\Vert \tilde{c}_{q-k}\Psi\right\Vert ^{2}}\label{eq:FirstCommutatorBound}\\
 & +\max_{q\in L_{k}}\left|\left\langle e_{q},\varphi_{k,p}\right\rangle \right|\sqrt{\sum_{q\in\mleft(L_{k}-k\mright)}\frac{1}{\vert\left|q+k\right|^{2}-\zeta\vert}}\sqrt{\sum_{q\in\mleft(L_{k}-k\mright)}\vert\left|q+k\right|^{2}-\zeta\vert\left\Vert \tilde{c}_{q+k}\Psi\right\Vert ^{2}}\nonumber \\
 & \leq\sqrt{C_{\epsilon}k_{F}^{1+\epsilon}\max_{q\in L_{k}}\left|\left\langle e_{q},\varphi_{k,p}\right\rangle \right|^{2}\left\langle \Psi,H_{\mathrm{kin}}^{\prime}\Psi\right\rangle }.\nonumber 
\end{align}
For $\left[b_{-k}\mleft(\psi_{-k,-p}\mright),D_{l}\right]^{\ast}\tilde{c}_{p\mp l}$
we note that from the calculation of equation (\ref{eq:bastDlcommutator})
\begin{align}
\left[b_{-k}\mleft(\psi_{-k,-p}\mright),D_{l}\right]^{\ast} & =-\left[b_{-k}^{\ast}\mleft(\psi_{-k,-p}\mright),D_{l}^{\ast}\right]=-\left[b_{-k}^{\ast}\mleft(\psi_{-k,-p}\mright),D_{-l}\right]\nonumber \\
 & =\sum_{q\in L_{k}}1_{B_{F}^{c}}\mleft(q-l\mright)\left\langle e_{-q},\psi_{-k,-p}\right\rangle \tilde{c}_{-q+k}^{\ast}\tilde{c}_{-q+l}^{\ast}\label{eq:SecondCommutatorCalculation}\\
 & +\sum_{q\in\mleft(L_{k}-k\mright)}1_{B_{F}}\mleft(q+l\mright)\left\langle e_{-q-k},\psi_{-k,-p}\right\rangle \tilde{c}_{-q-k}^{\ast}\tilde{c}_{-q-l}^{\ast}\nonumber 
\end{align}
as $D_{l}^{\ast}=D_{-l}$. Now, note that either of these sums are in
fact of the $b^{\ast}\mleft(\cdot\mright)$ form, since the $q$ summation
ranges and indicator functions force one momenta to lie inside $B_{F}$
and the other outside. We can take advantage of this to estimate $\left|\left[b_{-k}\mleft(\psi_{-k,-p}\mright),D_{l}\right]^{\ast}\tilde{c}_{p\mp l}\right|^{2}$
by commutation, as (considering the first term for definiteness)
\begin{align}
 & \left|\sum_{q\in L_{k}}1_{B_{F}^{c}}\mleft(q-l\mright)\left\langle e_{-q},\psi_{-k,-p}\right\rangle \tilde{c}_{-q+k}^{\ast}\tilde{c}_{-q+l}^{\ast}\tilde{c}_{p\mp l}\right|^{2}=\left|\sum_{q\in L_{k}}1_{B_{F}^{c}}\mleft(q-l\mright)\left\langle \psi_{-k,-p},e_{-q}\right\rangle \tilde{c}_{-q+l}\tilde{c}_{-q+k}\tilde{c}_{p\mp l}\right|^{2}\nonumber \\
 & +\sum_{q,q'\in L_{k}}1_{B_{F}^{c}}\mleft(q-l\mright)1_{B_{F}^{c}}\mleft(q'-l\mright)\left\langle e_{-q},\psi_{-k,-p}\right\rangle \left\langle \psi_{-k,-p},e_{-q'}\right\rangle \tilde{c}_{p\mp l}^{\ast}\left[\tilde{c}_{-q+l}\tilde{c}_{-q+k},\tilde{c}_{-q'+k}^{\ast}\tilde{c}_{-q'+l}^{\ast}\right]\tilde{c}_{p\mp l}\nonumber \\
 & \leq C_{\epsilon}k_{F}^{1+\epsilon}\max_{q\in L_{k}}\left|\left\langle e_{-q},\psi_{-k,-p}\right\rangle \right|^{2}H_{\mathrm{kin}}^{\prime}+\sum_{q\in L_{k}}1_{B_{F}^{c}}\mleft(q-l\mright)\left|\left\langle e_{-q},\psi_{-k,-p}\right\rangle \right|^{2}\tilde{c}_{p\mp l}^{\ast}\tilde{c}_{p\mp l}\\
 & \leq C_{\epsilon}k_{F}^{1+\epsilon}\max_{q\in L_{k}}\left|\left\langle e_{-q},\psi_{-k,-p}\right\rangle \right|^{2}H_{\mathrm{kin}}^{\prime}+\left\Vert \psi_{-k,-p}\right\Vert ^{2}\tilde{c}_{p\mp l}^{\ast}\tilde{c}_{p\mp l}\nonumber 
\end{align}
where the first bound follows as in equation (\ref{eq:FirstCommutatorBound})
whereas the second follows from the $b^{\ast}\mleft(\cdot\mright)$
form since
\begin{align}
\left[\tilde{c}_{-q+l}\tilde{c}_{-q+k},\tilde{c}_{-q'+k}^{\ast}\tilde{c}_{-q'+l}^{\ast}\right] & =\tilde{c}_{-q+l}\left\{ \tilde{c}_{-q+k},\tilde{c}_{-q'+k}^{\ast}\right\} \tilde{c}_{-q'+l}^{\ast}-\tilde{c}_{-q'+k}^{\ast}\left\{ \tilde{c}_{-q+l},\tilde{c}_{-q'+l}^{\ast}\right\} \tilde{c}_{-q+k}\\
 & =\delta_{q,q'}\mleft(1-\tilde{c}_{-q+k}^{\ast}\tilde{c}_{-q+k}-\tilde{c}_{-q+l}^{\ast}\tilde{c}_{-q+l}\mright).\nonumber 
\end{align}
The same argument applies to the second term of equation (\ref{eq:SecondCommutatorCalculation})
and the proposition follows.

$\hfill\square$

The first error terms can then be estimated:
\begin{prop}
\label{prop:EC451Bound}For any $\epsilon>0$ it holds as $k_{F}\rightarrow\infty$ that
\[
\pm\tilde{\mathcal{E}}_{\mathcal{C},4,1}, \,\, \pm\tilde{\mathcal{E}}_{\mathcal{C},5,1}\leq C_{\epsilon}k_{F}^{1-\beta+\epsilon}\mleft(\sum_{k\in S}\sqrt{\sum_{p\in M_{k}}\tilde{\eta}_{k,p}^{2}}\mright)\sqrt{\sum_{k\in \mathbb{Z}^3_*}\hat{V}_{k}^{2}}H_{\mathrm{kin}}^{\prime}
\]
for a constant $C_{\epsilon}>0$ depending only on $\epsilon$.
\end{prop}

\textbf{Proof:} By the first bound above we can for any $\Psi\in D\mleft(H_{\mathrm{kin}}^{\prime}\mright)$
estimate
\begin{align}
\left|\left\langle \Psi,\tilde{\mathcal{E}}_{\mathcal{C},4,1}\Psi\right\rangle \right| & \leq\sum_{k\in\mathbb{Z}_{\ast}^{3}}\sum_{l\in S}\sum_{p\in M_{k}\cap M_{l}}\left|\tilde{\eta}_{l,p}\right|\left\Vert \left[b_{k}^{\ast}\mleft(\varphi_{k,p}\mright),D_{l}\right]^{\ast}\tilde{c}_{p\mp l}\Psi\right\Vert \left\Vert \tilde{c}_{p\mp k}\Psi\right\Vert \nonumber \\
 & \leq\sqrt{C_{\epsilon}k_{F}^{1+\epsilon}\left\langle \Psi,H_{\mathrm{kin}}^{\prime}\Psi\right\rangle }\sum_{l\in S}\sum_{k\in\mathbb{Z}_{\ast}^{3}}\sum_{p\in M_{k}\cap M_{l}}\left|\tilde{\eta}_{l,p}\right|\max_{q\in L_{k}}\left|\left\langle e_{q},\varphi_{k,p}\right\rangle \right|\left\Vert \tilde{c}_{p\mp k}\Psi\right\Vert \\
 & \leq\sqrt{C_{\epsilon}k_{F}^{1+\epsilon}\left\langle \Psi,H_{\mathrm{kin}}^{\prime}\Psi\right\rangle }\sum_{l\in S}\sqrt{\sum_{k\in\mathbb{Z}_{\ast}^{3}}\sum_{p\in M_{k}\cap M_{l}}\tilde{\eta}_{l,p}^{2}\left\Vert \tilde{c}_{p\mp k}\Psi\right\Vert ^{2}}\sqrt{\sum_{k\in\mathbb{Z}_{\ast}^{3}}\sum_{p\in M_{k}}\max_{q\in L_{k}}\left|\left\langle e_{q},\varphi_{k,p}\right\rangle \right|^{2}}\nonumber \\
 & \leq\mleft(\sum_{l\in S}\sqrt{\sum_{p\in M_{l}}\tilde{\eta}_{l,p}^{2}}\mright)\sqrt{C_{\epsilon}k_{F}^{1+\epsilon}\sum_{k\in\mathbb{Z}_{\ast}^{3}}\sum_{p\in M_{k}}\max_{q\in L_{k}}\left|\left\langle e_{q},\varphi_{k,p}\right\rangle \right|^{2}}\sqrt{\left\langle \Psi,H_{\mathrm{kin}}^{\prime}\Psi\right\rangle \left\langle \Psi,\mathcal{N}_{E}\Psi\right\rangle }\nonumber 
\end{align}
which upon using that $\mathcal{N}_{E}\leq H_{\mathrm{kin}}^{\prime}$
and recalling $\sum_{p\in M_{k}}\max_{q\in L_{k}}\left|\left\langle e_{q},\varphi_{k,p}\right\rangle \right|^{2}\leq Ck_{F}^{1-2\beta}\hat{V}_{k}^{2}$
gives the first bound. For the second we likewise have
\begin{align}
\left|\left\langle \Psi,\tilde{\mathcal{E}}_{\mathcal{C},5,1}\Psi\right\rangle \right| & \leq\sqrt{C_{\epsilon}k_{F}^{1+\epsilon}\left\langle \Psi,H_{\mathrm{kin}}^{\prime}\Psi\right\rangle }\sum_{k\in\mathbb{Z}_{\ast}^{3}}\sum_{l\in S}\sum_{p\in M_{k}\cap M_{l}}\left|\tilde{\eta}_{l,p}\right|\max_{q\in L_{k}}\left|\left\langle e_{-q},\psi_{-k,-p}\right\rangle \right|\left\Vert \tilde{c}_{p\mp k}\Psi\right\Vert \nonumber \\
 & +\sqrt{2}\sum_{k\in\mathbb{Z}_{\ast}^{3}}\sum_{l\in S}\sum_{p\in M_{k}\cap M_{l}}\left|\tilde{\eta}_{l,p}\right|\left\Vert \psi_{-k,-p}\right\Vert \left\Vert \tilde{c}_{p\mp l}\Psi\right\Vert \left\Vert \tilde{c}_{p\mp k}\Psi\right\Vert 
\end{align}
and the first can be estimated as we did the previous one, whereas
the second obeys
\begin{align}
 & \quad\,\sum_{k\in\mathbb{Z}_{\ast}^{3}}\sum_{l\in S}\sum_{p\in M_{k}\cap M_{l}}\left|\tilde{\eta}_{l,p}\right|\left\Vert \psi_{-k,-p}\right\Vert \left\Vert \tilde{c}_{p\mp l}\Psi\right\Vert \left\Vert \tilde{c}_{p\mp k}\Psi\right\Vert \nonumber \\
 & \leq\sum_{l\in S}\sqrt{\sum_{k\in\mathbb{Z}_{\ast}^{3}}\sum_{p\in M_{k}\cap M_{l}}\tilde{\eta}_{l,p}^{2}\left\Vert \tilde{c}_{p\mp k}\Psi\right\Vert ^{2}}\sqrt{\sum_{k\in\mathbb{Z}_{\ast}^{3}}\sum_{p\in M_{k}\cap M_{l}}\left\Vert \psi_{-k,-p}\right\Vert ^{2}\left\Vert \tilde{c}_{p\mp l}\Psi\right\Vert ^{2}}\\
 & \leq\sum_{l\in S}\sqrt{\sum_{p\in M_{l}}\tilde{\eta}_{l,p}^{2}}\sqrt{\sum_{k\in\mathbb{Z}_{\ast}^{3}}\max_{p\in M_{k}}\left\Vert \psi_{k,p}\right\Vert ^{2}}\left\langle \Psi,\mathcal{N}_{E}\Psi\right\rangle \nonumber 
\end{align}
and we recall that $\max_{p\in M_{k}}\left\Vert \psi_{k,p}\right\Vert ^{2}\leq Ck_{F}^{1-2\beta}\hat{V}_{k}^{2}$.

$\hfill\square$

Recalling equation (\ref{eq:cpkDlCommutator}), we see that $\tilde{\mathcal{E}}_{\mathcal{C},4,2}$
can be written as
\begin{align}
\tilde{\mathcal{E}}_{\mathcal{C},4,2} & =\sum_{k\in\mathbb{Z}_{\ast}^{3}}\sum_{l\in S}\sum_{p\in M_{k}\cap M_{l}}\tilde{\eta}_{l,p}\tilde{c}_{p\mp l}^{\ast}b_{k}^{\ast}\mleft(\varphi_{k,p}\mright)\left[\tilde{c}_{p\mp k},D_{l}\right]\\
 & =\mp\sum_{k\in\mathbb{Z}_{\ast}^{3}}\sum_{l\in S}\sum_{p\in M_{k}\cap M_{l}}1_{B_{F}^{\circ}}\mleft(p\mp k\mp l\mright)\tilde{\eta}_{l,p}b_{k}^{\ast}\mleft(\varphi_{k,p}\mright)\tilde{c}_{p\mp l}^{\ast}\tilde{c}_{p\mp k\mp l}\nonumber 
\end{align}
since $\left[\tilde{c}^{\ast},b^{\ast}\mleft(\cdot\mright)\right]=0$.
Now, $\left[\tilde{c}^{\ast},b\mleft(\cdot\mright)\right]\neq0$, but
we can nonetheless write $\tilde{\mathcal{E}}_{\mathcal{C},5,2}$
in the similar form
\begin{align}
\tilde{\mathcal{E}}_{\mathcal{C},5,2} & =\mp\sum_{k\in\mathbb{Z}_{\ast}^{3}}\sum_{l\in S}\sum_{p\in M_{k}\cap M_{l}}1_{B_{F}^{\circ}}\mleft(p\mp k\mp l\mright)\tilde{\eta}_{l,p}\tilde{c}_{p\mp l}^{\ast}b_{-k}\mleft(\psi_{-k,-p}\mright)\tilde{c}_{p\mp k\mp l}\\
 & =\mp\sum_{k\in\mathbb{Z}_{\ast}^{3}}\sum_{l\in S}\sum_{p\in M_{k}\cap M_{l}}1_{B_{F}^{\circ}}\mleft(p\mp k\mp l\mright)\tilde{\eta}_{l,p}b_{-k}\mleft(\psi_{-k,-p}\mright)\tilde{c}_{p\mp l}^{\ast}\tilde{c}_{p\mp k\mp l}\nonumber 
\end{align}
as equation (\ref{eq:c-pkblpsiCommutator}) implies that
\begin{equation}
\left[\tilde{c}_{p\mp l}^{\ast},b_{-k}\mleft(\psi_{-k,-p}\mright)\right]=\pm1_{M_{-k}}\mleft(p\mp k\mp l\mright)\left\langle \psi_{-k,-p},e_{p\mp k'\mp l}\right\rangle \tilde{c}_{p\mp k\mp l}\label{eq:cpastbpsiCommutator}
\end{equation}
and the two indicator functions for $p\mp k\mp l$ have disjoint support.

We now bound these terms:
\begin{prop}
\label{prop:EC452Bound}It holds as $k_{F}\rightarrow\infty$ that
\begin{align*}
\pm\tilde{\mathcal{E}}_{\mathcal{C},4,2} & \leq Ck_{F}^{1-\beta}\mleft(\sum_{k\in S}\sqrt{\sum_{p\in M_{k}}\tilde{\eta}_{k,p}^{2}}\mright)\sqrt{\sum_{k\in\mathbb{Z}_{\ast}^{3}}\hat{V}_{k}^{2}}H_{\mathrm{kin}}^{\prime}\\
\pm\tilde{\mathcal{E}}_{\mathcal{C},5,2} & \leq Ck_{F}^{1-\beta}\mleft(\sum_{k\in S}\sqrt{\sum_{p\in M_{k}}\tilde{\eta}_{k,p}^{2}}\mright)\sqrt{\sum_{k\in\mathbb{Z}_{\ast}^{3}}\hat{V}_{k}^{2}\min\left\{ \left|k\right|,k_{F}\right\} }\mleft(H_{\mathrm{kin}}^{\prime}+k_{F}\mright)
\end{align*}
for a constant $C>0$ independent of all quantities.
\end{prop}

\textbf{Proof:} For any $\Psi\in D\mleft(H_{\mathrm{kin}}^{\prime}\mright)$
we can estimate
\begin{align}
\left|\left\langle \Psi,\tilde{\mathcal{E}}_{\mathcal{C},4,2}\Psi\right\rangle \right| & \leq\sum_{k\in\mathbb{Z}_{\ast}^{3}}\sum_{l\in S}\sum_{p\in M_{k}\cap M_{l}}1_{B_{F}^{\circ}}\mleft(p\mp k\mp l\mright)\left|\tilde{\eta}_{l,p}\right|\left\Vert b_{k}\mleft(\varphi_{k,p}\mright)\Psi\right\Vert \left\Vert \tilde{c}_{p\mp l}^{\ast}\tilde{c}_{p\mp k\mp l}\Psi\right\Vert \nonumber \\
 & \leq\sum_{l\in S}\sqrt{\sum_{k\in\mathbb{Z}_{\ast}^{3}}\sum_{p\in M_{k}}\left\Vert b_{k}\mleft(\varphi_{k,p}\mright)\Psi\right\Vert ^{2}}\sqrt{\sum_{k\in\mathbb{Z}_{\ast}^{3}}\sum_{p\in M_{k}\cap M_{l}}1_{B_{F}^{\circ}}\mleft(p\mp k\mp l\mright)\tilde{\eta}_{l,p}^{2}\left\Vert \tilde{c}_{p\mp k\mp l}\Psi\right\Vert ^{2}}\\
 & \leq\sqrt{\sum_{k\in\mathbb{Z}_{\ast}^{3}}\sum_{p\in M_{k}}\left\Vert b_{k}\mleft(\varphi_{k,p}\mright)\Psi\right\Vert ^{2}}\sum_{l\in S}\sqrt{\sum_{p\in M_{l}}\tilde{\eta}_{l,p}^{2}\left\langle \Psi,\mathcal{N}_{E}\Psi\right\rangle }\nonumber 
\end{align}
and similarly
\begin{equation}
\left|\left\langle \Psi,\tilde{\mathcal{E}}_{\mathcal{C},5,2}\Psi\right\rangle \right|\leq\sqrt{\sum_{k\in\mathbb{Z}_{\ast}^{3}}\sum_{p\in M_{k}}\left\Vert b_{k}^{\ast}\mleft(\psi_{-k,-p}\mright)\Psi\right\Vert ^{2}}\sum_{l\in S}\sqrt{\sum_{p\in M_{l}}\tilde{\eta}_{l,p}^{2}\left\langle \Psi,\mathcal{N}_{E}\Psi\right\rangle }.
\end{equation}
Now, as in the Propositions \ref{prop:EB2Bound} and \ref{prop:EB4Bound}
it holds that
\begin{align}
\sum_{k\in\mathbb{Z}_{\ast}^{3}}\sum_{p\in M_{k}}\left|b_{k}\mleft(\varphi_{k,p}\mright)\right|^{2} & \leq Ck_{F}^{2\mleft(1-\beta\mright)}\sum_{k\in\mathbb{Z}_{\ast}^{3}}\hat{V}_{k}^{2}H_{\mathrm{kin}}^{\prime}\\
\sum_{k\in\mathbb{Z}_{\ast}^{3}}\sum_{p\in M_{k}}\left|b_{k}^{\ast}\mleft(\psi_{-k,-p}\mright)\right|^{2} & \leq Ck_{F}^{2\mleft(1-\beta\mright)}\sum_{k\in\mathbb{Z}_{\ast}^{3}}\hat{V}_{k}^{2}\min\left\{ \left|k\right|,k_{F}\right\} \mleft(H_{\mathrm{kin}}^{\prime}+k_{F}\mright)\nonumber 
\end{align}
from which the claim follows.

$\hfill\square$

By equation (\ref{eq:cpastbpsiCommutator}), the final error term
is
\begin{align}\label{eq:cE_C_6_def}
\mathcal{\tilde{E}}_{\mathcal{C},6} & =\mp \sum_{k\in\mathbb{Z}_{\ast}^{3}}\sum_{l\in S}\sum_{p\in M_{k}\cap M_{l}}1_{M_{-k}}\mleft(p\mp k\mp l\mright)\left\langle \psi_{-k,-p},e_{p\mp k'\mp l}\right\rangle \tilde{\eta}_{l,p}\tilde{c}_{p\mp k\mp l}\tilde{c}_{p\mp k}D_{l}\\
 & =\mp \sum_{k\in\mathbb{Z}_{\ast}^{3}}\sum_{l\in S}\sum_{p\in M_{k}\cap M_{l}}1_{M_{-k}}\mleft(p\mp k\mp l\mright)\left\langle \psi_{-k,-p},e_{p\mp k'\mp l}\right\rangle \tilde{\eta}_{l,p}D_{l}\tilde{c}_{p\mp k\mp l}\tilde{c}_{p\mp k}\nonumber 
\end{align}
where we could commute $D_{l}$ to the left due to the indicator function
of the commutator
\begin{equation}
\left[\tilde{c}_{p\mp k},D_{l}\right]=\mp1_{B_{F}^{\circ}}\mleft(p\mp k\mp l\mright)\tilde{c}_{p\mp k\mp l}
\end{equation}
and, as is readily computed,
\begin{equation}
\left[\tilde{c}_{p\mp k\mp l},D_{l}\right]=\begin{cases}
+1_{B_{F}^{c}}\mleft(p-k\mright)\tilde{c}_{p-k} & \text{ for } M_{k}=L_{k}\\
-1_{B_{F}}\mleft(p+k\mright)\tilde{c}_{p+k} & \text{ for } M_{k}=L_{k}-k
\end{cases}
\end{equation}
which vanishes for $p\in M_{k}$. Here recall  \eqref{eq:MkPkPhikDefinition} for the definition of $M_k$. The term $\mathcal{\tilde{E}}_{\mathcal{C},6}$
can be controlled as follows:
\begin{prop}
\label{prop:EC6Bound}
For any $\epsilon>0$ it holds as $k_{F}\rightarrow\infty$ that
\[
\pm\tilde{\mathcal{E}}_{\mathcal{C},6}\leq C_{\epsilon}k_{F}^{1-\beta+\epsilon}\mleft(\sum_{k\in S}\sqrt{\sum_{p\in M_{k}}\tilde{\eta}_{k,p}^{2}}\mright)\sqrt{\sum_{k\in\mathbb{Z}_{\ast}^{3}}\hat{V}_{k}^{2}}H_{\mathrm{kin}}^{\prime}
\]
for a constant $C_{\epsilon}>0$ depending only on $\epsilon$.
\end{prop}

\textbf{Proof:} For any $\Psi\in D\mleft(H_{\mathrm{kin}}^{\prime}\mright)$
we have (using Proposition \ref{prop:DkDkKineticEstimate} and $D^\ast_{l} = D_{-l}$)
\begin{align}
\left|\left\langle \Psi,\tilde{\mathcal{E}}_{\mathcal{C},6}\Psi\right\rangle \right| & \leq\sum_{k\in\mathbb{Z}_{\ast}^{3}}\sum_{l\in S}\sum_{p\in M_{k}\cap M_{l}}1_{M_{-k}}\mleft(p\mp k\mp l\mright)\left|\left\langle \psi_{-k,-p},e_{p\mp k'\mp l}\right\rangle \right|\left|\tilde{\eta}_{l,p}\right|\left\Vert D_{l}^{\ast}\Psi\right\Vert \left\Vert \tilde{c}_{p\mp k\mp l}\tilde{c}_{p\mp k}\Psi\right\Vert \nonumber \\
 & \leq\sqrt{C_{\epsilon}k_{F}^{1+\epsilon}\left\langle \Psi,H_{\mathrm{kin}}^{\prime}\Psi\right\rangle }\sum_{l\in S}\sum_{k\in\mathbb{Z}_{\ast}^{3}}\sum_{p\in M_{k}\cap M_{l}}\max_{q\in L_{k}}\left|\left\langle e_{-q},\psi_{-k,-p}\right\rangle \right|\left|\tilde{\eta}_{l,p}\right|\left\Vert \tilde{c}_{p\mp k}\Psi\right\Vert \\
 & \leq\sqrt{C_{\epsilon}k_{F}^{1+\epsilon}\left\langle \Psi,H_{\mathrm{kin}}^{\prime}\Psi\right\rangle }\sum_{l\in S}\sqrt{\sum_{k\in\mathbb{Z}_{\ast}^{3}}\sum_{p\in M_{k}\cap M_{l}}\max_{q\in L_{k}}\left|\left\langle e_{-q},\psi_{-k,-p}\right\rangle \right|^{2}}\sqrt{\sum_{k\in\mathbb{Z}_{\ast}^{3}}\sum_{p\in M_{k}\cap M_{l}}\tilde{\eta}_{l,p}^{2}\left\Vert \tilde{c}_{p\mp k}\Psi\right\Vert ^{2}}\nonumber \\
 & \leq\sqrt{C_{\epsilon}k_{F}^{1+\epsilon}\sum_{k\in\mathbb{Z}_{\ast}^{3}}\sum_{p\in M_{k}}\max_{q\in L_{k}}\left|\left\langle e_{q},\psi_{k,p}\right\rangle \right|^{2}}\sum_{l\in S}\sqrt{\sum_{p\in M_{l}}\tilde{\eta}_{l,p}^{2}}\sqrt{\left\langle \Psi,H_{\mathrm{kin}}^{\prime}\Psi\right\rangle \left\langle \Psi,\mathcal{N}_{E}\Psi\right\rangle }.\nonumber 
\end{align}
Recalling $\sum_{p\in M_{k}}\max_{q\in L_{k}}\left|\left\langle e_{q},\psi_{k,p}\right\rangle \right|^{2}\leq Ck_{F}^{1-2\beta}\hat{V}_{k}^{2}$
and using $\mathcal{N}_{E}\leq H_{\mathrm{kin}}^{\prime}$ we have
the claim.

$\hfill\Square$

We can now conclude the main result of this section:

\noindent
\textbf{Proof of Theorem \ref{them:ECBound}:} By definition of $\tilde{\eta}_{k,p}$, the sum $\sum_{p\in M_{k}}\tilde{\eta}_{k,p}^{2}$
is (as $h_{k}\leq E_{k}$)
\begin{align}
\sum_{p\in M_{k}}\tilde{\eta}_{k,p}^{2} & =\sum_{p\in L_{k}}\vert\left|p\right|^{2}-k_{F}^{2}\vert\left|\left\langle e_{p},\eta_{k}\right\rangle \right|^{2}\leq2\sum_{p\in L_{k}}\lambda_{k,p}\left|\left\langle e_{p},\eta_{k}\right\rangle \right|^{2}\\
 & =2\left\langle \eta_{k},h_{k}\eta_{k}\right\rangle \leq2\left\langle \eta_{k},E_{k}\eta_{k}\right\rangle \nonumber 
\end{align}
for both $M_{k}=L_{k}$ and $M_{k}=L_{k}-k$ (the only difference
being $\vert\left|p\right|^{2}-k_{F}^{2}\vert\rightarrow\vert\left|p-k\right|^{2}-k_{F}^{2}\vert$
in the first line).

Combining the computation in (\ref{eq:etaEketa}) and Proposition \ref{prop:SimpleLuneRiemannSums}, we have 
\begin{equation}
\left\langle \eta_{k},E_{k}\eta_{k}\right\rangle =\frac{\hat{V}_{k}k_{F}^{-\beta}}{2\mleft(2\pi\mright)^{3}}\frac{\left\langle v_{k},h_{k}^{-1}v_{k}\right\rangle }{1+2\left\langle v_{k},h_{k}^{-1}v_{k}\right\rangle }\leq Ck_{F}^{-\beta}\hat{V}_{k}\left\langle v_{k},h_{k}^{-1}v_{k}\right\rangle \leq Ck_{F}^{1-2\beta}\hat{V}_{k}^{2}
\end{equation}
from which it is seen that all the bounds of the Propositions \ref{prop:EC1Bound},
\ref{prop:EC23Bound}, \ref{prop:EC451Bound}, \ref{prop:EC452Bound}
and \ref{prop:EC6Bound} can be controlled in the claimed manner.

$\hfill\square$

\section{\label{sec:EstimationoftheRemainingTerms}Estimation of the Remaining
Terms}

The results of the previous sections can be summarized in the inequality
\begin{equation}
H_{N}\geq E_{\mathrm{FS}}+E_{\mathrm{corr},\mathrm{bos}}+E_{\mathrm{corr},\mathrm{ex}}+\mathcal{E}_{\mathrm{B}}+\mathcal{E}_{\mathcal{C}}+2^{-1}\mleft(2\pi\mright)^{-3}k_{F}^{-\beta}\mleft(\mathcal{E}_{S}+\mathcal{E}_{\mathbb{Z}_{\ast}^{3}\backslash S}\mright)
\end{equation}
where $\mathcal{E}_{S}$ and $\mathcal{E}_{\mathbb{Z}_{\ast}^{3}\backslash S}$
are given by
\begin{align}
\mathcal{E}_{S} & =\sum_{k\in S}\hat{V}_{k}\mleft(\mleft(1-\frac{2\left\langle v_{k},h_{k}^{-1}v_{k}\right\rangle }{1+2\left\langle v_{k},h_{k}^{-1}v_{k}\right\rangle }\mright)D_{k}^{\ast}D_{k}-\sum_{p\in L_{k}}\mleft(\tilde{c}_{p}^{\ast}\tilde{c}_{p}+\tilde{c}_{p-k}^{\ast}\tilde{c}_{p-k}\mright)\mright)\\
\mathcal{E}_{\mathbb{Z}_{\ast}^{3}\backslash S} & =\sum_{k\in\mathbb{Z}_{\ast}^{3}\backslash S}\hat{V}_{k}\mleft(4\,\mathrm{Re}\mleft(B_{k}^{\ast}D_{k}\mright)+D_{k}^{\ast}D_{k}-\sum_{p\in L_{k}}\mleft(\tilde{c}_{p}^{\ast}\tilde{c}_{p}+\tilde{c}_{p-k}^{\ast}\tilde{c}_{p-k}\mright)\mright)\nonumber 
\end{align}
with $\mathcal{E}_{\mathrm{B}}$ and $\mathcal{E}_{\mathcal{C}}$
obeying the estimates of Theorems \ref{them:EBEstimate} and \ref{them:ECBound}, for any fixed symmetric set $S \in \mathbb{Z}^3_\ast$.

In this section we conclude the proof of Theorem \ref{them:MainTheorem}
by estimating these final error terms. To state the main results of
this section we define
\begin{align}
\mathcal{E}_{\mathcal{Q},4} & =2\sum_{p\in B_{F}}\mleft(\sum_{k\in\mathbb{Z}_{\ast}^{3}\backslash S}1_{B_{F}}\mleft(p+k\mright)\hat{V}_{k}\mright)\tilde{c}_{p}^{\ast}\tilde{c}_{p}-2\sum_{p\in B_{F}^{c}}\mleft(\sum_{k\in\mathbb{Z}_{\ast}^{3}\backslash S}1_{B_{F}}\mleft(p-k\mright)\hat{V}_{k}\mright)\tilde{c}_{p}^{\ast}\tilde{c}_{p}\\
\mathcal{E}_{\mathcal{Q},5} & =\sum_{k\in\mathbb{Z}_{\ast}^{3}\backslash S}\hat{V}_{k}\sum_{p,q\in A\cap\mleft(A+k\mright)}\tilde{c}_{p}^{\ast}\tilde{c}_{q-k}^{\ast}\tilde{c}_{q}\tilde{c}_{p-k}\nonumber 
\end{align}
where $A=\left\{ p\in B_{F}^{c}\mid\left|p\right|>2k_{F}\right\} $.
The estimates are as follows:
\begin{thm}
\label{them:FinalErrorBounds}Let $V$ obey Assumption \ref{Assumption:Potential}. Then for any $\epsilon>0$ it holds that
\begin{align*}
\mathcal{E}_{S} & \geq-2\sum_{k\in S}\hat{V}_{k}H_{\mathrm{kin}}^{\prime}\\
\mathcal{E}_{\mathbb{Z}_{\ast}^{3}\backslash S}-\mathcal{E}_{\mathcal{Q},4}-\mathcal{E}_{\mathcal{Q},5} & \geq-C_{\epsilon}k_{F}^{1+\epsilon}\sqrt{\sum_{k\in\mathbb{Z}_{\ast}^{3}\backslash S}\hat{V}_{k}^{2}}H_{\mathrm{kin}}^{\prime}\\
\mathcal{E}_{\mathcal{Q},4} & \geq - \mleft(C_V^\prime k_F^{\frac{2}{3}} + 2\sum_{k\in S}\hat{V}_{k}\mright) H_{\mathrm{kin}}^{\prime}
\end{align*}
for constants $C_{\epsilon}, C_V^\prime>0$ depending only on $\epsilon$ and $C_V$, respectively. Furthermore,
for any $S'\subset\mathbb{Z}_{\ast}^{3}$ containing $S$ and $\overline{B}\mleft(0,3k_{F}\mright)\cap\mathbb{Z}^{3}$
,
\[
\mathcal{E}_{\mathcal{Q},5}\geq-C_{\epsilon}^{\prime}\mleft(k_{F}^{-2}\sum_{k\in S'}\hat{V}_{k}+k_{F}^{3}\mleft(\sum_{k\in\mathbb{Z}_{\ast}^{3}\backslash S'}\frac{\hat{V}_{k}}{\left|k\right|^{2}}+\sqrt{\sum_{k\in\mathbb{Z}_{\ast}^{3}\backslash S'}\hat{V}_{k}^{2}\left|k\right|^{-\mleft(1-\epsilon\mright)}}\mright)\mright)H_{\mathrm{kin}}^{\prime}
\]
for a constant $C_{\epsilon}^{\prime}>0$ depending only on $\epsilon$.
\end{thm}

\subsubsection*{Estimation of $\mathcal{E}_{S}$}

The bound for $\mathcal{E}_{S}$ is almost immediate from the observation
that
\begin{align}
\mathcal{E}_{S} & =\sum_{k\in S}\hat{V}_{k}\mleft(\mleft(1-\frac{2\left\langle v_{k},h_{k}^{-1}v_{k}\right\rangle }{1+2\left\langle v_{k},h_{k}^{-1}v_{k}\right\rangle }\mright)D_{k}^{\ast}D_{k}-\sum_{p\in L_{k}}\mleft(\tilde{c}_{p}^{\ast}\tilde{c}_{p}+\tilde{c}_{p-k}^{\ast}\tilde{c}_{p-k}\mright)\mright)\\
 & \geq-\sum_{k\in S}\hat{V}_{k}\sum_{p\in L_{k}}\mleft(\tilde{c}_{p}^{\ast}\tilde{c}_{p}+\tilde{c}_{p-k}^{\ast}\tilde{c}_{p-k}\mright)\nonumber 
\end{align}
which leads to the following:
\begin{prop}
\label{prop:ESBound}It holds that
\[
\mathcal{E}_{S}\geq-2\sum_{k\in S}\hat{V}_{k}H_{\mathrm{kin}}^{\prime}.
\]
\end{prop}

\textbf{Proof:} From the above inequality we rearrange the sums to
see that
\begin{align}
\sum_{k\in S}\hat{V}_{k}\sum_{p\in L_{k}}\mleft(\tilde{c}_{p}^{\ast}\tilde{c}_{p}+\tilde{c}_{p-k}^{\ast}\tilde{c}_{p-k}\mright) & =\sum_{p\in B_{F}^{c}}\mleft(\sum_{k\in S}1_{L_{k}}\mleft(p\mright)\hat{V}_{k}\mright)\tilde{c}_{p}^{\ast}\tilde{c}_{p}+\sum_{p\in B_{F}}\mleft(\sum_{k\in S}1_{L_{k}-k}\mleft(p\mright)\hat{V}_{k}\mright)\tilde{c}_{p}^{\ast}\tilde{c}_{p}\\
 & \leq\sum_{k\in S}\hat{V}_{k}\mleft(\sum_{p\in B_{F}^{c}}\tilde{c}_{p}^{\ast}\tilde{c}_{p}+\sum_{p\in B_{F}}\tilde{c}_{p}^{\ast}\tilde{c}_{p}\mright)\leq2\sum_{k\in S}\hat{V}_{k}\mathcal{N}_{E}\nonumber 
\end{align}
and the claim now follows from the fact that $\mathcal{N}_{E}\leq H_{\mathrm{kin}}^{\prime}$.

$\hfill\square$

\subsection{Preliminary Analysis of Large $k$ Terms}

We split the large $k$ terms into a cubic part and a quartic part
as $\mathcal{E}_{\mathbb{Z}_{\ast}^{3}\backslash S}=4\,\mathrm{Re}\mleft(\mathcal{E}_{\mathcal{C},\mathbb{Z}_{\ast}^{3}\backslash S}\mright)+\mathcal{E}_{\mathcal{Q}}$
where
\begin{equation}
\mathcal{E}_{\mathcal{C},\mathbb{Z}_{\ast}^{3}\backslash S}=\sum_{k\in\mathbb{Z}_{\ast}^{3}\backslash S}\hat{V}_{k}B_{k}^{\ast}D_{k},\quad\mathcal{E}_{\mathcal{Q}}=\sum_{k\in\mathbb{Z}_{\ast}^{3}\backslash S}\hat{V}_{k}\mleft(D_{k}^{\ast}D_{k}-\sum_{p\in L_{k}}\mleft(\tilde{c}_{p}^{\ast}\tilde{c}_{p}+\tilde{c}_{p-k}^{\ast}\tilde{c}_{p-k}\mright)\mright),
\end{equation}
and we recall that $B_{k}=\sum_{p\in L_{k}}\tilde{c}_{p-k}\tilde{c}_{p}$
and $D_{k}=D_{1,k}+D_{2,k}$ for
\begin{equation}
D_{1,k}=+\sum_{p\in B_{F}^{c}\cap\mleft(B_{F}^{c}-k\mright)}\tilde{c}_{p}^{\ast}\tilde{c}_{p+k},\quad D_{2,k}=-\sum_{p\in B_{F}\cap\mleft(B_{F}+k\mright)}\tilde{c}_{p}^{\ast}\tilde{c}_{p-k}.
\end{equation}
We split the cubic terms $\mathcal{E}_{\mathcal{C},\mathbb{Z}_{\ast}^{3}\backslash S}$
further into a $D_{1,k}$ part and a $D_{2,k}$ part as
\begin{align}
\mathcal{E}_{\mathcal{C},\mathbb{Z}_{\ast}^{3}\backslash S} & =\sum_{k\in\mathbb{Z}_{\ast}^{3}\backslash S}\hat{V}_{k}\sum_{p\in L_{k}}\tilde{c}_{p}^{\ast}\tilde{c}_{p-k}^{\ast}D_{1,k}-\sum_{k\in\mathbb{Z}_{\ast}^{3}\backslash S}\hat{V}_{k}\sum_{p\in\mleft(L_{k}-k\mright)}\tilde{c}_{p}^{\ast}\tilde{c}_{p+k}^{\ast}D_{2,k}\\
 & =:\mathcal{E}_{\mathcal{C},\mathbb{Z}_{\ast}^{3}\backslash S}^{\mleft(1\mright)}-\mathcal{E}_{\mathcal{C},\mathbb{Z}_{\ast}^{3}\backslash S}^{\mleft(2\mright)}\nonumber 
\end{align}
and for the quartic terms we note that
\begin{align}
D_{1,k}^{\ast}D_{1,k} & =+\sum_{p\in B_{F}^{c}\cap\mleft(B_{F}^{c}+k\mright)}\tilde{c}_{p}^{\ast}D_{1,k}\tilde{c}_{p-k}+\sum_{p\in B_{F}^{c}\cap\mleft(B_{F}^{c}+k\mright)}\tilde{c}_{p}^{\ast}\tilde{c}_{p}\\
D_{2,k}^{\ast}D_{2,k} & =-\sum_{p\in B_{F}\cap\mleft(B_{F}-k\mright)}\tilde{c}_{p}^{\ast}D_{2,k}\tilde{c}_{p+k}+\sum_{p\in B_{F}\cap\mleft(B_{F}-k\mright)}\tilde{c}_{p}^{\ast}\tilde{c}_{p}\nonumber 
\end{align}
and (since e.g. $D_{1,k}^{\ast}=D_{1,-k}$ and $[D_{1,k},D_{2,l}]=0$)
\begin{equation}
\sum_{k\in\mathbb{Z}_{\ast}^{3}\backslash S}\hat{V}_{k}\mleft(D_{1,k}^{\ast}D_{2,k}+D_{2,k}^{\ast}D_{1,k}\mright)=-2\sum_{k\in\mathbb{Z}_{\ast}^{3}\backslash S}\hat{V}_{k}\sum_{p\in B_{F}\cap\mleft(B_{F}-k\mright)}\tilde{c}_{p}^{\ast}D_{1,k}\tilde{c}_{p+k}
\end{equation}
whence $\mathcal{E}_{\mathcal{Q}}$ can be decomposed as
\begin{equation}
\mathcal{E}_{\mathcal{Q}}=\mathcal{E}_{\mathcal{Q},1}-\mathcal{E}_{\mathcal{Q},2}-2\,\mathcal{E}_{\mathcal{Q},3}+\mathcal{E}_{\mathcal{Q},4}
\end{equation}
where
\begin{align}
\mathcal{E}_{\mathcal{Q},1} & =\sum_{k\in\mathbb{Z}_{\ast}^{3}\backslash S}\hat{V}_{k}\sum_{p\in B_{F}^{c}\cap\mleft(B_{F}^{c}+k\mright)}\tilde{c}_{p}^{\ast}D_{1,k}\tilde{c}_{p-k}\nonumber \\
\mathcal{E}_{\mathcal{Q},2} & =\sum_{k\in\mathbb{Z}_{\ast}^{3}\backslash S}\hat{V}_{k}\sum_{p\in B_{F}\cap\mleft(B_{F}-k\mright)}\tilde{c}_{p}^{\ast}D_{2,k}\tilde{c}_{p+k}\\
\mathcal{E}_{\mathcal{Q},3} & =\sum_{k\in\mathbb{Z}_{\ast}^{3}\backslash S}\hat{V}_{k}\sum_{p\in B_{F}\cap\mleft(B_{F}-k\mright)}\tilde{c}_{p}^{\ast}D_{1,k}\tilde{c}_{p+k}\nonumber 
\end{align}
and we noted that
\begin{align}
 & \quad\,\sum_{k\in\mathbb{Z}_{\ast}^{3}\backslash S}\hat{V}_{k}\mleft(\sum_{p\in B_{F}^{c}\cap\mleft(B_{F}^{c}+k\mright)}\tilde{c}_{p}^{\ast}\tilde{c}_{p}+\sum_{p\in B_{F}\cap\mleft(B_{F}-k\mright)}\tilde{c}_{p}^{\ast}\tilde{c}_{p}-\sum_{p\in L_{k}}\mleft(\tilde{c}_{p}^{\ast}\tilde{c}_{p}+\tilde{c}_{p-k}^{\ast}\tilde{c}_{p-k}\mright)\mright)\\
 & =\sum_{k\in\mathbb{Z}_{\ast}^{3}\backslash S}\hat{V}_{k}\mleft(\sum_{p\in B_{F}^{c}}\mleft(1_{B_{F}^{c}}\mleft(p-k\mright)-1_{B_{F}}\mleft(p-k\mright)\mright)\tilde{c}_{p}^{\ast}\tilde{c}_{p}+\sum_{p\in B_{F}}\mleft(1_{B_{F}}\mleft(p+k\mright)-1_{B_{F}^{c}}\mleft(p+k\mright)\mright)\tilde{c}_{p}^{\ast}\tilde{c}_{p}\mright)\nonumber \\
 & =2\sum_{k\in\mathbb{Z}_{\ast}^{3}\backslash S}\hat{V}_{k}\mleft(-\sum_{p\in B_{F}^{c}}1_{B_{F}}\mleft(p-k\mright)\tilde{c}_{p}^{\ast}\tilde{c}_{p}+\sum_{p\in B_{F}}1_{B_{F}}\mleft(p+k\mright)\tilde{c}_{p}^{\ast}\tilde{c}_{p}\mright)=\mathcal{E}_{\mathcal{Q},4}\nonumber 
\end{align}
since $\sum_{p\in B_{F}^{c}}\tilde{c}_{p}^{\ast}\tilde{c}_{p}=\mathcal{N}_{E}=\sum_{p\in B_{F}}\tilde{c}_{p}^{\ast}\tilde{c}_{p}$.

Now we decompose $\mathcal{E}_{\mathcal{Q},1}$ further. The decomposition is needed to create a large kinetic gap for the
points in $A_2$ while keeping the total number of points in $A_1$ manageable. More specifically, we need to
isolate the terms that make up $\mathcal{E}_{{\rm Q},5}$ since these can not be estimated as the rest can. Defining
(as in Proposition \ref{prop:DkDkKineticEstimate})
\begin{equation}
A_{1}=\left\{ p\in B_{F}^{c}\mid\left|p\right|\leq2k_{F}\right\} ,\quad A_{2}=\left\{ p\in B_{F}^{c}\mid\left|p\right|>2k_{F}\right\} ,
\end{equation}
and
\begin{equation}
D_{0,k}=\sum_{p\in A_{2}\cap\mleft(A_{2}-k\mright)}\tilde{c}_{p}^{\ast}\tilde{c}_{p+k}
\end{equation}
we split the sum of $\mathcal{E}_{\mathcal{Q},1}$ into 4 parts depending
on whether $p$ and $p-k$ are in $A_{1}$ or $A_{2}$:
\begin{align}
\sum_{p\in B_{F}^{c}\cap\mleft(B_{F}^{c}+k\mright)}\tilde{c}_{p}^{\ast}D_{1,k}\tilde{c}_{p-k} & =\sum_{p\in A_{1}\cap\mleft(A_{1}+k\mright)}\tilde{c}_{p}^{\ast}D_{1,k}\tilde{c}_{p-k}+\sum_{p\in A_{2}\cap\mleft(A_{2}+k\mright)}\tilde{c}_{p}^{\ast}D_{1,k}\tilde{c}_{p-k}\\
 & +\sum_{p\in A_{2}\cap\mleft(A_{1}+k\mright)}\tilde{c}_{p}^{\ast}D_{1,k}\tilde{c}_{p-k}+\mleft(\sum_{p\in\mleft(A_{1}-k\mright)\cap A_{2}}\tilde{c}_{p}^{\ast}D_{1,-k}\tilde{c}_{p+k}\mright)^{\ast}.\nonumber 
\end{align}
The second sum of this equation can be written in terms of $D_{0,k}$
as
\begin{align}
\sum_{p\in A_{2}\cap\mleft(A_{2}+k\mright)}\tilde{c}_{p}^{\ast}D_{1,k}\tilde{c}_{p-k} & =\sum_{q\in B_{F}^{c}\cap\mleft(B_{F}^{c}-k\mright)}\tilde{c}_{q}^{\ast}D_{0,-k}\tilde{c}_{q+k}\nonumber \\
 & =\sum_{q\in A_{1}\cap\mleft(A_{1}-k\mright)}\tilde{c}_{q}^{\ast}D_{0,-k}\tilde{c}_{q+k}+\sum_{q\in A_{2}\cap\mleft(A_{2}-k\mright)}\tilde{c}_{q}^{\ast}D_{0,-k}\tilde{c}_{q+k}\\
 & +\sum_{q\in A_{2}\cap\mleft(A_{1}-k\mright)}\tilde{c}_{q}^{\ast}D_{0,-k}\tilde{c}_{q+k}+\mleft(\sum_{q\in\mleft(A_{1}+k\mright)\cap A_{2}}\tilde{c}_{q}^{\ast}D_{0,k}\tilde{c}_{q-k}\mright)^{\ast}\nonumber 
\end{align}
so all in all (substituting also $k\rightarrow-k$ in some sums to
group terms together)
\begin{equation}
\mathcal{E}_{\mathcal{Q},1}=\mathcal{E}_{\mathcal{Q},1}^{1}+2\,\mathrm{Re}\mleft(\mathcal{E}_{\mathcal{Q},1}^{2}\mright)+\mathcal{E}_{\mathcal{Q},0}^{1}+2\,\mathrm{Re}\mleft(\mathcal{E}_{\mathcal{Q},0}^{2}\mright)+\mathcal{E}_{\mathcal{Q},5}
\end{equation}
where for $a=1,2$
\begin{equation}
\mathcal{E}_{\mathcal{Q},j}^{a}=\sum_{k\in\mathbb{Z}_{\ast}^{3}\backslash S}\hat{V}_{k}\sum_{p\in A_{a}\cap\mleft(A_{1}+k\mright)}\tilde{c}_{p}^{\ast}D_{j,k}\tilde{c}_{p-k},\quad j=0,1,
\end{equation}
and we recognized that
\begin{equation}
\sum_{k\in\mathbb{Z}_{\ast}^{3}\backslash S}\hat{V}_{k}\sum_{q\in A_{2}\cap\mleft(A_{2}-k\mright)}\tilde{c}_{q}^{\ast}D_{0,-k}\tilde{c}_{q+k}=\sum_{k\in\mathbb{Z}_{\ast}^{3}\backslash S}\hat{V}_{k}\sum_{p,q\in A_{2}\cap\mleft(A_{2}+k\mright)}\tilde{c}_{q}^{\ast}\tilde{c}_{p-k}^{\ast}\tilde{c}_{p}\tilde{c}_{q-k}=\mathcal{E}_{\mathcal{Q},5}.
\end{equation}

\subsubsection*{Schematic Forms}

By the decompositions above we see that to obtain the second estimate
of Theorem \ref{them:FinalErrorBounds}, i.e. that on $\mathcal{E}_{\mathbb{Z}_{\ast}^{3}\backslash S}-\mathcal{E}_{\mathcal{Q},4}-\mathcal{E}_{\mathcal{Q},5}$,
it suffices to estimate the sums
\begin{equation}
\mathcal{E}_{\mathcal{C},\mathbb{Z}_{\ast}^{3}\backslash S}^{(1)}=\sum_{k\in\mathbb{Z}_{\ast}^{3}\backslash S}\hat{V}_{k}\sum_{p\in L_{k}}\tilde{c}_{p}^{\ast}\tilde{c}_{p-k}^{\ast}D_{1,k},
\quad\mathcal{E}_{\mathcal{C},\mathbb{Z}_{\ast}^{3}\backslash S}^{(2)}=\sum_{k\in\mathbb{Z}_{\ast}^{3}\backslash S}\hat{V}_{k}\sum_{p\in\mleft(L_{k}-k\mright)}\tilde{c}_{p}^{\ast}\tilde{c}_{p+k}^{\ast}D_{2,k},
\end{equation}
and
\begin{align}
\mathcal{E}_{\mathcal{Q},j}^{a} & =\sum_{k\in\mathbb{Z}_{\ast}^{3}\backslash S}\hat{V}_{k}\sum_{p\in A_{a}\cap\mleft(A_{1}+k\mright)}\tilde{c}_{p}^{\ast}D_{j,k}\tilde{c}_{p-k}\nonumber \\
\mathcal{E}_{\mathcal{Q},2} & =\sum_{k\in\mathbb{Z}_{\ast}^{3}\backslash S}\hat{V}_{k}\sum_{p\in B_{F}\cap\mleft(B_{F}-k\mright)}\tilde{c}_{p}^{\ast}D_{2,k}\tilde{c}_{p+k}\\
\mathcal{E}_{\mathcal{Q},3} & =\sum_{k\in\mathbb{Z}_{\ast}^{3}\backslash S}\hat{V}_{k}\sum_{p\in B_{F}\cap\mleft(B_{F}-k\mright)}\tilde{c}_{p}^{\ast}D_{1,k}\tilde{c}_{p+k}\nonumber 
\end{align}
for $a=1,2$ and $j=0,1$.

We can summarize these in two schematic forms: $\mathcal{E}_{\mathcal{C},\mathbb{Z}_{\ast}^{3}\backslash S}^{(1)}$
and $\mathcal{E}_{\mathcal{C},\mathbb{Z}_{\ast}^{3}\backslash S}^{(2)}$
are both of the form
\begin{equation}
\tilde{\mathcal{E}}_{\mathcal{C},\mathbb{Z}_{\ast}^{3}\backslash S}=\sum_{k\in\mathbb{Z}_{\ast}^{3}\backslash S}\hat{V}_{k}\sum_{p\in\mathbb{Z}^{3}}1_{S_{k}}\mleft(p\mright)\tilde{c}_{p}^{\ast}\tilde{c}_{p\mp k}^{\ast}D_{j,k}
\end{equation}
where
\begin{equation}
\mleft(S_{k},\tilde{c}_{p\mp k}^{\ast},D_{j,k}\mright)=\begin{cases}
\mleft(L_{k},\tilde{c}_{p-k}^{\ast},D_{1,k}\mright) & \text{for \ensuremath{\mathcal{E}_{\mathcal{C},\mathbb{Z}_{\ast}^{3}\backslash S}^{(1)}}}\\
\mleft(L_{k}-k,\tilde{c}_{p+k}^{\ast},D_{2,k}\mright) & \text{for }\mathcal{E}_{\mathcal{C},\mathbb{Z}_{\ast}^{3}\backslash S}^{(2)}
\end{cases},
\end{equation}
while $\mathcal{E}_{\mathcal{Q},j}^{a}$, $\mathcal{E}_{\mathcal{Q},2}$
and $\mathcal{E}_{\mathcal{Q},3}$ are all of the form
\begin{equation}
\tilde{\mathcal{E}}_{\mathcal{Q}}=\sum_{k\in\mathbb{Z}_{\ast}^{3}\backslash S}\hat{V}_{k}\sum_{p\in\mathbb{Z}^{3}}1_{S_{k}}\mleft(p\mright)\tilde{c}_{p}^{\ast}D_{j,k}\tilde{c}_{p\mp k}
\end{equation}
where
\begin{equation}
\mleft(S_{k},\tilde{c}_{p\mp k},D_{j,k}\mright)=\begin{cases}
\mleft(A_{a}\cap\mleft(A_{1}+k\mright),\tilde{c}_{p-k},D_{j,k}\mright) & \text{for \ensuremath{\mathcal{E}_{\mathcal{Q},j}^{a}}}\\
\mleft(B_{F}\cap\mleft(B_{F}-k\mright),\tilde{c}_{p+k},D_{2,k}\mright) & \text{for }\mathcal{E}_{\mathcal{Q},2}\\
\mleft(B_{F}\cap\mleft(B_{F}-k\mright),\tilde{c}_{p+k},D_{1,k}\mright) & \text{for }\mathcal{E}_{\mathcal{Q},3}
\end{cases}.
\end{equation}
It consequently suffices to estimate these schematic forms. Noting
that $\tilde{\mathcal{E}}_{\mathcal{C},\mathbb{Z}_{\ast}^{3}\backslash S}$
can be written as
\begin{equation}
\tilde{\mathcal{E}}_{\mathcal{C},\mathbb{Z}_{\ast}^{3}\backslash S}=\sum_{p\in\mathbb{Z}^{3}}\tilde{c}_{p}^{\ast}\mleft(\sum_{k\in\mathbb{Z}_{\ast}^{3}\backslash S}1_{S_{k}}\mleft(p\mright)\hat{V}_{k}\tilde{c}_{p\mp k}^{\ast}D_{j,k}\mright)
\end{equation}
we can for any $\Psi\in D\mleft(H_{\mathrm{kin}}^{\prime}\mright)$
estimate
\begin{align}
\left|\left\langle \Psi,\tilde{\mathcal{E}}_{\mathcal{C},\mathbb{Z}_{\ast}^{3}\backslash S}\Psi\right\rangle \right| & \leq\sqrt{\sum_{p\in\mathbb{Z}^{3}}\vert\left|p\right|^{2}-\zeta\vert\left\Vert \tilde{c}_{p}\Psi\right\Vert ^{2}}\sqrt{\sum_{p\in\mathbb{Z}^{3}}\frac{1}{\vert\left|p\right|^{2}-\zeta\vert}\left\Vert \sum_{k\in\mathbb{Z}_{\ast}^{3}\backslash S}1_{S_{k}}\mleft(p\mright)\hat{V}_{k}\tilde{c}_{p\mp k}^{\ast}D_{j,k}\Psi\right\Vert ^{2}}\nonumber \\
 & \leq\sqrt{\left\langle \Psi,H_{\mathrm{kin}}^{\prime}\Psi\right\rangle \sum_{p\in\mathbb{Z}^{3}}\vert\left|p\right|^{2}-\zeta\vert^{-1}\left\langle \Psi,T_{p}^{\mathcal{C}}\Psi\right\rangle }\label{eq:ECLargeEstimate}
\end{align}
where $T_{p}^{\mathcal{C}}=\left|\sum_{k\in\mathbb{Z}_{\ast}^{3}\backslash S}1_{S_{k}}\mleft(p\mright)\hat{V}_{k}\tilde{c}_{p\mp k}^{\ast}D_{j,k}\right|^{2}$,
and similarly
\begin{equation}
\left|\left\langle \Psi,\tilde{\mathcal{E}}_{\mathcal{Q}}\Psi\right\rangle \right|\leq\sqrt{\left\langle \Psi,H_{\mathrm{kin}}^{\prime}\Psi\right\rangle \sum_{p\in\mathbb{Z}^{3}}\vert\left|p\right|^{2}-\zeta\vert^{-1}\left\langle \Psi,T_{p}^{\mathcal{Q}}\Psi\right\rangle }\label{eq:EQEstimate}
\end{equation}
for $T_{p}^{\mathcal{Q}}=\left|\sum_{k\in\mathbb{Z}_{\ast}^{3}\backslash S}1_{S_{k}}\mleft(p\mright)\hat{V}_{k}D_{j,k}\tilde{c}_{p\mp k}\right|^{2}$.

\subsection{Estimation of $T_{p}^{\mathcal{C}}$ and $T_{p}^{\mathcal{Q}}$}

By expansion and (anti-)commutation we can write $T_{p}^{\mathcal{C}}$
as
\begin{align}
T_{p}^{\mathcal{C}} & =\sum_{k,l\in\mathbb{Z}_{\ast}^{3}\backslash S}1_{S_{k}}\mleft(p\mright)1_{S_{l}}\mleft(p\mright)\hat{V}_{k}\hat{V}_{l}D_{j,k}^{\ast}\tilde{c}_{p\mp k}\tilde{c}_{p\mp l}^{\ast}D_{j,l}\nonumber \\
 & =-\sum_{k,l\in\mathbb{Z}_{\ast}^{3}\backslash S}1_{S_{k}}\mleft(p\mright)1_{S_{l}}\mleft(p\mright)\hat{V}_{k}\hat{V}_{l}D_{j,k}^{\ast}\tilde{c}_{p\mp l}^{\ast}\tilde{c}_{p\mp k}D_{j,l}+\sum_{k\in\mathbb{Z}_{\ast}^{3}\backslash S}1_{S_{k}}\mleft(p\mright)\hat{V}_{k}^{2}D_{j,k}^{\ast}D_{j,k}\\
 & =-\left|\sum_{k\in\mathbb{Z}_{\ast}^{3}\backslash S}1_{S_{k}}\mleft(p\mright)\hat{V}_{k}D_{j,k}^{\ast}\tilde{c}_{p\mp k}\right|^{2}+\sum_{k\in\mathbb{Z}_{\ast}^{3}\backslash S}1_{S_{k}}\mleft(p\mright)\hat{V}_{k}^{2}D_{j,k}^{\ast}D_{j,k}\nonumber \\
 & -\sum_{k,l\in\mathbb{Z}_{\ast}^{3}\backslash S}1_{S_{k}}\mleft(p\mright)1_{S_{l}}\mleft(p\mright)\hat{V}_{k}\hat{V}_{l}\tilde{c}_{p\mp l}^{\ast}\left[D_{j,k}^{\ast},D_{j,l}\right]\tilde{c}_{p\mp k}\nonumber 
\end{align}
where we also used that $\left[\tilde{c}_{p\pm k},D_{j,l}\right]=0$
in this particular case, since for $\mathcal{E}_{\mathcal{C},\mathbb{Z}_{\ast}^{3}\backslash S}^{(1)}$
the momenta $p\mp k=p-k\in B_{F}$ but $D_{j,l}=D_{1,l}$ only involves
momenta in $B_{F}^{c}$, and vice versa for $\mathcal{E}_{\mathcal{C},\mathbb{Z}_{\ast}^{3}\backslash S}^{(2)}$.

Now, the first term of the right-hand side of the equation above is
manifestly negative, so we have the bound
\begin{align}
T_{p}^{\mathcal{C}} & \leq\sum_{k\in\mathbb{Z}_{\ast}^{3}\backslash S}1_{S_{k}}\mleft(p\mright)\hat{V}_{k}^{2}D_{j,k}^{\ast}D_{j,k}-\sum_{k,l\in\mathbb{Z}_{\ast}^{3}\backslash S}1_{S_{k}}\mleft(p\mright)1_{S_{l}}\mleft(p\mright)\hat{V}_{k}\hat{V}_{l}\tilde{c}_{p\mp l}^{\ast}\left[D_{j,k}^{\ast},D_{j,l}\right]\tilde{c}_{p\mp k}\\
 & =:T_{p}^{\mathcal{C},1}-T_{p}^{\mathcal{C},2}.\nonumber 
\end{align}
We computed the commutator $[D_{j,k}^{\ast},D_{j,l}]$ in equation
(\ref{eq:DjkDjlCommutator}), with the result
\begin{equation}
\left[D_{j,k}^{\ast},D_{j,l}\right]=\sum_{q\in B_{F}^{\circ}\cap\mleft(B_{F}^{\circ}\mp k\mright)\cap\mleft(B_{F}^{\circ}\mp l\mright)}\tilde{c}_{q\pm k}^{\ast}\tilde{c}_{q\pm l}-\sum_{q\in B_{F}^{\circ}\cap\mleft(B_{F}^{\circ}\pm k\mright)\cap\mleft(B_{F}^{\circ}\pm l\mright)}\tilde{c}_{q\mp l}^{\ast}\tilde{c}_{q\mp k}.
\end{equation}
Performing the substitution $q\mapsto q\pm k\pm l$ in the second sum, we
can also write this as
\begin{align}
\left[D_{j,k}^{\ast},D_{j,l}\right] & =\sum_{q\in\mleft(B_{F}^{\circ}\mp k\mright)\cap\mleft(B_{F}^{\circ}\mp l\mright)}\mleft(1_{B_{F}^{\circ}}\mleft(q\mright)-1_{B_{F}^{\circ}}\mleft(q\pm k\pm l\mright)\mright)\tilde{c}_{q\pm k}^{\ast}\tilde{c}_{q\pm l}\\
 & =\sum_{q\in\mleft(B_{F}^{\circ}\mp k\mright)\cap\mleft(B_{F}^{\circ}\mp l\mright)}\mleft(1_{B_{F}^{\circ}}\mleft(q\mright)1_{(B_{F}^{\circ})^{c}}\mleft(q\pm k\pm l\mright)-1_{(B_{F}^{\circ})^{c}}\mleft(q\mright)1_{B_{F}^{\circ}}\mleft(q\pm k\pm l\mright)\mright)\tilde{c}_{q\pm k}^{\ast}\tilde{c}_{q\pm l}\nonumber 
\end{align}
where we used the indicator function identity $1_{A}\mleft(x\mright)-1_{A}\mleft(y\mright)=1_{A}\mleft(x\mright)1_{A^{c}}\mleft(y\mright)-1_{A^{c}}\mleft(x\mright)1_{A}\mleft(y\mright)$.
Writing out the possible choices of $B_{F}^{\circ}=B_{F}$ or $B_{F}^{c}$, it is straightforward to see that there holds the alternative identity
\begin{align}
\mp\left[D_{j,k}^{\ast},D_{j,l}\right] & =\sum_{q\in B_{F}\cap\mleft(B_{F}^{\circ}\mp k\mright)\cap\mleft(B_{F}^{\circ}\mp l\mright)}1_{B_{F}^{c}}\mleft(q\pm k\pm l\mright)\tilde{c}_{q\pm k}^{\ast}\tilde{c}_{q\pm l}\label{eq:DjkDjlCommutatorAlternate}\\
 & -\sum_{q\in B_{F}\cap\mleft(B_{F}^{\circ}\pm k\mright)\cap\mleft(B_{F}^{\circ}\pm l\mright)}1_{B_{F}^{c}}\mleft(q\mp k\mp l\mright)\tilde{c}_{q\mp l}^{\ast}\tilde{c}_{q\mp k}.\nonumber 
\end{align}
Using this identity we can now estimate $\sum_{p\in\mathbb{Z}^{3}}\vert\left|p\right|^{2}-\zeta\vert^{-1}T_{p}^{\mathcal{C}}$
as follows:
\begin{prop}
\label{prop:TCpBound}For any $\epsilon>0$ it holds as $k_{F}\rightarrow\infty$ that
\[
\sum_{p\in\mathbb{Z}^{3}}\frac{1}{\vert\left|p\right|^{2}-\zeta\vert}T_{p}^{\mathcal{C}}\leq C_{\epsilon}k_{F}^{2+\epsilon}\sum_{k\in\mathbb{Z}_{\ast}^{3}\backslash S}\hat{V}_{k}^{2}H_{\mathrm{kin}}^{\prime}
\]
for a constant $C_{\epsilon}>0$ depending only on $\epsilon$.
\end{prop}

\textbf{Proof:} $T_{p}^{\mathcal{C},1}$ can immediately be bounded
as
\begin{equation}
T_{p}^{\mathcal{C},1}=\sum_{k\in\mathbb{Z}_{\ast}^{3}\backslash S}1_{S_{k}}\mleft(p\mright)\hat{V}_{k}^{2}D_{j,k}^{\ast}D_{j,k}\leq C_{\epsilon}k_{F}^{1+\epsilon}\mleft(\sum_{k\in\mathbb{Z}_{\ast}^{3}\backslash S}1_{S_{k}}\mleft(p\mright)\hat{V}_{k}^{2}\mright)H_{\mathrm{kin}}^{\prime}
\end{equation}
by Proposition \ref{prop:DkDkKineticEstimate}, and by equation (\ref{eq:DjkDjlCommutatorAlternate})
we can for any $\Psi\in D\mleft(H_{\mathrm{kin}}^{\prime}\mright)$
estimate $\left\langle \Psi,T_{p}^{\mathcal{C},2}\Psi\right\rangle $
as
\begin{align}\label{eq:TpC2-new}
\left|\left\langle \Psi,T_{p}^{\mathcal{C},2}\Psi\right\rangle \right| & \le \sum_{k,l\in \mathbb{Z}^3_*\backslash S} \,\, \sum_{q\in B_{F}\cap\mleft(B_{F}^{\circ}\mp k\mright)\cap\mleft(B_{F}^{\circ}\mp l\mright)} 1_{S_k}(p) 1_{S_l}(p) 1_{B_{F}^{c}} \mleft(q\pm k\pm l\mright)\hat V_k \hat V_l   \| \tilde{c}_{q\pm k} \Psi\| \|\tilde{c}_{q\pm l}\Psi\| \nonumber\\
 &+\sum_{k,l\in \mathbb{Z}^3_*\backslash S} \,\, \sum_{q\in B_{F}\cap\mleft(B_{F}^{\circ}\mp k\mright)\cap\mleft(B_{F}^{\circ}\mp l\mright)} 1_{S_{-k}}(p) 1_{S_{-l}}(p) 1_{B_{F}^{c}} \mleft(q\pm k\pm l\mright)\hat V_k \hat V_l   \| \tilde{c}_{q\pm k} \Psi\| \|\tilde{c}_{q\pm l}\Psi\| \\ 
&\leq \sum_{q\in B_{F}}\mleft(\sum_{k\in\mathbb{Z}_{\ast}^{3}\backslash S}1_{S_{k}}\mleft(p\mright)1_{B_{F}^{\circ}\mp k}\mleft(q\mright)\hat{V}_{k}\left\Vert \tilde{c}_{q\pm k}\Psi\right\Vert \mright)^{2} + \sum_{q\in B_{F}}\mleft(\sum_{k\in\mathbb{Z}_{\ast}^{3}\backslash S}1_{S_{-k}}\mleft(p\mright)1_{B_{F}^{\circ}\mp k}\mleft(q\mright)\hat{V}_{k}\left\Vert \tilde{c}_{q\pm k}\Psi\right\Vert \mright)^{2}. \nonumber 
\end{align}
The two sums on the right-hand side of \eqref{eq:TpC2-new} can be treated the same way: Considering the sum involving $1_{S_{k}}$ for definiteness we have
\begin{align}
&\quad\,\sum_{q\in B_{F}}\mleft(\sum_{k\in\mathbb{Z}_{\ast}^{3}\backslash S}1_{S_{k}}\mleft(p\mright)1_{B_{F}^{\circ}\mp k}\mleft(q\mright)\hat{V}_{k}\left\Vert \tilde{c}_{q\pm k}\Psi\right\Vert \mright)^{2}  \nonumber\\
& \leq  \sum_{q\in B_{F}}\mleft(\sum_{k\in\mathbb{Z}_{\ast}^{3}\backslash S}1_{S_{k}}\mleft(p\mright)\hat{V}_{k}^{2}\frac{1_{B_{F}^{\circ}\mp k}\mleft(q\mright)}{\vert\left|q\mp k\right|^{2}-\zeta\vert}\mright)\mleft(\sum_{k\in\mathbb{Z}_{\ast}^{3}\backslash S}1_{B_{F}^{\circ}\mp k}\mleft(q\mright)\vert\left|q\mp k\right|^{2}-\zeta\vert\left\Vert \tilde{c}_{q\mp k}\Psi\right\Vert ^{2}\mright)\\
 & \leq \sum_{k\in\mathbb{Z}_{\ast}^{3}\backslash S}1_{S_{k}}\mleft(p\mright)\hat{V}_{k}^{2}\sum_{q\in B_{F}}\frac{1_{B_{F}^{\circ}\mp k}\mleft(q\mright)}{\vert\left|q\mp k\right|^{2}-\zeta\vert}\left\langle \Psi,H_{\mathrm{kin}}^{\prime}\Psi\right\rangle \nonumber\\
 &\leq C_{\epsilon}k_{F}^{1+\epsilon}\sum_{k\in\mathbb{Z}_{\ast}^{3}\backslash S}1_{S_{k}}\mleft(p\mright)\hat{V}_{k}^{2}\left\langle \Psi,H_{\mathrm{kin}}^{\prime}\Psi\right\rangle \nonumber 
\end{align}


where Proposition \ref{lemma:GeneralKineticSum} could be applied
due to the condition $q\in B_{F}$ in the sum. Combining these estimates
and applying Proposition \ref{lemma:GeneralKineticSum} once more,
we conclude that
\begin{align}
\sum_{p\in\mathbb{Z}^{3}}\frac{1}{\vert\left|p\right|^{2}-\zeta\vert}T_{p}^{\mathcal{C}} & \leq C_{\epsilon}k_{F}^{1+\epsilon}\sum_{k\in\mathbb{Z}_{\ast}^{3}\backslash S}\hat{V}_{k}^{2}\sum_{p\in\mathbb{Z}^{3}}\frac{1_{S_{k}}\mleft(p\mright)}{\vert\left|p\right|^{2}-\zeta\vert}H_{\mathrm{kin}}^{\prime}\\
 & \leq C_{\epsilon}^{\prime}k_{F}^{2+\epsilon^{\prime}}\sum_{k\in\mathbb{Z}_{\ast}^{3}\backslash S}\hat{V}_{k}^{2}H_{\mathrm{kin}}^{\prime}.\nonumber 
\end{align}
$\hfill\square$

As we did for $T_{p}^{\mathcal{\mathcal{C}}}$, we expand $T_{p}^{\mathcal{Q}}$
and commute for the identity
\begin{align}
T_{p}^{\mathcal{Q}} & =\sum_{k,l\in\mathbb{Z}_{\ast}^{3}\backslash S}1_{S_{k}}\mleft(p\mright)1_{S_{l}}\mleft(p\mright)\hat{V}_{k}\hat{V}_{l}\tilde{c}_{p\mp k}^{\ast}D_{j,k}^{\ast}D_{j,l}\tilde{c}_{p\mp l}\nonumber \\
 & =\sum_{k,l\in\mathbb{Z}_{\ast}^{3}\backslash S}1_{S_{k}}\mleft(p\mright)1_{S_{l}}\mleft(p\mright)\hat{V}_{k}\hat{V}_{l}\mleft(D_{j,l}\tilde{c}_{p\mp k}^{\ast}+\left[\tilde{c}_{p\mp k}^{\ast},D_{j,l}\right]\mright)\mleft(\tilde{c}_{p\mp l}D_{j,k}^{\ast}+\left[D_{j,k}^{\ast},\tilde{c}_{p\mp l}\right]\mright)\nonumber \\
 & +\sum_{k,l\in\mathbb{Z}_{\ast}^{3}\backslash S}1_{S_{k}}\mleft(p\mright)1_{S_{l}}\mleft(p\mright)\hat{V}_{k}\hat{V}_{l}\tilde{c}_{p\mp k}^{\ast}\left[D_{j,k}^{\ast},D_{j,l}\right]\tilde{c}_{p\mp l}\nonumber \\
 & =-\left|\sum_{k\in\mathbb{Z}_{\ast}^{3}\backslash S}1_{S_{k}}\mleft(p\mright)\hat{V}_{k}\tilde{c}_{p\mp k}^{\ast}D_{j,k}^{\ast}\right|^{2}+\sum_{k\in\mathbb{Z}_{\ast}^{3}\backslash S}1_{S_{k}}\mleft(p\mright)\hat{V}_{k}^{2}D_{j,k}D_{j,k}^{\ast}\\
 & +\sum_{k,l\in\mathbb{Z}_{\ast}^{3}\backslash S}1_{S_{k}}\mleft(p\mright)1_{S_{l}}\mleft(p\mright)\hat{V}_{k}\hat{V}_{l}\tilde{c}_{p\mp k}^{\ast}\left[D_{j,k}^{\ast},D_{j,l}\right]\tilde{c}_{p\mp l}\nonumber \\
 & +2\,\mathrm{Re}\sum_{k,l\in\mathbb{Z}_{\ast}^{3}\backslash S}1_{S_{k}}\mleft(p\mright)1_{S_{l}}\mleft(p\mright)\hat{V}_{k}\hat{V}_{l}D_{j,l}\tilde{c}_{p\mp k}^{\ast}\left[D_{j,k}^{\ast},\tilde{c}_{p\mp l}\right]\nonumber \\
 & +\sum_{k,l\in\mathbb{Z}_{\ast}^{3}\backslash S}1_{S_{k}}\mleft(p\mright)1_{S_{l}}\mleft(p\mright)\hat{V}_{k}\hat{V}_{l}\left[D_{j,l}^{\ast},\tilde{c}_{p\mp k}\right]^{\ast}\left[D_{j,k}^{\ast},\tilde{c}_{p\mp l}\right],\nonumber 
\end{align}
which yields the inequality
\begin{equation}
T_{p}^{\mathcal{Q}}\leq T_{p}^{\mathcal{Q},1}+T_{p}^{\mathcal{Q},2}+2\,\mathrm{Re}\mleft(T_{p}^{\mathcal{Q},3}\mright)+T_{p}^{\mathcal{Q},4}
\end{equation}
where
\begin{align}
T_{p}^{\mathcal{Q},1} & =\sum_{k\in\mathbb{Z}_{\ast}^{3}\backslash S}1_{S_{k}}\mleft(p\mright)\hat{V}_{k}^{2}D_{j,k}D_{j,k}^{\ast}\nonumber \\
T_{p}^{\mathcal{Q},2} & =\sum_{k,l\in\mathbb{Z}_{\ast}^{3}\backslash S}1_{S_{k}}\mleft(p\mright)1_{S_{l}}\mleft(p\mright)\hat{V}_{k}\hat{V}_{l}\tilde{c}_{p\mp k}^{\ast}\left[D_{j,k}^{\ast},D_{j,l}\right]\tilde{c}_{p\mp l}\\
T_{p}^{\mathcal{Q},3} & =\sum_{k,l\in\mathbb{Z}_{\ast}^{3}\backslash S}1_{S_{k}}\mleft(p\mright)1_{S_{l}}\mleft(p\mright)\hat{V}_{k}\hat{V}_{l}D_{j,l}\tilde{c}_{p\mp k}^{\ast}\left[D_{j,k}^{\ast},\tilde{c}_{p\mp l}\right]\nonumber \\
T_{p}^{\mathcal{Q},4} & =\sum_{k,l\in\mathbb{Z}_{\ast}^{3}\backslash S}1_{S_{k}}\mleft(p\mright)1_{S_{l}}\mleft(p\mright)\hat{V}_{k}\hat{V}_{l}\left[D_{j,l}^{\ast},\tilde{c}_{p\mp k}\right]^{\ast}\left[D_{j,k}^{\ast},\tilde{c}_{p\mp l}\right].\nonumber 
\end{align}
We can then estimate $\sum_{p\in\mathbb{Z}^{3}}\vert\left|p\right|^{2}-\zeta\vert^{-1}T_{p}^{\mathcal{Q}}$
in a similar fashion:
\begin{prop}
\label{prop:TQpBound}For any $\epsilon>0$ it holds as $k_{F}\rightarrow\infty$ that
\[
\sum_{p\in\mathbb{Z}^{3}}\frac{1}{\vert\left|p\right|^{2}-\zeta\vert}T_{p}^{\mathcal{Q}}\leq C_{\epsilon}k_{F}^{2+\epsilon}\sum_{k\in\mathbb{Z}_{\ast}^{3}\backslash S}\hat{V}_{k}^{2}H_{\mathrm{kin}}^{\prime}
\]
for a constant $C_{\epsilon}>0$ depending only on $\epsilon$.
\end{prop}

\textbf{Proof:} Exactly as in the previous proposition we see that
the bound
\begin{equation}
T_{p}^{\mathcal{Q},1},\,T_{p}^{\mathcal{Q},2}\leq C_{\epsilon}k_{F}^{1+\epsilon}\mleft(\sum_{k\in\mathbb{Z}_{\ast}^{3}\backslash S}1_{S_{k}}\mleft(p\mright)\hat{V}_{k}^{2}\mright)H_{\mathrm{kin}}^{\prime}
\end{equation}
holds: The $T_{p}^{\mathcal{Q},1}$ bound follows since $D_{j,k}D_{j,k}^{\ast}=D_{j,-k}^{\ast}D_{j,-k}$
and Proposition \ref{prop:DkDkKineticEstimate} is also valid for
$D_{0,k}^{\ast}D_{0,k}$ (indeed, this is the final equality of the
proposition), and the $T_{p}^{\mathcal{Q},2}$ bound follows since
it is readily computed that $[D_{0,k}^{\ast},D_{0,l}]$ can also be
written in the form
\begin{align}
-\left[D_{0,k}^{\ast},D_{0,l}\right] & =\sum_{q\in A_{2}^{c}\cap\mleft(A_{2}-k\mright)\cap\mleft(A_{2}-l\mright)}1_{A_{2}}\mleft(q+k+l\mright)\tilde{c}_{q+k}^{\ast}\tilde{c}_{q+l}\\
 & -\sum_{q\in A_{2}^{c}\cap\mleft(A_{2}+k\mright)\cap\mleft(A_{2}+l\mright)}1_{A_{2}}\mleft(q-k-l\mright)\tilde{c}_{q-l}^{\ast}\tilde{c}_{q-k}\nonumber 
\end{align}
and the fact that $A_{2}^{c}=\overline{B}\mleft(0,2k_{F}\mright)\cap\mathbb{Z}^{3}$
ensures that Proposition \ref{lemma:GeneralKineticSum} still applies.

For $T_{p}^{\mathcal{Q},3}$ (and $T_{p}^{\mathcal{Q},4}$) we calculate
\begin{align}
\left[D_{j,k}^{\ast},\tilde{c}_{p\mp l}\right] & =\pm\sum_{q\in B_{F}^{\circ}\cap\mleft(B_{F}^{\circ}\mp k\mright)}\left[\tilde{c}_{q\pm k}^{\ast}\tilde{c}_{q},\tilde{c}_{p\mp l}\right]=\mp\sum_{q\in B_{F}^{\circ}\cap\mleft(B_{F}^{\circ}\mp k\mright)}\delta_{q\pm k,p\mp l}\tilde{c}_{q}\label{eq:DjkcplCommutator}\\
 & =\mp1_{B_{F}^{\circ}}\mleft(p\mp l\mright)1_{B_{F}^{\circ}}\mleft(p\mp k\mp l\mright)\tilde{c}_{p\mp k\mp l},\nonumber 
\end{align}

for $j=1,2$ and likewise

\begin{align}
\left[D_{0,k}^{\ast},\tilde{c}_{p - l}\right] & =- 1_{A_2}\mleft(p - l\mright)1_{A_2}\mleft(p - k - l\mright)\tilde{c}_{p - k - l}
\end{align}

for $j=0$, i.e. this case agrees with the $j=1$ case up to the substitution $B_F^\circ \rightarrow A_2$,
so $T_{p}^{\mathcal{Q},3}$ can be written as
\begin{align*}
T_{p}^{\mathcal{Q},3} & =\mp\sum_{k,l\in\mathbb{Z}_{\ast}^{3}\backslash S}1_{S_{k}}\mleft(p\mright)1_{S_{l}}\mleft(p\mright)1_{B_{F}^{\circ}}\mleft(p\mp l\mright)1_{B_{F}^{\circ}}\mleft(p\mp k\mp l\mright)\hat{V}_{k}\hat{V}_{l}D_{j,l}\tilde{c}_{p\mp k}^{\ast}\tilde{c}_{p\mp k\mp l}\\
 & =\mp\sum_{l\in\mathbb{Z}_{\ast}^{3}\backslash S}1_{S_{l}}\mleft(p\mright)1_{B_{F}^{\circ}}\mleft(p\mp l\mright)\hat{V}_{l}D_{j,l}\mleft(\sum_{k\in\mathbb{Z}_{\ast}^{3}\backslash S}1_{S_{k}}\mleft(p\mright)1_{B_{F}^{\circ}}\mleft(p\mp k\mp l\mright)\hat{V}_{k}\tilde{c}_{p\mp k}^{\ast}\tilde{c}_{p\mp k\mp l}\mright)
\end{align*}
which implies the estimate
\[
\left|\left\langle \Psi,T_{p}^{\mathcal{Q},3}\Psi\right\rangle \right|\leq\sqrt{\sum_{l\in\mathbb{Z}_{\ast}^{3}\backslash S}1_{S_{l}}\mleft(p\mright)\hat{V}_{l}^{2}\left\Vert D_{j,l}\Psi\right\Vert ^{2}}\sqrt{\sum_{l\in\mathbb{Z}_{\ast}^{3}\backslash S}1_{S_{l}}\mleft(p\mright)\left\Vert \sum_{k\in\mathbb{Z}_{\ast}^{3}\backslash S}1_{S_{k}}\mleft(p\mright)1_{B_{F}^{\circ}}\mleft(p\mp k\mp l\mright)\hat{V}_{k}\tilde{c}_{p\mp k}^{\ast}\tilde{c}_{p\mp k\mp l}\Psi\right\Vert ^{2}}.
\]
Again
\begin{equation}
\sum_{l\in\mathbb{Z}_{\ast}^{3}\backslash S}1_{S_{l}}\mleft(p\mright)\hat{V}_{l}^{2}\left\Vert D_{j,l}\Psi\right\Vert ^{2}\leq C_{\epsilon}k_{F}^{1+\epsilon}\sum_{l\in\mathbb{Z}_{\ast}^{3}\backslash S}1_{S_{l}}\mleft(p\mright)\hat{V}_{l}^{2}\left\langle \Psi,H_{\mathrm{kin}}^{\prime}\Psi\right\rangle 
\end{equation}
while the second factor obeys
\begin{align}
 & \quad\,\sum_{l\in\mathbb{Z}_{\ast}^{3}\backslash S}1_{S_{l}}\mleft(p\mright)\left\Vert \sum_{k\in\mathbb{Z}_{\ast}^{3}\backslash S}1_{S_{k}}\mleft(p\mright)1_{B_{F}^{\circ}}\mleft(p\mp k\mp l\mright)\hat{V}_{k}\tilde{c}_{p\mp k}^{\ast}\tilde{c}_{p\mp k\mp l}\Psi\right\Vert ^{2}\nonumber \\
 & \leq\sum_{l\in\mathbb{Z}_{\ast}^{3}\backslash S}1_{S_{l}}\mleft(p\mright)\mleft(\sum_{k\in\mathbb{Z}_{\ast}^{3}\backslash S}1_{S_{k}}\mleft(p\mright)\frac{1_{B_{F}^{\circ}}\mleft(p\mp k\mp l\mright)}{\vert\left|p\mp k\mp l\right|^{2}-\zeta\vert}\hat{V}_{k}^{2}\mright)\nonumber \\
 & \qquad\qquad\qquad\;\cdot\mleft(\sum_{k\in\mathbb{Z}_{\ast}^{3}\backslash S}1_{B_{F}^{\circ}}\mleft(p\mp k\mp l\mright)\vert\left|p\mp k\mp l\right|^{2}-\zeta\vert\left\Vert \tilde{c}_{p\mp k\mp l}\Psi\right\Vert ^{2}\mright)\\
 & \leq\sum_{k\in\mathbb{Z}_{\ast}^{3}\backslash S}1_{S_{k}}\mleft(p\mright)\hat{V}_{k}^{2}\sum_{l\in\mathbb{Z}_{\ast}^{3}\backslash S}1_{S_{l}}\mleft(p\mright)\frac{1_{B_{F}^{\circ}}\mleft(p\mp k\mp l\mright)}{\vert\left|p\mp k\mp l\right|^{2}-\zeta\vert}\left\langle \Psi,H_{\mathrm{kin}}^{\prime}\Psi\right\rangle \nonumber \\
 & \leq C_{\epsilon}k_{F}^{1+\epsilon}\sum_{k\in\mathbb{Z}_{\ast}^{3}\backslash S}1_{S_{k}}\mleft(p\mright)\hat{V}_{k}^{2}\left\langle \Psi,H_{\mathrm{kin}}^{\prime}\Psi\right\rangle \nonumber 
\end{align}
where we could use Proposition \ref{lemma:GeneralKineticSum} once
more since the summation over $l$ is restricted by the indicator
function $1_{S_{l}}\mleft(p\mright)$, with $S_{l}$ being either $B_{F}\cap\mleft(B_{F}-l\mright)\subset\mleft(B_{F}-l\mright)$
or $A_{a}\cap\mleft(A_{1}+l\mright)\subset\mleft(A_{1}+l\mright)$, since
$\left|B_{F}\right|,\left|A_{1}\right|\leq\left|\overline{B}\mleft(0,2k_{F}\mright)\cap\mathbb{Z}^{3}\right|$.

Combining the two estimates we get
\begin{equation}
\pm T_{p}^{\mathcal{Q},3}\leq C_{\epsilon}k_{F}^{1+\epsilon}\sum_{k\in\mathbb{Z}_{\ast}^{3}\backslash S}1_{S_{k}}\mleft(p\mright)\hat{V}_{k}^{2}H_{\mathrm{kin}}^{\prime}
\end{equation}
and equation (\ref{eq:DjkcplCommutator}) also yields
\begin{equation}
T_{p}^{\mathcal{Q},4}=\sum_{k,l\in\mathbb{Z}_{\ast}^{3}\backslash S}1_{S_{k}}\mleft(p\mright)1_{S_{l}}\mleft(p\mright)1_{B_{F}^{\circ}}\mleft(p\mp k\mright)1_{B_{F}^{\circ}}\mleft(p\mp l\mright)1_{B_{F}^{\circ}}\mleft(p\mp k\mp l\mright)\hat{V}_{k}\hat{V}_{l}\tilde{c}_{p\mp k\mp l}^{\ast}\tilde{c}_{p\mp k\mp l}
\end{equation}
which as the summand is symmetric in $k$ and $l$ can be estimated
by
\begin{align}
T_{p}^{\mathcal{Q},4} & \leq\sum_{k,l\in\mathbb{Z}_{\ast}^{3}\backslash S}1_{S_{k}}\mleft(p\mright)1_{S_{l}}\mleft(p\mright)1_{B_{F}^{\circ}}\mleft(p\mp k\mright)1_{B_{F}^{\circ}}\mleft(p\mp l\mright)1_{B_{F}^{\circ}}\mleft(p\mp k\mp l\mright)\hat{V}_{k}^{2}\tilde{c}_{p\mp k\mp l}^{\ast}\tilde{c}_{p\mp k\mp l}\\
 & \leq\sum_{k\in\mathbb{Z}_{\ast}^{3}\backslash S}1_{S_{k}}\mleft(p\mright)\hat{V}_{k}^{2}\mathcal{N}_{E}\leq\sum_{k\in\mathbb{Z}_{\ast}^{3}\backslash S}1_{S_{k}}\mleft(p\mright)\hat{V}_{k}^{2}H_{\mathrm{kin}}^{\prime}.\nonumber 
\end{align}
All in all this shows that $T_{p}^{\mathcal{Q}}\leq C_{\epsilon}k_{F}^{1+\epsilon}\sum_{k\in\mathbb{Z}_{\ast}^{3}\backslash S}1_{S_{k}}\mleft(p\mright)\hat{V}_{k}^{2}H_{\mathrm{kin}}^{\prime}$
and the claim now follows as in Proposition \ref{prop:TCpBound}.

$\hfill\square$

By the equations (\ref{eq:ECLargeEstimate}) and (\ref{eq:EQEstimate})
combined with these propositions we see that
\begin{equation}
\pm\mathcal{E}_{\mathcal{C},\mathbb{Z}_{\ast}^{3}\backslash S},\,\,\pm(\mathcal{E}_{\mathcal{Q}}-\mathcal{E}_{\mathcal{Q},4}-\mathcal{E}_{\mathcal{Q},5})\leq C_{\epsilon}k_{F}^{1+\epsilon}\sqrt{\sum_{k\in\mathbb{Z}_{\ast}^{3}\backslash S}\hat{V}_{k}^{2}}H_{\mathrm{kin}}^{\prime}
\end{equation}
which implies the second estimate of Theorem \ref{them:FinalErrorBounds}.

\subsection{Estimation of $\mathcal{E}_{\mathcal{Q},4}$ and $\mathcal{E}_{\mathcal{Q},5}$}

To estimate $\mathcal{E}_{\mathcal{Q},4}$ we will use the following:
\begin{prop}
\label{prop:InsideOutsideComparison}Let $\hat{V}_{k}$ obey Assumption
\ref{Assumption:Potential}. Then
\[
\inf_{q\in B_{F}}\sum_{k\in\mleft(B_{F}+q\mright)\backslash\left\{ 0\right\} }\hat{V}_{k}-\sup_{p\in B_{F}^{c}}\sum_{k\in\mleft(B_{F}+p\mright)}\hat{V}_{k}\geq-C_{V}^{\prime}k_{F}^{\frac{2}{3}},\quad k_{F}\rightarrow\infty,
\]
for a constant $C_{V}^{\prime}>0$ depending only on $C_{V}$.
\end{prop}

For the proof see appendix section \ref{subsec:InsideOutsideComparison}.

The estimate for $\mathcal{E}_{\mathcal{Q},4}$ is then as follows:
\begin{prop}
Let $\hat{V}_{k}$ obey Assumption \ref{Assumption:Potential}. Then
it holds that
\[
\mathcal{E}_{\mathcal{Q},4}\geq-\mleft(C_{V}^{\prime}k_{F}^{\frac{2}{3}}+2\sum_{k\in S}\hat{V}_{k}\mright)H_{\mathrm{kin}}^{\prime}
\]
for a constant $C_{V}^{\prime}>0$ depending only on $C_{V}$.
\end{prop}

\textbf{Proof:} We can write $\mathcal{E}_{\mathcal{Q},4}$ in the
form
\begin{align}
\mathcal{E}_{\mathcal{Q},4} & =2\sum_{p\in B_{F}}\mleft(\sum_{k\in\mathbb{Z}_{\ast}^{3}}1_{B_{F}}\mleft(p+k\mright)\hat{V}_{k}\mright)\tilde{c}_{p}^{\ast}\tilde{c}_{p}-2\sum_{p\in B_{F}^{c}}\mleft(\sum_{k\in\mathbb{Z}_{\ast}^{3}}1_{B_{F}}\mleft(p-k\mright)\hat{V}_{k}\mright)\tilde{c}_{p}^{\ast}\tilde{c}_{p}\\
 & -2\sum_{p\in B_{F}}\mleft(\sum_{k\in S}1_{B_{F}}\mleft(p+k\mright)\hat{V}_{k}\mright)\tilde{c}_{p}^{\ast}\tilde{c}_{p}+2\sum_{p\in B_{F}^{c}}\mleft(\sum_{k\in S}1_{B_{F}}\mleft(p-k\mright)\hat{V}_{k}\mright)\tilde{c}_{p}^{\ast}\tilde{c}_{p}\nonumber 
\end{align}
which, estimating the final line as in Proposition \ref{prop:ESBound},
implies that
\begin{equation}
\mathcal{E}_{\mathcal{Q},4}\geq2\sum_{p\in B_{F}}\mleft(\sum_{k\in\mleft(B_{F}+p\mright)\backslash\left\{ 0\right\} }\hat{V}_{k}\mright)\tilde{c}_{p}^{\ast}\tilde{c}_{p}-2\sum_{p\in B_{F}^{c}}\mleft(\sum_{k\in\mleft(B_{F}+p\mright)}\hat{V}_{k}\mright)\tilde{c}_{p}^{\ast}\tilde{c}_{p}-2\sum_{k\in S}\hat{V}_{k}H_{\mathrm{kin}}^{\prime}
\end{equation}
where we also absorbed the indicator functions into the summation
range (and substituted $k\rightarrow-k$ in the first sum). Now, clearly
\begin{align}
2\sum_{p\in B_{F}}\mleft(\sum_{k\in\mleft(B_{F}+p\mright)\backslash\left\{ 0\right\} }\hat{V}_{k}\mright)\tilde{c}_{p}^{\ast}\tilde{c}_{p} & \geq2\mleft(\inf_{q\in B_{F}}\sum_{k\in\mleft(B_{F}+q\mright)\backslash\left\{ 0\right\} }\hat{V}_{k}\mright)\sum_{q\in B_{F}}\tilde{c}_{q}^{\ast}\tilde{c}_{q}\\
-2\sum_{p\in B_{F}^{c}}\mleft(\sum_{k\in\mleft(B_{F}+p\mright)}\hat{V}_{k}\mright)\tilde{c}_{p}^{\ast}\tilde{c}_{p} & \geq-2\mleft(\sup_{p\in B_{F}^{c}}\sum_{k\in\mleft(B_{F}+p\mright)}\hat{V}_{k}\mright)\sum_{p\in B_{F}^{c}}\tilde{c}_{p}^{\ast}\tilde{c}_{p}\nonumber 
\end{align}
and by particle-hole symmetry either of the sums on the right-hand
side are in fact $\mathcal{N}_{E}$. Thus we can apply Proposition
\ref{prop:InsideOutsideComparison} to conclude
\begin{align}
\mathcal{E}_{\mathcal{Q},4} & \geq2\mleft(\inf_{q\in B_{F}}\sum_{k\in\mleft(B_{F}+q\mright)\backslash\left\{ 0\right\} }\hat{V}_{k}-\sup_{p\in B_{F}^{c}}\sum_{k\in\mleft(B_{F}+p\mright)}\hat{V}_{k}\mright)\mathcal{N}_{E}-2\sum_{k\in S}\hat{V}_{k}H_{\mathrm{kin}}^{\prime}\\
 & \geq-C_{V}^{\prime}k_{F}^{\frac{2}{3}}\mathcal{N}_{E}-2\sum_{k\in S}\hat{V}_{k}H_{\mathrm{kin}}^{\prime}\geq-\mleft(C_{V}^{\prime}k_{F}^{\frac{2}{3}}+2\sum_{k\in S}\hat{V}_{k}\mright)H_{\mathrm{kin}}^{\prime}.\nonumber 
\end{align}
$\hfill\square$

\subsubsection*{Estimation of $\mathcal{E}_{\mathcal{Q},5}$}

Finally we come to $\mathcal{E}_{\mathcal{Q},5}$, which we recall
is
\begin{equation}
\mathcal{E}_{\mathcal{Q},5}=\sum_{k\in\mathbb{Z}_{\ast}^{3}\backslash S}\hat{V}_{k}\sum_{p,q\in A_{2}\cap\mleft(A_{2}+k\mright)}\tilde{c}_{p}^{\ast}\tilde{c}_{q-k}^{\ast}\tilde{c}_{q}\tilde{c}_{p-k}
\end{equation}
where $A_{2}=\mathbb{Z}^{3}\backslash\overline{B}\mleft(0,2k_{F}\mright)$.
Noting as in Proposition \ref{prop:DkDkKineticEstimate} that
\begin{equation}\label{eq:cQ-5-normal order}
\sum_{p,q\in A_{2}\cap\mleft(A_{2}+k\mright)}\tilde{c}_{p}^{\ast}\tilde{c}_{q-k}^{\ast}\tilde{c}_{q}\tilde{c}_{p-k}=D_{0,k}^{\ast}D_{0,k}-\sum_{p\in A_{2}\cap\mleft(A_{2}+k\mright)}\tilde{c}_{p}^{\ast}\tilde{c}_{p}\geq-3^{-1}k_{F}^{-2}H_{\mathrm{kin}}^{\prime}
\end{equation}
we can for any $S'\subset\mathbb{Z}_{\ast}^{3}$ containing $S$ estimate
\begin{align}
\mathcal{E}_{\mathcal{Q},5} & =\sum_{k\in S'\backslash S}\hat{V}_{k}\sum_{p,q\in A_{2}\cap\mleft(A_{2}+k\mright)}\tilde{c}_{p}^{\ast}\tilde{c}_{q-k}^{\ast}\tilde{c}_{q}\tilde{c}_{p-k}+\sum_{k\in\mathbb{Z}_{\ast}^{3}\backslash S'}\hat{V}_{k}\sum_{p,q\in A_{2}\cap\mleft(A_{2}+k\mright)}\tilde{c}_{p}^{\ast}\tilde{c}_{q-k}^{\ast}\tilde{c}_{q}\tilde{c}_{p-k}\label{eq:Eq5SprimeSplit}\\
 & \geq-\frac{1}{3}k_{F}^{-2}\sum_{k\in S'}\hat{V}_{k}H_{\mathrm{kin}}^{\prime}+\sum_{k\in\mathbb{Z}_{\ast}^{3}\backslash S'}\hat{V}_{k}\sum_{p,q\in A_{2}\cap\mleft(A_{2}+k\mright)}c_{p}^{\ast}c_{q-k}^{\ast}c_{q}c_{p-k}\nonumber 
\end{align}
and the $k_{F}^{-2}$ factor ensures that we can take $S'$ to be
considerably larger than $S$ without worsening the overall estimate.
The remaining sum then not only involves exclusively momenta which
are large, but we can also assume $k$ to be large. In that case we
can make the following estimate:
\begin{prop}
For any $\epsilon>0$ and $S'$ containing $\overline{B}(0,3k_{F})\cap\mathbb{Z}^{3}$
it holds that
\[
\pm\sum_{k\in\mathbb{Z}_{\ast}^{3}\backslash S'}\hat{V}_{k}\sum_{p,q\in A_{2}\cap\mleft(A_{2}+k\mright)}c_{p}^{\ast}c_{q-k}^{\ast}c_{q}c_{p-k}\leq C_{\epsilon}\mleft(\sum_{k\in\mathbb{Z}_{\ast}^{3}\backslash S'}\frac{\hat{V}_{k}}{\left|k\right|^{2}}+\sqrt{\sum_{k\in\mathbb{Z}_{\ast}^{3}\backslash S'}\hat{V}_{k}^{2}\left|k\right|^{-\mleft(1-\epsilon\mright)}}\mright)\mathcal{N}_{E}H_{\mathrm{kin}}^{\prime}
\]
for a constant $C_{\epsilon}>0$ depending only on $\epsilon$.
\end{prop}

\textbf{Proof:} By the triangle inequality we have for all $k\in\mathbb{Z}_{\ast}^{3}\backslash S'\subset\mathbb{Z}_{\ast}^{3}\backslash\overline{B}\mleft(0,3k_{F}\mright)$
that
\begin{equation}
3^{-1}\left|k\right|\leq\left|k\right|-2k_{F}\leq\left|p\right|-k_{F}+\left|p-k\right|-k_{F}\leq\sqrt{\left|p\right|^{2}-k_{F}^{2}}+\sqrt{\left|p-k\right|^{2}-k_{F}^{2}}
\end{equation}
when $p\in A_{2}\cap\mleft(A_{2}+k\mright)$, so for any $\Psi\in D\mleft(H_{\mathrm{kin}}^{\prime}\mright)$
we can estimate
\begin{align}
 & \;\;\;\left|\sum_{k\in\mathbb{Z}_{\ast}^{3}\backslash S'}\hat{V}_{k}\sum_{p,q\in A_{2}\cap\mleft(A_{2}+k\mright)}\left\langle \Psi,c_{p}^{\ast}c_{q-k}^{\ast}c_{q}c_{p-k}\Psi\right\rangle \right|\nonumber \\
 & \leq\sum_{k\in\mathbb{Z}_{\ast}^{3}\backslash S'}\hat{V}_{k}\sum_{p,q\in A_{2}\cap\mleft(A_{2}+k\mright)}\left\Vert c_{q-k}c_{p}\Psi\right\Vert \left\Vert c_{q}c_{p-k}\Psi\right\Vert \\
 & \leq\sum_{k\in\mathbb{Z}_{\ast}^{3}\backslash S'}\hat{V}_{k}\sum_{p,q\in A_{2}\cap\mleft(A_{2}+k\mright)}\frac{\sqrt{\left|p\right|^{2}-k_{F}^{2}}+\sqrt{\left|p-k\right|^{2}-k_{F}^{2}}}{3^{-1}\left|k\right|}\left\Vert c_{q-k}c_{p}\Psi\right\Vert \left\Vert c_{q}c_{p-k}\Psi\right\Vert \nonumber \\
 & =6\sum_{k\in\mathbb{Z}_{\ast}^{3}\backslash S'}\frac{\hat{V}_{k}}{\left|k\right|}\sum_{p,q\in A_{2}\cap\mleft(A_{2}+k\mright)}\sqrt{\left|p\right|^{2}-k_{F}^{2}}\left\Vert c_{q-k}c_{p}\Psi\right\Vert \left\Vert c_{q}c_{p-k}\Psi\right\Vert \nonumber 
\end{align}
where we also made the substitutions $p\mapsto p+k$, $q\mapsto q+k$ and $k\mapsto -k$ in one sum to reduce to the same expression.

Now we split the $q$ summation into a $\left|q\right|\geq\left|k\right|$
and $\left|q\right|<\left|k\right|$ part. In the first case we can
estimate
\begin{align}
 & \quad\,\sum_{k\in\mathbb{Z}_{\ast}^{3}\backslash S'}\frac{\hat{V}_{k}}{\left|k\right|}\sum_{p,q\in A_{2}\cap\mleft(A_{2}+k\mright)}1_{B(0,\left|k\right|)^{c}}\mleft(q\mright)\sqrt{\left|p\right|^{2}-k_{F}^{2}}\left\Vert c_{q-k}c_{p}\Psi\right\Vert \left\Vert c_{q}c_{p-k}\Psi\right\Vert \nonumber \\
 & \leq\sum_{k\in\mathbb{Z}_{\ast}^{3}\backslash S'}\frac{\hat{V}_{k}}{\left|k\right|}\sum_{p,q\in A_{2}\cap\mleft(A_{2}+k\mright)}\frac{\left|q\right|-k_{F}}{\left|k\right|-k_{F}}\sqrt{\left|p\right|^{2}-k_{F}^{2}}\left\Vert c_{q-k}c_{p}\Psi\right\Vert \left\Vert c_{q}c_{p-k}\Psi\right\Vert \nonumber \\
 & \leq\frac{3}{2}\sum_{k\in\mathbb{Z}_{\ast}^{3}\backslash S'}\frac{\hat{V}_{k}}{\left|k\right|^{2}}\sum_{p,q\in A_{2}\cap\mleft(A_{2}+k\mright)}\sqrt{\left|p\right|^{2}-k_{F}^{2}}\sqrt{\left|q\right|^{2}-k_{F}^{2}}\left\Vert c_{q-k}c_{p}\Psi\right\Vert \left\Vert c_{q}c_{p-k}\Psi\right\Vert \\
 & \leq\frac{3}{2}\sum_{k\in\mathbb{Z}_{\ast}^{3}\backslash S'}\frac{\hat{V}_{k}}{\left|k\right|^{2}}\sum_{p,q\in A_{2}\cap\mleft(A_{2}+k\mright)}\mleft(\left|q\right|^{2}-k_{F}^{2}\mright)\left\Vert c_{q}c_{p-k}\Psi\right\Vert ^{2}\nonumber \\
 & \leq\frac{3}{2}\sum_{k\in\mathbb{Z}_{\ast}^{3}\backslash S'}\frac{\hat{V}_{k}}{\left|k\right|^{2}}\left\langle \Psi,\mathcal{N}_{E}H_{\mathrm{kin}}^{\prime}\Psi\right\rangle \nonumber 
\end{align}
as also $\left|k\right|-k_{F}>\frac{2}{3}\left|k\right|$ for $k\in\mathbb{Z}_{\ast}^{3}\backslash S'$.
Meanwhile, in the second case,
\begin{align}
 & \qquad\,\sum_{k\in\mathbb{Z}_{\ast}^{3}\backslash S'}\frac{\hat{V}_{k}}{\left|k\right|}\sum_{p,q\in A_{2}\cap\mleft(A_{2}+k\mright)}1_{B(0,\left|k\right|)}\mleft(q\mright)\sqrt{\left|p\right|^{2}-k_{F}^{2}}\left\Vert c_{q-k}c_{p}\Psi\right\Vert \left\Vert c_{q}c_{p-k}\Psi\right\Vert \nonumber \\
 & \leq\sqrt{\sum_{k\in\mathbb{Z}_{\ast}^{3}\backslash S'}\hat{V}_{k}^{2}\left|k\right|^{-\mleft(1-\epsilon\mright)}\sum_{p,q\in A_{2}\cap\mleft(A_{2}+k\mright)}\mleft(\left|q\right|^{2}-k_{F}^{2}\mright)\left\Vert c_{q}c_{p-k}\Psi\right\Vert ^{2}}\nonumber \\
 & \times \sqrt{\sum_{k\in\mathbb{Z}_{\ast}^{3}\backslash S'}\sum_{p,q\in A_{2}\cap\mleft(A_{2}+k\mright)}\frac{1_{B(0,\left|k\right|)}\mleft(q\mright)}{\left|k\right|^{1+\epsilon}}\frac{\left|p\right|^{2}-k_{F}^{2}}{\left|q\right|^{2}-k_{F}^{2}}\left\Vert c_{q-k}c_{p}\Psi\right\Vert ^{2}}\\
 & \leq\sqrt{\sum_{k\in\mathbb{Z}_{\ast}^{3}\backslash S'}\hat{V}_{k}^{2}\left|k\right|^{-\mleft(1-\epsilon\mright)}\left\langle \Psi,\mathcal{N}_{E}H_{\mathrm{kin}}^{\prime}\Psi\right\rangle }\nonumber \\
 & \times \sqrt{\sum_{q\in A_{2}}\frac{1}{\left|q\right|^{1+\epsilon}\mleft(\left|q\right|^{2}-k_{F}^{2}\mright)}\sum_{k\in\mathbb{Z}_{\ast}^{3}\backslash S'}\sum_{p\in A_{2}\cap\mleft(A_{2}+k\mright)}\mleft(\left|p\right|^{2}-k_{F}^{2}\mright)\left\Vert c_{q-k}c_{p}\Psi\right\Vert ^{2}}\nonumber \\
 & \leq\sqrt{\sum_{k\in\mathbb{Z}_{\ast}^{3}\backslash S'}\hat{V}_{k}^{2}\left|k\right|^{-\mleft(1-\epsilon\mright)}}\sqrt{\sum_{q\in A_{2}}\frac{1}{\left|q\right|^{1+\epsilon}\mleft(\left|q\right|^{2}-k_{F}^{2}\mright)}}\left\langle \Psi,\mathcal{N}_{E}H_{\mathrm{kin}}^{\prime}\Psi\right\rangle \nonumber 
\end{align}
and for $q\in A_{2}$, $k_{F}^{2}\leq\frac{1}{4}\left|q\right|^{2}$,
so
\begin{equation}
\sum_{q\in A_{2}}\frac{1}{\left|q\right|^{1+\epsilon}\mleft(\left|q\right|^{2}-k_{F}^{2}\mright)}\leq\frac{4}{3}\sum_{q\in A_{2}}\frac{1}{\left|q\right|^{3+\epsilon}}\leq C_{\epsilon}.
\end{equation}
$\hfill\square$

By inserting this bound in equation (\ref{eq:Eq5SprimeSplit}) and
using the trivial bound $\mathcal{N}_{E}H_{\mathrm{kin}}^{\prime}\leq\left|B_{F}\right|H_{\mathrm{kin}}^{\prime}\leq Ck_{F}^{3}H_{\mathrm{kin}}^{\prime}$
we arrive at the final estimate of Theorem \ref{them:FinalErrorBounds}.

\subsection{Proof of Theorem \ref{them:MainTheorem}}

We can now prove the first part of Theorem \ref{them:MainTheorem}:
\begin{prop}
Let $\frac{1}{6}\leq\beta\leq1$ and let $V$ obey Assumption \ref{Assumption:Potential}.
Then it holds as $k_{F}\rightarrow\infty$ that
\[
H_{N}\geq E_{\mathrm{FS}}+E_{\mathrm{corr},\mathrm{bos}}+E_{\mathrm{corr},\mathrm{ex}}+\mathcal{E}
\]
for an operator $\mathcal{E}$ obeying
\[
\mathcal{E}\geq-C_{V,\epsilon}k_{F}^{-\frac{1}{6}+2\mleft(1-\beta\mright)+\epsilon}\mleft(H_{\mathrm{kin}}^{\prime}+k_{F}\mright)
\]
for any $\epsilon>0$ where $C_{V,\epsilon}>0$ is a constant depending
only on $C_{V}$ and $\epsilon$.
\end{prop}

\textbf{Proof:} As remarked in the beginning of the section we have
by Theorem \ref{them:BigFactorization} that
\[
H_{N}\geq E_{\mathrm{FS}}+E_{\mathrm{corr},\mathrm{bos}}+E_{\mathrm{corr},\mathrm{ex}}+\mathcal{E}
\]
for $\mathcal{E}=\mathcal{E}_{\mathrm{B}}+\mathcal{E}_{\mathcal{C}}+2^{-1}\mleft(2\pi\mright)^{-3}k_{F}^{-\beta}\mleft(\mathcal{E}_{S}+\mathcal{E}_{\mathbb{Z}_{\ast}^{3}\backslash S}\mright)$.
By Theorems \ref{them:EBEstimate} and \ref{them:ECBound}, $\mathcal{E}_{\mathrm{B}}$
and $\mathcal{E}_{\mathcal{C}}$ obey the bounds
\begin{align}
\pm\mathcal{E}_{\mathrm{B}} & \leq C_{\epsilon}k_{F}^{2\mleft(1-\beta\mright)+\epsilon}\mleft(\sqrt{\sum_{k\in\mathbb{Z}_{\ast}^{3}\backslash S}\hat{V}_{k}^{2}}+k_{F}^{-\frac{1}{2}}\sum_{k\in S}\hat{V}_{k}\mright)\sqrt{\sum_{k\in\mathbb{Z}_{\ast}^{3}}\hat{V}_{k}^{2}\min\left\{ \left|k\right|,k_{F}\right\} }\mleft(H_{\mathrm{kin}}^{\prime}+k_{F}\mright)\nonumber\\&+C k_F^{3(1-\beta)} \sqrt{\sum_{k\in\mathbb{Z}_{\ast}^{3}}\hat{V}_{k}^{2}}\sum_{k\in\mathbb{Z}_{\ast}^{3}}\hat{V}_{k}^{2}\left|k\right|^{\frac{1}{2}} \\
\pm\mathcal{E}_{\mathcal{C}} & \leq C_{\epsilon}k_{F}^{2\mleft(1-\beta\mright)+\epsilon}\mleft(k_{F}^{-\frac{1}{2}}\sum_{k\in S}\hat{V}_{k}\mright)\mleft(\sqrt{\sum_{k\in\mathbb{Z}_{\ast}^{3}}\hat{V}_{k}^{2}\min\left\{ \left|k\right|,k_{F}\right\} }+k_{F}^{-\frac{1}{2}}\sum_{k\in S}\hat{V}_{k}\mright)\mleft(H_{\mathrm{kin}}^{\prime}+k_{F}\mright)\nonumber
\end{align}
and under Assumption \ref{Assumption:Potential} it holds that with
$S=\overline{B}(0,k_{F}^{1/3})\cap\mathbb{Z}_{\ast}^{3}$
\begin{equation}\label{eq:optimizing-cut-off-end}
\sqrt{\sum_{k\in\mathbb{Z}_{\ast}^{3}\backslash S}\hat{V}_{k}^{2}} \le C_{V}^{\prime}k_{F}^{-\frac{1}{6}},\qquad k_{F}^{-\frac{1}{2}}\sum_{k\in S}\hat{V}_{k}\leq C_{V}^{\prime}k_{F}^{-\frac{1}{6}},\qquad\sqrt{\sum_{k\in\mathbb{Z}_{\ast}^{3}}\hat{V}_{k}^{2}}\sum_{k\in\mathbb{Z}_{\ast}^{3}}\hat{V}_{k}^{2}\left|k\right|^{\frac{1}{2}}\leq C_{V}^{\prime},
\end{equation}
and
\begin{equation}
\sum_{k\in\mathbb{Z}_{\ast}^{3}}\hat{V}_{k}^{2}\min\left\{ \left|k\right|,k_{F}\right\} \leq C_{V}^{\prime}\log\mleft(k_{F}\mright)\leq C_{V,\epsilon}k_{F}^{\epsilon}
\end{equation}
for every $\epsilon>0$ which together imply a bound of the form
\begin{equation}\label{eq:final-error-665}
\mathcal{E}_{\mathrm{B}}+\mathcal{E}_{\mathcal{C}}\geq-C_{V,\epsilon}k_{F}^{-\frac{1}{6}+2\mleft(1-\beta\mright)+\epsilon}\mleft(H_{\mathrm{kin}}^{\prime}+k_{F}\mright)
\end{equation}
where we used the assumption $\beta\geq\frac{1}{6}$ to absorb the
$k_{F}^{3\mleft(1-\beta\mright)}$ term into the rest.

By Theorem \ref{them:FinalErrorBounds} it follows that with $S'=\overline{B}(0,k_{F}^{5/2})\cap\mathbb{Z}_{\ast}^{3}$
\begin{align}
\mathcal{E}_{S}+\mathcal{E}_{\mathbb{Z}_{\ast}^{3}\backslash S} & \geq-C_{\epsilon}\mleft(k_{F}^{1+\epsilon}\sqrt{\sum_{k\in\mathbb{Z}_{\ast}^{3}\backslash S}\hat{V}_{k}^{2}}+\sum_{k\in S}\hat{V}_{k}+C_{V}^{\prime}k_{F}^{\frac{2}{3}}\mright)H_{\mathrm{kin}}^{\prime}\nonumber \\
 & -C_{\epsilon}^{\prime}\mleft(k_{F}^{-2}\sum_{k\in S'}\hat{V}_{k}+k_{F}^{3}\mleft(\sum_{k\in\mathbb{Z}_{\ast}^{3}\backslash S'}\hat{V}_{k}\left|k\right|^{-2}+\sqrt{\sum_{k\in\mathbb{Z}_{\ast}^{3}\backslash S'}\hat{V}_{k}^{2}\left|k\right|^{-\mleft(1-\epsilon\mright)}}\mright)\mright)H_{\mathrm{kin}}^{\prime}\\
 & \geq-C_{V,\epsilon}k_{F}^{-\frac{1}{6}+1+\epsilon}H_{\mathrm{kin}}^{\prime}\nonumber 
\end{align}
so all in all
\begin{equation}
\mathcal{E}=\mathcal{E}_{\mathrm{B}}+\mathcal{E}_{\mathcal{C}}+\frac{k_{F}^{-\beta}}{2\mleft(2\pi\mright)^{3}}\mleft(\mathcal{E}_{S}+\mathcal{E}_{\mathbb{Z}_{\ast}^{3}\backslash S}\mright)\geq-C_{V,\epsilon}k_{F}^{-\frac{1}{6}+2\mleft(1-\beta\mright)+\epsilon}\mleft(H_{\mathrm{kin}}^{\prime}+k_{F}\mright)
\end{equation}
since $\beta\leq1$.

$\hfill\square$

As remarked in the introduction, we can by this result conclude the
inequality
\begin{equation}
\mleft(1-o\mleft(1\mright)\mright)H_{\mathrm{kin}}^{\prime}\leq2\mleft(H_{N}-E_{\mathrm{FS}}\mright)-\tilde{E}_{\mathrm{corr},\mathrm{bos}}+Ck_{F},\quad k_{F}\rightarrow\infty,
\end{equation}
when $\beta>\frac{11}{12}$ (to ensure $-\frac{1}{6}+2\mleft(1-\beta\mright)+\epsilon<0$
for some $\epsilon$), where
\begin{equation}
\tilde{E}_{\mathrm{corr},\mathrm{bos}}=\frac{1}{\pi}\sum_{k\in\mathbb{Z}_{\ast}^{3}}\int_{0}^{\infty}F\mleft(\frac{2\hat{V}_{k}k_{F}^{-\beta}}{\mleft(2\pi\mright)^{3}}\sum_{p\in L_{k}}\frac{\lambda_{k,p}}{\lambda_{k,p}^{2}+t^{2}}\mright)dt,\quad F\mleft(x\mright)=\log\mleft(1+x\mright)-x.
\end{equation}
The second part of Theorem \ref{them:MainTheorem} is now an immediate
consequence of the following:
\begin{prop}
Let $V$ obey Assumption \ref{Assumption:Potential}. Then for any
$\epsilon>0$ it holds that
\[
-\tilde{E}_{\mathrm{corr},\mathrm{bos}}\leq C_{V,\epsilon}k_{F}^{3-2\beta+\epsilon},\quad k_{F}\rightarrow\infty,
\]
for a constant $C_{V,\epsilon}>0$ depending only on $C_{V}$ and
$\epsilon$.
\end{prop}

\textbf{Proof:} By the inequality $\log\mleft(1+x\mright)\geq x-\frac{1}{2}x^{2}$,
valid for all $x\geq0$, we see that
\begin{align}
-\tilde{E}_{\mathrm{corr},\mathrm{bos}} & \leq\frac{1}{\pi}\sum_{k\in\mathbb{Z}_{\ast}^{3}}\int_{0}^{\infty}\frac{1}{2}\mleft(\frac{2\hat{V}_{k}k_{F}^{-\beta}}{\mleft(2\pi\mright)^{3}}\sum_{p\in L_{k}}\frac{\lambda_{k,p}}{\lambda_{k,p}^{2}+t^{2}}\mright)^{2}dt\nonumber \\
 & =\frac{4k_{F}^{-2\beta}}{\mleft(2\pi\mright)^{7}}\sum_{k\in\mathbb{Z}_{\ast}^{3}}\hat{V}_{k}^{2}\sum_{p,q\in L_{k}}\int_{0}^{\infty}\frac{\lambda_{k,p}}{\lambda_{k,p}^{2}+t^{2}}\frac{\lambda_{k,q}}{\lambda_{k,q}^{2}+t^{2}}dt\\
 & =\frac{k_{F}^{-2\beta}}{\mleft(2\pi\mright)^{6}}\sum_{k\in\mathbb{Z}_{\ast}^{3}}\hat{V}_{k}^{2}\sum_{p,q\in L_{k}}\frac{1}{\lambda_{k,p}+\lambda_{k,q}}\nonumber 
\end{align}
where we applied the integral identity $\int_{0}^{\infty}\frac{a}{a^{2}+t^{2}}\frac{b}{b^{2}+t^{2}}dt=\frac{\pi}{2}\mleft(a+b\mright)^{-1}$,
valid for all $a,b>0$. Now by Proposition \ref{prop:SimpleLuneRiemannSums}, 
\begin{equation}
\sum_{p,q\in L_{k}}\frac{1}{\lambda_{k,p}+\lambda_{k,q}}\leq\left|L_{k}\right|\sum_{p\in L_{k}}\frac{1}{\lambda_{k,p}}\leq Ck_{F}^{3}\min\left\{ \left|k\right|,k_{F}\right\},  
\end{equation}
and as noted above, $\sum_{k\in\mathbb{Z}_{\ast}^{3}}\hat{V}_{k}^{2}\min\left\{ \left|k\right|,k_{F}\right\} \leq C_{V,\epsilon}k_{F}^{\epsilon}$
under Assumption \ref{Assumption:Potential}, from which the claim
follows.

$\hfill\square$

\appendix
\section{Appendix}

\subsection{\label{subsec:KineticSumEstimates}Kinetic Sum Estimates}

We will use the following well-known estimate for the number of lattice points
on a sphere (see for instance \cite[Section 2]{BouRudSar-17}) which is a consequence of the fact that the number of 2D integer points on a circle of radius $R$ is $O(R^\eps)$ for every $\eps>0$. 
\begin{prop}
\label{prop:LatticePointSphereEstimate}For any $n\in\mathbb{N}$
and $\epsilon>0$ it holds that
\[
r_{3}\mleft(n\mright):=|\{p\in\mathbb{Z}^{3}\mid\left|p\right|^{2}=n\}|\leq C_{\epsilon}n^{\frac{1}{2}+\epsilon}
\]
for a constant $C_{\epsilon}>0$ depending only on $\epsilon$.
\end{prop}

We note that
\begin{equation}
\zeta=\frac{1}{2}\mleft(\inf_{p\in B_{F}^{c}}\left|p\right|^{2}+\sup_{q\in B_{F}}\left|q\right|^{2}\mright)
\end{equation}
obeys $|\zeta-k_{F}^{2}|\leq k_{F}+1$ (\cite[eq. A.90]{ChrHaiNam-23a})
and crucially
\begin{equation}
\vert\left|p\right|^{2}-\zeta\vert\geq\frac{1}{2},\quad\forall p\in\mathbb{Z}^{3}.\label{eq:ZetaLowerBound}
\end{equation}

Now we are ready to give the proof of Lemma \ref{lemma:GeneralKineticSum}.

\textbf{Proof:} By rearrangement it holds for sufficiently large $k_{F}$
that
\begin{align}
\sum_{p\in A}\frac{1}{\vert\left|p\right|^{2}-\zeta\vert} & \leq\sum_{p\in\overline{B}\left(0,2k_{F}\right)\cap\mathbb{Z}^{3}}\frac{1}{\vert\left|p\right|^{2}-\zeta\vert}=\sum_{p\in\overline{B}\left(0,2k_{F}\right)\cap\mathbb{Z}_{\ast}^{3}}\frac{1}{\vert\left|p\right|^{2}-\zeta\vert}+\zeta^{-1}\\
 & \leq\sum_{p\in\overline{B}\left(0,2k_{F}\right)\cap\mathbb{Z}_{\ast}^{3}}\frac{1}{\vert\left|p\right|^{2}-\zeta\vert}+2k_{F}^{-2},\quad k_{F}\rightarrow\infty. \nonumber
\end{align}
Here we used $\vert\left|p\right|^{2}-\zeta\vert^{-1} \leq \vert\left|q\right|^{2}-\zeta\vert^{-1}$ for all $p\notin\overline{B}\mleft(0,2k_{F}\mright)$ and $q\in\overline{B}\mleft(0,2k_{F}\mright)$, which follows from the facts that 
 $\vert\left|p\right|^{2}-\zeta\vert^{-1}$ is radially increasing
for $p\in B_{F}$, radially decreasing for $p\in B_{F}^{c}$, and
\begin{equation}
\vert\left|p\right|^{2}-\zeta\vert^{-1}\leq\mleft(4k_{F}^{2}-\zeta\mright)^{-1}\leq\mleft(3k_{F}^{2}-k_{F}-1\mright)^{-1}\leq\frac{1}{2}k_{F}^{-2},\quad k_{F}\rightarrow\infty,
\end{equation}
for $p\notin\overline{B}\mleft(0,2k_{F}\mright)$ while
\begin{equation}
\vert\left|q\right|^{2}-\zeta\vert^{-1}\geq\zeta^{-1}\geq\mleft(k_{F}^{2}+k_{F}+1\mright)^{-1}\geq\frac{1}{2}k_{F}^{-2},\quad k_{F}\rightarrow\infty,
\end{equation}
for $q\in B_{F}$. 

The sum can now be written as
\begin{equation}
\sum_{p\in\overline{B}\mleft(0,2k_{F}\mright)\cap\mathbb{Z}_{\ast}^{3}}\frac{1}{\vert\left|p\right|^{2}-\zeta\vert}=\sum_{n=1}^{\left\lfloor 4k_{F}^{2}\right\rfloor }\frac{r_{3}\mleft(n\mright)}{\vert n-\zeta\vert}=\sum_{n=1}^{m}\frac{r_{3}\mleft(n\mright)}{\zeta-n}+\sum_{n=m'}^{\left\lfloor 4k_{F}^{2}\right\rfloor }\frac{r_{3}\mleft(n\mright)}{n-\zeta}
\end{equation}
where
\begin{equation}
m=\sup_{q\in B_{F}}\left|q\right|^{2},\quad m'=\inf_{p\in B_{F}^{c}}\left|p\right|^{2}.
\end{equation}
We can use Proposition \ref{prop:LatticePointSphereEstimate} and
the fact that $t\mapsto\sqrt{t}\mleft(\zeta-t\mright)^{-1}$ is increasing
for $t\in\mleft(0,\zeta\mright)$ to estimate the first sum as
\begin{equation}
\sum_{n=1}^{m}\frac{r_{3}\mleft(n\mright)}{\zeta-n}\leq C_{\epsilon}\sum_{n=1}^{m}\frac{n^{\frac{1}{2}+\frac{1}{2}\epsilon}}{\zeta-n}\leq C_{\epsilon}k_{F}^{\epsilon}\mleft(\frac{\sqrt{m}}{\zeta-m}+\sum_{n=1}^{m-1}\frac{\sqrt{n}}{\zeta-n}\mright)\leq C_{\epsilon}k_{F}^{\epsilon}\mleft(2\sqrt{m}+\int_{1}^{m}\frac{\sqrt{t}}{\zeta-t}\,dt\mright)
\end{equation}
where we also used that equation (\ref{eq:ZetaLowerBound}) implies
that $\mleft(\zeta-m\mright)^{-1}\leq2$. The integral obeys
\begin{align}
\int_{1}^{m}\frac{\sqrt{t}}{\zeta-t}\,dt & \leq\int_{0}^{m}\frac{2(\sqrt{t})^{2}}{\zeta-(\sqrt{t})^{2}}(\sqrt{t})'dt=\int_{0}^{\sqrt{m}}\frac{2t^{2}}{\zeta-t^{2}}dt=\sqrt{\zeta}\log\mleft(\frac{\sqrt{\zeta}+\sqrt{m}}{\sqrt{\zeta}-\sqrt{m}}\mright)-2\sqrt{m}\\
 & =\sqrt{\zeta}\log\mleft(\frac{\mleft(\sqrt{\zeta}+\sqrt{m}\mright)^{2}}{\zeta-m}\mright)-2\sqrt{m}\leq\sqrt{\zeta}\log\mleft(8 \zeta \mright)-2\sqrt{m}\nonumber 
\end{align}
whence
\begin{equation}
\sum_{n=1}^{m}\frac{r_{3}\mleft(n\mright)}{\zeta-n}\leq C_{\epsilon}k_{F}^{\epsilon}\sqrt{\zeta}\log\mleft(8\zeta\mright)\leq C_{\epsilon}^{\prime}k_{F}^{1+\epsilon'}.
\end{equation}
For the other sum we can similarly estimate
\begin{equation}
\sum_{n=m'}^{\left\lfloor 4k_{F}^{2}\right\rfloor }\frac{r_{3}\mleft(n\mright)}{n-\zeta}\leq C_{\epsilon}k_{F}^{\epsilon}\mleft(\frac{\sqrt{m'}}{m'-\zeta}+\sum_{n=m'+1}^{\left\lfloor 4k_{F}^{2}\right\rfloor }\frac{\sqrt{n}}{n-\zeta}\mright)\leq C_{\epsilon}k_{F}^{\epsilon}\mleft(2\sqrt{m'}+\int_{m'}^{\left\lfloor 4k_{F}^{2}\right\rfloor }\frac{\sqrt{t}}{t-\zeta}\,dt\mright)
\end{equation}
as $t\mapsto\sqrt{t}\mleft(t-\zeta\mright)^{-1}$ is decreasing on $\mleft(\zeta,\infty\mright)$.
This integral can be bounded as
\begin{align}
\int_{m'}^{\left\lfloor 4k_{F}^{2}\right\rfloor }\frac{\sqrt{t}}{t-\zeta}\,dt & \leq\int_{\sqrt{m'}}^{2k_{F}}\frac{2t^{2}}{t^{2}-\zeta}\,dt=4k_{F}-2\sqrt{m'}+\int_{\sqrt{m'}}^{2k_{F}}\frac{2\zeta}{t^{2}-\zeta}\,dt\nonumber \\
 & =4k_{F}-2\sqrt{m'}+\sqrt{\zeta}\mleft(\log\mleft(\frac{2k_{F}-\sqrt{\zeta}}{2k_{F}+\sqrt{\zeta}}\mright)-\log\mleft(\frac{\sqrt{m'}-\sqrt{\zeta}}{\sqrt{m'}+\sqrt{\zeta}}\mright)\mright)\\
 & \leq4k_{F}-2\sqrt{m'}+\sqrt{\zeta}\log\mleft(8m'\mright)\nonumber 
\end{align}
whence
\begin{equation}
\sum_{n=m'}^{\left\lfloor 4k_{F}^{2}\right\rfloor }\frac{r_{3}\mleft(n\mright)}{n-\zeta}\leq C_{\epsilon}k_{F}^{\epsilon}\mleft(4k_{F}+\sqrt{\zeta}\log\mleft(8m'\mright)\mright)\leq C_{\epsilon}^{\prime}k_{F}^{1+\epsilon'}.
\end{equation}
Combining the estimates yields the claim.

$\hfill\square$

\subsection{\label{subsec:One-BodyOperatorEstimates}One-Body Operator Estimates}

Let $\mleft(V,\left\langle \cdot,\cdot\right\rangle \mright)$ be an
$n$-dimensional Hilbert space, $h:V\rightarrow V$ be a positive
self-adjoint operator with eigenbasis $\mleft(x_{i}\mright)_{i=1}^{n}$
and eigenvalues $\mleft(\lambda_{i}\mright)_{i=1}^{n}$, and let $v\in V$
be a vector with $\left\langle x_{i},v\right\rangle \geq0$ for all
$1\leq i\leq n$.

We define
\begin{equation}
E=(h^{\frac{1}{2}}\mleft(h+2P_{v}\mright)h^{\frac{1}{2}})^{\frac{1}{2}}=(h^{2}+2P_{h^{\frac{1}{2}}v})^{\frac{1}{2}}
\end{equation}
where $P_{w}=\left|w\right\rangle \left\langle w\right|$ for any
$w\in V$, and in terms of this further define
\begin{equation}
C=\frac{1}{2}\mleft(h^{-\frac{1}{2}}E^{\frac{1}{2}}+h^{\frac{1}{2}}E^{-\frac{1}{2}}\mright),\quad S=\frac{1}{2}\mleft(h^{-\frac{1}{2}}E^{\frac{1}{2}}-h^{\frac{1}{2}}E^{-\frac{1}{2}}\mright).
\end{equation}
Note that $E^{\frac{1}{2}}$ is the fourth root of a rank one perturbation.
As in \cite[Proposition 9.9]{ChrHaiNam-23a}, the Sherman-Morrison
formula
\begin{equation}
\mleft(A+gP_{w}\mright)^{-1}=A^{-1}-\frac{g}{1+g\left\langle w,A^{-1}w\right\rangle }P_{A^{-1}w}
\end{equation}
and the integral identity $a^{\frac{1}{4}}=\frac{2\sqrt{2}}{\pi}\int_{0}^{\infty}\mleft(1-t^{4}\mleft(a+t^{4}\mright)^{-1}\mright)dt$,
$a\geq0$, yields the following characterization of such operator
roots:
\begin{prop}
\label{prop:FourthRootofRankOnePerturbation}Let $A:V\rightarrow V$
be a positive self-adjoint operator. Then for any $w\in V$ it holds
that
\begin{align*}
\mleft(A+P_{w}\mright)^{\frac{1}{4}} & =A^{\frac{1}{4}}+\frac{2\sqrt{2}}{\pi}\int_{0}^{\infty}\frac{t^{4}}{1+\left\langle w,\mleft(A+t^{4}\mright)^{-1}w\right\rangle }P_{\mleft(A+t^{4}\mright)^{-1}w}dt\\
\mleft(A+P_{w}\mright)^{-\frac{1}{4}} & =A^{-\frac{1}{4}}-\frac{2\sqrt{2}}{\pi}\int_{0}^{\infty}\frac{t^{4}}{1+\left\langle w,A^{-1}\mleft(A^{-1}+t^{4}\mright)^{-1}w\right\rangle t^{4}}P_{A^{-1}\mleft(A^{-1}+t^{4}\mright)^{-1}w}dt.
\end{align*}
\end{prop}

This implies the following:
\begin{prop}
For all $1\leq i,j\leq n$ it holds that
\begin{align*}
\frac{1}{1+2\left\langle v,h^{-1}v\right\rangle }\frac{2\sqrt{\lambda_{i}\lambda_{j}}}{\sqrt{\lambda_{i}}+\sqrt{\lambda_{j}}}\frac{\left\langle x_{i},v\right\rangle \left\langle v,x_{j}\right\rangle }{\lambda_{i}+\lambda_{j}} & \leq\left\langle x_{i},\mleft(E^{\frac{1}{2}}-h^{\frac{1}{2}}\mright)x_{j}\right\rangle \leq\frac{2\sqrt{\lambda_{i}\lambda_{j}}}{\sqrt{\lambda_{i}}+\sqrt{\lambda_{j}}}\frac{\left\langle x_{i},v\right\rangle \left\langle v,x_{j}\right\rangle }{\lambda_{i}+\lambda_{j}}\\
\frac{1}{1+2\left\langle v,h^{-1}v\right\rangle }\frac{2}{\sqrt{\lambda_{i}}+\sqrt{\lambda_{j}}}\frac{\left\langle x_{i},v\right\rangle \left\langle v,x_{j}\right\rangle }{\lambda_{i}+\lambda_{j}} & \leq\left\langle x_{i},\mleft(h^{-\frac{1}{2}}-E^{-\frac{1}{2}}\mright)x_{j}\right\rangle \leq\frac{2}{\sqrt{\lambda_{i}}+\sqrt{\lambda_{j}}}\frac{\left\langle x_{i},v\right\rangle \left\langle v,x_{j}\right\rangle }{\lambda_{i}+\lambda_{j}}.
\end{align*}
\end{prop}

\textbf{Proof:} Taking $A=h^{2}$ and $w=\sqrt{2}h^{\frac{1}{2}}v$
in Proposition \ref{prop:FourthRootofRankOnePerturbation}, we have
\begin{align}
E^{\frac{1}{2}} & =h^{\frac{1}{2}}+\frac{4\sqrt{2}}{\pi}\int_{0}^{\infty}\frac{t^{4}}{1+2\left\langle v,h\mleft(h^{2}+t^{4}\mright)^{-1}v\right\rangle }P_{h^{\frac{1}{2}}\mleft(h^{2}+t^{4}\mright)^{-1}v}dt\\
E^{-\frac{1}{2}} & =h^{-\frac{1}{2}}-\frac{4\sqrt{2}}{\pi}\int_{0}^{\infty}\frac{t^{4}}{1+2\left\langle v,h^{-1}\mleft(h^{-2}+t^{4}\mright)^{-1}v\right\rangle t^{4}}P_{h^{-\frac{3}{2}}\mleft(h^{-2}+t^{4}\mright)^{-1}v}dt\nonumber 
\end{align}
so
\begin{align}
\left\langle x_{i},\mleft(E^{\frac{1}{2}}-h^{\frac{1}{2}}\mright)x_{j}\right\rangle  & =\frac{4\sqrt{2}}{\pi}\int_{0}^{\infty}\frac{t^{4}}{1+2\left\langle v,h\mleft(h^{2}+t^{4}\mright)^{-1}v\right\rangle }\frac{\sqrt{\lambda_{i}}\left\langle x_{i},v\right\rangle }{\lambda_{i}^{2}+t^{4}}\frac{\sqrt{\lambda_{j}}\left\langle v,x_{j}\right\rangle }{\lambda_{j}^{2}+t^{4}}dt\\
\left\langle x_{i},\mleft(h^{-\frac{1}{2}}-E^{-\frac{1}{2}}\mright)x_{j}\right\rangle  & =\frac{4\sqrt{2}}{\pi}\int_{0}^{\infty}\frac{t^{4}}{1+2\left\langle v,h^{-1}\mleft(h^{-2}+t^{4}\mright)^{-1}v\right\rangle t^{4}}\frac{\lambda_{i}^{-\frac{3}{2}}\left\langle x_{i},v\right\rangle }{\lambda_{i}^{-2}+t^{4}}\frac{\lambda_{j}^{-\frac{3}{2}}\left\langle v,x_{j}\right\rangle }{\lambda_{j}^{-2}+t^{4}}dt\nonumber 
\end{align}
and the estimates now follow from the fact that
\begin{equation}
0\leq\left\langle v,h\mleft(h^{2}+t^{4}\mright)^{-1}v\right\rangle ,\left\langle v,h^{-1}\mleft(h^{-2}+t^{4}\mright)^{-1}v\right\rangle t^{4}\leq\left\langle v,h^{-1}v\right\rangle 
\end{equation}
for all $t\geq0$, as well as the integral identities (for $a,b>0$)
\begin{align}
\int_{0}^{\infty}\frac{a^{-\frac{3}{2}}}{a^{-2}+t^{4}}\frac{b^{-\frac{3}{2}}}{b^{-2}+t^{4}}t^{4}dt = \int_{0}^{\infty}\frac{1}{a^{2}+t^{4}}\frac{1}{b^{2}+t^{4}}t^{4}dt  =\frac{\pi}{2\sqrt{2}}\frac{1}{\sqrt{a}+\sqrt{b}}\frac{1}{a+b}. 
\end{align}
$\hfill\square$

This leads to the following bounds for $C$ and $S$:
\begin{prop}
For all $1\leq i,j\leq n$ it holds that
\begin{align*}
\left|\left\langle x_{i},\mleft(C-1\mright)x_{j}\right\rangle \right|,\left|\left\langle x_{i},Sx_{j}\right\rangle \right| & \leq\frac{\left\langle x_{i},v\right\rangle \left\langle v,x_{j}\right\rangle }{\lambda_{i}+\lambda_{j}}\\
\left|\left\langle x_{i},Sx_{j}\right\rangle -\frac{\left\langle x_{i},v\right\rangle \left\langle v,x_{j}\right\rangle }{\lambda_{i}+\lambda_{j}}\right| & \leq2\left\langle v,h^{-1}v\right\rangle \frac{\left\langle x_{i},v\right\rangle \left\langle v,x_{j}\right\rangle }{\lambda_{i}+\lambda_{j}}.
\end{align*}
\end{prop}

\textbf{Proof:} From the definition of $C$ we have
\begin{equation}
\left\langle x_{i},\mleft(C-1\mright)x_{j}\right\rangle =\frac{1}{2}\mleft(\lambda_{i}^{-\frac{1}{2}}\left\langle x_{i},\mleft(E^{\frac{1}{2}}-h^{\frac{1}{2}}\mright)x_{j}\right\rangle -\lambda_{i}^{\frac{1}{2}}\left\langle x_{i},\mleft(h^{-\frac{1}{2}}-E^{-\frac{1}{2}}\mright)x_{j}\right\rangle \mright)
\end{equation}
and by the proposition
\begin{align}
0\leq\lambda_{i}^{-\frac{1}{2}}\left\langle x_{i},\mleft(E^{\frac{1}{2}}-h^{\frac{1}{2}}\mright)x_{j}\right\rangle  & \leq\frac{2\sqrt{\lambda_{j}}}{\sqrt{\lambda_{i}}+\sqrt{\lambda_{j}}}\frac{\left\langle x_{i},v\right\rangle \left\langle v,x_{j}\right\rangle }{\lambda_{i}+\lambda_{j}}\leq2\frac{\left\langle x_{i},v\right\rangle \left\langle v,x_{j}\right\rangle }{\lambda_{i}+\lambda_{j}}\\
0\leq\lambda_{i}^{\frac{1}{2}}\left\langle x_{i},\mleft(h^{-\frac{1}{2}}-E^{-\frac{1}{2}}\mright)x_{j}\right\rangle  & \leq\frac{2\sqrt{\lambda_{i}}}{\sqrt{\lambda_{i}}+\sqrt{\lambda_{j}}}\frac{\left\langle x_{i},v\right\rangle \left\langle v,x_{j}\right\rangle }{\lambda_{i}+\lambda_{j}}\leq2\frac{\left\langle x_{i},v\right\rangle \left\langle v,x_{j}\right\rangle }{\lambda_{i}+\lambda_{j}}\nonumber 
\end{align}
whence the claim for $C-1$. For $S$ we have
\begin{align}
\left\langle x_{i},Sx_{j}\right\rangle  & =\frac{1}{2}\mleft(\lambda_{i}^{-\frac{1}{2}}\left\langle x_{i},\mleft(E^{\frac{1}{2}}-h^{\frac{1}{2}}\mright)x_{j}\right\rangle +\lambda_{i}^{\frac{1}{2}}\left\langle x_{i},\mleft(h^{-\frac{1}{2}}-E^{-\frac{1}{2}}\mright)x_{i}\right\rangle \mright)\\
 & \leq\frac{1}{2}\mleft(\frac{2\sqrt{\lambda_{j}}}{\sqrt{\lambda_{i}}+\sqrt{\lambda_{j}}}\frac{\left\langle x_{i},v\right\rangle \left\langle v,x_{j}\right\rangle }{\lambda_{i}+\lambda_{j}}+\frac{2\sqrt{\lambda_{i}}}{\sqrt{\lambda_{i}}+\sqrt{\lambda_{j}}}\frac{\left\langle x_{i},v\right\rangle \left\langle v,x_{j}\right\rangle }{\lambda_{i}+\lambda_{j}}\mright)=\frac{\left\langle x_{i},v\right\rangle \left\langle v,x_{j}\right\rangle }{\lambda_{i}+\lambda_{j}}\nonumber 
\end{align}
hence the general bound for $S$, and also
\begin{equation}
\left\langle x_{i},Sx_{j}\right\rangle \geq\frac{1}{1+2\left\langle v,h^{-1}v\right\rangle }\frac{\left\langle x_{i},v\right\rangle \left\langle v,x_{j}\right\rangle }{\lambda_{i}+\lambda_{j}}
\end{equation}
whence
\begin{equation}
\left|\left\langle x_{i},Sx_{j}\right\rangle -\frac{\left\langle x_{i},v\right\rangle \left\langle v,x_{j}\right\rangle }{\lambda_{i}+\lambda_{j}}\right|\leq\mleft(1-\frac{1}{1+2\left\langle v,h^{-1}v\right\rangle }\mright)\frac{\left\langle x_{i},v\right\rangle \left\langle v,x_{j}\right\rangle }{\lambda_{i}+\lambda_{j}}\leq2\left\langle v,h^{-1}v\right\rangle \frac{\left\langle x_{i},v\right\rangle \left\langle v,x_{j}\right\rangle }{\lambda_{i}+\lambda_{j}}.
\end{equation}
$\hfill\square$

Proposition \ref{prop:OneBodyOperatorEstimates} now follows by the
substitutions $\lambda_{i}\rightarrow\lambda_{k,p}$, $\left\langle x_{i},v\right\rangle \rightarrow\sqrt{\frac{\hat{V}_{k}k_{F}^{-\beta}}{2\mleft(2\pi\mright)^{3}}}$
and using that
\begin{equation}
\left\langle v_{k},h_{k}^{-1}v_{k}\right\rangle =\frac{\hat{V}_{k}k_{F}^{-\beta}}{2\mleft(2\pi\mright)^{3}}\sum_{p\in L_{k}}\lambda_{k,p}^{-1}\leq C\hat{V}_{k}k_{F}^{1-\beta}.
\end{equation}

\subsection{\protect\label{subsec:InsideOutsideComparison}Proof of Proposition
\ref{prop:InsideOutsideComparison}}

We will apply the following elementary lemmas in the proof:
\begin{lem}
\label{lemma:ExtensionResult}Let $\hat{V}_{k}$ obey Assumption \ref{Assumption:Potential}.
Then there exists a continuous monotone decreasing function $f:\left[0,\infty\right)\rightarrow\left[0,C_{V}\right]$
such that $f(0)=C_V$ and $\hat{V}_{k}=f(\left|k\right|)$ for all $k\in\mathbb{Z}_{\ast}^{3}$.
\end{lem}

\begin{lem}
\label{lemma:LensCountingResult}Let $z\in\mathbb{R}^{3}$ and $R_{1},R_{2}>0$
be given. Then it holds that
\[
\left|\left|\overline{B}(0,R_{1})\cap\overline{B}(z,R_{2})\cap\mathbb{Z}^{3}\right|-\mathrm{Vol}\mleft(\overline{B}(0,R_{1})\cap\overline{B}(z,R_{2})\mright)\right|\leq C\mleft(1+R_{1}^{2}+R_{2}^{2}\mright)
\]
for a constant $C>0$ independent of all quantities.
\end{lem}

The first essentially allows us to assume that $\hat{V}_{k}$ is defined
for all $k\in\mathbb{R}^{3}$. The proof follows by linear interpolation
in the radial direction. 
The second lattice point counting result follows from the standard argument from the (three-dimensional) Gauss circle problem, using the inclusion 
\begin{equation}\big(\overline{B}(0,(R_1-\sqrt{3}/2)_+)\cap \overline{B}(z,(R_2-\sqrt{3}/2)_+) \big) \subset \Omega \subset \big( \overline{B} (0,R_1+\sqrt{3}/2)\cap \overline{B}(z,R_2+\sqrt{3}/2)\big),
\end{equation}
where $\Omega$ is the union of unit cubes centered at lattice points in $\overline{B}(0,R_1)\cap \overline{B}(z,R_2)\cap\mathbb{Z}^{3}$ (the constant $\sqrt{3}/2$ arising as the distance from the center of a unit cube to its corners).

We can now prove Proposition \ref{prop:InsideOutsideComparison}.

\textbf{Proof:} By Lemma \ref{lemma:ExtensionResult} we can write
\begin{equation}
\sum_{k\in\mleft(B_{F}+q\mright)\backslash\left\{ 0\right\} }\hat{V}_{k}-\sum_{k\in\mleft(B_{F}+p\mright)}\hat{V}_{k}=\sum_{k\in\mleft(B_{F}+q\mright)}f\mleft(\left|k\right|\mright)-\sum_{k\in\mleft(B_{F}+p\mright)}f\mleft(\left|k\right|\mright)-C_{V}
\end{equation}
so it suffices to show that $\sum_{k\in\mleft(B_{F}+q\mright)}f\mleft(\left|k\right|\mright)-\sum_{k\in\mleft(B_{F}+p\mright)}f\mleft(\left|k\right|\mright)\geq-C_{V}^{\prime}k_{F}^{\frac{2}{3}}$
for any $p\in B_{F}^{c}$ and $q\in B_{F}$.

Consider first the case $\left|p\right|>3k_{F}$: In this case every
element of $(B_{F}+q)$ has a smaller norm than every element of $(B_{F}+p)$,
and since both sets have the same cardinality the monotonicity of
$f$ implies the strong inequality $\sum_{k\in\mleft(B_{F}+q\mright)}f\mleft(\left|k\right|\mright)\geq\sum_{k\in\mleft(B_{F}+p\mright)}f\mleft(\left|k\right|\mright)$
in this case. Going forward we can thus assume that $\left|p\right|\leq3k_{F}$,
hence that $\mleft(B_{F}+p\mright),\,\mleft(B_{F}+q\mright)\subset\overline{B}(0,4k_{F})$.

Now we can for any $\epsilon>0$ estimate
\begin{align}
 & \quad\,\sum_{k\in\mleft(B_{F}+q\mright)}f\mleft(\left|k\right|\mright)-\sum_{k\in\mleft(B_{F}+p\mright)}f\mleft(\left|k\right|\mright)\nonumber \\
 & \geq\sum_{k\in\mleft(B_{F}+q\mright)}\min\{f\mleft(\left|k\right|\mright),\epsilon\}-\sum_{k\in\mleft(B_{F}+p\mright)}1_{\{f(\left|k\right|)\leq\epsilon\}}\mleft(k\mright)f\mleft(\left|k\right|\mright)-\sum_{k\in\mleft(B_{F}+p\mright)}1_{\{f(\left|k\right|)>\epsilon\}}\mleft(k\mright)f\mleft(\left|k\right|\mright)\label{eq:SumDifferenceDecomposition}\\
 & \geq\sum_{k\in\mleft(B_{F}+q\mright)}\min\{f\mleft(\left|k\right|\mright),\epsilon\}-\sum_{k\in\mleft(B_{F}+p\mright)}\min\{f\mleft(\left|k\right|\mright),\epsilon\}-\sum_{k\in\mleft(B_{F}+p\mright)}1_{\{f(\left|k\right|)>\epsilon\}}\mleft(k\mright)f\mleft(\left|k\right|\mright).\nonumber 
\end{align}
For the final sum we note that by Assumption \ref{Assumption:Potential}
\begin{equation}
\{k\in\mathbb{Z}^{3}\mid f(\left|k\right|)>\epsilon\}=\{k\in\mathbb{Z}^{3}\mid\hat{V}_{k}>\epsilon\}\subset\{k\in\mathbb{Z}^{3}\mid C_{V}\left|k\right|^{-2}>\epsilon\}=B(0,C_{V}^{\frac{1}{2}}\epsilon^{-\frac{1}{2}})\cap\mathbb{Z}^{3}
\end{equation}
so (employing Assumption \ref{Assumption:Potential} once more)
\begin{equation}
\sum_{k\in\mleft(B_{F}+p\mright)}1_{\{f(\left|k\right|)>\epsilon\}}\mleft(k\mright)f\mleft(\left|k\right|\mright)\leq C_{V}+\sum_{k\in B(0,C_{V}^{1/2}\epsilon^{-1/2})\cap\mathbb{Z}_{\ast}^{3}}C_{V}\left|k\right|^{-2}\leq C_{V}^{\prime}(1+\epsilon^{-\frac{1}{2}})\label{eq:DecompositionLowkBound}
\end{equation}
For the difference between the sums involving $\min\{f\mleft(\left|k\right|\mright),\epsilon\}$
we will work with a slightly modified version of $f$: We set $f_{\delta}\mleft(t\mright):=f\mleft(t\mright)+\delta\mleft(1+t\mright)^{-2}$.
Then $f_{\delta}$ is a \textit{strictly} monotone decreasing function, hence
is a bijection between $\left[0,\infty\right)$ and $\left(0,C_{V}+\delta\right]$,
for any $\delta>0$. Furthermore $\min\{f\mleft(\left|k\right|\mright),\epsilon\}=\lim_{\delta\rightarrow0^{+}}\min\{f_{\delta}\mleft(\left|k\right|\mright),\epsilon\}$
for any $k\in\mathbb{R}^{3}$, so this modification has no impact
in the limit $\delta\rightarrow0^{+}$.

By the layer-cake decomposition we can express $\min\{f_{\delta}\mleft(\left|k\right|\mright),\epsilon\}$
as
\begin{equation}
\min\{f_{\delta}\mleft(\left|k\right|\mright),\epsilon\}=\int_{0}^{\epsilon}1_{\{t\leq f_{\delta}\mleft(\left|k\right|\mright)\}}\mleft(t\mright)dt=\int_{0}^{\epsilon}1_{\{\left|k\right|\leq f_{\delta}^{-1}\mleft(t\mright)\}}\mleft(t\mright)dt=\int_{0}^{\epsilon}1_{\overline{B}(0,f_{\delta}^{-1}\mleft(t\mright))}\mleft(k\mright)dt
\end{equation}
(where one should understand $\overline{B}(0,f_{\delta}^{-1}\mleft(t\mright))=\emptyset$
if $t>C_{V}+\delta$). Consequently
\begin{align}
\sum_{k\in\mleft(B_{F}+p\mright)}\min\{f\mleft(\left|k\right|\mright),\epsilon\} & =\sum_{k\in\mleft(B_{F}+p\mright)}\int_{0}^{\epsilon}1_{\overline{B}(0,f_{\delta}^{-1}\mleft(t\mright))}\mleft(k\mright)dt=\int_{0}^{\epsilon}\sum_{k\in\mathbb{Z}^{3}}1_{\overline{B}(0,f_{\delta}^{-1}\mleft(t\mright))\cap\overline{B}(p,k_{F})}\mleft(k\mright)dt\\
 & =\int_{0}^{\epsilon}\left|\overline{B}(0,f_{\delta}^{-1}\mleft(t\mright))\cap\overline{B}(p,k_{F})\cap\mathbb{Z}^{3}\right|dt\nonumber 
\end{align}
and similarly for $\sum_{k\in\mleft(B_{F}+q\mright)}\min\{f\mleft(\left|k\right|\mright),\epsilon\}$.
Then
\begin{align}
 & \sum_{k\in\mleft(B_{F}+q\mright)}\min\{f_{\delta}\mleft(\left|k\right|\mright),\epsilon\}-\sum_{k\in\mleft(B_{F}+p\mright)}\min\{f_{\delta}\mleft(\left|k\right|\mright),\epsilon\}\nonumber \\
 & =\int_{0}^{\epsilon}\left|\overline{B}(0,f_{\delta}^{-1}\mleft(t\mright))\cap\overline{B}(q,k_{F})\cap\mathbb{Z}^{3}\right|-\left|\overline{B}(0,f_{\delta}^{-1}\mleft(t\mright))\cap\overline{B}(p,k_{F})\cap\mathbb{Z}^{3}\right|dt\\
 & =\int_{f_{\delta}\mleft(4k_{F}\mright)}^{\epsilon}\left|\overline{B}(0,f_{\delta}^{-1}\mleft(t\mright))\cap\overline{B}(q,k_{F})\cap\mathbb{Z}^{3}\right|-\left|\overline{B}(0,f_{\delta}^{-1}\mleft(t\mright))\cap\overline{B}(p,k_{F})\cap\mathbb{Z}^{3}\right|dt\nonumber 
\end{align}
where we could replace the lower integration bound by $f_{\delta}\mleft(4k_{F}\mright)$
due to the condition $\mleft(B_{F}+p\mright),\,\mleft(B_{F}+q\mright)\subset\overline{B}(0,4k_{F})$
implying that the integrand vanishes beyond this range (the integrand
also vanishes for the same reason if it should happen that $\epsilon<f_{\delta}\mleft(4k_{F}\mright)$,
so we can assume this is not the case). Lemma \ref{lemma:LensCountingResult}
now lets us estimate
\begin{align}
 & \sum_{k\in\mleft(B_{F}+q\mright)}\min\{f_{\delta}\mleft(\left|k\right|\mright),\epsilon\}-\sum_{k\in\mleft(B_{F}+p\mright)}\min\{f_{\delta}\mleft(\left|k\right|\mright),\epsilon\}\\
 & \geq\int_{f_{\delta}\mleft(4k_{F}\mright)}^{\epsilon}\mathrm{Vol}\mleft(\overline{B}(0,f_{\delta}^{-1}\mleft(t\mright))\cap\overline{B}(q,k_{F})\mright)-\mathrm{Vol}\mleft(\overline{B}(0,f_{\delta}^{-1}\mleft(t\mright))\cap\overline{B}(p,k_{F})\mright)-C\mleft(1+k_{F}^{2}+f_{\delta}^{-1}\mleft(t\mright)^{2}\mright)dt\nonumber \\
 & \geq-C\epsilon\sup_{t\in\mleft(f_{\delta}\mleft(4k_{F}\mright),\epsilon\mright)}\mleft(1+k_{F}^{2}+f_{\delta}^{-1}\mleft(t\mright)^{2}\mright)=-C\epsilon\mleft(1+k_{F}^{2}+\mleft(4k_{F}\mright)^{2}\mright)\geq-C\epsilon\mleft(1+k_{F}^{2}\mright)\nonumber 
\end{align}
where we used that $\mathrm{Vol}\mleft(\overline{B}(0,f_{\delta}^{-1}\mleft(t\mright))\cap\overline{B}(q,k_{F})\mright)\geq\mathrm{Vol}\mleft(\overline{B}(0,f_{\delta}^{-1}\mleft(t\mright))\cap\overline{B}(p,k_{F})\mright)$
since $\left|q\right|\leq\left|p\right|$.

Taking the limit $\delta\rightarrow0^{+}$, recalling equations (\ref{eq:SumDifferenceDecomposition})
and (\ref{eq:DecompositionLowkBound}) and choosing $\epsilon=k_{F}^{-\frac{4}{3}}$
now yields the claim.

$\hfill\square$

\end{document}